\documentclass[twocolumn]{emulateapj}
\usepackage{amsmath}
\usepackage{multirow} 
\usepackage{nicefrac}
\usepackage[none]{hyphenat}
\usepackage{enumitem}

\def\Rsun{R$_{\odot}$}
\def\Msun{M$_{\odot}$}

\begin{document}

\title{Mind your P{\scriptsize s} and Q{\scriptsize s}: the Interrelation between \\ Period (P) and Mass-ratio (Q) Distributions of Binary Stars}

\author{Maxwell Moe\altaffilmark{1,2} \& Rosanne Di Stefano\altaffilmark{3}}

\altaffiltext{1}{Steward Observatory, University of Arizona, 933~N.~Cherry~Ave.,~Tucson,~AZ 85721,~USA; moem@email.arizona.edu}

\altaffiltext{2}{Einstein Fellow}

\altaffiltext{3}{Harvard-Smithsonian Center for Astrophysics, 60~Garden~St., Cambridge, MA 02138, USA}

\begin{abstract}
  We compile observations of early-type binaries identified via spectroscopy, eclipses, long-baseline interferometry, adaptive optics, common proper motion, etc.  Each observational technique is sensitive to companions across a narrow parameter space of orbital periods $P$ and mass ratios $q$~=~$M_{\rm comp}$/$M_1$.  After combining the samples from the various surveys and correcting for their respective selection effects, we find the properties of companions to O-type and B-type main-sequence (MS) stars differ among three regimes.  First, at short orbital periods $P$~$\lesssim$~20~days (separations $a$~$\lesssim$~0.4~AU), the binaries have small eccentricities $e$~$\lesssim$~0.4, favor modest mass ratios $\langle q \rangle$~$\approx$~0.5, and exhibit a small excess of twins $q$~$>$~0.95. Second, the companion frequency peaks at intermediate periods log $P$\,(days)~$\approx$~3.5 ($a$~$\approx$~10~AU),  where the binaries have mass ratios weighted toward small values $q$~$\approx$~0.2\,-\,0.3 and follow a Maxwellian ``thermal'' eccentricity distribution.  Finally, companions with long orbital periods log\,$P$\,(days)~$\approx$~5.5\,-\,7.5 ($a$~$\approx$~200\,-\,5,000~AU) are outer tertiary components in hierarchical triples, and have a mass ratio distribution across $q$~$\approx$~0.1\,-\,1.0 that is nearly consistent with random pairings drawn from the initial mass function.  We discuss these companion distributions and properties in the context of binary star formation and evolution.  We also reanalyze the binary statistics of solar-type MS primaries, taking into account that (30\,$\pm$\,10)\% of single-lined spectroscopic binaries likely contain white dwarf companions instead of low-mass stellar secondaries. The mean frequency of stellar companions with $q$~$>$~0.1 and log\,$P$\,(days) $<$ 8.0 per primary increases from 0.50\,$\pm$\,0.04 for solar-type MS primaries to 2.1\,$\pm$\,0.3 for O-type MS primaries.  We fit joint probability density functions $f(M_1,q,P,e)$~$\neq$~$f(M_1)f(q)f(P)f(e)$ to the corrected distributions, which can be incorporated into binary population synthesis studies. 
\end{abstract}

\keywords{binaries: close, general; stars: massive, formation, evolution, statistics}

\section{Introduction}

 Spectral type B (3\,\Msun~$\lesssim$~$M_1$~$\lesssim$~16\,\Msun) and O ($M_1$~$\gtrsim$~16\,\Msun) main-sequence (MS) primaries with closely orbiting stellar companions can evolve to produce  X-ray binaries \citep{Verbunt1993}, millisecond pulsars \citep{Lorimer2008}, Type Ia \citep{Wang2012} and possibly Type Ib/c \citep{Yoon2010} supernovae, Algols \citep{vanRensbergen2011}, short \citep{Nakar2007} and perhaps long \citep{Izzard2004} gamma ray bursts, accretion induced collapse \citep{Ivanova2004}, and sources of gravitational waves \citep{Schneider2001}. It is therefore important to constrain the binary statistics of massive stars in order to fully characterize the rates and properties of these channels of binary evolution.  The close binary fraction, i.e. the fraction of primaries with stellar companions at separations $a$~$\lesssim$~1~AU, increases dramatically between M-type and O-type MS stars \citep[][ and references therein]{Abt1983, Duquennoy1991,Fischer1992,Raghavan2010,Moe2014}.  In fact, most massive stars with $M_1$~$\gtrsim$~15\,\Msun\ will interact with a stellar companion before they explode as core-collapse supernovae \citep{Sana2012}.  However, the interrelations among binary properties, e.g. primary mass, mass ratio, orbital period, eccentricity, age, metallicity, and environment, are only beginning to be accurately quantified. See \citet{Duchene2013} for a recent review.  

\begin{figure*}[t!]
\centerline{
\includegraphics[trim=0.0cm 1.93cm 0.0cm 0.35cm, clip=true, width=6.7in]{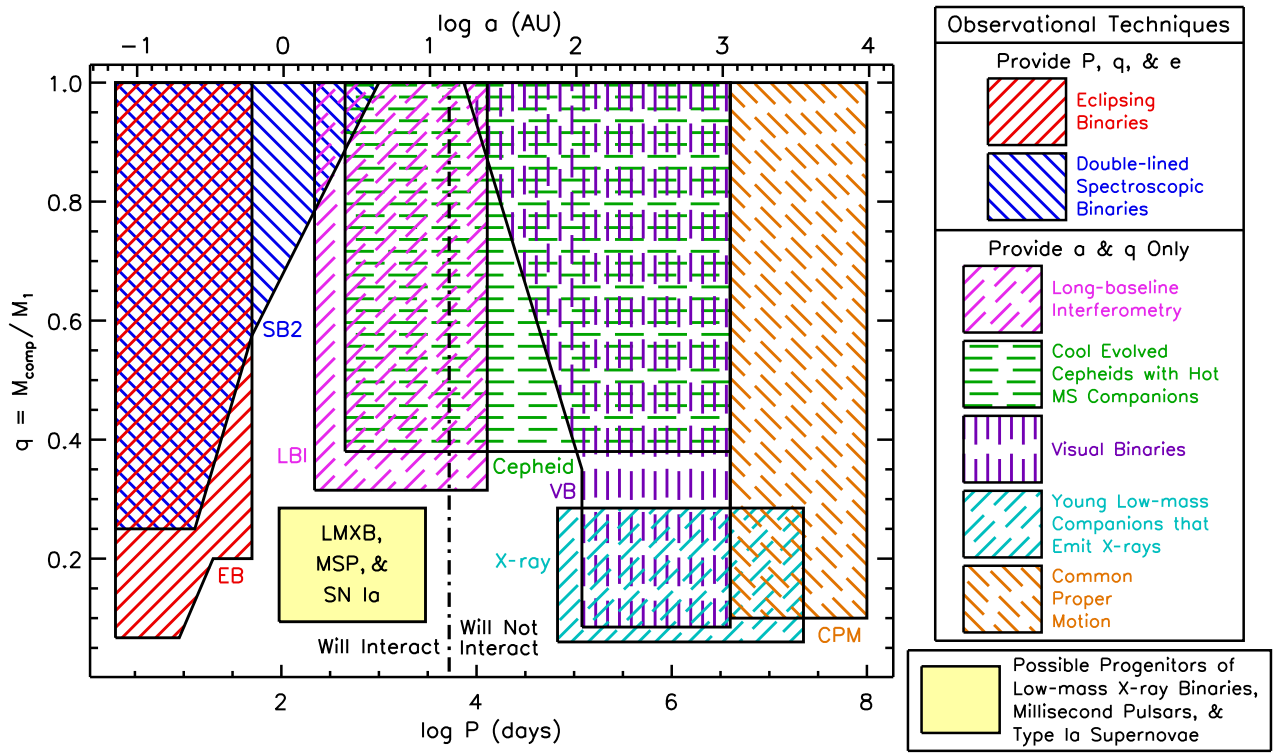}}
\caption{Schematic diagram of the various observational techniques (regions enclosed with solid and dashed lines) used to identify companions to early-type primaries as a function of orbital period $P$ and mass ratio $q$~=~$M_{\rm comp}$/$M_1$.  We compare the approximate parameter space of detection abilities for double-lined spectroscopic binaries (blue), eclipsing binaries (red), long-baseline interferometry (magenta), Cepheids (green), visual binaries (purple), X-ray emission (aqua), and common proper motion (orange). In this diagram, we show only observational techniques where the orbital period $P$ (or orbital separation $a$) and mass ratio $q$ can be estimated, and the nature of the companion is reliably known to be a non-degenerate pre-MS or MS star.  Assuming an average eccentricity $\langle e \rangle$ $\approx$ 0.5, only binaries with log\,$P$\,(days)~$\lesssim$~3.7 (left of dot-dashed line) will substantially interact via Roche-lobe overflow as the primary evolves toward the giant phase (see \S11).  Certain channels of low-mass X-ray binaries, millisecond pulsars, and Type Ia supernovae are expected to derive from early-type MS primaries with low-mass companions $q$~$\approx$~0.1\,-\,0.3 at initially moderate orbital periods $P$~$\approx$~100\,-\,3,000~days (filled yellow region).  We emphasize that only recent observations of eclipsing binaries \citep[][red]{Moe2014,Moe2015}, companions identified via long-baseline interferometry \citep[][magenta]{Rizzuto2013,Sana2014}, and binaries with intermediate-mass Cepheid primaries \citep[][green]{Evans2013,Evans2015} come close to probing this portion of the parameter space.}
\end{figure*}

  The precise distributions of binary properties enlightens our understanding of binary star formation.  For example, a mass-ratio distribution that is consistent with random pairings drawn from the initial mass function (IMF) would suggest the companions formed relatively independently from the primaries  \citep{Abt1990,Tout1991,McDonald1995}.  Alternatively, correlated component masses, which are expected and generally observed for close binaries \citep{Tokovinin2000,Raghavan2010,Sana2012}, indicate coevolution during the pre-MS phase via physical processes such as fragmentation, fission, competitive accretion, and/or mass transfer through Roche-lobe overflow \citep{Bonnell1992, Clarke1996, Kroupa1995,Bate1997,Kratter2006,Kouwenhoven2009,Marks2011,Bate2012}.  As another example, an eccentricity distribution that is weighted toward large values implies a Maxwellian ``thermal'' orbital velocity distribution \citep{Ambartsumian1937,Kroupa2008}.  Such a population would suggest the binaries formed through dynamical interactions, possibly through tidal or disk capture, perturbations in a dense cluster, triple-star secular evolution / Kozai cycles, and/or three-body exchanges \citep{Heggie1975,Pringle1989,Turner1995,Kroupa1995,Kiseleva1998,Naoz2014}.  Meanwhile, circularized orbits demonstrate tidal evolution on the MS and/or pre-MS \citep{Zahn1977,Zahn1989b,Meibom2005}.  

The statistical distributions of zero-age MS binaries are also utilized as input parameters in population synthesis studies of binary star evolution \citep{Hurley2002,Belczynski2008}.  The predicted rates and properties of certain channels of binary evolution are highly dependent on the adopted MS binary statistics \citep{Fryer1999,Kiel2006,Davis2010,Claeys2014}.  Moreover, MS binary distributions, such as the period and mass-ratio distributions, may be significantly correlated with each other \citep{Abt1990}.  Separately adjusting the input MS binary distributions to the extremes may still not encompass the true nature of the binary population.

Companions to massive stars have been detected through a variety of methods, including spectroscopy \citep{Sana2012}, eclipses \citep{Moe2015}, long-baseline interferometry \citep{Rizzuto2013}, adaptive optics \citep{Shatsky2002}, common-proper motion \citep{Abt1990}, etc.  Each observational technique is sensitive to companions across a certain interval of orbital periods $P$ and mass ratios $q$~$\equiv$~$M_{\rm comp}$/$M_1$ (see Fig.~1).  Despite significant advances in the observational instruments and methods, the properties of low-mass companions ($q$~$\lesssim$~0.3) around early-type stars at intermediate orbital periods ($P$~$\approx$~50\,-\,10$^4$~days) remain elusive.  This region is especially interesting because low-mass X-ray binaries and millisecond pulsars that form in the galactic field \citep{Kalogera1998,Kiel2006} as well as  certain channels of Type Ia supernovae \citep{Whelan1973,Ruiter2011} may evolve from extreme mass-ratio binaries at initially intermediate orbital periods (Fig.~1). Although we investigate all portions of the binary parameter space, we are especially concerned with determining accurate companion statistics at intermediate orbital periods.  

In order to use the properties of MS binaries to understand their formation and evolution, we must have a clear and complete description of young MS binaries. The primary goal of this study is to fit mathematical functions to the intrinsic distributions and properties of a zero-age MS population of binary stars.  We anchor our fits according to dozens of binary star samples observed through a variety of techniques.  However, each survey is affected by observational biases and incompleteness, and so we must first account for each survey's particular selection effects when measuring the intrinsic distribution functions. In addition, because each survey is sensitive to a narrow, specific interval of primary mass, orbital period, and mass ratio, we must combine the statistics from each sample in a self-consistent, homogeneous manner.  Only then can we understand binary star populations as a whole and reliably measure the intrinsic interrelations between the binary star properties. 

\renewcommand{\arraystretch}{1.5}
\setlength{\tabcolsep}{2.5pt}
\begin{figure*}[t!]\footnotesize
{\small {\bf Table 1:} Binary properties (top), the parameters that model their distributions (middle), and multiplicity statistics (bottom).} \\
\vspace*{-0.3cm}
\begin{center}
\begin{tabular}{|c|c|l|}
\hline
  Property  &  Range  &  Definition \\
\hline
$M_1$ (\Msun) &  0.8 $<$ $M_1$ $<$ 40 & Zero-age MS mass of the primary, i.e., the initially most massive component of the system \\
\hline
$q$ = $M_{\rm comp}$/$M_1$ & 0.1 $<$ $q$ $<$ 1.0 & Mass ratio of companion to primary \\
\hline
$P$ (days) & 0.2 $<$ log\,$P$ $<$ 8.0 & Orbital period of companion with respect to the primary \\
\hline
$e$ & 0.0 $<$ $e$ $<$ $e_{\rm max}(P)$ & Orbital eccentricity up to maximum value $e_{\rm max}$ (Eqn.~3) without Roche-lobe filling \\
\hline
\hline
  Parameter  &  Domain  &  Definition \\
\hline 
$f_{\rm logP;q>0.3}$($M_1$,$P$) & 0.3 $<$ $q$ $<$ 1.0 & Frequency of companions with $q$ $>$ 0.3 per decade of orbital period \\
\hline
{\large $\gamma$}$_{\rm smallq}$($M_1$,$P$) & 0.1 $<$ $q$ $<$ 0.3 & Power-law slope of the mass-ratio distribution $p_q$ $\propto$ $q^{\gamma_{\rm smallq}}$ across 0.1 $<$ $q$ $<$ 0.3 (Eqn.~2) \\
\hline
{\large $\gamma$}$_{\rm largeq}$($M_1$,$P$) & 0.3 $<$ $q$ $<$ 1.0 & Power-law slope of the mass-ratio distribution $p_q$ $\propto$ $q^{\gamma_{\rm largeq}}$ across 0.3 $<$ $q$ $<$ 1.0 (Eqn.~2) \\
\hline
${\cal F}_{\rm twin}$($M_1$,$P$) & 0.95 $<$ $q$ $<$ 1.00 & Excess fraction of twins with $q$ $>$ 0.95 relative to companions with $q$ $>$ 0.3 \\
\hline
$\eta$($M_1$,$P$) &  0.0 $<$ $e$ $<$ $e_{\rm max}(P)$ &  Power-law slope of the eccentricity distribution $p_e$ $\propto$ $e^{\eta}$ across 0.0 $<$ $e$ $<$ $e_{\rm max}$ \\
\hline
\hline
Statistic  &  Domain  &  Definition \\
\hline 
$f_{\rm mult;q>0.3}$($M_1$) & 0.3 $<$ $q$ $<$ 1.0  & Multiplicity frequency; average frequency of companions with $q$ $>$ 0.3 per primary (Eqn. 1) \\
\hline 
$f_{\rm mult;q>0.1}$($M_1$) & 0.1 $<$ $q$ $<$ 1.0  & Total multiplicity frequency; average frequency of companions with $q$ $>$ 0.1 per primary \\
\hline                      
${\cal F}_{n;q>0.1}$($M_1$) &  0~$\le$~$n$~$\le$~3 & Fraction of primaries that have $n$ companions with $q$ $>$ 0.1 (Eqns.~28\,-\,29) \\
\hline
\end{tabular}
\end{center}
\end{figure*}

We organize the rest of this paper as follows.  In \S2, we define the parameters that describe the statistics and distributions of MS binary stars.  We then review the various observational methods for detecting early-type binaries.  We analyze spectroscopic binaries~(\S3), eclipsing binaries~(\S4), binaries discovered via long-baseline interferometry and sparse aperture masking~(\S5), binaries containing Cepheid primaries that evolved from B-type MS stars~(\S6), and visual binaries identified through adaptive optics, lucky imaging, speckle interferometry, space-based observations, and common proper motion~(\S7).  For each observational technique and sample of early-type binaries, we account for their respective selection effects to recover the intrinsic binary statistics.  To extend the baseline toward smaller masses, we parameterize the multiplicity statistics of solar-type MS primaries in~\S8.  In~\S9, we combine the statistics of the corrected binary populations to measure the interrelations between primary mass, multiplicity frequency, mass ratio, orbital period, and eccentricity.  We discuss the implications for binary star formation (\S10) and evolution (\S11).  Finally, we summarize our main results in \S12.

\section{Definitions}

  In the following sections, we discuss the various observational techniques for identifying stellar companions to late-B ($M_1$~$\approx$~3\,-\,5\,\Msun), mid-B ($M_1$~$\approx$~5\,-\,9\,\Msun), early-B ($M_1$~$\approx$~9\,-\,16\,\Msun) and O-type ($M_1$~$\gtrsim$~16\,\Msun) MS stars.  Unless otherwise stated, we use the stellar relations provided in \citet[][and references therein]{Pecaut2013} to estimate the primary mass $M_1$ from the spectral type.  For each observational sample, we recover the intrinsic frequency of companions across the specified period range, the intrinsic distribution of mass ratios $q$, and, if possible, the intrinsic distribution of eccentricities $e$.  Our aim is to measure five parameters, which we designate as $f_{\rm logP;q>0.3}$, {\large$\gamma$}$_{\rm smallq}$, {\large$\gamma$}$_{\rm largeq}$, ${\cal F}_{\rm twin}$, and $\eta$ (see definitions below), that describe the binary statistics and distributions. We summarize the binary properties and the parameters that describe their distributions in Table~1.

Our statistical parameters model the binary properties of a zero-age MS population.  We note the binary properties may differ during the earlier pre-MS phase when the components are still forming, which we analyze in more detail in \S10.   A primary with mass $M_1$ is the initially most massive and brightest component of a system.   An observed population of stars in the galactic field, however, has a distribution of ages.  The brightest component of an observed system may therefore be the secondary if the  primary has since evolved into a faint compact remnant.  In \S8 and \S11, we show that $\approx$(5\,-\,30)\% of stars which appear to be the primaries are actually the original secondaries of evolved binaries.  The precise value depends on the spectral type and location of the stars under consideration.  In particular, (20\,$\pm$\,10)\% of early-type MS stars in a volume-limited sample are the initial secondaries of evolved binary star systems (\S11; see also \citealt{deMink2014}).  We account for this correction factor of ${\cal C}_{\rm evol}$ = 1.2\,$\pm$\,0.1 due to binary evolution in \S3\,-\,7, and we further justify this value in \S11.  For early-type stars targeted in young open clusters with ages $\tau$ $\lesssim$ 3 Myr, the effect of binary evolution is negligible \citep{Sana2012}. For such young populations, we assume ${\cal C}_{\rm evol}$ = 1.0.  We emphasize this correction factor ${\cal C}_{\rm evol}$ describes only the evolution of binary star properties from the zero-age MS to an older stellar population that may contain compact remnants.  This factor does not include the evolution of binary star properties during the earlier pre-MS phase, which we discuss in \S10. 

Many of the observational techniques we consider are sensitive to companions with $q$~$>$~0.3 (see Fig.~1), and so the density of binaries in this portion of the parameter space is more reliably known.   We therefore choose to anchor our definitions according to binaries with $q$~=~0.3\,-\,1.0 where the statistics are more complete and accurate.  We also consider binaries with $q$~=~0.1\,-\,0.3, but utilize different variables to describe these small mass-ratio systems where the statistics are less certain.

 We define $f_{\rm logP;q>0.3}$\,($M_1$,\,$P$) as the frequency of companions per decade of orbital period with mass ratios $q$~$\equiv$~$M_{\rm comp}$/$M_1$~$>$~0.3.   For example, if a sample of 100 primaries with $M_1$~=~10\,\Msun\ have 15 companions with masses $M_{\rm comp}$~$=$~3\,-\,10\,\Msun\ and periods $P$~=~100\,-\,1,000~days, then $f_{\rm logP;q>0.3}$\,($M_1$\,=\,10\,\Msun,\,log\,$P$\,=\,2.5)~=~0.15. For a given mass $M_1$, the frequency $f_{\rm logP;q>0.3}$\,($P$) provides the period distribution of companions with $q$ $>$ 0.3. Note that $f_{\rm logP;q>0.3}$\,($P$) = constant is simply \"{O}pik's law \citep{Opik1924,Abt1983}, i.e. a uniform distribution with respect to logarithmic period.   Integration of $f_{\rm logP;q>0.3}$  gives the multiplicity frequency:
\begin{equation}
f_{\rm mult;q>0.3}\,(M_1) = \int_{0.2}^{8.0} f_{\rm logP;q>0.3}\,(M_1,\,P)\,d{\rm log}P,
\end{equation}
\noindent  i.e. the mean number of companions with $q$~$>$~0.3 per primary.  We investigate stellar companions with $P$~$\gtrsim$~1.6~days that are not Roche-lobe filling and binaries with $P$~$<$~10$^8$~days that are gravitationally bound according to their common proper motion (see Fig.~1).  Triple stars in which both companions directly orbit the primary star contribute two orbits to the multiplicity frequency.  The multiplicity frequency $f_{\rm mult;q>0.3}$ can exceed unity if a primary star contains, on average, more than one stellar companion with $q$ $>$ 0.3.  

In the present study, we do not fully differentiate between single stars, binaries, and triples (but see \S9.4). We instead tabulate the corrected total frequency of MS companions $f_{\rm logP;q>0.3}$\,($M_1$,\,$P$), where the MS primary with mass $M_1$ is the most massive component of the system and $P$ is the orbital period of the stellar companion with respect to the primary. Current observations of massive stars are not sensitive to companions in certain portions of the $P$ and $q$ parameter space (see Fig.~1), and so we must correct for incompleteness to self-consistently derive $f_{\rm logP;q>0.3}$.  By correcting for incompleteness, we cannot evaluate the properties of multiple stars without making assumptions.  Namely, we cannot determine whether a companion that escapes detection is in a true binary or is the inner or outer component of a hierarchical triple.

For example, suppose that Survey~1 detects 50 companions with $q$~$>$~0.1 and 0.2~$<$~log\,$P$\,(days)~$<$~4.0 in a Sample~A of 100 B-type primaries.  After accounting for incompleteness and selection effects, we may estimate the corrected frequency of companions with $q$~$>$~0.1 and 0.2~$<$~log\,$P$~$<$~4.0 to be $\approx$\,0.70, slightly larger than the 50\% raw detection rate.  Suppose also that Survey~2 identifies 20 companions with $q$~$>$~0.1 and 4.0~$<$~log\,$P$~$<$~8.0 in a different Sample~B of 50 B-type primaries.  We may find after correcting for incompleteness that the intrinsic frequency of wide companions with $q$~$>$~0.1 and 4.0~$<$~log\,$P$~$<$~8.0 is $\approx$\,0.50.  By combining these two surveys, all we can definitively calculate is that the corrected multiplicity frequency is $f_{\rm mult;q>0.1}$ = 0.70\,+\,0.50 = 1.20.  Although a certain fraction of systems with B-type primaries must be in triples and/or higher-order multiples to accommodate an average multiplicity frequency greater than unity, we cannot evaluate the precise proportions of singles, binaries, triples, and quadruples.  The parameters $f_{\rm logP;q>0.3}$, $f_{\rm mult;q>0.3}$, and $f_{\rm mult;q>0.1}$ therefore model the corrected total frequency of all companions, including companions observed in binaries, triples, quadruples, etc. as well as companions not yet detected but expected to exist after accounting for incompleteness.  

 For massive stars, we assume in \S9.4 that the distribution of singles, binaries, triples, and quadruples follow a Poisson distribution, as is observed for solar-type systems \citep{Kraus2011,Duchene2013}. Only by making this assumption for massive stars can we estimate their single star fraction ${\cal F}_{n=0;q>0.1}$, i.e., the fraction of primaries that do not have any companions with $q$~$>$~0.1 and 0.2~$<$~log\,$P$\,(days)~$<$~8.0, their binary star fraction ${\cal F}_{n=1;q>0.1}$, their triple star fraction ${\cal F}_{n=2;q>0.1}$, etc. (see Table~1).  However, the inferred multiplicity fractions ${\cal F}_{n;q>0.1}$ rely heavily on this assumption, and so the multiplicity frequencies $f_{\rm logP;q>0.3}$, $f_{\rm mult;q>0.3}$, $f_{\rm mult;q>0.1}$ are more robust and our preferred statistics.

Next, the parameters {\large $\gamma$}$_{\rm smallq}$\,($M_1$,\,$P$), {\large $\gamma$}$_{\rm largeq}$\,($M_1$,\,$P$), and ${\cal F}_{\rm twin}$\,($M_1$,\,$P$) describe the mass-ratio probability distribution $p_q$ (see Fig.~2).  We adopt a broken power-law probability distribution $p_q$~$\propto$~$q^{\gamma}$ with slopes:

\begin{align}
  {\mbox {\large $\gamma$}} = &~{\mbox {\large $\gamma$}}_{\rm smallq}~~{\rm for}~0.1\,<\,q\,<\,0.3, \nonumber \\
           &~{\mbox {\large $\gamma$}}_{\rm largeq}~~{\rm for}~0.3\,<\,q~<\,1.0,
\end{align}

\noindent where the mass-ratio probability distribution $p_q$ is continuous at $q$ = 0.3. If the mass-ratio probability distribution can be described by a single power-law distribution $p_q$~$\propto$~$q^{\gamma}$ across $q$~=~0.1\,-\,1.0, then {\large $\gamma$} = {\large $\gamma$}$_{\rm smallq}$ = {\large $\gamma$}$_{\rm largeq}$.  Note that {\large $\gamma$}~=~0 is a uniform mass-ratio probability distribution while {\large $\gamma$}~=~$-$2.35 implies random pairings drawn from a Salpeter IMF. A more realistic IMF flattens below $M$~$\lesssim$~0.5\,\Msun, and so random pairings from the true IMF no longer provides {\large $\gamma$}~=~$-$2.35 in this low-mass regime.  \citet{Tout1991} and \citet{Kouwenhoven2009} examine binary star pairing algorithms based on more realistic IMFs, and we perform our own Monte Carlo pairings when examining the mass-ratio distribution of solar-type binaries in \S8.5.  Nevertheless, for the majority of this study, we examine companions $q$~$>$~0.1 to massive stars $M_1$~$>$~5\,\Msun\ ($M_2$~$>$~0.5\,\Msun) across a narrowly selected interval $\delta M_1$/$M_1$~$<$~0.3 of primary masses.  For such massive star populations, random pairings from the IMF indeed implies {\large $\gamma$}~=~$-$2.35.

 Certain observational techniques can detect extreme mass-ratio binaries $q$ $\approx$ 0.05\,-\,0.10 \citep{Shatsky2002,Abt1990,Moe2014,Hinkley2015}, which we exclude when quantifying our multiplicity statistics.  Alternatively, other observational methods are sensitive to only companions with $q$ $\gtrsim$ 0.3 \citep[][see Fig.~1]{Rizzuto2013,Evans2013,Sana2014}.  For these surveys, we cannot determine {\large $\gamma$}$_{\rm smallq}$.  Nonetheless, we can still measure the power-law component {\large $\gamma$}$_{\rm largeq}$ and companion frequency $f_{\rm logP;q>0.3}$ across large mass ratios $q$ = 0.3\,-\,1.0.

\begin{figure}[t!]
\centerline{
\includegraphics[trim=0.0cm 0.1cm 0.0cm 0.3cm, clip=true, width=3.5in]{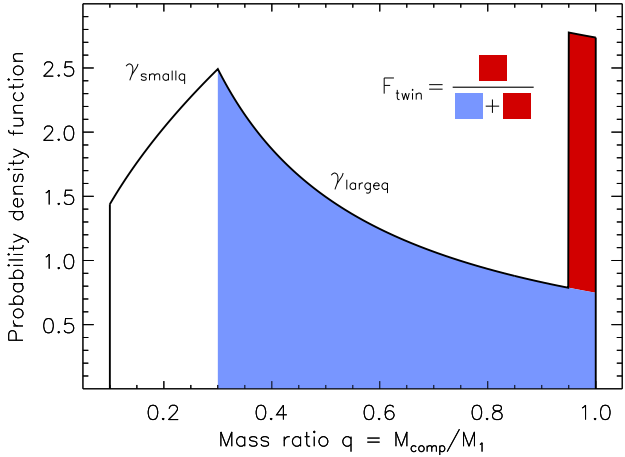}}
\caption{In general, three independent parameters are required to sufficiently model the mass-ratio probability distribution $p_q$: (1) the power-law slope {\large $\gamma$}$_{\rm smallq}$ across small mass ratios $q$~=~0.1\,-\,0.3, (2) the power-law slope {\large $\gamma$}$_{\rm largeq}$ across large mass ratios $q$~=~0.3\,-\,1.0, and (3) an {\it excess} fraction ${\cal F}_{\rm twin}$ of twin components with $q$~=~0.95\,-\,1.00 relative to the underlying power-law component {\large $\gamma$}$_{\rm largeq}$ across $q$~=~0.3\,-\,1.0.  In this particular example, {\large $\gamma$}$_{\rm smallq}$~=~0.5, {\large $\gamma$}$_{\rm largeq}$~=~$-$1.0, ${\cal F}_{\rm twin}$ = 0.10, and the probability distribution function is normalized such that $\int_{0.3}^{1.0} p_q\,$d$q$~=~1.  Throughout this study, we examine dozens of binary samples that span specific, narrow intervals of primary mass $M_1$ and orbital period $P$.  For each sample, we correct for selection effects and measure the parameters {\large $\gamma$}$_{\rm smallq}$\,($M_1$,\,$P$), {\large $\gamma$}$_{\rm largeq}$\,($M_1$,\,$P$), and ${\cal F}_{\rm twin}$\,($M_1$,\,$P$) that describe the intrinsic mass-ratio distribution for the given $M_1$ and $P$.}
\end{figure}

Because we define the companion frequency $f_{\rm logP;q>0.3}$ according to binaries with $q$ $>$ 0.3, the power-law component {\large $\gamma$}$_{\rm smallq}$ mainly serves as a statistic that describes the frequency of binaries with $q$~=~0.1\,-\,0.3.  For example, suppose a sample of ${\cal N}_{\rm prim}$ = 100 primaries contain ${\cal N}_{\rm largeq}$ = 28 companions uniformly distributed across $q$ = 0.3\,-\,1.0 ({\large $\gamma$}$_{\rm largeq}$ = 0.0) and with orbital periods $P$ = 10\,-\,1,000 days.  This results in $f_{\rm logP;q>0.3}$ = ${\cal N}_{\rm largeq}$/${\cal N}_{\rm prim}$/$\Delta$log$P$ = (28\,$\pm$\,$\sqrt{28}$)/100/(3\,$-$\,1) = 0.14\,$\pm$\,0.03, where we propagate the uncertainties from Poisson statistics.  Suppose also the same sample contains ${\cal N}_{\rm smallq}$~=~8 additional companions with $q$~=~0.1\,-\,0.3 across the same period interval.  This number implies the uniform mass-ratio distribution extends all the way down to $q$ = 0.1, i.e., {\large $\gamma$}$_{\rm smallq}$ = 0.0\,$\pm$\,0.7, where we propagate the Poisson uncertainties in ${\cal N}_{\rm smallq}$. If instead the number ${\cal N}_{\rm smallq}$ = 6 is smaller, for example, then the power-law component {\large $\gamma$}$_{\rm smallq}$ = 0.7\,$\pm$\,0.8 would be larger.  Alternatively, if there are more binaries with small mass ratios, e.g., ${\cal N}_{\rm smallq}$ = 10, then the slope {\large $\gamma$}$_{\rm smallq}$ = $-$0.5\,$\pm$\,0.6 must be negative to account for the higher frequency of low-mass companions.  

 Some authors advocate the use of binomial statistics when analyzing binary star properties \citep{Burgasser2003,Kroupa2011}.  This is appropriate only when measuring the binary star {\it fraction} ${\cal F}_{n \ge 1;q>0.1}$, i.e. the fraction of primaries that have at least one companion.  In the current study, we directly measure the multiplicity {\it frequencies} $f_{\rm logP;q>0.3}$, $f_{\rm mult;q>0.3}$, and $f_{\rm mult;q>0.1}$, where we allow for the possibility that primaries can have more than one companion.  In these cases of measuring the frequency of companions per primary, Poisson statistics are most appropriate.

   For some populations of close binaries, there is a clear narrow peak in the mass-ratio probability distribution at $q$~$\gtrsim$~0.95 \citep{Tokovinin2000, Pinsonneault2006, Raghavan2010}.  We therefore define ${\cal F}_{\rm twin}$\,($M_1$,\,$P$) as the excess fraction of twins with $q$ $>$ 0.95 relative to the underlying power-law component {\large $\gamma$}$_{\rm largeq}$.  For example, if 90\% of the binaries with $q$~$>$~0.3 are uniformly distributed across 0.3~$<$~$q$~$<$~1.0 and the remaining 10\% are evenly distributed across 0.95~$<$~$q$~$<$~1.0, then {\large $\gamma$}$_{\rm largeq}$~=~0.0 and ${\cal F}_{\rm twin}$~=~0.1.  The parameter ${\cal F}_{\rm twin}$ is therefore the {\it excess} fraction, not the total fraction,  of twin components with $q$~$>$~0.95 (see Fig.~2).

Finally, $\eta$\,($M_1$,\,$P$) describes the eccentricity probability distribution $p_e$ $\propto$ $e^{\eta}$ according to a power-law. Note that $\eta$ = 1 is a Maxwellian ``thermal'' eccentricity distribution \citep{Ambartsumian1937,Heggie1975,Kroupa2008}. For a given $M_1$ and $P$, we consider the eccentricity distribution across the interval $e$~=~0.0\,-\,$e_{\rm max}$, where the upper limit is:

\begin{equation}
 e_{\rm max}\,(P) = 1 - \Big(\frac{P}{2~{\rm days}}\Big)^{-\nicefrac{2}{3}}~~{\rm for~}P~>~2~{\rm days}.
\end{equation}

\noindent This relation guarantees the binary components have Roche-lobe fill-factors $\lesssim$70\% at periastron. Some binaries may initially have $e$ $>$ $e_{\rm max}$ and nearly fill their Roche lobes at periastron, but their orbits will rapidly evolve toward smaller eccentricities due to tides. We assume all binaries with $P$ $\le$ 2 days are circularized, which is consistent with both observations and tidal theory of early-type binaries \citep{Zahn1975,Abt1990,Sana2012}.

According to the above definitions, we have implicitly assumed that the distributions of mass ratios $q$ and eccentricities $e$ are independent.  For both solar-type binaries \citep{Raghavan2010} and early-type binaries \citep{Moe2015}, the mass ratios $q$ and eccentricities $e$ are not generally observed to be significantly correlated with each other.  The only possible exception is that the excess twin population ${\cal F}_{\rm twin}$ among short-period solar-type binaries have systematically smaller eccentricities than their non-twin counterparts \citep{Halbwachs2003}.  This may provide insight into the binary twin formation process, but may also be due to tidal evolution whereby binaries with larger, more massive companions have shorter circularization timescales \citep{Zahn1977,Hut1981}. For all other parameter combinations, the above definitions allow for possible correlations between the binary physical properties and their distributions.  

Only a small fraction of visual early-type binaries have measured orbital eccentricities \citep{Abt1990,Sana2014}.  \citet{Harrington1977} demonstrate that it is relatively more difficult to measure the orbital elements of visual binaries which have large eccentricities $e$~$\gtrsim$~0.7. A subsample of visual binaries that have reliable orbital solutions is therefore biased toward smaller eccentricities. \citet{Tokovinin2015} confirm this observational selection bias for a volume-limited sample of solar-type wide binaries.  They find that binaries in various visual orbit catalogs are weighted toward smaller eccentricities compared to those in their complete sample.  For binaries identified through spectroscopic radial velocity variations and eclipses, the eccentricities can always be readily measured. Only spectroscopic (\S3) and eclipsing (\S4) binaries can currently be utilized to quantify an unbiased eccentricity distribution for early-type binaries (see Fig.~1).  In each of the following sections, we measure $f_{\rm logP;q>0.3}$\,($M_1$,\,$P$), {\large $\gamma$}$_{\rm largeq}$\,($M_1$,\,$P$), ${\cal F}_{\rm twin}$\,($M_1$,\,$P$), and, if possible, {\large $\gamma$}$_{\rm smallq}$\,($M_1$,\,$P$) and $\eta$\,($M_1$\,$P$).

\section{Spectroscopic Binaries}

\subsection{Sample Selection}

Multi-epoch spectroscopic radial velocity observations are capable of detecting companions to massive MS stars with the shortest orbital periods \citep{Wolff1978,Garmany1980, Levato1987,Abt1990,Sana2012,Chini2012,Kobulnicky2014}. The mass ratio of a double-lined spectroscopic binary (SB2) can be directly measured from the observed ratio of velocity semi-amplitudes $q$~=~$M_2$/$M_1$ = $K_1$/$K_2$.  The orbital eccentricity $e$ of an SB2 is derived by fitting the radial velocities as a function of orbital phase.

\renewcommand{\arraystretch}{1.5}
\setlength{\tabcolsep}{8pt}
\begin{figure*}[t!]\footnotesize
{\small {\bf Table 2:} Statistics based on four surveys containing early-type spectroscopic binaries with log\,$P$\,(days)~=~0.3\,-\,2.7.}  \\
\vspace*{-0.3cm}
\begin{center}
\begin{tabular}{|c|c|c|c|}
\hline
   Survey & Primary Mass & Period Interval & Statistic \\
\hline
\multirow{5}{*} {All Four} & \multirow{5}{*} {$\langle M_1 \rangle$ =  16\,$\pm$\,8\,\Msun} & log\,$P$\,(days) = 0.75\,$\pm$\,0.25 & $\eta$ = $-$0.3\,$\pm$\,0.2 \\
\cline{3-4}
                         &           & log\,$P$\,(days) = 1.85\,$\pm$\,0.85                  & $\eta$ = 0.6\,$\pm$\,0.3 \\
\cline{3-4}
                         &           & log\,$P$\,(days) = 0.8\,$\pm$\,0.5                    & {\large $\gamma$}$_{\rm largeq}$ = $-$0.3\,$\pm$\,0.3 \\
\cline{3-4}
                         &           & \multirow{2}{*} {log\,$P$\,(days) = 2.0\,$\pm$\,0.7}  & {\large $\gamma$}$_{\rm largeq}$ = $-$1.6\,$\pm$\,0.5 \\
\cline{4-4}
                         &           &                                                       & ${\cal F}_{\rm twin}$ $<$ 0.05 \\
\hline
\citet{Levato1987} \& & \multirow{2}{*} {$\langle M_1 \rangle$ = 6\,$\pm$\,2\,\Msun} & \multirow{4}{*} {log\,$P$\,(days) = 0.8\,$\pm$\,0.5} & \multirow{2}{*}  { ${\cal F}_{\rm twin}$ = 0.17\,$\pm$\,0.09} \\
\citet{Abt1990}  & & & \\
\cline{1-2}
\cline{4-4}
\citet{Kobulnicky2014} \& & \multirow{2}{*} {$\langle M_1 \rangle$ = 20\,$\pm$\,7\,\Msun} &  & \multirow{2}{*} { ${\cal F}_{\rm twin}$ = 0.08\,$\pm$\,0.04} \\
\citet{Sana2012}  & & & \\
\hline
\citet{Levato1987},  & \multirow{3}{*} {$\langle M_1 \rangle$ = 8\,$\pm$\,3\,\Msun} & \multirow{4}{*} {log\,$P$\,(days) = 0.8\,$\pm$\,0.5}  & \multirow{3}{*}  {{\large $\gamma$}$_{\rm smallq}$ = $-$0.5\,$\pm$\,0.8} \\
\citet{Abt1990}, \& & & & \\
\citet{Kobulnicky2014} & & & \\
\cline{1-2}
\cline{4-4}
\citet{Sana2012} & $\langle M_1 \rangle$ = 28\,$\pm$\,8\,\Msun\ & & {\large $\gamma$}$_{\rm smallq}$ = 0.6\,$\pm$\,0.8 \\
\hline
\citet{Levato1987} & $\langle M_1 \rangle$ = 5\,$\pm$\,2\,\Msun\ & \multirow{4}{*} {log\,$P$\,(days) = 0.8\,$\pm$\,0.5} & $f_{\rm logP;q>0.3}$ = 0.07\,$\pm$\,0.04 \\
\cline{1-2}
\cline{4-4}
\citet{Abt1990} & $\langle M_1 \rangle$ = 7\,$\pm$\,2\,\Msun\ &  & $f_{\rm logP;q>0.3}$ = 0.10\,$\pm$\,0.04 \\
\cline{1-2}
\cline{4-4}
\citet{Kobulnicky2014} & $\langle M_1 \rangle$ = 12\,$\pm$\,3\,\Msun\ & & $f_{\rm logP;q>0.3}$ = 0.12\,$\pm$\,0.05 \\
\cline{1-2}
\cline{4-4}
\citet{Sana2012} & $\langle M_1 \rangle$ = 28\,$\pm$\,8\,\Msun\ & & $f_{\rm logP;q>0.3}$ = 0.24\,$\pm$\,0.06 \\
\hline
\citet{Abt1990} \&  & \multirow{2}{*} {$\langle M_1 \rangle$ = 9\,$\pm$\,3\,\Msun} & \multirow{2}{*} {log\,$P$\,(days) = 1.8\,$\pm$\,0.5} & \multirow{2}{*} {$f_{\rm logP;q>0.3}$ = 0.08\,$\pm$\,0.05} \\
\citet{Kobulnicky2014} & & & \\
\hline 
\citet{Sana2012} & $\langle M_1 \rangle$ = 28\,$\pm$\,8\,\Msun\ & log\,$P$\,(days) = 2.0\,$\pm$\,0.7 & $f_{\rm logP;q>0.3}$ = 0.12\,$\pm$\,0.06 \\
\hline

\end{tabular}
\end{center}
\end{figure*}

\begin{figure}[t!]
\centerline{
\includegraphics[trim=0.4cm 0.0cm 0.4cm 0.0cm, clip=true, width=3.25in]{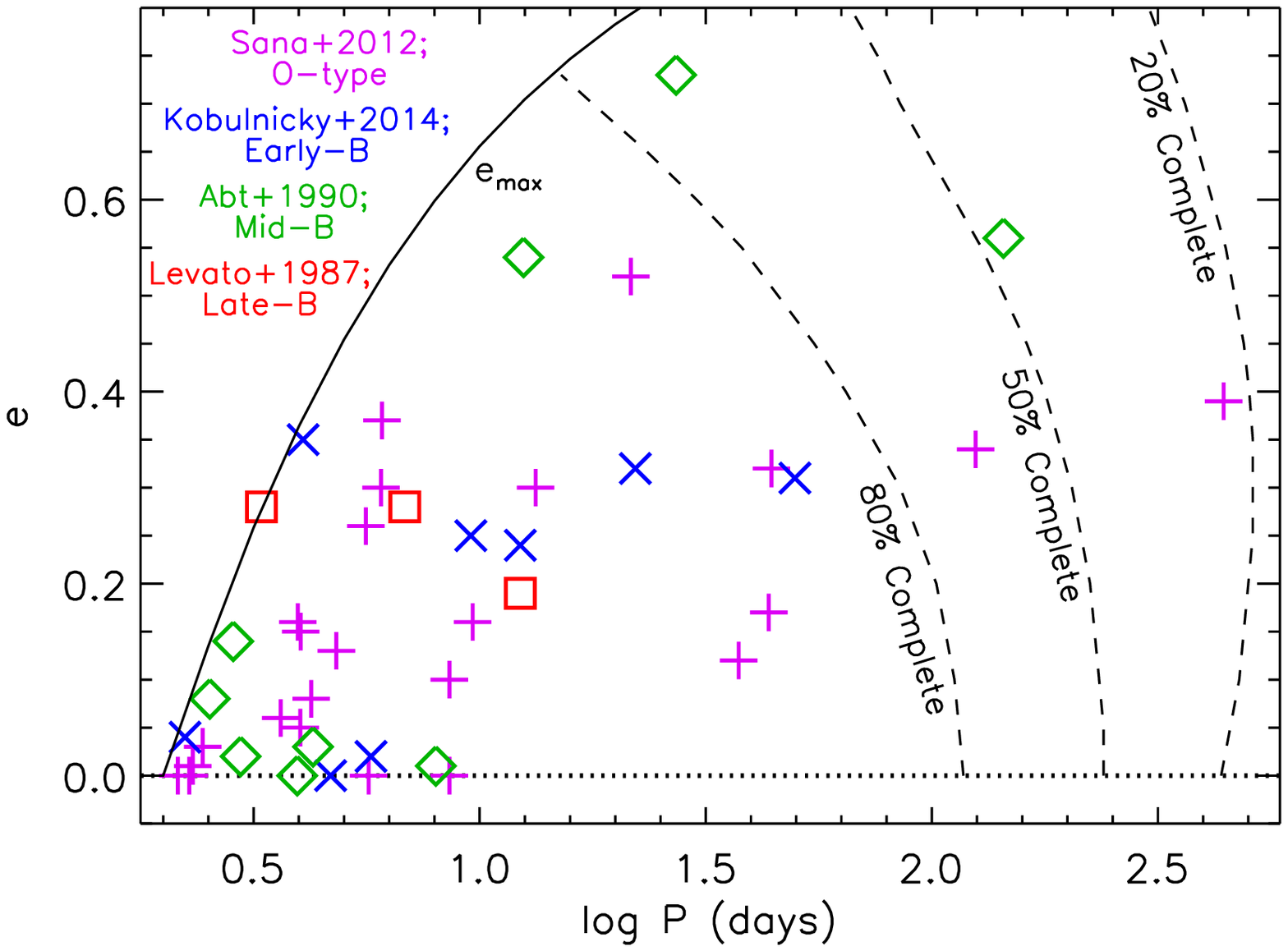}}
\caption{Eccentricities $e$ of 44 SB2s as a function of orbital period $P$ from four spectroscopic surveys: \citet[][magenta pluses]{Sana2012}, \citet[][blue crosses]{Kobulnicky2014}, \citet[][green diamonds]{Abt1990}, and \citet[][red squares]{Levato1987}. We display the maximum expected eccentricity $e_{\rm max}$($P$) according to Eqn.~3 (black line).  Assuming $M_1$ = 10\,\Msun, $q$ = 0.4, random orientations, and the median cadence and sensitivity of the spectroscopic surveys, we show completeness levels of 80\%, 50\%, and 20\% (dashed).  Note that many SB2s with $P$ $<$ 10 days are circularized ($e$ = 0; dotted), while all early-type SB2s with $P$~$>$~10~days are in eccentric orbits. }
\end{figure}

We initially analyze 44 SB2s with orbital periods $P$~=~2\,-\,500~days from four surveys of early-type stars: \citet[][81~B-type primaries; $\langle M_1 \rangle$~$\approx$~5\,\Msun; 3~SB2s]{Levato1987}, \citet[][109~B2\,-\,B5 primaries; $\langle M_1 \rangle$~$\approx$~7\,\Msun; 9~SB2s]{Abt1990}, \citet[][83~B0\,-\,B2 primaries; $\langle M_1 \rangle$~$\approx$~12\,\Msun; 8~SB2s]{Kobulnicky2014}, and \citet[][71~O-type primaries; $\langle M_1 \rangle$~$\approx$~28\,\Msun; 24~SB2s]{Sana2012}.  We note the \citet{Kobulnicky2014} survey also includes O-type stars, but we consider only their 83 systems with early-B primaries in our current analysis.

\begin{figure}[t!]
\centerline{
\includegraphics[trim=0.8cm 0.2cm 0.8cm 0.1cm, clip=true, width=3.15in]{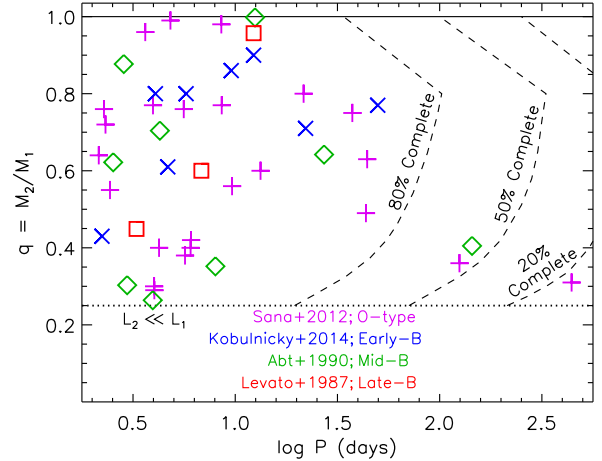}}
\caption{Similar to Fig.~3, but for the mass ratios $q$~=~$M_2$/$M_1$ as a function of orbital period $P$.  Early-type SB2s with MS components can reveal only companions with $q$ $>$ 0.25 (above dotted line). Assuming $M_1$ = 10\,\Msun, $e$ = 0.5$e_{\rm max}$, and the median cadence and sensitivity of the spectroscopic surveys, we show completeness levels of 80\%, 50\%, and 20\% (dashed).  Short-period SB2s with $P$ $<$ 20 days span the entire observable mass-ratio interval $q$ $\approx$ 0.25\,-\,1.0.  Alternatively, long-period systems with $P$~=~100\,-\,500~days include only SB2s with $q$~$\approx$~0.3\,-\,0.4, even though the spectroscopic surveys are substantially incomplete in this corner of the parameter space. }
\end{figure}

We list the multiplicity statistics based on the four samples in Table~2.  In Figs.~3 and~4, we display the eccentricities $e$ and mass ratios $q$, respectively, of the 44 SB2s as a function of orbital period. \citet{Sana2012} and \citet{Kobulnicky2014} identified additional SB2s with $q$~$>$~0.55 and $P$~$>$~500~days, which confirms that spectroscopic observations can reveal moderate mass-ratio binaries at intermediate orbital periods.  However, spectroscopic binaries become increasingly less complete and biased toward moderate $q$ at longer $P$ (see \S3.2 and Fig.~4), so we limit our sample selection to SB2s with $P$~$<$ 500 days when discussing these four surveys.

\renewcommand{\arraystretch}{1.65}
\setlength{\tabcolsep}{8pt}
\begin{figure*}[t!]\footnotesize
{\small {\bf Table 3:} Binary statistics based on 23 early-type SB2s with log\,$P$\,(days)~=~1.3\,$\pm$\,0.3 contained in the SB9 catalog.}  \\
\vspace*{-0.45cm}
\begin{center}
\begin{tabular}{|c|c|c|}
\hline
  Sample   &  Primary Mass & Statistic \\
\hline
\multirow{2}{*} {Spectral Types O9\,-\,B3} & \multirow{2}{*} {$\langle M_1 \rangle$ = 14\,$\pm$\,4\,\Msun} & $\eta$ = 0.9\,$\pm$\,0.4 \\
\cline{3-3}
  &  & {\large $\gamma$}$_{\rm largeq}$ = $-$0.6\,$\pm$\,0.9 \\
\hline
\multirow{2}{*} {Spectral Types B5\,-\,B9.5} & \multirow{2}{*} {$\langle M_1 \rangle$ = 4.5\,$\pm$\,1.5\,\Msun} & $\eta$ = $-$0.3\,$\pm$\,0.2 \\
\cline{3-3}
  &  & {\large $\gamma$}$_{\rm largeq}$ = $-$1.0\,$\pm$\,0.8 \\
\hline
\end{tabular}
\end{center}
\end{figure*}
\renewcommand{\arraystretch}{1.0}

We also examine the 23 detached SB2s with primary spectral types O and B, luminosity classes III-V, orbital periods $P$ = 8\,-\,40 days, and measured mass ratios $q$~=~$K_1$/$K_2$ and eccentricities $e$ from the Ninth Catalog of Spectroscopic Binaries \citep[SB9;][]{Pourbaix2004}.  This sample includes 10 systems with O9-B3 primaries ($\langle M_1 \rangle $~$\approx$~14\,\Msun) and 13 systems with B5-B9.5 primaries ($\langle M_1 \rangle $~$\approx$~4.5\,\Msun).  We report the multiplicity statistics based on these 23 SB2s in Table~3.  The spectroscopic binaries contained in the SB9 catalog are compiled from a variety of surveys and samples, and so the cadence and sensitivity of the spectroscopic observations are not as homogeneous. We therefore consider only the SB2s with $P$ $<$ 40 days, which are relatively complete regardless of the instruments utilized. We cannot infer the intrinsic frequency of companions per primary based on the SB9 catalog.  We can nonetheless utilize the sample of 23 early-type SB2s to measure the eccentricity and mass-ratio distributions.

For SB2s, the secondary must be comparable in luminosity to the primary in order for both components to be visible in the combined spectrum.   Because MS stars follow a steep mass-luminosity relation, SB2s with early-type MS primaries can reveal only moderate mass ratios $q$~$>$~0.25 (Figs.~1~\&~4).    For single-lined spectroscopic binaries (SB1s) with low-luminosity companions, a lower limit to the mass ratio can be estimated from the observed reflex motion of the primary.  A statistical mass-ratio distribution can be recovered for SB1s by assuming the intrinsic binary population has random orientations \citep{Mazeh1992}.  However, SB1s with early-type MS primaries may not necessarily have low-mass A-K type stellar companions.  Instead, many SB1s with O-type and B-type primaries may contain 1\,-\,3\,\Msun\ stellar remnants such as white dwarfs, neutron stars, or even black holes \citep{Wolff1978,Garmany1980}. It is imperative to never implicitly assume that early-type SB1s contain two MS components. In the context of spectroscopic binaries with massive primaries, only SB2s provide a definitive uncontaminated census of unevolved companions \citep{Moe2014}.  At the very least, SB1s can provide a self-consistency check and an upper limit to the frequency of low-mass stellar companions.

\subsection{Corrections for Incompleteness}

The ability to detect spectroscopic binaries not only depends on the resolution of the spectrograph, the signal to noise ratio of the spectra,  and the cadence of the observations, but also on the physical properties of the binary.  Early-type stars, including those in binaries with $P$ $\gtrsim$ 10 days where tidal synchronization is inefficient, are rotationally broadened by $\langle v_{\rm rot} \rangle$~$\approx$~100\,-\,200~km~s$^{-1}$ \citep{Abt2002,Levato2013}.  The primary's orbital velocity semi-amplitude must therefore be $K_1$~$\gtrsim$~10~km~s$^{-1}$ in order for the orbital reflex motion of the primary to be detectable \citep{Levato1987,Abt1990,Sana2012,Kobulnicky2014}.  For SB2s, the atmospheric absorption features of both the primary and secondary components need to be distinguishable.  Due to blending of the broad absorption features, early-type SB2s require an even higher threshold of $K_1$~$\gtrsim$~40~km~s$^{-1}$ to be observed.   The primary's velocity semi-amplitude $K_1$  decreases toward wider separations $a$, smaller mass ratios $q$, and lower inclinations $i$.  Lower mass companions at longer orbital periods are more likely to be missed in the spectroscopic binary surveys.  Finally, highly eccentric binaries spend only a small fraction of time near periastron while exhibiting appreciable radial velocity variations.  Depending on the cadence of the spectroscopic observations, eccentric binaries can be either more or less complete compared to binaries in circular orbits. 

To measure the detection efficiencies and correct for incompleteness, we utilize a Monte Carlo technique to generate a large population of binaries with different primary masses $M_1$, mass ratios $q$, and orbital configurations $P$ and $e$.   For each binary, we select from a randomly generated set of orientations, i.e., an inclination $i$ distribution such that cos\,$i$~=~[0,\,1] is uniform and a distribution of periastron angles $\omega$~=~[0$^{\rm o}$,\,360$^{\rm o}$] that is also uniform.  The velocity semi-amplitude $K_1$ criterion used in previous studies does not adequately describe the detection efficiencies of eccentric binaries.  For each binary, we instead synthesize radial velocity measurements $v_{1,r}$ and  $v_{2,r}$ at 20 random epochs, which is the median cadence of the spectroscopic binary surveys \citep{Levato1987,Abt1990,Sana2012,Kobulnicky2014}. For simplicity, we assume the systemic velocity of the binary is zero.  In an individual spectrum of a early-type star, the radial velocities can be centroided to an accuracy of $\approx$\,2\,-\,3~km~s$^{-1}$.  However, atmospheric variations limit the true sensitivity across multiple epochs to $\delta v_{1,r}$ $\approx$\,3\,-\,5~km~s$^{-1}$ \citep{Levato1987,Abt1990,Sana2012,Kobulnicky2014}.   In order for a simulated binary to have an orbital solution that can be fitted, and therefore an eccentricity and mass ratio that can be measured,  we require a minimum number of radial velocity measurements of the primary $v_{1,r}$ to significantly differ from the systemic velocity.   We impose that at least 5 of the 20 measurements satisfy $|v_{1,r}|$~$\gtrsim$ (3\,-\,5)\,$\delta v_{1,r}$~$\approx$~15~km~s$^{-1}$ in order to provide a precise and unique orbital solution for the primary. 

The rotationally broadened spectral features of both the primary and secondary must also be distinguishable to be cataloged as an SB2.  At the very least, the primary's absorption features must not only shift during the orbit, but also have velocity profiles that visibly change due to the moving absorption lines from the orbiting secondary \citep{DeBecker2006,Sana2012}. For $q$~$<$~0.8, we require at least 3 of the 20 measurements to satisfy $|v_{1,r} - v_{2,r}|$~$\gtrsim$ $\langle v_{\rm rot} \rangle /2$ $\approx$ 75~km~s$^{-1}$.   This velocity threshold is appropriate for both O-type and B-type primaries because the mean rotational velocities $\langle v_{\rm rot} \rangle$~$\approx$~150~km~s$^{-1}$ do not significantly vary between these two spectral types \citep{Abt2002,Levato2013}.  For twin binaries with $q$ $=$ 1.0, the absorption lines of both the primary and secondary are of comparable depths, and so it is more difficult to distinguish the two components if their absorption features are significantly blended \citep{Sana2011}.  For these twin systems, we require at least 3 of the 20 measurements to satisfy $|v_{1,r} - v_{2,r}|$~$>$ $\langle v_{\rm rot} \rangle$ = 150~km~s$^{-1}$.  We linearly interpolate our velocity difference thresholds between 75~km~s$^{-1}$ at $q$ = 0.8 and 150~km~s$^{-1}$ at $q$ = 1.0.

\begin{figure}[t!]
\centerline{
\includegraphics[trim=0.0cm 0.2cm 0.0cm 0.1cm, clip=true, width=3.45in]{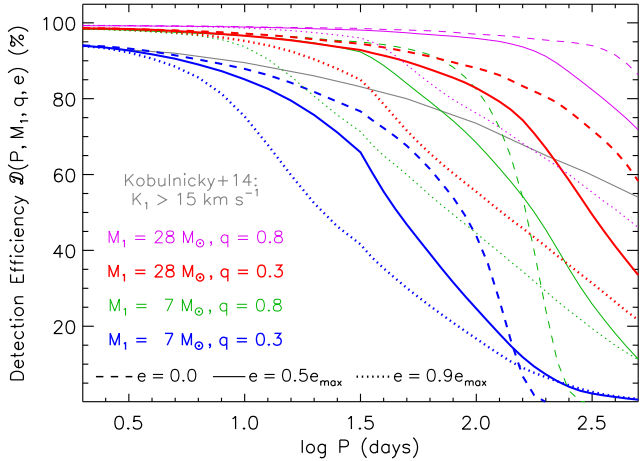}}
\caption{The detection efficiencies ${\cal D}$($P$,\,$M_1$,\,$q$,\,$e$) of SB2s assuming random orientations and the sensitivity and cadence of the spectroscopic observations.  As a function of orbital period $P$, we display the completeness levels for $M_1$ = 28 \Msun\ (O-type) primaries with $q$ = 0.8 (thin magenta) and $q$~=~0.3 (thick red) companions as well as  $M_1$ = 7 \Msun\ (B-type) primaries with $q$~=~0.8 (thin green) and $q$~=~0.3 (thick blue) companions.  For each combination of $M_1$ and $q$, we compare the detection efficiencies for circular orbits (dashed), intermediate eccentricities $e$ = 0.5$e_{\rm max}$ (solid), and large eccentricities $e$ = 0.9$e_{\rm max}$ (dotted). SB2s at $P$~$<$~10~days are relatively complete, while the detection efficiencies of longer period systems considerably vary and critically depend on the primary mass, mass ratio, and eccentricity.  We also display the completeness levels as calculated by \citet[][grey]{Kobulnicky2014}, who assumed a sensitivity of $K_1$~$=$~15~km~s$^{-1}$ for both SB1s and SB2s and that the binary population has a flat mass-ratio distribution ({\large $\gamma$} = 0.0) and a flat eccentricity distribution ($\eta$~=~0.0).}
\end{figure}

For our simulated population, we keep track of the fraction of binaries that satisfy the above criteria.  In this manner, we calculate the detection efficiencies ${\cal D}$($P$,\,$M_1$,\,$q$,\,$e$) of SB2s as a function of the physical properties of the binary.  In Fig.~5, we display the detection efficiencies ${\cal D}$ as a function of orbital period $P$ for various combinations of $M_1$, $q$, and $e$.  The sample of SB2s are relatively complete at short orbital periods $P$~$<$~10~days, while the longer period systems have detection efficiencies considerably less than unity.  As expected, binaries with small mass ratios $q$ $\approx$ 0.3 are less complete than systems with large mass ratios $q$ $\approx$ 0.8.  Because of Kepler's laws,   SB2s with lower mass mid-B and late-B primaries are less complete than SB2s with more massive O-type and early-B primaries.  Eccentric binaries with $P$ = 10\,-\,200 days have smaller detection efficiencies than their counterparts in circular orbits due to the finite cadence of observations.  At the longest orbital periods $P$ $\gtrsim$ 200 days, however, only eccentric binaries near periastron produce detectable radial velocity variations, and so ${\cal D}$ becomes larger with increasing eccentricity.

We also compare in Fig.~5 our detection efficiencies ${\cal D}$ to the completeness levels as computed by \citet[][dashed line in their Fig.~26]{Kobulnicky2014}.  For this calculation, \citet{Kobulnicky2014} assumed a sensitivity threshold of $K_1$ = 15 km s$^{-1}$ for both SB1s and SB2s, and that the binary population has an underlying flat mass-ratio distribution ({\large $\gamma$} = 0) and flat eccentricity distribution ($\eta$ = 0).  Although these choices of {\large $\gamma$} = 0 and $\eta$ = 0 are consistent with observations of short-period spectroscopic binaries where the observations are relatively complete (see below), the longer period systems may have different mass-ratio and eccentricity distributions.  

To correct for incompleteness, it is better to calculate the detection efficiency ${\cal D}$($P$,\,$M_1$,\,$q$,\,$e$) for each individual SB2.  The binary's relative contribution to the total sample is then determined by the statistical weight $w$~=~1/${\cal D}$.  In other words, for every one system observed with detection efficiency ${\cal D}$($P$,\,$M_1$,\,$q$,\,$e$), there are $w$~=~1/${\cal D}$ total systems with similar properties in the intrinsic population after considering selection effects. Given the small sample sizes, this approach is sufficient for recovering the intrinsic binary statistics from the observed SB2 population.  

Our calculated detection efficiencies ${\cal D}$ have some level of uncertainty, especially considering the uncertainties in our adopted detection criteria (e.g., five observations with $|v_{1,r}|$ $>$ 15 km s$^{-1}$ and three observations with $|v_{1,r} - v_{2,r}|$ $>$ 75\,-\,150 km s$^{-1}$, depending on $q$).  We vary our detection criteria within reasonable limits and estimate the uncertainty in ${\cal D}$ to be $\approx$10\,-\,15\%.  We propagate this uncertainty into the statistical weights $w$~=~1/${\cal D}$.  For example, if ${\cal D}$ $\approx$ 0.6, then $w$ $\approx$ 1.7\,$\pm$\,0.4, while for a smaller detection efficiency ${\cal D}$ = 0.3, the uncertainty in the weight $w$~=~3.3$_{-0.9}^{+2.3}$ becomes larger and asymmetric. 

In Fig.~3, we compare the observed SB2s to the detection efficiencies ${\cal D}$($P$,\,$M_1$,\,$q$,\,$e$) as a function of $P$ and $e$ while assuming a primary mass $M_1$ = 10 \Msun\ and mass ratio $q$ = 0.4.  Similarly, in Fig.~4, we compare the observed SB2s to the detection efficiencies ${\cal D}$ as a function of $P$ and $q$ while assuming an eccentricity $e$~=~0.5$e_{\rm max}$ and the same primary mass $M_1$~=~10\,\Msun.  These completeness levels are for illustration purposes only, as we calculate ${\cal D}$($P$,\,$M_1$,\,$q$,\,$e$) for each system in the full four-dimensional parameter space.  For example, the longest period SB2 in Figs. 3\,-\,4 (i.e., the system with the O-type primary, log\,$P$ $\approx$ 2.6, $q$ $\approx$ 0.3, $e$~$\approx$~0.4) has a detection efficiency ${\cal D}$ $\approx$ 0.35 and statistical weight $w$~$\approx$~2.9$_{-0.7}^{+1.5}$.  Meanwhile, the second longest period system (i.e., the SB2 with the mid-B primary, log\,$P$~$\approx$~2.2, $q$~$\approx$~0.4, $e$ $\approx$ 0.6) has an even smaller detection efficiency ${\cal D}$ $\approx$ 0.24 and therefore larger weight $w$ $\approx$ 4.2$_{-1.4}^{+4.2}$.  We compute $w$ for each of the 44 early-type SB2s from the four surveys and for the 23 early-type SB2s from the SB9 catalog.  By weighting each observed SB2 by $w$, we can now calculate the intrinsic eccentricity and mass-ratio distributions.

\subsection{Eccentricity Distribution}

To investigate the intrinsic eccentricity distribution as a function of orbital period, we divide the SB2s from the four surveys into short-period ($P$ = 3\,-\,10 days) and long-period ($P$ = 10\,-\,500 days) subsamples.   We consider only systems with $e$ $<$ $e_{\rm max}$, i.e.,  $e$~$<$~0.3 and $e$~$<$~0.6 for the short-period and long-period subsamples, respectively.   In Fig.~6, we display the cumulative distribution of eccentricities for these two subsamples of SB2s after correcting for incompleteness. The length of each vertical step in the cumulative distribution is proportional to the statistical weight $w$ of the SB2 it represents.

\begin{figure}[t!]
\centerline{
\includegraphics[trim=3.6cm 0.2cm 4.7cm 0.1cm, clip=true, width=3.3in]{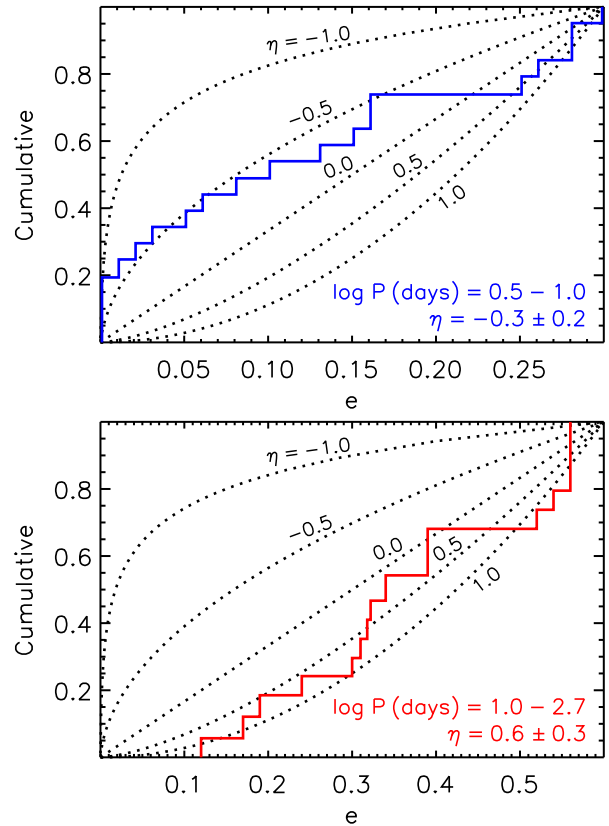}}
\caption{After accounting for incompleteness, we show the corrected cumulative distributions of eccentricities $e$ for short-period (top; blue) and long-period (bottom; red) early-type SB2s from the combined sample of four spectroscopic surveys.  The long-period subsample not only contains SB2s with larger eccentricities, but is also weighted toward larger values of $\eta$.}
\end{figure}

Our short-period subsample of 20 SB2s with $P$~=~3\,-\,10~days and $e$ $<$ 0.3 is relatively complete with statistical weights all below $w$~$<$~1.4.  For this population, we measure the power-law exponent of the eccentricity distribution to be $\eta$~=~$-$0.3\,$\pm$\,0.2  (see Table~2 and top panel of Fig.~6.).  The uncertainty in $\eta$ derives from the quadrature sum of the uncertainties from Poisson sample statistics and the uncertainties in the statistical weights $w$.  Our result of  $\eta$~=~$-$0.3\,$\pm$\,0.2 is consistent with the measurement of $\eta$~=~$-$0.4\,$\pm$\,0.2 by \citet{Sana2012}, whose O-type spectroscopic binary sample is dominated by short-period systems.

For our long-period subsample of 13 SB2s with $P$~=~10\,-\,500~days and $e$ $<$ 0.6, we measure  $\eta$~=~0.6\,$\pm$\,0.3 after correcting for selection effects (Table 2 and bottom panel of Fig.~6). Although the statistical weights $w$~=~1.2\,-\,4.2 of the SB2s in this long-period subsample are relatively large and uncertain, our measurement of $\eta$~=~0.6\,$\pm$\,0.3 is still robust.  For example, if we were to set all the weights to $w$ = 1, we would still measure a positive exponent $\eta$ $=$ 0.2. Regardless of how we correct for selection effects, early-type SB2s at intermediate orbital periods are weighted toward larger values of $\eta$ relative to the short-period systems.

The sample of SB2s from the four spectroscopic surveys is not large enough to measure changes in $\eta$ as a function of $M_1$.  We therefore investigate the 23 early-type SB2s with $P$ = 8\,-\,40 days from the SB9 catalog.  To easily compare the eccentricity distributions, we analyze the distributions of $e$/$e_{\rm max}$, where $e_{\rm max}$($P$) is determined for each SB2 according to Eqn.~3. Because we include only SB2s with $P$~$<$~40~days from the SB9 catalog, the detection efficiencies are ${\cal D}$ $>$ 60\% for all our systems, i.e., the individual weights are $w$ $<$ 1.7.

After correcting for incompleteness, we display in Fig.~7 the cumulative distributions of $e$/$e_{\rm max}$ for the 10 early (O9\,-\,B3) and 13 late (B5\,-\,B9.5) SB2s we selected from the SB9 catalog. Early-type SB2s with more massive primaries are clearly weighted toward larger values of $\eta$.  For the O9\,-\,B3 subsample, we measure $\eta$~=~0.9\,$\pm$\,0.4.  Meanwhile, for the B5\,-\,B9.5 subsample, we measure $\eta$~=~$-$0.3\,$\pm$\,0.2 (Table~3). The observed differences between these two distributions may suggest more massive binaries form with systematically larger eccentricities.  Alternatively, both early-B and late-B binaries may initially be born with $\eta$~$\approx$~0.9, but the long-lived late-B binaries have had more time to tidally evolve toward smaller eccentricities.  Based on the early-type SB2 observations alone, we cannot differentiate which of these two scenarios is the most likely explanation.  

\begin{figure}[t!]
\centerline{
\includegraphics[trim=0.1cm 0.0cm 0.0cm 0.0cm, clip=true, width=3.55in]{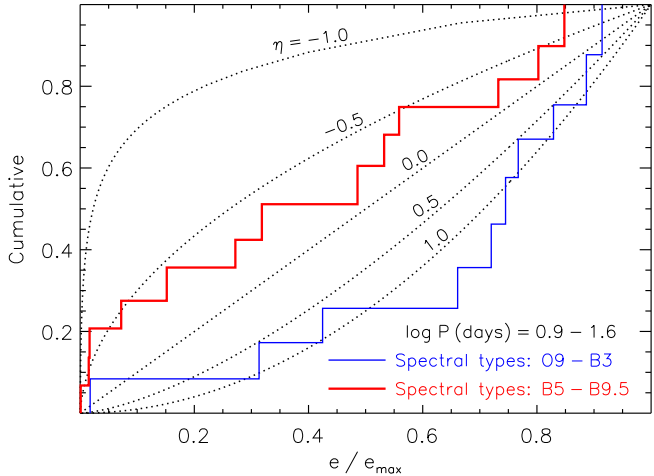}}
\caption{After accounting for selection effects, we compare the corrected cumulative distributions of $e$/$e_{\rm max}$ for the O9\,-\,B3 (thin blue) and B5\,-\,B9.5 (thick red) subsamples of SB2s with $P$~=~8\,-\,40~days from the SB9 catalog. Long-lived late-B SB2s are weighted toward smaller values of $\eta$, either because they are initially born with and/or have tidally evolved toward smaller eccentricities.}
\end{figure}

\subsection{Mass-ratio Distribution}

  In a magnitude-limited survey, binaries with equally-bright twin components are probed across larger distances compared to single stars and binaries with faint companions.  This Malmquist bias, typically called the \"{O}pik effect in the context of binary stars, can lead to an artificial peak near unity in the mass ratio distribution \citep{Opik1923}.  Fortunately, the four spectroscopic binary surveys have already accounted for the \"{O}pik effect, either by targeting early-type stars in open clusters / stellar associations with fixed distances or by removing distant twin binaries that do not reside in a volume-limited sample.  We can therefore weight each observed SB2 by their respective values of $w$ to correct for incompleteness.  

After accounting for selection effects, we show in Fig.~8 the corrected cumulative distributions of mass ratios for the 32 short-period ($P$ = 2\,-\,20 days) and 10 long-period ($P$ = 20\,-\,500 days) SB2s with $q$~$>$~0.3 from the four spectroscopic surveys. Assuming the mass-ratio distribution can be described by a power-law across 0.3~$<$~$q$~$<$~1.0, we measure {\large $\gamma$}$_{\rm largeq}$~$=$~0.1\,$\pm$\,0.3 for the short-period subsample of early-type SB2s.  This is consistent with the result of {\large $\gamma$} = $-$0.1\,$\pm$\,0.6 by \citet{Sana2012}, whose O-type binary sample is dominated by short-period systems with $P$~=~2\,-\,20~days.

A simple power-law distribution, however, does not fully describe the data.  Allowing for an excess fraction of twin components with $q$ $>$ 0.95, we measure {\large $\gamma$}$_{\rm largeq}$~=~$-$0.3\,$\pm$\,0.3 and ${\cal F}_{\rm twin}$~=~0.11\,$\pm$\,0.04 for our short-period subsample (see Fig.~8).  For early-type binaries with $P$ = 2\,-\,20 days, the relative density ${\cal N}$/$\Delta q$ = 5/(1.00\,$-$\,0.95) $\approx$ 100 of twin companions with $q$ $=$ 0.95\,-\,1.00 is larger than the density ${\cal N}$/$\Delta q$ = 5/(0.95\,$-$\,0.80) $\approx$ 30 of companions with $q$ = 0.80\,-\,0.95 at the $\approx$2.6$\sigma$ significance level (see Fig.~4).  We emphasize that this twin excess is real considering the four spectroscopic binary surveys have already accounted for the \"{O}pik effect. Nonetheless, the excess twin fraction of ${\cal F}_{\rm twin}$~$=$~0.11 is quite small; the remaining 89\% of companions with $P$~=~2\,-\,20~days and $q$~$>$~0.3 follow a power-law distribution with {\large $\gamma$}$_{\rm largeq}$~$\approx$~$-$0.3 across the broad interval $q$~=~0.3\,-\,1.0.

\begin{figure}[t!]
\centerline{
\includegraphics[trim=0.4cm 0.0cm 0.4cm 0.0cm, clip=true, width=3.55in]{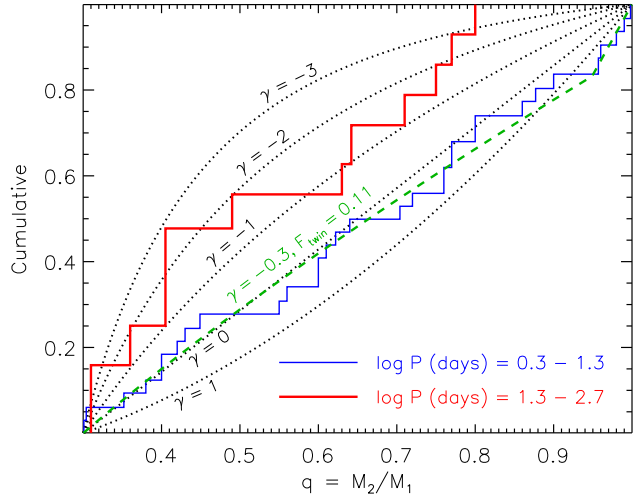}}
\caption{After accounting for incompleteness, we show the corrected cumulative distributions of mass ratios $q$ = $M_2$/$M_1$ for short-period (thin blue) and long-period (thick red) early-type SB2s from the combined sample of four spectroscopic surveys.  The short-period subsample is best fit by a two-parameter model (dashed green) with a power-law component of {\large $\gamma$}$_{\rm largeq}$~$\approx$~$-$0.3 and a small excess twin fraction of ${\cal F}_{\rm twin}$~$\approx$~0.11.  There is no indication of an excess twin population in the long-period subsample, which is adequately described by a power-law distribution with {\large $\gamma$}$_{\rm largeq}$~$=$~$-$1.6 that is weighted toward small mass ratios.}
\end{figure}

 Based on the four individual SB2 surveys, we do not measure any statistically significant trends in {\large $\gamma$}$_{\rm largeq}$ as a function of spectral type.  We therefore adopt the average value {\large $\gamma$}$_{\rm largeq}$~=~$-$0.3\,$\pm$\,0.3 for early-type short-period SB2s (Table~2). The short-period SB2s do, however, reveal a change in the excess twin fraction ${\cal F}_{\rm twin}$ as a function of primary mass.  After combining the 23 O-type and early-B SB2s with $q$ $>$ 0.3 and $P$~=~2\,-\,20~days, only 3 have $q$ $>$ 0.95 (i.e., 3/23~$\approx$~13\%). Using a maximum likelihood method, we measure the excess twin fraction to be  ${\cal F}_{\rm twin}$~=~0.08\,$\pm$\,0.04 for SB2s with $P$~=~2\,-\,20~days and O/early-B primaries (Table 2).  Of the nine short-period SB2s with $q$~$>$~0.3 and mid/late B-type primaries, two have $q$ $>$ 0.95 (i.e., 2/9~$\approx$~22\%).  After accounting for incompleteness toward small mass ratios, we measure ${\cal F}_{\rm twin}$ = 0.17\,$\pm$\,0.09 for these SB2s with lower mass primaries (Table~2).  According to these measurements, there is an indication that the excess twin fraction increases among short-period binaries as the primary mass decreases.  

After correcting for selection effects, it is evident from Fig.~8 that the long-period subsample of 10 early-type SB2s is weighted toward smaller mass ratios.  We measure an excess twin fraction that is consistent with zero, i.e. the 1$\sigma$ upper limit is  ${\cal F}_{\rm twin}$~$<$~0.05 (Table~2).  Note the completeness levels for twin systems $q$ $\approx$ 1.0 at intermediate orbital periods is smaller than the detection efficiencies for $q$ $\approx$ 0.8 systems (see Fig.~4).  This is because $q$ $\approx$ 1.0 binaries have similar absorption features and therefore must have larger radial velocity differences $|v_{\rm 1,r}$ $-$ $v_{\rm 2,r}|$ $>$ 150~km~s$^{-1}$ to be distinguishable \citep[\S3.2,][]{Sana2011}.  Nonetheless, the completeness levels of twin $q$ $=$ 1.0 binaries across log~$P$ = 1.3\,-\,2.7 are sufficiently large than an excess twin population would still be detectable.  The deficit of $q$ = 0.8\,-\,1.0 systems at intermediate orbital periods log~$P$ = 1.3\,-\,2.7 is therefore intrinsic to the population of early-type binaries.  

We also measure the power-law slope {\large $\gamma$}$_{\rm largeq}$~=~$-$1.6\,$\pm$\,0.5 of the mass-ratio distribution to be weighted toward small values for the long-period SB2s (Table~2).  The large uncertainty is mainly due to the small sample size, not the uncertainties in the correction factors $w$.  For example, if we were to set the weights $w$~=~1 for all ten long-period SB2s, we would still measure {\large $\gamma$}$_{\rm largeq}$ $=$ $-$0.8.  The detection efficiencies of binaries with $q$ = 0.3 are certainly smaller than systems with $q$ = 0.6 (\S3.2), and so {\large $\gamma$}$_{\rm largeq}$ $<$ $-$0.8 is an upper limit.  Hence, early-type SB2s become weighted toward smaller mass ratios $q$ $\approx$ 0.2\,-\,0.4 with increasing orbital period. This trend is already seen in the observed population of early-type SB2s (Fig.~4).  By correcting for incompleteness, the intrinsic population of binaries with longer orbital periods are even further skewed toward smaller mass ratios.  Our result is consistent with the conclusions of \citet{Abt1990}, who also find that early-B spectroscopic binaries become weighted toward smaller mass ratios with increasing separation. In addition, \citet{Kobulnicky2014} speculate there may be comparatively more binaries with extreme mass ratios at intermediate orbital periods that are hiding below the spectroscopic detection limits.

We next investigate the mass-ratio distributions of the SB2s with $P$ = 8\,-\,40 days we selected from the SB9 catalog. Unlike the four spectroscopic binary surveys, the SB9 catalog is still affected by the \"{O}pik effect.  We therefore consider only the SB2s from the SB9 catalog with $q$ $<$ 0.9 in order to remove any bias toward binaries with equally-bright components.  Although we cannot quantify the excess twin fraction ${\cal F}_{\rm twin}$, we can still measure the power-law component {\large $\gamma$}$_{\rm largeq}$ of the SB2s in the SB9 catalog.  

\begin{figure}[t!]
\centerline{
\includegraphics[trim=0.0cm 0.1cm 0.0cm 0.2cm, clip=true, width=3.3in]{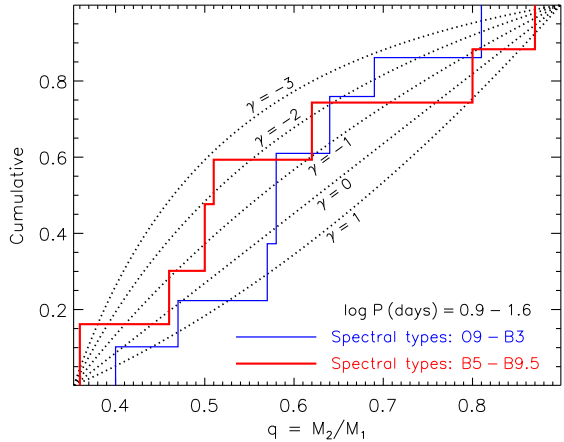}}
\caption{After accounting for selection effects, we compare the corrected cumulative distributions of mass ratios $q$ = $M_2$/$M_1$ for the O9\,-\,B3 (thin blue) and B5\,-\,B9.5 (thick red) subsamples of SB2s with $P$~=~8\,-\,40~days from the SB9 catalog. For these orbital periods, the O-type/early-B ({\large $\gamma$} $\approx$ $-$0.6) and late-B ({\large $\gamma$} $\approx$ $-$1.0) subsamples have mass-ratio distributions that are consistent with each other.}
\end{figure}

After correcting for incompleteness, we display in Fig.~9 the cumulative distribution of mass ratios across $q$~=~0.35\,-\,0.90 for the O9-B3 and B5-B9.5 subsamples from the SB9 catalog.  The sample sizes are quite small, and so we cannot discern any statistically significant differences between the two distributions.  We measure {\large $\gamma$}$_{\rm largeq}$~=~$-$0.6\,$\pm$\,0.9 for the O9\,-\,B3 subsample and {\large $\gamma$}$_{\rm largeq}$~=~$-$1.0\,$\pm$\,0.8 for the B5\,-\,B9.5 subsample (Table~3).  The measured values of {\large $\gamma$}$_{\rm largeq}$ $\approx$ $-$1.0\,-\,$-$0.6 for these SB2s at intermediate orbital periods are between the values of {\large $\gamma$}$_{\rm largeq}$ $\approx$ $-$0.3 and {\large $\gamma$}$_{\rm largeq}$ $\approx$ $-$1.6 we measured above for the short-period and long-period SB2 subsamples, respectively.

\subsection{Companion Frequencies}

Now that we have measured the SB2 detection efficiencies, we can calculate the intrinsic frequency $f_{\rm logP;q>0.3}$\,($M_1$,\,$P$) of companions with $q$~$>$~0.3 per decade of orbital period.   In the \citet{Sana2012} sample of ${\cal N}_{\rm prim}$~=~71 O-type primaries, there are ${\cal N}_{\rm SB2}$~=~17~SB2s with $q$~$\ge$~0.3 and $P$~=~2\,-\,20~days. The detection efficiencies of these 17 O-type short-period SB2s are nearly 100\% (\S3.2), and so the correction factor due to incompleteness is ${\cal C}_{\rm largeq}$~=~$\sum_{j=1}^{{\cal N}_{\rm SB2}} w_j/{\cal N}_{\rm SB2}$~=~1.0\,$\pm$\,0.1.   In addition, all 71 O-type primaries in the \citet{Sana2012} sample are in young open clusters, and so the correction factor ${\cal C}_{\rm evol}$ = 1.0 due to binary evolution is negligible (see \S2 and \S11).  The intrinsic frequency of companions with $q$~$>$~0.3 per decade of orbital period is $f_{\rm logP;q>0.3}$ = ${\cal N}_{\rm SB2}$\,${\cal C}_{\rm largeq}$\,${\cal C}_{\rm evol}$\,/\,${\cal N}_{\rm prim}$\,/\,$\Delta$log$P$ = (17\,$\pm$\,$\sqrt{17}$)$\times$(1.0\,$\pm$\,0.1)$\times$1.0/71/(1.3\,$-$\,0.3) = 0.24\,$\pm$\,0.06 for O-type primaries and short orbital periods (Table~2).  In other words, (24\,$\pm$\,6)\% of O-type zero-age MS primaries have a companion with $P$ = 2\,-\,20 days and $q$~$>$~0.3.

 We perform similar calculations for the samples of short-period companions to B-type MS primaries.   For B-type primaries, however, the correction factor ${\cal C}_{\rm largeq}$ for incompleteness is slightly larger than unity because the detection efficiencies of SB2s with lower mass primaries are smaller (see \S3.2).  We find  ${\cal C}_{\rm largeq}$~=~1.3\,$\pm$\,0.1, 1.4\,$\pm$\,0.2, and 1.5\,$\pm$\,0.2 for the early-B, mid-B, and late-B subsamples, respectively.  The  \citet{Levato1987} and \citet{Abt1990} samples are volume limited, and \citet{Kobulnicky2014} targeted systems in the Cyg OB2 association, which has a distribution of ages $\tau$ = 1\,-\,7 Myr \citep{Wright2015}.  We must therefore also account for the correction factor ${\cal C}_{\rm evol}$ = 1.2\,$\pm$\,0.1 due to binary evolution (\S2 and \S11).  Following the same approach as above, we measure $f_{\rm logP;q>0.3}$ = 0.12\,$\pm$\,0.05, 0.10\,$\pm$\,0.04, and 0.07\,$\pm$\,0.04 for the early-B, mid-B, and late-B subsamples, respectively, at short orbital periods (Table~2).  The intrinsic frequency of companions with $q$~$>$~0.3 and $P$~=~2\,-\,20~days is $\approx$\,4 times larger for O-type primaries compared to late-B primaries (see also \S9).  This trend is consistent with the conclusions of \citet{Chini2012}, who find the observed spectroscopic binary fraction increases by a factor of $\approx$\,5\,-\,7 between B9 and O5 primaries (see their Fig.~3).

In addition to the 17 O-type SB2s with $q$~$\ge$~0.3 and $P$~=~2\,-\,20~days, \citet{Sana2012} identified one SB2 with $q$ = 0.25\,-\,0.30 and three SB1s across the same period interval.  Considering the sensitivity of the spectroscopic binary observations and the young nature of the \citet{Sana2012} O-type sample, we expect these ${\cal N}_{\rm smallq}$~=~4 binaries to contain low-mass stellar companions with $q$~=~0.1\,-\,0.3.  Given ${\cal N}_{\rm largeq}$~=~17, ${\cal N}_{\rm smallq}$~=~4, {\large $\gamma$}$_{\rm largeq}$~=~$-$0.3, ${\cal F}_{\rm twin}$~=~0.08, and our definitions according to \S2, then {\large $\gamma$}$_{\rm smallq}$~=~0.6\,$\pm$\,0.8 (Table~2).  The relative density ${\cal N}_{\rm smallq}$/$\Delta q$ = 4/(0.3\,$-$\,0.1)~=~20 of $q$~=~0.1\,-\,0.3 binaries is slightly smaller than the density ${\cal N}_{\rm largeq}$/$\Delta q$ = 17/(1.0\,$-$\,0.3)~=~24 of binaries with $q$~=~0.3\,-\,1.0.  The mass-ratio probability distribution must flatten and possibly turn over below $q$~$\lesssim$~0.3.

Similarly, in addition to the 15 SB2s with $q$ $\ge$ 0.3 and $P$~=~2\,-\,20 days, the three samples of B-type spectroscopic binaries contain one additional SB2 with $q$ = 0.25\,-\,0.30 and 32 SB1s across the same period interval \citep{Levato1987,Abt1990,Kobulnicky2014}.  At these short orbital periods, the ratio of SB1s to SB2s in the combined B-type sample (32/16 = 2.0) is an order of magnitude larger than the ratio in the O-type sample (3/18 = 0.2).  Unlike the O-type SB1s, the 32 B-type SB1s do not all contain low-mass stellar companions with $q$ = 0.1\,-\,0.3 due to three effects we discuss below.

First, not all B-type binaries with $q$ = 0.3\,-\,1.0 will appear as SB2s, even at these shortest of orbital periods (see blue curve in Fig.~5). Although $q$ = 0.3\,-\,1.0 binaries may not be distinguishable as SB2s, they will still be detectable as SB1s.  Given the correction factors ${\cal C}_{\rm largeq}$~=~1.3\,-\,1.5 due to incompleteness of short-period B-type SB2s, we expect a total of ${\cal N}_{\rm largeq}$~=~22 binaries with $q$~$\ge$~0.3 and $P$~=~2\,-\,20~days in the combined B-type sample.  This implies 22\,$-$\,15~=~7 of the 32 SB1s actually contain stellar companions with $q$~$>$~0.3.  

Second, the sample of 25 remaining B-type SB1s is contaminated by stellar remnants.  In \S8, we show that $\approx$30\% of close, faint, stellar-mass companions to solar-type stars are actually white dwarfs instead of M-dwarfs.  We expect similar percentages of contamination by stellar remnants for the B-type stars in our sample (see below and \S11).  We estimate that $\approx$9 of the B-type SB1s probably contain compact remnant companions.  The remaining $\approx$16 B-type SB1s have stellar companions with $q$ $<$ 0.3.  

Finally, the B-type spectroscopic surveys include SB1s with small velocity semi-amplitudes $K_1$~$\lesssim$~20~km~s$^{-1}$, some as small as $K_1$~$\approx$~6~km~s$^{-1}$ \citep{Levato1987,Abt1990,Kobulnicky2014}.  If the orientations are sufficiently close to edge-on, these small-amplitude SB1s may be extreme mass-ratio stellar binaries with $q$~$<$~0.1.  Among the three B-type spectroscopic binary surveys, we identify 11 SB1s with $P$ = 2\,-\,20 days and sufficiently small velocity semi-amplitudes such that the binary could have $q$~$<$~0.1.  Assuming random orientations, we estimate that $\approx$7 of these SB1s may indeed have extreme mass ratios $q$~$<$~0.1.    

After accounting for the three effects described above, we estimate there are ${\cal N}_{\rm smallq}$ = 9\,$\pm$\,4 binaries with $q$~=~0.1\,-\,0.3 and $P$ = 2\,-\,20 days in the combined B-type sample.  Given ${\cal N}_{\rm largeq}$ = 22, then the power-law component across small mass ratios is {\large $\gamma$}$_{\rm smallq}$~=~$-$0.5\,$\pm$\,0.8 for B-type binaries with short orbital periods (Table~2).  This is  consistent with the value {\large $\gamma$}$_{\rm smallq}$~=~0.6\,$\pm$\,0.8 we measured above for O-type primaries across the same period interval.

In our calculation of {\large $\gamma$}$_{\rm smallq}$~=~$-$0.5\,$\pm$\,0.8 for B-type binaries, we accounted for contamination by $\approx$9 compact remnant companions in the SB1 sample.  Suppose instead that we neglected this effect and assumed there are ${\cal N}_{\rm smallq}$ = 18 stellar companions with $q$~=~0.1\,-\,0.3 and $P$~=~2\,-\,20~days in the combined B-type sample.  We would then measure {\large $\gamma$}$_{\rm smallq}$~=~$-$1.9\,$\pm$\,0.5, which is discrepant with the O-type sample at the 2.7$\sigma$ significance level. Either the relative frequency of short-period companions with extreme mass ratios is significantly larger for B-type stars compared to O-type stars, or a sizable fraction of B-type SB1s contain compact remnant companions.  We favor the latter scenario for a three reasons.  First, the power-law component {\large $\gamma$}$_{\rm largeq}$ across large mass ratios does not vary between O-type and B-type short-period binaries (\S3.4).  We would naturally expect that {\large $\gamma$}$_{\rm smallq}$ does not significantly vary either.  Second, other observational techniques demonstrate that {\large $\gamma$}$_{\rm smallq}$ is larger than or comparable to {\large $\gamma$}$_{\rm largeq}$ (see \S4\,-\,8). In other words, the mass-ratio probability distribution is always observed to flatten, not steepen, below $q$~$<$~0.3.   Finally, we utilize observations in \S11 to predict that a non-negligible fraction of early-type stars in a volume-limited survey contain compact remnant companions.  We therefore conclude that {\large $\gamma$}$_{\rm smallq}$~=~$-$0.5\,$\pm$\,0.8 for short-period B-type binaries.  This self-consistency analysis indicates that $\approx$30\% of an older population of B-type SB1s contain compact remnant companions.

We next utilize the ${\cal N}_{\rm SB2}$ = 6 SB2s with O-type primaries and log\,$P$\,(days) = 1.3\,-\,2.7 to estimate $f_{\rm logP;q>0.3}$ at intermediate orbital periods. We measure the correction factor for incompleteness of $q$ $>$ 0.3 companions to be ${\cal C}_{\rm largeq}$ $=$ $\sum_{j=1}^{{\cal N}_{\rm SB2}} w_j$/${\cal N}_{\rm SB2}$ = 2.0\,$\pm$\,0.5. The intrinsic frequency of companions with $q$ $>$ 0.3 per decade of orbital period is $f_{\rm logP;q>0.3}$ = ${\cal N}_{\rm SB2}$\,${\cal C}_{\rm largeq}$\,${\cal C}_{\rm evol}$/${\cal N}_{\rm prim}$/$\Delta$log$P$ = (6\,$\pm$\,$\sqrt{6}$)$\times$(2.0\,$\pm$\,0.5)$\times$1.0/71/(2.7\,$-$\,1.3) = 0.12\,$\pm$\,0.06 for O-type stars and intermediate orbital periods (Table~2).  For O-type primaries, the intrinsic frequency of companions with $q$ $>$ 0.3 decreases with increasing logarithmic orbital period.  This is consistent with the results of \citet{Sana2012}, who find that the logarithmic period distribution of massive binaries are skewed toward shorter periods.

Finally, we combine the ${\cal N}_{\rm SB2}$ = 4 SB2s with $P$~=~20\,-\,200~days from the samples of ${\cal N}_{\rm prim}$ = 109\,+\,83 = 192 early-B and mid-B primaries. We measure a large correction factor ${\cal C}_{\rm largeq}$ $=$ $\sum_{j=1}^{{\cal N}_{\rm SB2}} w_j$/${\cal N}_{\rm SB2}$ = 3.0\,$\pm$\,1.0 because spectroscopic binary surveys of B-type primaries are significantly incomplete at intermediate orbital periods (see Fig.~5). The intrinsic frequency of companions with $q$ $>$ 0.3 per decade of orbital period is $f_{\rm logP;q>0.3}$  = ${\cal N}_{\rm SB2}$\,${\cal C}_{\rm largeq}$\,${\cal C}_{\rm evol}$/${\cal N}_{\rm prim}$/$\Delta$log$P$ = (4\,$\pm$\,$\sqrt{4}$)(3.0$\pm$1.0)(1.2\,$\pm$\,0.1)/192/(2.3\,$-$\,1.3) = 0.08\,$\pm$\,0.05 for early/mid B-type stars at intermediate orbital periods (Table 2).  This is consistent with the frequency $f_{\rm logP;q>0.3}$ $\approx$ 0.10\,-\,0.12 at shorter periods log\,$P$\,(days) $\approx$ 0.8, indicating the period distribution of companions to early/mid B-type stars approximately obeys \"{O}pik's law across $P$~=~2\,-\,200~days (see \S9).  

\subsection{Direct Spectral Detection}

We recently became aware of a new direct spectral detection method for identifying binary stars with close to intermediate separations and small mass ratios \citep{Gullikson2016a}.  \citet{Gullikson2016b}  obtained large signal-to-noise ratio SNR\,$\sim$\,100\,-\,1,000, high-resolution $R$~=~80,000 spectra of 341 bright ($V$~$<$~6\,mag) A-type and B-type MS primaries with moderate to large rotation velocities $v_{\rm rot}$\,sin\,$i$~$>$~80\,km\,s$^{-1}$.  By exploiting the differences in spectral aborption widths between the rapidly rotating primaries and low-mass companions ($M_2$~$\lesssim$~1.5\Msun, $T_{\rm eff}$ $\lesssim$ 6,500K) that quickly spin down to $v_{\rm rot}$\,sin\,$i$~$\approx$~15\,km\,s$^{-1}$,  \citet{Gullikson2016b} can distinguish the two components in a single-epoch spectrum.  Technically, \citet{Gullikson2016b} cross-correlate the observed spectrum with model spectra of cooler companions in order to measure the spectral type of the secondary.  Although the orbital periods cannot be measured with a single-epoch spectrum, this technique is sensitive to companions with projected separations $\rho$~$\lesssim$~2$''$ that lie within the slit or optical fiber.  Given the median distance $d$~$\approx$~100\,pc to the 341 systems in their sample, the binaries have separations $a$~$\lesssim$~200 AU.  In total, \citet{Gullikson2016b} find 64 companions within $a$~$\lesssim$~200~AU of the 341 primaries. 

 \citet{Gullikson2016b} also estimate their detection efficiencies as a function of mass ratio $q$ (see their Fig.~6).  The completeness estimates peak at $\approx$90\% near $q$~$\approx$~0.3.  Toward smaller mass ratios $q$ $<$ 0.15, the detection efficiencies plummet because the secondaries become substantially fainter and cannot be distinguished from the more luminous primaries in a combined spectrum.  At large mass ratios $q$ $>$ 0.6, the detection efficiencies steadily decrease because the absorption profiles of the secondaries become too similar to those of the primaries.

Utilizing their estimated detection efficiencies, \citet{Gullikson2016b} recover the intrinsic mass-ratio distribution of A-type and B-type binaries with $a$~$\lesssim$~200\,AU.  They find the mass-ratio distribution peaks at $q$~=~0.3.  Fitting our two-component power-law to the histogram distribution in Fig.~7 of \citet{Gullikson2016b}, we estimate {\large $\gamma$}$_{\rm smallq}$~$\approx$~0.5 and {\large $\gamma$}$_{\rm largeq}$~$\approx$~$-$1.9.

The \citet{Gullikson2016b} sample covers a broad range of primary masses $M_1$~$\approx$~1.6\,-\,10\,\Msun\ and orbital separations $a$~$\approx$~0.05\,-\,200\,AU.  Because the intrinsic binary statistics can significantly vary within such a large parameter space, the inferred mass-ratio distribution may be biased.  We therefore cull the \citet{Gullikson2016b} sample according to primary mass and orbital separation as follows.  Due to the IMF and short lifetimes of early-B stars, the \citet{Gullikson2016b} sample is dominated by A-type and late-B stars.  We select the ${\cal N}_{\rm prim}$~=~281 primaries with MS spectral types B6\,-\,A7, which correspond to primary masses $M_1$~=~1.6\,-\,4.6\,\Msun. From this subset of  ${\cal N}_{\rm prim}$~=~281 primaries,  \citet{Gullikson2016b} identify 46 companions.  We remove the five spectroscopic binaries (HIP\,5348, 13165, 16611, 76267, 100221) contained in the SB9 catalog \citep{Pourbaix2004} with known orbital periods $P$~$<$~20~days.  We also remove the four wide binaries (HIP\,12706, 79199, 84606, 115115) in the Washington Double Star catalog \citep[WDS][]{Mason2009} with listed projected separations $\rho$~$>$~0.7$''$.  The SB9 catalog is relatively complete shortward of $P$~$<$~20~days (see above), while the WDS catalog contains most bright visual binaries with $q$~$>$~0.2 and $\rho$ $>$ 0.7$''$ (see \S7).  The 37 binaries remaining in our subsample most likely have orbital periods $P$~$>$~20~days and projected separations $\rho$~$<$~0.7$''$.  Given the median distance $d$ = 90 pc to the 281 A-type and late-B primaries in our subsample, the 37 companions span log\,$P$\,(days)~=~1.3\,-\,4.9.  

\citet{Gullikson2016b} note that their predicted detection efficiencies for large mass ratio $q$~$>$~0.6 binaries may be overestimated.  Utilizing an artificial injection and recovery algorithm, \citet{Gullikson2016b} predict a detection efficiency of $\approx$80\% for hot companions with masses comparable to the primaries.  In practice, they detect only 15 of the 25 known binaries (60\%) with hot companions in the SB9 and WDS catalogs.  Aware of this discrepancy,  \citet{Gullikson2016b} multiply their predicted detection efficiencies by a factor of $f$~=~0.8 to roughly match the empirical detection rate of $\approx$60\%.  However, the SB9 and WDS catalogs are themselves incomplete, even for binaries with $q$~$>$~0.6 and especially for binaries with intermediate separations.  The ratio 15/25 = 60\% is therefore an upper limit to the true detection efficiency for hot companions.  

\renewcommand{\arraystretch}{1.6}
\setlength{\tabcolsep}{8pt}
\begin{figure}[t!]\footnotesize
{\small {\bf Table 4:} Binary statistics based on the \citet{Gullikson2016b} survey of A-type / late-B primaries.}  \\
\vspace*{-0.45cm}
\begin{center}
\begin{tabular}{|c|c|}
\hline
 Primary Mass / Period Interval & Statistic \\
\hline
 $M_1$ = 3.1\,$\pm$\,1.5\,\Msun;    & {\large $\gamma$}$_{\rm smallq}$ = 0.7\,$\pm$\,0.8 \\
\cline{2-2}
 log\,$P$\,(days) = 3.1\,$\pm$\,1.8 &  {\large $\gamma$}$_{\rm largeq}$ = $-$1.0\,$\pm$\,0.5  \\
\cline{2-2}
                                    & $f_{\rm logP;q>0.3}$ = 0.06\,$\pm$\,0.02 \\
\hline
\end{tabular}
\end{center}
\end{figure}
\renewcommand{\arraystretch}{1.0}

We correct for selection effects as follows.  For binaries with 0.15~$<$~$q$~$<$~0.60, we adopt the completeness levels as displayed in Fig.~6 of \citet{Gullikson2016b}.  Toward larger mass ratios, we interpolate the completeness levels between the \citet{Gullikson2016b} value of $\approx$83\% at $q$ = 0.60 and and a smaller value of 60\% at $q$ = 0.75. For $q$~$>$~0.75, the detection efficiencies are too uncertain and may become too small to be included in our statistical analysis.  We therefore consider only binaries with $q$~$<$~0.75, and we remove the one detected system (HIP\,14043) in our subsample with $q$~$>$~0.75.  In Fig.~10, we display the corrected cumulative distribution of mass ratios for the 36 remaining binaries in our subsample with 0.15~$<$~$q$~$<$~0.75.  The length of each vertical step in Fig.~10 is inversely proportional to its corresponding detection efficiency.

Across $q$~=~0.15\,-\,0.75, the binaries in our subsample cannot be described by a single power-law mass-ratio distribution.  This is consistent with the conclusions of \citet{Gullikson2016b}.  Using a maximum likelihood method, we fit our two-component power-law model to the distribution in Fig.~10.  We measure {\large $\gamma$}$_{\rm smallq}$ = 0.7\,$\pm$\,0.8 and {\large $\gamma$}$_{\rm largeq}$ = $-$1.0\,$\pm$\,0.5 (Table~4).  Although not as narrowly peaked as the mass-ratio distribution presented in Fig.~7 of \citet{Gullikson2016b}, we confirm their main result that the intrinsic mass-ratio distribution is weighted toward $q$~$=$~0.3 with a flattening and possible turnover below $q$ $<$ 0.3.  This is a robust result considering we select a subsample from the \citet{Gullikson2016b} data set, apply different correction factors for incompleteness, and examine a narrower interval of mass ratios $q$~=~0.15\,-\,0.75.

\begin{figure}[t!]
\centerline{
\includegraphics[trim=0.0cm 0.1cm 0.0cm 0.2cm, clip=true, width=3.55in]{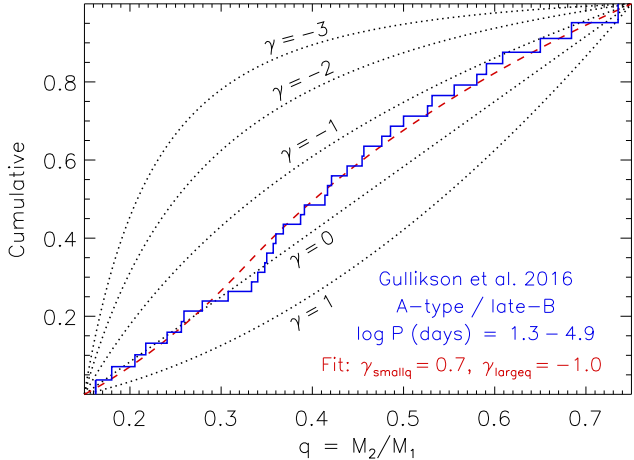}}
\caption{After accounting for incompleteness, we show the corrected cumulative distribution (blue) of mass ratios for the 36 binaries found in the \citet{Gullikson2016b} survey with B6-A7 primaries, orbital periods log\,$P$~=~1.3\,-\,4.9, and mass ratios $q$ = 0.15\,-\,0.75.  The intrinsic mass-ratio distribution cannot be described by a single power-law slope $\gamma$, which is consistent with the conclusions of \citet{Gullikson2016b}.  By fitting our broken two-component power-law distribution to the data, we measure {\large $\gamma$}$_{\rm smallq}$~=~0.7\,$\pm$\,0.8 across small mass ratios and {\large $\gamma$}$_{\rm largeq}$~=~$-$1.0\,$\pm$\,0.5 across large mass ratios (dashed red).}
\end{figure}

Of the 36 detected companions with $q$~=~0.15\,-\,0.75 in our subsample, 28 have large mass ratios $q$~$=$~0.30\,-\,0.75.  By weighting each system according to the inverse of their respective detection efficiency, we estimate a total of 34\,$\pm$\,9 companions with $q$ = 0.30\,-\,0.75.  The uncertainty derives from Poisson statistics and the uncertainties in the correction factors.  If the power-law component {\large $\gamma$}$_{\rm largeq}$~=~$-$1.0\,$\pm$\,0.5 extends all the way to $q$~=~1.0, then we expect 11\,$\pm$\,4 companions with $q$~=~0.75\,-\,1.00.  If instead there is an excess fraction ${\cal F}_{\rm twin}$~$\approx$~0.10 of twins,  as is indicated by observations of other samples of A-type MS binaries at closer (\S3.4) and wider (\S7.1) separations, then there would be 15\,$\pm$\,5 companions with $q$~=~0.75\,-\,1.00.  We adopt the average of these two estimates, and find a total of ${\cal N}_{\rm largeq}$~=~47\,$\pm$\,12 binaries with large mass ratios $q$~=~0.3\,-\,1.0.  We account for the correction factor ${\cal C}_{\rm evol}$~=~1.2\,$\pm$\,0.1 due to binary evolution, whereby (20\,$\pm$\,10)\% of A-type / late-B primaries in the magnitude-limited sample of \citet{Gullikson2016b} are actually the original secondaries of evolved binaries (see~\S11).  We measure a corrected frequency of companions with $q$~$>$~0.3 per decade of orbital period of $f_{\rm logP;q>0.3}$~=~${\cal N}_{\rm largeq} {\cal C}_{\rm evol}$/${\cal N}_{\rm prim}$/$\Delta$log$P$ = (47\,$\pm$\,11)(1.2\,$\pm$\,0.1)/281/(4.9\,$-$\,1.3) = 0.06\,$\pm$\,0.02.  We report the binary statistics based on the \citet{Gullikson2016b} survey in Table~4.  

\section{Eclipsing Binaries}

\renewcommand{\arraystretch}{1.6}
\setlength{\tabcolsep}{8pt}
\begin{figure*}[t!]\footnotesize
{\small {\bf Table 5:} Binary statistics based on our analysis of OGLE-III LMC EBs with B-type MS primaries and $P$~=~2\,-\,50~days.}  \\
\vspace*{-0.45cm}
\begin{center}
\begin{tabular}{|c|c|c|c|}
\hline
 Reference    & Primary Mass & Period Interval & Statistic \\
\hline
\multirow{3}{*} {\citet{Moe2013}} & \multirow{3}{*} {$\langle M_1 \rangle$ = 10\,$\pm$\,3\,\Msun}
  & \multirow{3}{*} {log\,$P$\,(days) = 0.8\,$\pm$\,0.5} & $f_{\rm logP;q>0.3}$ = 0.16\,$\pm$\,0.04 \\ 
\cline{4-4}
 & & & {\large $\gamma$}$_{\rm largeq}$ = $-$0.9\,$\pm$\,0.3 \\
\cline{4-4}
 & & & ${\cal F}_{\rm twin}$ = 0.12\,$\pm$\,0.06 \\
\hline
\citet{Moe2014} & $\langle M_1 \rangle$ = 10\,$\pm$\,4\,\Msun
  & log\,$P$\,(days) = 0.7\,$\pm$\,0.2 & {\large $\gamma$}$_{\rm smallq}$ = 0.2\,$\pm$\,0.8 \\
\hline
\multirow{7}{*} {\citet{Moe2015}} & \multirow{4}{*} {$\langle M_1 \rangle$ = 7\,$\pm$\,2\,\Msun}
  & \multirow{7}{*} {log\,$P$\,(days) = 1.5\,$\pm$\,0.2} & $\eta$ =  0.5\,$\pm$\,0.4 \\ 
\cline{4-4}
 & & & {\large $\gamma$}$_{\rm largeq}$ $=$ $-$1.1\,$\pm$\,0.3 \\
\cline{4-4}
 & & & ${\cal F}_{\rm twin}$ $<$ 0.04 \\
\cline{4-4}
 & & &  $f_{\rm logP;q>0.3}$ = 0.14\,$\pm$\,0.04 \\
\cline{4-4}
\cline{2-2}
 & \multirow{3}{*} {$\langle M_1 \rangle$ = 10\,$\pm$\,2\,\Msun} &  & $\eta$ = 0.8\,$\pm$\,0.3 \\
\cline{4-4}
 & & & {\large $\gamma$}$_{\rm largeq}$ $=$ $-$1.6\,$\pm$\,0.4 \\
\cline{4-4}
 & & & ${\cal F}_{\rm twin}$ $<$ 0.05 \\
\hline
\end{tabular}
\end{center}
\end{figure*}
\renewcommand{\arraystretch}{1.0}

Eclipsing binaries (EBs) with MS components are generally observed at short orbital periods $P$~$\lesssim$~50~days.  This is partially because of geometrical selection effects, but also due to the finite cadence of the observations \citep{Soderhjelm2005}.  EBs  offer an independent assessment of the close binary properties of massive stars.  Deep and wide-field surveys, such as the third phase of the Optical Gravitational Lensing Experiment \citep[OGLE-III][]{Graczyk2011}, have identified thousands of early-type EBs in the Magellanic Clouds.  Despite the geometrical selection effects, we can achieve EB samples at short and intermediate orbital periods that are 1\,-\,2 orders of magnitude larger than the spectroscopic binary samples.  Unlike early-type SB2s, which can be observed only if $q$ $>$ 0.25, binaries with $q$ $\approx$ 0.07\,-\,0.25 can produce detectable eclipses \citep{Moe2014}.  We present the parameter space of early-type EBs in terms of orbital periods $P$ and mass ratios $q$ as the red region in Fig.~1.

In several papers \citep{Moe2013,Moe2014,Moe2015}, we have analyzed OGLE-III EBs with B-type MS primaries in the Large Magellanic Cloud (LMC).  By utilizing the known distance to the LMC and the calibrated evolutionary tracks of B-type MS stars, we have measured the physical properties of OGLE-III LMC EBs based solely on the photometric light curves.  We also corrected for geometrical and evolutionary selection effects to recover the intrinsic multiplicity statistics.  The methods are thoroughly discussed in these three papers, and so we report only the main results pertinent to the present study (Table~5).  

In \citet{Moe2013}, we analyzed the eclipse depth and period distributions of EBs with early-B primaries and $P$ = 2\,-\,20 days.  After accounting for selection effects, we recovered the power-law component {\large $\gamma$}$_{\rm largeq}$ = $-$0.9\,$\pm$\,0.3 of the intrinsic mass-ratio distribution (Table~5).  In \citet{Moe2013}, we assumed this power-law slope continued all the way down to $q$ = 0.1, and so we measured a total binary fraction of (22\,$\pm$\,5)\% across $P$~=~2\,-\,20~days and $q$~=~0.1\,-\,1.0.  We also observed a small excess fraction $\approx$(7\,$\pm$\,5)\% of twin components relative to all companions with $q$ $>$ 0.1.  In the present study, we calculate our binary statistics ${\cal F}_{\rm twin}$ and $f_{\rm logP;q>0.3}$ according to binaries with $q$ $>$ 0.3.  Based on the EB data, the excess fraction of twins with $q$~$>$~0.95 relative to binaries with $q$~$>$~0.3 is ${\cal F}_{\rm twin}$~=~0.12\,$\pm$\,0.06 (Table~5).  Similarly, the fraction of early-B stars that have a companion with $P$ = 2\,-\,20 days and $q$ = 0.3\,-\,1.0 is (13\,$\pm$\,3)\%. Accounting for the correction factor ${\cal C}_{\rm evol}$ = (1.2\,$\pm$\,0.1) due to binary evolution within the OGLE-III LMC field of B-type MS stars, then the intrinsic frequency is $f_{\rm logP;q>0.3}$ = (0.13\,$\pm$\,0.03)(1.2\,$\pm$\,0.1)/(1.3\,$-$\,0.3) = 0.16\,$\pm$\,0.04 (Table 5).    At these short orbital periods, the early-B companion frequency $f_{\rm logP;q>0.3}$ $=$ 0.16\,$\pm$\,0.04 measured from EBs is between the O-type companion frequency $f_{\rm logP;q>0.3}$ $=$ 0.24\,$\pm$\,0.06 and the early/mid B-type companion frequencies $f_{\rm logP;q>0.3}$ $\approx$ 0.10\,-\,0.12 calculated from SB2s (\S3).  In addition, the parameters that describe the mass-ratio distribution across $q$~$>$~0.3 and $P$~=~2\,-\,20~days are consistent between the EB values ({\large $\gamma$}$_{\rm largeq}$ = $-$0.9\,$\pm$\,0.3 and ${\cal F}_{\rm twin}$ = 0.12\,$\pm$\,0.06) and SB2 values ({\large $\gamma$}$_{\rm largeq}$ = $-$0.3\,$\pm$\,0.3 and ${\cal F}_{\rm twin}$ = 0.11\,$\pm$\,0.04).

In \citet{Moe2014}, we identified and measured the physical properties of young EBs with early/mid B-type MS primaries, low-mass pre-MS companions ($q$ $\approx$ 0.07\,-\,0.36), and short orbital periods $P$~=~3.0\,-\,8.5 days.  We found that (1.5\,$\pm$\,0.5)\% of B-type MS stars have extreme mass-ratio companions with $q$~$=$~0.1\,-\,0.3 and $P$~=~3.0\,-\,8.5~days \citep[Table~5 and Fig.~12 in][]{Moe2014}. After accounting for the correction factor ${\cal C}_{\rm evol}$ = 1.2\,$\pm$\,0.1 due to binary evolution, then the frequency of small mass-ratio companions with $q$~=~0.1\,-\,0.3 per decade of orbital period is $f_{\rm logP;smallq}$ = (0.015\,$\pm$\,0.005)(1.2\,$\pm$\,0.1)/(log\,8.5\,$-$\,log\,3) = 0.04\,$\pm$\,0.02.  To calculate {\large $\gamma$}$_{\rm smallq}$ at these short orbital periods,  we average the statistics of $q$~$>$~0.3 companions from SB2s (\S3) and EBs above.  Given {\large $\gamma$}$_{\rm largeq}$~=~$-$0.6,  ${\cal F}_{\rm twin}$~=~0.12, and $f_{\rm logP;q>0.3}$~$=$~0.13 for early/mid B-type primaries, then $f_{\rm logP;smallq}$~=~0.04\,$\pm$\,0.02 implies {\large $\gamma$}$_{\rm smallq}$~=~0.2\,$\pm$\,0.8 (Table~5).  We emphasize this is the only measurement of {\large $\gamma$}$_{\rm smallq}$ for early-type binaries with short orbital periods where the companions are definitively known to be stellar in nature (see Fig.~1).  Our EB measurement of {\large $\gamma$}$_{\rm smallq}$~=~0.2\,$\pm$\,0.8 is between the values of {\large $\gamma$}$_{\rm smallq}$~=~0.6\,$\pm$\,0.8 and {\large $\gamma$}$_{\rm smallq}$~=~$-$0.5\,$\pm$\,0.8 we inferred from the number of O-type and B-type SB1s, respectively (\S3).

In \citet{Moe2015}, we analyzed the properties of EBs with B-type MS primaries and intermediate periods $P$ = 20\,-\,50 days.   For the entire population of EBs, which is dominated by relatively older mid-B primaries, we measured $\eta$~=~0.1\,$\pm$\,0.2 for the eccentricity distribution.  We also discovered the ages $\tau$ and eccentricities $e$ of the EBs are anticorrelated at a statistically significant level due to tidal evolution.  By selecting only the young systems with $\tau$~$<$~10~Myr that have not yet tidally evolved toward smaller eccentricities, we found that companions to early-B primaries at these orbital periods are initially born onto the zero-age MS with $\eta$~=~0.8\,$\pm$\,0.3 (Table~5). We measured a statistically significant anticorrelation between $\tau$ and $e$ for both early-B and mid-B subsamples.   We currently select the 29 EBs from \citet{Moe2015} with measured primary masses $M_1$~=~5\,-\,9\,\Msun, ages $\tau$~$<$~30~Myr, and eccentricities $e$~$<$~0.7.  Based on this subsample, we find binaries with mid-B primaries and intermediate orbital periods are born onto the zero-age MS with $\eta$~=~0.5\,$\pm$\,0.4 (Table~5).

After correcting for selection effects in \citet{Moe2015}, we measured {\large $\gamma$}$_{\rm largeq}$~=~$-$1.1\,$\pm$\,0.3 and ${\cal F}_{\rm twin}$~$<$~0.04 for mid-B primaries and intermediate orbit periods (Table~5).  The data also indicate that EBs with slightly more massive primaries favor smaller mass ratios, i.e.,  {\large $\gamma$}$_{\rm largeq}$~=~$-$1.6\,$\pm$\,0.4 and ${\cal F}_{\rm twin}$~$<$~0.05 (Table~5).  As found for SB2s, early-type binaries with longer orbital periods $P$~$>$~20~days favor smaller mass ratios, i.e., smaller values of {\large $\gamma$}$_{\rm largeq}$ and ${\cal F}_{\rm twin}$.  

 In \citet{Moe2015}, we also estimated that (6.7\,$\pm$\,2.2)\% of mid-B primaries have companions with $q$ $>$ 0.2 and $P$ = 20\,-\,50 days. Based on the measured intrinsic mass-ratio distribution \citep[Fig.~18 in][]{Moe2015}, 72\% of these systems have companions with $q$ $>$ 0.3.  In other words, (4.8\,$\pm$\,1.3)\% of B-type MS stars have a companion with $q$ $>$ 0.3 and $P$ = 20\,-\,50 days.  After accounting for the correction factor ${\cal C}_{\rm evol}$ = 1.2\,$\pm$\,0.1 due to binary evolution within the OGLE-III LMC field, then the intrinsic zero-age MS companion frequency is $f_{\rm logP;q>0.3}$ = (0.048\,$\pm$\,0.013)(1.2\,$\pm$\,0.1)/(log\,50~$-$~log\,20) = 0.14\,$\pm$\,0.04 for mid-B primaries and intermediate orbital periods (Table~5).  This is slightly larger than but consistent with the measurements of $f_{\rm logP;q>0.3}$ = 0.08\,-\,0.10 based on observations of mid-B SB2s (\S3).

\section{Long-Baseline Interferometry}

Long-baseline interferometry (LBI) can reveal binary companions at extremely small angular separations $\approx$5\,-\,100 mas \citep{Rizzuto2013,Sana2014}. Given the typical distances $d$ $\approx$ 0.1\,-\,2~kpc to early-type MS stars, these angular separations correspond to physical projected separations $\rho$~$\approx$~1.5\,-\,30AU, i.e. intermediate orbital periods 2.3~$\lesssim$~log\,$P$\,(days)~$\lesssim$~4.3.  LBI is limited by the brightness contrasts $\Delta$m~$\lesssim$~4~mag between the binary components, and so the secondaries must be comparable in luminosity to the primaries in order to be detected.  Unlike SB2s, which become biased toward moderate $q$ with increasing $P$, the sensitivity of LBI  is nearly constant with respect to orbital separation (see Fig.~4 in \citealt{Rizzuto2013} and Fig.~7 in \citealt{Sana2014}). LBI can therefore provide an unbiased sample of companions with $q$~$\gtrsim$~0.3 and 2.3 $\lesssim$ log\,$P$\,(days) $\lesssim$ 4.3 (magenta region in our Fig.~1).

\subsection{Early-B Primaries}

For a sample of ${\cal N}_{\rm prim}$ = 58 B0-B5 MS primaries in the Sco-Cen OB association ($d$ $\approx$ 130 pc), \citet{Rizzuto2013} used LBI to identify 24 companions with angular separations 7\,-\,130 mas.  They measured the brightness contrasts $\Delta$m of the binary components at wavelengths $\lambda$ = 550\,-\,800\,nm, and then estimated the mass ratios $q$ from $\Delta$m  according to stellar evolutionary tracks.  \citet{Rizzuto2013} report ${\cal N}_{\rm comp}$~=~18 companions  with $q$~$\ge$~0.3 and projected orbital separations $\rho$~=~1.5\,-\,30 AU, which correspond to orbital periods 2.3~$\lesssim$~log\,$P$\,(days)~$\lesssim$~4.3.  This subsample is relatively complete across the specified mass-ratio and period interval.  In Fig.~11, we display the cumulative distribution of mass ratios for these 18 systems.  The mass-ratio distribution of the 18 binaries are fitted to high accuracy by a single power-law distribution with {\large $\gamma$}$_{\rm largeq}$~=~$-$2.4\,$\pm$\,0.4. The upper limit on the excess twin fraction is ${\cal F}_{\rm twin}$ $<$ 0.03 (Table 6). Companions to early-B primaries at intermediate orbital periods are weighted toward extreme mass ratios. For $q$ $\gtrsim$ 0.3, the binary mass ratios are surprisingly consistent with random pairings drawn from a Salpeter IMF ({\large $\gamma$}~=~$-$2.35). 

\begin{figure}[t!]
\centerline{
\includegraphics[trim=0.5cm 0.15cm 0.5cm 0.2cm, clip=true, width=3.4in]{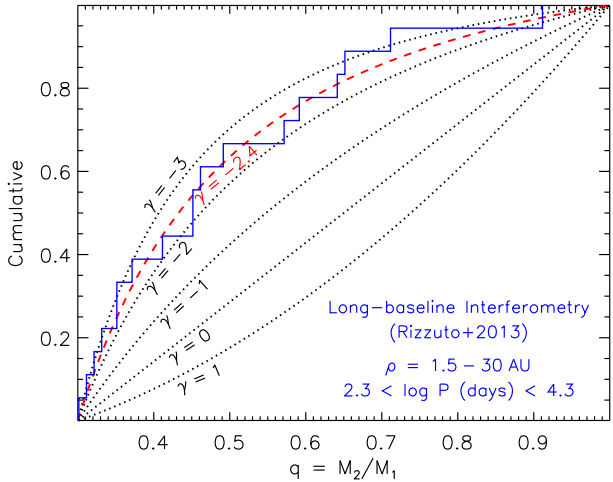}}
\caption{Cumulative distribution of mass ratios $q$ for the 18 companions (blue) to 58 early-B MS stars with $q$ $\ge$ 0.3 and projected separations $\rho$~=~1.5\,-\,30~AU (2.3 $\lesssim$ log $P$ $\lesssim$ 4.3) identified through long-baseline interferometry \citep{Rizzuto2013}.  In this parameter space, the companions are relatively complete and described by a power-law mass-ratio distribution with {\large $\gamma$}$_{\rm largeq}$~~=~$-$2.4\,$\pm$\,0.4 (dashed red).  For $q$ $>$ 0.3, this is surprisingly consistent with random pairings drawn from a Salpeter IMF ({\large $\gamma$}~=~$-$2.35).}
\end{figure}

 \citet{Rizzuto2013} do not directly state that their binaries discovered through long-baseline interferometry strongly favor small mass ratios.  They instead compile observations of short-period spectroscopic and long-period visual companions to the 58 early-B MS stars in their sample.  They then report a mass-ratio distribution with {\large $\gamma$}~$\approx$~$-$0.5 that is averaged across all orbital periods.  We emphasize that the binary distributions of $P$ and $q$ are not necessarily uncorrelated. LBI offers a unique perspective into the binary properties of massive stars at intermediate periods, and should therefore be treated independently.  

We wish to evaluate the robustness of our measurement of {\large $\gamma$}$_{\rm largeq}$~=~$-$2.4\,$\pm$\,0.4, and so we estimate the systematic uncertainties in deriving $q$ from $\Delta$m.  We calculate our own values of $q$ from the measured brightness contrasts $\Delta$m reported in \citet{Rizzuto2013} by incorporating different stellar evolutionary tracks, ages, and atmospheric parameters.  We also apply this method to the O-type LBI binary sample of \citet{Sana2014}, who report only the brightness contrasts $\Delta$m (see \S5.2).

In Fig.~12, we compare the brightness contrasts $\Delta$m to the mass ratios $q$ calculated by \citet{Rizzuto2013} for the 20 binaries in their sample with $\rho$~=~1.5\,-\,30~AU.  This subsample includes the 18 systems with $q$ $\ge$ 0.3 incorporated above as well as two additional systems with $q$ $\approx$ 0.26\,-\,0.29 near the detection limit. \citet{Rizzuto2013} report an uncertainty of 10\% in the mass ratios, which we indicate in our Fig.~12.   Eighteen of the 20 binaries have relatively unevolved MS primaries with spectral types B0\,-\,B5 and luminosity classes IV\,-\,V. Two systems, $\epsilon$-Cen and $\kappa$-Sco, have $\approx$B1III spectral types  and are therefore on the upper MS or giant branch (shown in our Fig.~12 as the two systems with diamond symbols).  Given the same brightness contrast $\Delta$m,  binaries with older primaries on the upper MS have more massive companions.

To determine our own values of $q$ from the brightness contrasts $\Delta$m, we utilize the solar-metallicity pre-MS and MS stellar evolutionary tracks from \citet{Tognelli2011} and \citet{Bertelli2009}, respectively.  We consider two primary masses, $M_1$~=~12\,\Msun\ and $M_1$~=~6\,\Msun, which are representative of B1V and B5V MS stars, respectively.  We model the brightness contrasts at two different ages, $\tau$~=~5~Myr and $\tau$~=~16~Myr, which span the estimated ages of the stellar subgroups within the Sco-Cen OB association \citep{Rizzuto2013}.  We incorporate the bolometric corrections and color indices from \citet{Pecaut2013} to calculate the red R$_{\rm C}$-band magnitudes from the stellar luminosities and effective temperatures.  In Fig.~12, we plot our derived brightness contrasts $\Delta$R$_{\rm C}$ as a function of $q$ for the four different combinations of primary mass and age.  The four curves are quite similar to each other except for two minor aspects.  First, at ages $\tau$~=~5~Myr, secondaries with $M_2$~$\lesssim$~2\,\Msun\ are still on the pre-MS.  This explains why the blue dashed curve corresponding to our model with $M_1$~=~6\,\Msun\ and $\tau$~=~5~Myr is non-monotonic near $q$~=~0.3 ($M_2$~$\approx$~2\,\Msun).  Second, our model with a $M_1$~=~12\Msun\ primary is near the tip of the MS at $\tau$~=~16~Myr (green curve).  For this older model, a binary with a given mass ratio $q$ is observed to have a slightly larger brightness contrast $\Delta$m.  In general, our models are consistent with the mass ratios reported by \citet{Rizzuto2013}.  In fact, the two systems with $\approx$B1III primaries better match our model with $M_1$~=~12\,\Msun\ and $\tau$~=~16~Myr (green curve in Fig.~12).

\begin{figure}[t!]
\centerline{
\includegraphics[trim=0.5cm 0.0cm 0.5cm 0.1cm, clip=true, width=3.4in]{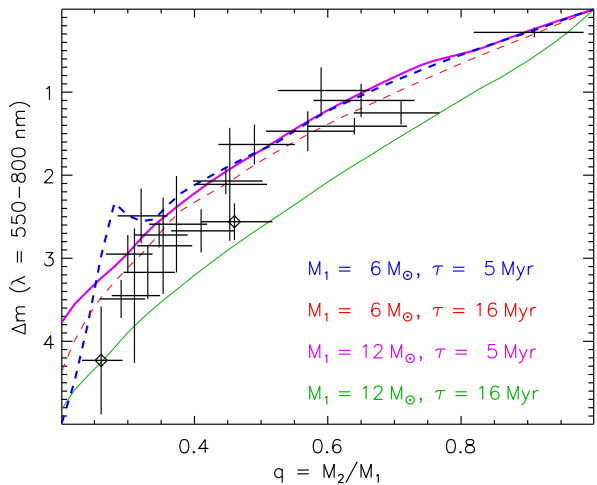}}
\caption{Measured mass ratios $q$ = $M_2$/$M_1$ and brightness contrasts $\Delta$m (mag) at $\lambda$ = 550\,-\,800\,nm for the 20 companions to early-B primaries with projected separations $\rho$ $=$ 1.5\,-\,30 AU as reported by \citet{Rizzuto2013}.  Eighteen of the binaries have MS primaries with luminosity classes of IV-V, while two systems have primaries with luminosity classes of III (shown with diamond symbols). We model the brightness contrasts $\Delta$R$_{\rm c}$ in the red band for a primary mass $M_1$ = 6\,\Msun\ (dashed) with ages $\tau$~=~5~Myr (thick blue) and $\tau$~=~16~Myr (thin red) and for a primary mass $M_1$~=~12\,\Msun\ (solid) with ages $\tau$~=~5~Myr (thick magenta) and $\tau$~=~16~Myr (thin green). Our models are consistent with the mass ratios provided in \citet{Rizzuto2013}.  }
\end{figure}

 To estimate the systematic uncertainties in {\large $\gamma$}$_{\rm largeq}$, we utilize the red line in Fig.~12 to calculate our own values of $q$.  Based on this model, we find the mass ratios $q$ are slightly smaller than those reported by \citet{Rizzuto2013}. We find 16 systems (instead of 18) with $q$~$>$~0.3.  By fitting a power-law mass-ratio distribution to these 16 binaries, we measure {\large $\gamma$}$_{\rm largeq}$ = $-$2.2\,$\pm$\,0.5.  This nearly matches the previous result of {\large $\gamma$}$_{\rm largeq}$ = $-$2.4\,$\pm$\,0.4 from using the mass ratios $q$ directly provided by \citet{Rizzuto2013}.  The systematic uncertainty is smaller than the measurement uncertainty, and so we adopt the average value {\large $\gamma$}$_{\rm largeq}$ = $-$2.3\,$\pm$\,0.5 of the two methods (Table~6). Binaries with early-B primaries and intermediate orbital periods are indeed weighted toward extreme mass ratios.  

We also average the number ${\cal N}_{\rm largeq}$ = 17 of binaries with $q$ $>$ 0.3 and 2.3~$<$~log\,$P$\,(days)~$<$~4.3 based on the two methods for measuring $q$ in the \citet{Rizzuto2013} sample.  After accounting for the correction factor ${\cal C}_{\rm evol}$ = 1.2\,$\pm$\,0.1 due to binary evolution, then the intrinsic companion frequency is $f_{\rm logP;q>0.3}$ = ${\cal N}_{\rm largeq}$\,${\cal C}_{\rm evol}$/${\cal N}_{\rm prim}$/$\Delta$log$P$ = (17\,$\pm$\,$\sqrt{17}$)(1.2\,$\pm$\,0.1)/58/(4.3\,$-$\,2.3) = 0.18\,$\pm$\,0.05 for early-B binaries and intermediate orbital periods (Table~6).  This is slightly larger than but consistent with our measurements of $f_{\rm logP;q>0.3}$ = 0.12\,-\,0.16 for binaries with early-B primaries and shorter orbital periods based on observations of SB2s (\S3) and EBs (\S4).

\subsection{O-type Primaries}

\citet{Sana2014} recently surveyed $\approx$100 O-type stars with near-infrared magnitudes $H$ $<$ 7.5 ($\lambda$~$\approx$~1.6\,$\mu$m) in the southern hemisphere using both long-baseline interferometry (LBI) and sparse aperture masking (SAM) techniques.  The combination of these observational methods provide a relatively complete sample of companions with angular separations 2\,-\,60~mas and brightness contrasts $\Delta$H~$<$~4.0~mag (see Fig.~7 in \citealt{Sana2014}). \citet{Sana2014} also resolved additional companions at wider separations $\gtrsim$0.3$''$ with adaptive optics, which we examine in \S7.  

\renewcommand{\arraystretch}{1.6}
\setlength{\tabcolsep}{10pt}
\begin{figure}[t!]\footnotesize
{\small {\bf Table 6:} Companion statistics based on long-baseline interferometric observations of early-B stars \citep{Rizzuto2013} and O-type stars \citep{Sana2014}.}  \\
\vspace*{-0.45cm}
\begin{center}
\begin{tabular}{|c|c|}
\hline
 Reference and Primary Mass /   &  \\
 Period Interval & Statistic \\
\hline
 \citet{Rizzuto2013}; & {\large $\gamma$}$_{\rm largeq}$  = $-$2.3\,$\pm$\,0.5 \\
\cline{2-2}
  $\langle M_1 \rangle$ = 10\,$\pm$\,3\,\Msun;  & ${\cal F}_{\rm twin}$ $<$ 0.03 \\
\cline{2-2}
 log\,$P$\,(days) = 3.3\,$\pm$\,1.0 & $f_{\rm logP;q>0.3}$ = 0.18\,$\pm$\,0.05 \\
\hline
\citet{Sana2014}; & {\large $\gamma$}$_{\rm largeq}$  = $-$1.4\,$\pm$\,0.4 \\
\cline{2-2}
  $\langle M_1 \rangle$ = 28\,$\pm$\,8\,\Msun;  & ${\cal F}_{\rm twin}$ $<$ 0.03 \\
\cline{2-2}
 log\,$P$\,(days) = 3.6\,$\pm$\,1.1 & $f_{\rm logP;q>0.3}$ = 0.19\,$\pm$\,0.05 \\
\hline

\end{tabular}
\end{center}
\end{figure}
\renewcommand{\arraystretch}{1.0}

 It is difficult to reliably measure the mass ratios and correct for incompleteness for binaries with supergiant primaries.  In our current analysis, we consider only the 56 O-type primaries in the \citet{Sana2014} main sample that were observed by both LBI and SAM methods and have luminosity classes II.5\,-\,V (see their Fig.~1).  Their survey is magnitude-limited, so we must correct for the \"{O}pik effect / Malmquist bias toward binaries with equally bright components.  We remove the two detected binaries (HD\,93222 and HD\,123590) with observed total magnitudes $H$ $\approx$ 7.2\,-\,7.5 and brightness contrasts $\Delta$H~$\lesssim$~0.3~mag.  These two systems would be fainter than the $H$ = 7.5 limit if we were to consider the luminosity from the primaries alone.  Our culled sample from the \citet{Sana2014} survey contains ${\cal N}_{\rm prim}$ = 54 O-type MS primaries.  

From this subsample of 54 O-type MS primaries, \citet{Sana2014} identified 25 companions with angular separations 2\,-\,60 mas and brightness contrasts $\Delta$H~$<$~4.0~mag.    Given the typical distances $d$~=~1\,-\,2~kpc to the O-type stars with luminosity classes II.5\,-\,V in the \citet{Sana2014} sample (see their Fig.~3), the angular separations correspond to projected separations $\rho$~$\approx$~3\,-\,90~AU, i.e., 2.5~$<$~log\,$P$\,(days)~$<$~4.7.  In Fig.~13, we show the measured brightness contrasts $\Delta$H and uncertainties for the 25 binaries as reported in \citet{Sana2014}.  For the few systems with multiple measurements of $\Delta$H, we adopt a weighted average and uncertainty.  

\begin{figure}[t!]
\centerline{
\includegraphics[trim=0.6cm 0.1cm 0.5cm 0.2cm, clip=true, width=3.15in]{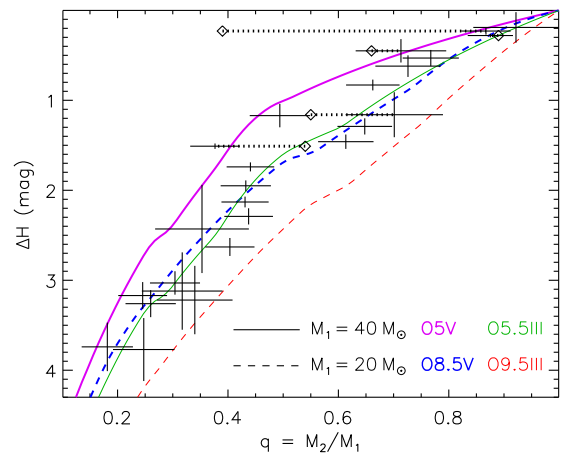}}
\caption{Near-infrared brightness contrast $\Delta$H versus binary mass ratio $q$.  We model $\Delta$H for $M_1$~=~40\,\Msun\ (solid) and $M_1$~=~20\,\Msun\ (dashed) primaries.  For dwarf stars, these masses correspond to O5V (thick magenta) and O8.5V (thick dashed blue) primaries.  As giants, they will appear as O5.5III (thin green) and O9.5III (thin dashed red).  We display the 25 O-type binaries from \citet{Sana2014} with $\Delta$H~$<$~4.0~mag, angular separations 2\,-\,60~mas, and primary luminosity classes II.5-V.  We utilize our models to measure the mass ratios $q$ according to the observed $\Delta$H and the spectral types and luminosity classes of the primaries. Five of the binaries are SB2s, where we link with dotted lines the dynamical mass ratios (diamonds) to our measurements inferred from the brightness contrasts.  }
\end{figure}

We employ a method similar to that described in \S5.1 to measure the binary mass ratios from the observed brightness contrasts $\Delta$H. For our 25 O-type binaries, we utilize the calibrated relations for galactic O-type stars in  \citet{Martins2005} to estimate the primary mass, effective temperature, and absolute M$_{\rm V}$ magnitude according to the primary's spectral type and luminosity class.  We then calculate the near-infrared magnitude $H$ according to the temperature-dependent color indices $(V-H)$ reported in \citet{Pecaut2013}.  If the secondary is also an O-type star with $M_2$~$\gtrsim$~16\,\Msun, then we use this same technique to estimate its own value of $H$.  For these O-type + O-type binaries, we assume the secondary is alway a MS star with luminosity class V if $q$~$<$~0.6.  For $q$~$>$~0.6, we smoothly interpolate the secondary's luminosity class between V at $q$~=~0.6 and the luminosity class of that of the primary at $q$~=~1.0.  If the secondary is a B-type star with $M_2$~$\lesssim$~16\,\Msun, we interpolate the solar-metallicity stellar evolutionary tracks of \citet{Bertelli2009} to determine the secondary's near-infrared magnitude $H$.  For these O-type + B-type binaries, we adopt an age appropriate for the spectral type and luminosity class of the primary.  

In Fig.~13, we display our modeled brightness contrasts $\Delta$H as a function of mass ratio $q$ for $M_1$ = 20\,\Msun\ (dashed) and $M_1$ = 40\,\Msun\ (solid) primaries.  For dwarf stars, these masses correspond to O8.5V (thick dashed blue) and O5V (thick solid magenta), respectively. The magenta model corresponding to the O5V primary is flatter than the dashed blue model of the O8.5V primary.  This is because the MS mass-luminosity relation flattens toward larger masses.  For example, a 40 \Msun\ MS star is $\Delta$H~$\approx$~1.0~mag brighter than a 20\,\Msun\ MS star (value of magenta curve at $q$~=~0.5). Meanwhile, a 20 \Msun\ MS star is $\Delta$H $\approx$ 1.7 mag brighter than a 10\,\Msun\ MS star (value of dashed blue curve at $q$~=~0.5).  The $M_1$~=~40\,\Msun\ primary increases in brightness by $\Delta$H~$\approx$~0.6~mag as it becomes an O5.5III star (thin green line in Fig.~13).  This is why the green and magenta solid curves differ by $\Delta$H~$\approx$~0.6~mag at $q$~$\lesssim$~0.5.  Similarly, the $M_1$~=~20\,\Msun\ primary will increase in brightness by $\Delta$H $\approx$ 1.0 mag as it evolves into an O9.5III giant (thin dashed red).

For the 25 binaries we have selected from \citet{Sana2014}, we determine the mass ratios $q$ from our models according to the listed brightness contrasts $\Delta$H and spectral types and luminosity classes of the primaries.  We propagate the measurement uncertainties in $\Delta$H as well as errors of $\approx$0.5 in both the spectral subtypes and luminosity classes.  We display our solutions for the mass ratios of the 25 systems in Fig.~13.  Twenty-three of the binaries are between our O5V (magenta) and O9.5III (dashed red) models. The two remaining systems have $\approx$O4V primaries and lie just above the magenta curve.  

Five of the 25 O-type binaries resolved with LBI/SAM in our subsample are also long-period SB2s with independent measurements of the mass ratio \citep{Sana2014}.  The SB2s with dynamical mass ratio measurements are HD\,54662 \citep[$q$~=~0.39,][]{Boyajian2007}, HD\,150136 \citep[$q$~=~0.54,][]{Mahy2012}, HD\,152246 \citep[$q$~=~0.89,][]{Nasseri2014}, HD\,152314 \citep[$q$~=~0.55,][]{Sana2012}, and HD\,164794 \citep[$q$~=~0.66,][]{Rauw2012}.  We display the five spectroscopic mass ratios as diamond symbols in Fig.~13.  Three of the five SB2s have dynamical mass ratios consistent with our values.  For these three systems, we adopt the SB2 dynamical mass ratios because they are measured to higher precision.  

For one of the two remaining systems, HD\,150136, we measure $q$~=~0.38\,$\pm$\,0.05 according to the moderate brightness contrast $\Delta$H~=~1.5~mag.  Meanwhile,  \citet{Mahy2012} report $q$~$\approx$~0.54 based on the SB2 dynamics.  This apparent discrepancy is because HD\,15136 is a triple system.  The tertiary component resolved with LBI orbits an inner binary of spectral types O3-3.5V ($M_1$~$\approx$~64\,\Msun) and O5.5-6V ($M_2$~$\approx$~40\,\Msun) in a $P$~=~2.7~day orbit \citep{Mahy2012}.  The additional luminosity from the inner companion increases the brightness contrast $\Delta$H between the inner binary and tertiary, which biases our mass ratio measurement toward smaller values.  To derive a dynamical mass of $M_3$~$\approx$~35~\Msun, \citet{Mahy2012} assume the tertiary component is coplaner with the inner binary. Utilizing the observed spectral type O6.5-7V of the tertiary alone \citep{Mahy2012}, the mass is $M_3$~$\approx$~27\,\Msun\ according to the O-type stellar relations provided in \citet{Martins2005}. This implies a mass ratio of $q$~=~$M_3$/$M_1$~$\approx$~0.42.  We adopt $q$ = 0.48 for the tertiary companion in HD\,150136, which is halfway between the dynamical coplaner estimate of $q$~$\approx$~0.54 and the measurement of $q$~=~0.42 based on the observed spectral types.  

Finally, for HD\,54662, we measure $q$ $\approx$ 0.87 based on the small brightness contrast $\Delta$H~=~0.2~mag, while \citet{Boyajian2007} report an SB2 dynamical mass ratio of $q$~=~$K_1$/$K_2$~=~0.39.  The spectroscopic absorption features of the binary components in HD\,54662 are significantly blended \citep{Boyajian2007}, and so the velocity semi-amplitudes $K_1$ and $K_2$ are rather uncertain. Moreover, \citet{Boyajian2007} fit O6.5V and O9V spectral types to the binary components with an optical flux ratio of $F_2$/$F_1$ $\approx$ 0.5.  These spectral types and optical brightness contrast imply a much larger mass ratio of  $q$ $\approx$ 0.7 \citep{Martins2005}.  For HD\,54662, we adopt this spectroscopic measurement of $q$ = 0.70, which is between the near-infrared brightness contrast measurement of $q$ = 0.89 and the uncertain dynamical measurement of $q$ = 0.39.  

\begin{figure}[t!]
\centerline{
\includegraphics[trim=0.5cm 0.0cm 0.5cm 0.2cm, clip=true, width=3.45in]{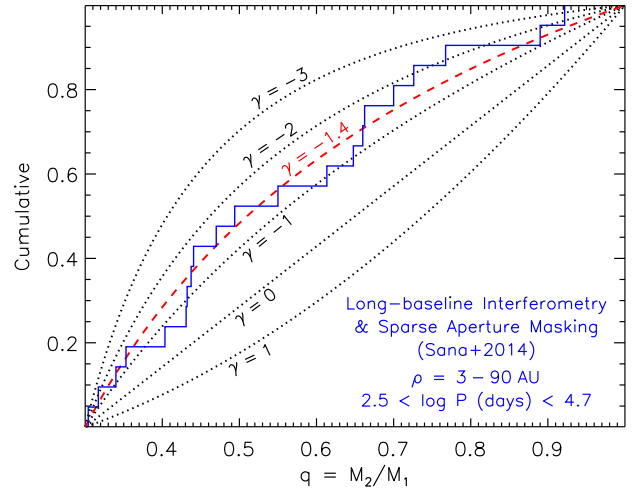}}
\caption{Cumulative distribution of mass ratios $q$ for the 21 companions (blue) to 54 O-type MS stars with $q$~$\ge$~0.3 and projected separations $\rho$~=~3\,-\,90~AU (2.5~$\lesssim$~log\,$P$~$\lesssim$~4.7) identified through long-baseline interferometry and sparse aperture masking \citep{Sana2014}.  For this relatively complete subsample, we measure the power-law component of the mass-ratio probability distribution across $q$~=~0.3\,-\,1.0 to be {\large $\gamma$}$_{\rm largeq}$~=~$-$1.4\,$\pm$\,0.4 (dashed red). This confirms that slightly wider companions to O-type stars are weighted toward smaller mass ratios $q$ $\approx$ 0.3\,-\,0.5.}
\end{figure}

For the 20 companions resolved at intermediate orbital periods without spectroscopic mass measurements, we adopt the mass ratios measured from the brightness contrasts $\Delta$H.  Given the detection limit  $\Delta$H~$<$~4.0~mag, companions to O-type primaries with luminosity classes III-V are relatively complete for $q$~$>$~0.3 (see Fig.~13).   For the ${\cal N}_{\rm prim}$~=~54 O-type primaries with luminosity classes II.5\,-\,V in \citet{Sana2014}, we find ${\cal N}_{\rm largeq}$ = 21 companions with $q$~$>$~0.3 and 2.5~$<$~log\,$P$\,(days)~$<$~4.7. Because we limited our sample to O-type stars with luminosity classes II.5\,-\,V, the correction factor ${\cal C}_{\rm evol}$ = 1.1\,$\pm$\,0.1 due to binary evolution is relatively small.  The intrinsic frequency of companions with $q$ $>$ 0.3 per decade of orbital period is $f_{\rm logP;q>0.3}$ = ${\cal N}_{\rm largeq} {\cal C}_{\rm evol}$/${\cal N}_{\rm prim}$/$\Delta$log$P$ = (21\,$\pm$\,$\sqrt{21}$)(1.1\,$\pm$\,0.1)/54/(4.7\,$-$\,2.5) = 0.19\,$\pm$\,0.05 for O-type stars and intermediate orbital periods (Table~6).  This is between the values $f_{\rm logP;q>0.3}$ = 0.12\,-\,0.24 we measured for O-type SB2s (\S3).

We display the cumulative distribution of mass ratios for the 21 binaries with $q$~$>$~0.3 in Fig.~14.  We fit the mass-ratio probability distribution, and measure {\large $\gamma$}$_{\rm largeq}$~=~$-$1.4\,$\pm$\,0.4 and ${\cal F}_{\rm twin}$~$<$~0.03 (Table~6).  Although not as steep as the slope {\large $\gamma$}$_{\rm largeq}$~=~$-$2.3\,$\pm$\,0.5 measured for LBI companions to early-B primaries (\S5.1), the mass-ratio distribution of LBI/SAM O-type binaries is still weighted toward small mass ratios.  In fact, the O-type LBI/SAM measurement of {\large $\gamma$}$_{\rm largeq}$~=~$-$1.4\,$\pm$\,0.4 nearly matches the O-type long-period SB2 measurement of {\large $\gamma$}$_{\rm largeq}$~=~$-$1.6\,$\pm$\,0.5  (\S3). These two independent measurements of {\large $\gamma$}$_{\rm largeq}$~$\approx$~$-$1.5 for intermediate-period O-type binaries are smaller than the value {\large $\gamma$}$_{\rm largeq}$~=~$-$0.3\,$\pm$\,0.3 we measured at shorter periods $P$~$<$~20~days (see \S3).  The LBI/SAM observations of O-type stars confirm that companions at slightly wider separations favor smaller mass ratios.  

\section{Cepheids}

The majority of Cepheid giant variables evolve from mid-B MS stars with $\langle M_1 \rangle$ $\approx$ 4\,-\,8\,\Msun\ \citep{Turner1996,Evans2013}.  B-type MS stars with close stellar companions at log\,$P$\,(days) $\lesssim$ 2.6, i.e., $P$~$\lesssim$~1~yr, will fill their Roche lobes before they can expand into the instability strip \citep{Evans2013}.  Many of the mid-B primaries with close binary companions are unlikely to evolve into Cepheid variables. The Cepheid population can, however, offer invaluable insight into the frequency and properties of companions to intermediate-mass stars at longer orbital periods log\,$P$~$\gtrsim$~2.6. Although the orbits may have tidally evolved toward smaller eccentricities, the masses of detached binaries with Cepheid primaries and MS companions have not significantly changed from their original zero-age MS values \citep{Evans2013}.  Unlike their B-type MS progenitors, which have rotationally and pressure broadened spectra (see \S3), Cepheid giants have narrow absorption lines. Companions that produce small velocity semi-amplitudes $K_1$~$\approx$~2~km~s$^{-1}$ can be detected once the primary evolves into a Cepheid \citep{Evans2015}.  Spectroscopic surveys of Cepheid primaries are therefore more sensitive toward companions with smaller masses and longer orbital periods.

\citet{Evans2013} took advantage of the temperature differences between cool Cepheid giants and hot late-B/early-A companions that are still on the MS.  For a magnitude-limited sample of ${\cal N}_{\rm Cepheid}$ = 76 Cepheids, they compiled all known massive companions with $M_2$ $\gtrsim$ 2\,\Msun\ and $T_2$ $\gtrsim$ 10,000 K that exhibit a UV excess.  \citet{Evans2013} measured the masses $M_1$ of the primaries according to a mass-luminosity relation for Cepheids, and the masses $M_2$ of the hot MS companions from their UV spectral features.  This technique is sensitive to companions with $q$ $\gtrsim$ 0.3 that are hot enough to produce a UV excess, regardless of the orbital separation.   \citet{Evans2013} also utilized spectroscopic and photometric follow-up observations to estimate the orbital periods of the binaries in their sample.  They find 16 companions with 2.7~$<$~log\,$P$\,(days)~$<$~6.5 and $q$~$\ge$~0.35, which is relatively complete in this parameter space (green region in our Fig.~1).

\subsection{Wide Companions}

We initially examine the ${\cal N}_{\rm longP}$~=~8 long-period companions with 4.1 $<$ log\,$P$ $<$ 6.5 and $q$ $>$ 0.3 in the \citet{Evans2013} sample.  We display in Fig.~15 the cumulative distribution of mass ratios $q$ for these 8 wide binaries.  This subsample is relatively complete across the specified mass-ratio interval.  We measure {\large $\gamma$}$_{\rm largeq}$ = $-$2.1\,$\pm$\,0.5 and ${\cal F}_{\rm twin}$ $<$ 0.07 (Table~7).  We find that wide companions to intermediate-mass stars have mass ratios $q$ $\gtrsim$ 0.3 consistent with random pairings drawn from a Salpeter IMF ({\large $\gamma$} = $-$2.35).

\begin{figure}[t!]
\centerline{
\includegraphics[trim=0.5cm 0.1cm 0.5cm 0.2cm, clip=true, width=3.4in]{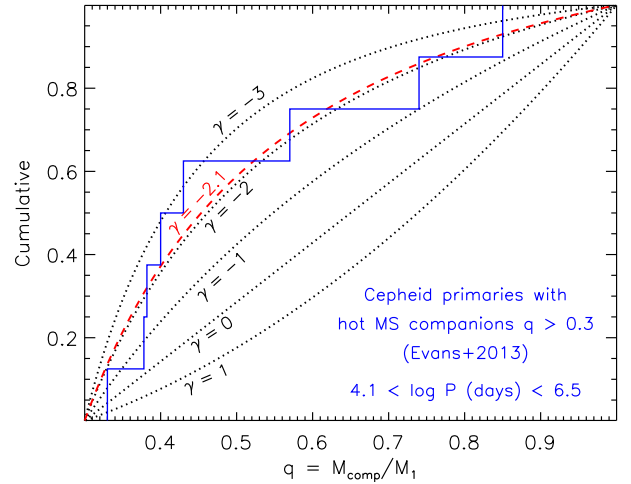}}
\caption{Cumulative distribution of mass ratios $q$ for the 8 wide companions (blue) to 76 Cepheid primaries ($\langle M_1 \rangle$~=~6\,$\pm$\,2\,\Msun) with log\,$P$\,(days) = 4.1\,-\,6.5 and $q$~$>$~0.3 as listed in \citet{Evans2013}.  Hot MS companions to cool Cepheid primaries produce a detectable UV excess if $q$~$\gtrsim$~0.3. This subsample is therefore relatively complete, where we measure the power-law component of the mass-ratio probability distribution to be {\large $\gamma$}$_{\rm largeq}$~=~$-$2.1\,$\pm$\,0.5 (dashed red).  Cepheids, which evolved from mid-B MS primaries, have wide companions with mass ratios $q$~$>$~0.3 that are consistent with random pairings drawn from a Salpeter IMF ({\large $\gamma$} = $-$2.35).}
\end{figure}

To calculate $f_{\rm logP;q>0.3}$,  we account for the fact that many mid-B MS primaries with close companions log\,$P$\,(days) $<$ 2.6 are unlikely to evolve into Cepheids. In the previous sections, we have measured $f_{\rm logP;q>0.3}$~$\approx$~0.08\,-\,0.14 for mid-B MS primaries and 0.2~$<$~log\,$P$~$<$~2.6.  We estimate that $f_{\rm logP;q>0.3}$\,$\times$\,$\Delta$log$P$~$\approx$~(0.11\,$\pm$\,0.3)\,$\times$\,(2.6\,$-$\,0.2) $\approx$ (20\,-\,30)\% of mid-B MS stars will interact with a binary companion with $q$ $>$ 0.3 and log $P$ $<$ 2.6. Depending on {\large$\gamma$}$_{\rm smallq}$, an additional (5\,-\,15)\% of mid-B stars will interact with a small mass-ratio $q$ = 0.1\,-\,0.3 companion across log\,$P$ = 0.2\,-\,2.6.  By adding these two statistics, we find that (35\,$\pm$\,10)\% of mid-B MS primaries have companions with $q$~$>$~0.1 and log\,$P$ $<$ 2.6 (see also \S9).  Approximately one-third of these close binaries will probably merge, particularly those with smaller mass ratios and/or shorter separations \citep[][\S11]{Hurley2002,Belczynski2008}.   These merger products can continue to evolve into Cepheid giants. Alternatively, the other two-thirds of close binaries  will undergo stable mass transfer or survive common envelope (CE) evolution \citep{Hurley2002,Belczynski2008}, thereby preventing the primary from evolving into a Cepheid. In total, we estimate that (25\,$\pm$\,15)\% of mid-B primaries will interact with a close stellar companion in such a manner that it does not evolve into a Cepheid.  The remaining ${\cal F}_{\rm Cepheid}$~=~0.75\,$\pm$\,0.15 of mid-B MS primaries are capable of evolving into Cepheid variables. 

Because only a subset ${\cal F}_{\rm Cepheid}$~=~0.75\,$\pm$\,0.15 of mid-B MS stars evolve into Cepheids, the frequency of wide companions to B-type MS stars is smaller than the frequency of wide companions to Cepheids.  We also account for the correction factor ${\cal C}_{\rm evol}$ = 1.2\,$\pm$\,0.1, whereby (20\,$\pm$\,10)\% of observed Cepheids are actually the original secondaries of evolved binaries.  We calculate $f_{\rm logP;q>0.3}$~=~${\cal N}_{\rm longP}$\,${\cal F}_{\rm Cepheid}$\,${\cal C}_{\rm evol}$/${\cal N}_{\rm Cepheid}$/$\Delta$log$P$ = (8\,$\pm$\,$\sqrt{8}$)(0.75\,$\pm$\,0.15)(1.2\,$\pm$\,0.1)/76/(6.5\,$-$\,4.1) = 0.04\,$\pm$\,0.02 for mid-B MS stars and long orbital periods (Table~7).

\renewcommand{\arraystretch}{1.55}
\setlength{\tabcolsep}{10pt}
\begin{figure}[t!]\footnotesize
{\small {\bf Table 7:} Companion statistics based on observations of Cepheid variables that evolved from mid-B MS primaries ($\langle M_1 \rangle$ = 6\,$\pm$\,2\,\Msun).}  \\
\vspace*{-0.45cm}
\begin{center}
\begin{tabular}{|c|c|}
\hline
 Reference and    &  \\
 Period Interval & Statistic \\
\hline
 \citet{Evans2013}; & {\large $\gamma$}$_{\rm largeq}$  = $-$2.1\,$\pm$\,0.5 \\
\cline{2-2}
  log\,$P$\,(days) = 5.3\,$\pm$\,1.2  & ${\cal F}_{\rm twin}$ $<$ 0.07 \\
\cline{2-2}
                                      & $f_{\rm logP;q>0.3}$ = 0.04\,$\pm$\,0.02 \\
\hline
           & {\large $\gamma$}$_{\rm largeq}$  = $-$1.7\,$\pm$\,0.5 \\
\cline{2-2}
\citet{Evans2015}; & {\large $\gamma$}$_{\rm smallq}$  $=$ 0.2\,$\pm$\,0.9 \\
\cline{2-2}
  log\,$P$\,(days) = 3.1\,$\pm$\,0.5   & ${\cal F}_{\rm twin}$ $<$ 0.06 \\
\cline{2-2}
                                      & $f_{\rm logP;q>0.3}$ = 0.17\,$\pm$\,0.08 \\
\hline
\end{tabular}
\end{center}
\end{figure}
\renewcommand{\arraystretch}{1.0}

\subsection{Companions with Intermediate Orbital Periods}

We next examine the companions to Cepheids at intermediate orbital periods.  In a recent follow-up paper, \citet{Evans2015} identified all known spectroscopic binary companions  to Cepheids, including those that did not necessarily exhibit a UV excess.  Given $\langle M_1 \rangle$ $\approx$ 6\,\Msun\ and the sensitivity $K_1$ $\approx$ 2 km s$^{-1}$ of their radial velocity measurements, the \citet{Evans2015} sample is relatively complete for $q$~$>$~0.1 and for $P$~$\approx$~1\,-\,10 years, i.e. 2.6~$<$~log\,$P$\,(days)~$<$~3.6 (see their Figs.~4~\&~5).  At slightly longer orbital periods $P$~$>$~10~years, the \citet{Evans2015} survey most likely missed low-mass companions due to the limited sensitivity and cadence of the spectroscopic observations.  We analyze the 17 companions reported in Table~8 of \citet{Evans2015} that have $P$~=~1\,-\,10~years and measured values or limits on the mass ratios.

Of the 17 intermediate-period companions to Cepheids, ${\cal N}_{\rm largeq}$ = 10 have measured mass ratios $q$ $>$ 0.3 based on the observed UV excess from the hot MS companions.  We plot the cumulative distribution of mass ratios for these 10 systems in Fig.~16, and measure {\large $\gamma$}$_{\rm largeq}$~=~$-$1.7\,$\pm$\,0.5 and ${\cal F}_{\rm twin}$~$<$~0.06 (Table~7).  Like the wide binaries, companions to Cepheids at intermediate orbital periods have mass ratios across $q$~$=$~0.3\,-\,1.0 weighted toward small values $q$ $\approx$ 0.3\,-\,0.5. 

Of the seven remaining companions to Cepheids at intermediate orbital periods, only one has a measured mass ratio $q$ = 0.27 near the sensitivity limit. The other six systems do not have detectable UV excesses.  Assuming these six SB1s have MS companions, the lack of UV excess provides an upper limit for the mass ratios $q$ $\lesssim$ 0.3 (see Table~8 in \citealt{Evans2015}).  Alternatively, the companions may not exhibit a UV excess because they are faint compact remnants.  In either case, the six long-period SB1s  must have $q$ $\gtrsim$ 0.1 given the sensitivity $K_1$~=~2~km~s$^{-1}$ of the spectroscopic radial velocity observations.  In fact, assuming random orientations, we expect one additional system with $q$~=~0.1\,-\,0.3 to have escaped detection \citep[see Fig.~4 in][]{Evans2015}.  Assuming ${\cal N}_{\rm smallq}$~=~8, i.e., all observed and suspected companions with $q$~=~0.1\,-\,0.3 are stellar in nature, and given ${\cal N}_{\rm largeq}$~=~10, {\large $\gamma$}$_{\rm largeq}$~=~$-$1.7, and ${\cal F}_{\rm twin}$~=~0.0, then {\large $\gamma$}$_{\rm smallq}$~=$-$0.1\,$\pm$\,0.7.  If instead  $\approx$30\% of the six observed SB1s contain compact remnant companions, as found for other populations of SB1s (see \S3\,-\,4, \S8, and \S11), then ${\cal N}_{\rm smallq}$~=~6.  This smaller number of $q$~=~0.1\,-\,0.3 stellar companions implies a positive slope {\large $\gamma$}$_{\rm smallq}$~=~0.6\,$\pm$\,0.8. We adopt the average {\large $\gamma$}$_{\rm smallq}$~=~0.2\,$\pm$\,0.9 of these two values (Table~7), where we propagate the uncertainty in the number of compact remnant companions.

  Regardless of how we account for selection effects, binary evolution, and contamination by compact remnants, the slope of the mass-ratio distribution across $q$ = 0.1\,-\,0.3 must be {\large $\gamma$}$_{\rm smallq}$~$>$~$-$1.4 at the 2$\sigma$ confidence level.  Although some observations of early-type intermediate-period binaries indicate the power-law component of the mass-ratio distribution across $q$~=~0.3\,-\,1.0 is consistent with random pairings drawn from a Salpeter IMF ({\large $\gamma$}$_{\rm largeq}$~$\approx$~$-$2.35), this steep slope cannot continue all the way down to $q$~=~0.1.  Instead, the mass-ratio distribution must flatten and possibly turnover below $q$ $<$ 0.3.  As a whole, the population of early-type binaries with intermediate orbital periods is inconsistent with random pairings drawn from the IMF.  

  \begin{figure}[t!]
\centerline{
\includegraphics[trim=0.5cm 0.1cm 0.5cm 0.2cm, clip=true, width=3.4in]{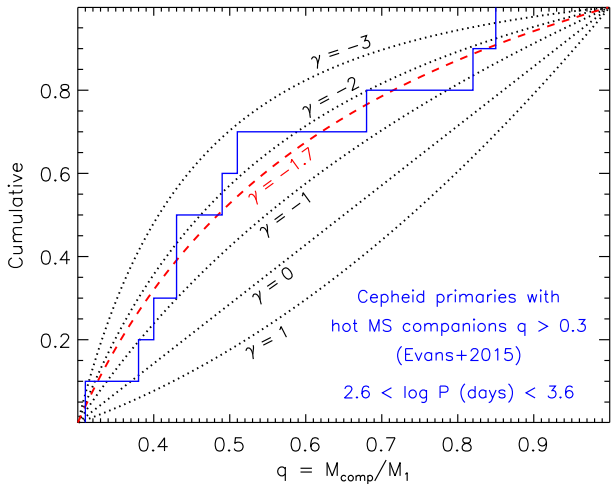}}
\caption{Similar to Fig.~15, but for the 10 companions (blue) to Cepheids with $q$ $>$ 0.3 and intermediate orbital periods 2.6~$<$~log\,$P$\,(days)~$<$~3.6 \citep{Evans2015}. We measure {\large $\gamma$}$_{\rm largeq}$~=~$-$1.7\,$\pm$\,0.5 across $q$~=~0.3\,-\,1.0 (dashed red), which is slightly larger than but consistent with our measurement at long orbital periods ({\large $\gamma$}$_{\rm largeq}$~=~$-$2.1\,$\pm$\,0.5).}
\end{figure}

Finally, we determine the intrinsic companion frequency $f_{\rm logP;q>0.3}$ based on observations of Cepheids with MS companions at intermediate separations. According to Table~6 in \citet{Evans2015}, ${\cal N}_{\rm Cepheid}$ = 31 Cepheids brighter than $V$ $<$ 8.0 mag were extensively monitored for spectroscopic radial velocity variations.  Of these primaries, ${\cal N}_{\rm largeq}$~=~6 have companions with orbital periods $P$~=~1\,-\,10~years (2.6~$<$~log\,$P$\,(days)~$<$~3.6) and with UV excesses that demonstrate $q$~$>$~0.3.  We again account for the fraction ${\cal F}_{\rm Cepheid}$~=~0.75\,$\pm$\,0.15 of mid-B MS primaries that evolve into Cepheids and for the correction factor ${\cal C}_{\rm evol}$~=~1.2\,$\pm$\,0.1 due to evolved secondaries that are observed as Cepheids. We measure $f_{\rm logP;q>0.3}$ = ${\cal N}_{\rm largeq} {\cal F}_{\rm Cepheid} {\cal C}_{\rm evol}$/${\cal N}_{\rm Cepheid}$/$\Delta$log$P$ = (6\,$\pm$\,$\sqrt{6}$)(0.75\,$\pm$\,0.15)(1.2\,$\pm$\,0.1)/31/(3.6\,$-$\,2.6) = 0.17\,$\pm$\,0.08 for mid-B primaries and intermediate orbital periods (Table~7).  This is higher than the measurements $f_{\rm logP;q>0.3}$~=~0.04\,-\,0.14 at shorter (\S3\,-\,4) and longer (\S6.1 and \S7) orbital periods, indicating the period distribution of mid-B binaries peaks at 3~$\lesssim$~log\,$P$\,(days)~$\lesssim$~5 (see also \S9).

\section{Visual Binaries}

Because O-type and B-type MS stars have low space densities, we must study these primaries over large distances $d$ $\gtrsim$ 100 pc to achieve an adequate sample size.  Companions to early-type stars can therefore be visually resolved only at larger orbital separations $a$ $\gtrsim$ 20\,AU, i.e., $P$~$\gtrsim$~10$^4$~days, even with speckle interferometry \citep{Mason1998,Preibisch1999,Mason2009}, adaptive optics \citep{Duchene2001,Shatsky2002,Sana2014}, lucky imaging \citep{Peter2012}, and space-based observations \citep{Caballero2014,Aldoretta2015}.   MS binaries with large brightness contrasts $\Delta$m~$\gtrsim$~4~mag, and therefore small mass ratios $q$~=~$M_{\rm comp}$/$M_1$~$\lesssim$~0.3, can only be detected at even longer orbital periods $P$~$\gtrsim$~10$^5$~days (see purple region in our Fig.~1).   At wide separations $a$~$\gtrsim$~1,000 AU, i.e., angular separations $\rho$~$\gtrsim$~5$''$, confusion with background and foreground stars becomes non-negligible.  Continuous astrometric observations can help confirm that wide visual binaries (VBs) are gravitationally bound according to their common proper motion (CPM; orange region in our Fig.~1).  However, it is also possible that two young individual stars are only loosely associated because they recently formed in the same cluster \citep{Abt2000}.  It is difficult to select a window of angular separations that is complete toward low-mass companions while simultaneously not significantly biased toward optical doubles.  
  
In addition to CPM, the spectral energy distributions of VBs can help confirm their physical association.    For example, \citet{Duchene2001} and \citet{Shatsky2002} utilized the observed near-infrared colors to differentiate optical doubles from physically associated pairs that share the same age, distance, and dust reddening.  As another example, late-B MS stars are typically X-ray quiet, while young and magnetically active G-M MS and pre-MS stars can emit X-rays \citep{Evans2011}.  Late-B MS stars that appear to be X-ray bright probably have young low-mass companions with $q$ $\approx$ 0.05\,-\,0.40 \citep{Hubrig2001,Stelzer2003,Evans2011}.  Indeed, (43\,$\pm$\,6)\% of X-ray bright late-B and early-A MS stars were resolved with adaptive optics to have low-mass companions at angular separations $\rho$~=~0.3$''$\,-\,26$''$ \citep{DeRosa2011}, i.e.  4.7~$\lesssim$~log\,$P$\,(days)~$\lesssim$~7.4 (aqua region in our Fig.~1).  The remaining $\approx$57\% of the X-ray bright late-B/early-A stars most likely contain low-mass companions with log~$P$~$<$~4.7 that cannot be spatially resolved.  Unfortunately, the precise orbital periods of these putative extreme mass-ratio binaries have not yet been measured.  We therefore do not know the intrinsic frequency of low-mass, X-ray emitting companions to late-B stars as a function of orbital period.  

In the following, we examine the statistics of VB companions to A-type, B-type, and O-type stars.  We avoid the separation-contrast bias by analyzing only the systems with sufficiently wide separations such that the observations are complete down to $q$ = 0.3 or, if possible, $q$ = 0.1.  For B-type and O-type binaries, we consider only companions with separations $\rho$~$\lesssim$~2$''$\,-\,6$''$, depending on the survey, to ensure with high probability ${\cal P}$ $>$ 95\% that the binary components are gravitationally bound.  In this manner, we can directly measure $f_{\rm logP;q>0.3}$, {\large $\gamma$}$_{\rm largeq}$, ${\cal F}_{\rm twin}$, and, if possible, {\large $\gamma$}$_{\rm smallq}$  without having to correct for incompleteness or selection effects.

\subsection{A-type Primaries}

To fill in the gap between solar-type primaries ($\langle M_1 \rangle$~=~1.0\,$\pm$\,0.2\,\Msun; see \S8) and B-type primaries ($M_1$~=~3\,-\,16\,\Msun), we examine the \citet{DeRosa2014} VB survey of A-type primaries ($\langle M_1 \rangle$~=~2.0\,$\pm$\,0.3\,\Msun).  They searched ${\cal N}_{\rm prim,AO}$ = 363 A-type stars within 75~pc for visual companions utilizing adaptive optics in the near-infrared H and K bands. \citet{DeRosa2014} also report the mass ratios $q$ of all their detected VBs according to the observed brightness contrasts.  Their adaptive optics survey is complete down to $q$ = 0.3 ($\Delta$K~$\lesssim$~4.0~mag) for angular separations $\rho$~$\gtrsim$~0.3$''$, and complete down to $q$~=~0.1 ($\Delta$K~$\lesssim$~6.5~mag) beyond $\rho$~$\gtrsim$~0.9$''$ (see their Figs.~7-8).  At wide separations $\rho$~$>$~8$''$, there is a $>$5\% probability that VBs with $\Delta$K~$\approx$~6~mag are optical doubles based on the observed background stellar densities \citep{DeRosa2014}. At even wider separations $\rho$~$>$~15$''$, the adaptive optics survey is incomplete given the limited field of view.

\citet{DeRosa2014} also analyzed ${\cal N}_{\rm prim,CPM}$~=~228 A-type stars in digitized photographic plates in search for astrometric CPM binaries. They utilized stellar isochrones and infrared colors to confirm that CPM pairs share the same distance and dust reddening.  A significant majority of A-type stars in the \citet{DeRosa2014} sample have ages $\tau$ $>$ 100 Myr (see their Fig.~3).    Unlike CPM companions to young B-type and O-type primaries, which may be spurious associations  \citep{Abt2000}, CPM companions to systematically older A-type stars are most likely gravitationally bound and dynamically stable.  

Based on the observational selection effects, we choose three subsamples from the  \citet{DeRosa2014} adaptive optics survey and one subsample from their CPM survey.   All four of these subsamples are relatively complete and unbiased within the specified mass-ratio and separation intervals (see Table~8). In the following, we measure $f_{\rm logP;q>0.3}$, {\large $\gamma$}$_{\rm largeq}$, ${\cal F}_{\rm twin}$, and, if possible, {\large $\gamma$}$_{\rm smallq}$ for each of these four subsamples.

We first select the ${\cal N}_{\rm largeq}$~=~12 VBs in the \citet{DeRosa2014} adaptive optics survey with $q$~$>$~0.3 and small angular separations $\rho$~=~0.3$''$\,-\,0.9$''$ ($\langle a_{\rm proj}\rangle$~$\approx$~25~AU; log\,$P$~$\approx$~4.3\,-\,4.9). The A-type stars are relatively old ($\tau$ $\gtrsim$ 100 Myr; \citealt{DeRosa2014}), and so (20\,$\pm$\,10)\% are likely to contain compact remnant companions (see \S11).  We account for this correction factor ${\cal C}_{\rm evol}$~=~1.2\,$\pm$\,0.1 due to stellar evolution within binary systems.  The intrinsic frequency of companions with $q$ $>$ 0.3 per decade of orbital period is $f_{\rm logP;q>0.3}$ = ${\cal N}_{\rm largeq} {\cal C}_{\rm evol}$/${\cal N}_{\rm prim,AO}$/$\Delta$log$P$ = (12\,$\pm$\,$\sqrt{12}$)(1.2\,$\pm$\,0.1)/363/(4.9\,$-$\,4.3) = 0.07\,$\pm$\,0.02 (Table~8).  

\renewcommand{\arraystretch}{1.6}
\setlength{\tabcolsep}{3pt}
\begin{figure}[t!]\footnotesize
{\small {\bf Table 8:} Companion statistics based on visually resolved companions to A-type MS stars ($\langle M_1 \rangle$ = 2.0\,$\pm$\,0.3\,\Msun) in the \citet{DeRosa2014} adaptive optics and common proper motion surveys.}  \\
\vspace*{-0.45cm}
\begin{center}
\begin{tabular}{|c|c|c|c|c|}
\hline
 Period Interval & $f_{\rm logP;q>0.3}$ & ${\cal F}_{\rm twin}$ & {\large $\gamma$}$_{\rm largeq}$ & {\large $\gamma$}$_{\rm smallq}$ \\
\hline
 4.3\,$<$\,log\,$P$\,$<$\,4.9 & 0.07\,$\pm$\,0.02 & 0.05\,$\pm$\,0.05 & $-$1.2\,$\pm$\,0.5 & - \\
\hline
 4.9\,$<$\,log\,$P$\,$<$\,6.3 & 0.06\,$\pm$\,0.01 & 0.07\,$\pm$\,0.04 & $-$2.1\,$\pm$\,0.4 & $-$0.8\,$\pm$\,0.4 \\
\hline
 6.3\,$<$\,log\,$P$\,$<$\,6.7 & 0.04\,$\pm$\,0.02 &    $<$\,0.11      & $-$2.5\,$\pm$\,0.7 & - \\
\hline
 7.2\,$<$\,log\,$P$\,$<$\,8.2 & 0.03\,$\pm$\,0.01 &    $<$\,0.10      & $-$2.2\,$\pm$\,0.6 & $-$1.1\,$\pm$\,0.5 \\
\hline
\end{tabular}
\end{center}
\end{figure}
\renewcommand{\arraystretch}{1.0}

\begin{figure}[t!]
\centerline{
\includegraphics[trim=0.5cm 0.1cm 0.5cm 0.1cm, clip=true, width=3.5in]{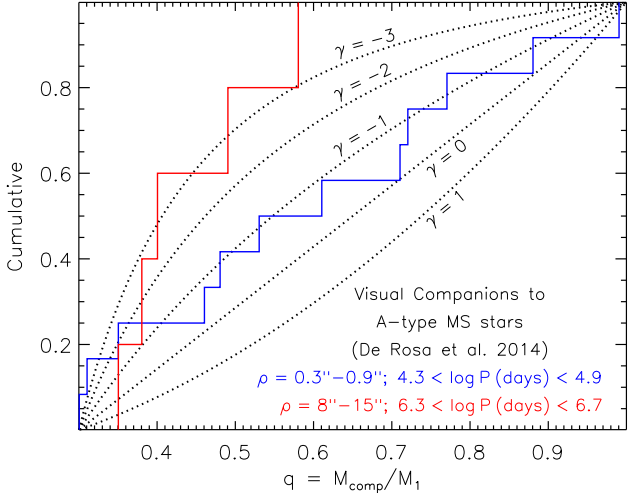}}
\caption{Cumulative distribution of mass ratios $q$ $>$ 0.3 for visually resolved companions to A-type MS stars identified in the \citet{DeRosa2014} adaptive optics survey.  We compare the 12 VBs with $q$ $>$ 0.3 and projected separations $\rho$ = 0.3$''$\,-\,0.9$''$ (blue) to the 5 VBs with $q$ $>$ 0.3 and  $\rho$ = 8$''$\,-\,15$''$ (red).  Both subsamples are relatively complete and unbiased, where we find that wider VBs favor smaller mass ratios.  Quantitatively,  we measure an excess twin fraction ${\cal  F}_{\rm twin}$~$\approx$~0.05 and power-law slope {\large $\gamma$}$_{\rm largeq}$~$\approx$~$-$1.2 for the VBs with shorter separations, and measure ${\cal  F}_{\rm twin}$~$\approx$~0.00 and {\large $\gamma$}$_{\rm largeq}$~$\approx$~$-$2.5 for the wider VB subsample.}
\end{figure}

We display the cumulative distribution of mass ratios $q$ for these 12 VBs in Fig.~17 (blue line). By fitting a single power-law across $q$~=~0.3\,-\,1.0, we measure $\gamma$~=~$-$0.9\,$\pm$\,0.4.  This is consistent with the value $\gamma$~=~$-$0.5 reported by \citet{DeRosa2014} for their population of close VBs with $a_{\rm proj}$~=~30\,-\,125~AU. One of the 12 VBs in our subsample has $q$~=~0.99, so there may be an excess fraction of twins.  Allowing for a non-zero excess twin fraction, we measure ${\cal F}_{\rm twin}$~=~0.05\,$\pm$\,0.05 and {\large $\gamma$}$_{\rm largeq}$~=~$-$1.2\,$\pm$\,0.5 (Table~8).

Second, we analyze the 54 VBs in the \citet{DeRosa2014} adaptive optics survey with $q$ $>$ 0.1 and separations $\rho$ = 0.9$''$\,-\,8.0$''$ ($\langle a_{\rm proj}\rangle$~$\approx$~150~AU; log\,$P$~$\approx$~4.9\,-\,6.3). Of these 54 VBs,   ${\cal N}_{\rm largeq}$~=~24 have large mass ratios $q$~=~0.3\,-\,1.0.  The intrinsic companion frequency at these slightly wider separations is $f_{\rm logP;q>0.3}$ = ${\cal N}_{\rm largeq} {\cal C}_{\rm evol}$/${\cal N}_{\rm prim,AO}$/$\Delta$log$P$ = (24\,$\pm$\,$\sqrt{24}$)(1.2\,$\pm$\,0.1)/363/(6.3\,$-$\,4.9) = 0.06\,$\pm$\,0.01 (Table~8).

In Fig.~18, we display the cumulative distribution of mass ratios $q$~=~0.1\,-\,1.0 for the 54 VBs.  Given such a large sample size, we can clearly see that a single power-law component $\gamma$ across all mass ratios $q$~=~0.1\,-\,1.0 does not adequately describe the data.  For instance, there is a small but statistically significant excess twin fraction.  Of the 24 VBs with $q$~$>$~0.3, two have $q$~$=$~0.99\,-\,1.00, implying ${\cal F}_{\rm twin}$~$\approx$~2/24~$\approx$~0.08.  The \citet{DeRosa2014} sample is volume limited within 75~pc, so this excess twin fraction cannot be due to the \"{O}pik effect / Malmquist bias.  Even after considering the twin systems with $q$~$>$~0.95, a single power-law does not fit the data across $q$ = 0.10\,-\,0.95.  For large mass ratios $q$~$\gtrsim$~0.3, the data favor a steeper slope of $\gamma$~$\approx$~$-$2 (see Fig.~18).  Meanwhile, the slope across small mass ratios $q$~$\approx$~0.1\,-\,0.3 trends toward  $\gamma$~$\approx$~$-$1.  Fitting our three-parameter model to the data, we measure an excess twin fraction ${\cal F}_{\rm twin}$ = 0.07\,$\pm$\,0.04, a slope {\large $\gamma$}$_{\rm largeq}$~=~$-$2.1\,$\pm$\,0.4 across large mass ratios $q$~=~0.3\,-\,1.0, and a slope {\large $\gamma$}$_{\rm smallq}$~=~$-$0.8\,$\pm$\,0.4 across small ratios $q$~=~0.1\,-\,0.3 (Table~8; dashed red line in Fig.~18).

\begin{figure}[t!]
\centerline{
\includegraphics[trim=0.5cm 0.1cm 0.5cm 0.1cm, clip=true, width=3.55in]{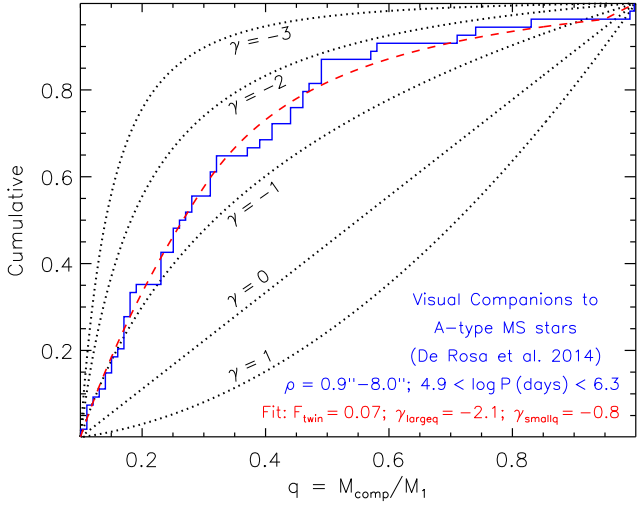}}
\caption{Cumulative distribution of mass ratios $q$ for the 54 VBs in the \citet{DeRosa2014} adaptive optics survey with $q$~$>$~0.1 and angular separations $\rho$~=~0.9$''$\,-\,8.0$''$ (blue).  The mass-ratio distribution of this large and relatively complete subsample cannot be adequately described by a single power-law $p_q$ $\propto$ $q^{\gamma}$ across 0.1~$<$~$q$~$<$~1.0.  We instead fit our three-parameter model (dashed red), where we measure an excess twin fraction of ${\cal F}_{\rm twin}$~=~0.07\,$\pm$\,0.04, a power-law slope {\large $\gamma$}$_{\rm largeq}$~=~$-$2.1\,$\pm$\,0.4 across large mass ratios $q$~$>$~0.3, and a power-law slope {\large $\gamma$}$_{\rm smallq}$~=~$-$0.8\,$\pm$\,0.4 across small ratios 0.1~$<$~$q$~$<$~0.3.}
\end{figure}

Third, we examine the ${\cal N}_{\rm largeq}$ = 5 VBs in the \citet{DeRosa2014} adaptive optics survey with $q$ $>$ 0.3 and wide separations $\rho$ = 8.0$''$\,-\,15.0$''$ ($\langle a_{\rm proj}\rangle$~$\approx$~600~AU; log\,$P$~$\approx$~6.3\,-\,6.7). We measure a frequency of companions with $q$ $>$ 0.3 per decade of orbital period of  $f_{\rm logP;q>0.3}$ = ${\cal N}_{\rm largeq} {\cal C}_{\rm evol}$/${\cal N}_{\rm prim,AO}$/$\Delta$log$P$ = (5\,$\pm$\,$\sqrt{5}$)(1.2\,$\pm$\,0.1)/363/(6.7\,$-$\,6.3) = 0.04\,$\pm$\,0.02 (Table~8). In Fig.~17, we display the cumulative distribution of mass ratios for these 5 VBs. The 5 A-type VBs with $q$~$>$~0.3 and wide separations $\rho$~=~8$''$\,-\,15$'$ (red line in Fig.~17) are weighted toward smaller mass ratios compared to the 12 A-type VBs with $q$~$>$~0.3 and smaller separations $\rho$~=~0.3$''$\,-\,0.9$''$ (blue line in Fig.~17).  For the 5 VBs with wider separations, we measure {\large $\gamma$}$_{\rm largeq}$~=~$-$2.5\,$\pm$\,0.7 and ${\cal F}_{\rm twin}$~$<$~0.11 (Table~8).  This is consistent with the value $\gamma$ = $-$2.3 reported by \citet{DeRosa2014} for their sample of wide VBs with $a_{\rm proj}$ = 125\,-\,800 AU.  

Finally, we choose the 11 VBs in the \citet{DeRosa2014} CPM survey with $q$ $>$ 0.1 and projected separations 2,000\,-\,8,000~AU (log\,$P$~$\approx$~7.2\,-\,8.2).  Of these systems, ${\cal N}_{\rm largeq}$ = 5 have large mass ratios $q$ $>$ 0.3.  The intrinsic companion frequency is  $f_{\rm logP;q>0.3}$ = ${\cal N}_{\rm largeq} {\cal C}_{\rm evol}$/${\cal N}_{\rm prim,CPM}$/$\Delta$log$P$ = (5\,$\pm$\,$\sqrt{5}$)(1.2\,$\pm$\,0.1)/228/(8.2\,$-$\,7.2) = 0.03\,$\pm$\,0.01 (Table~8). As expected, the frequency of companions per decade of orbital period gradually decreases toward the widest separations.

\begin{figure}[t!]
\centerline{
\includegraphics[trim=0.5cm 0.1cm 0.5cm 0.1cm, clip=true, width=3.55in]{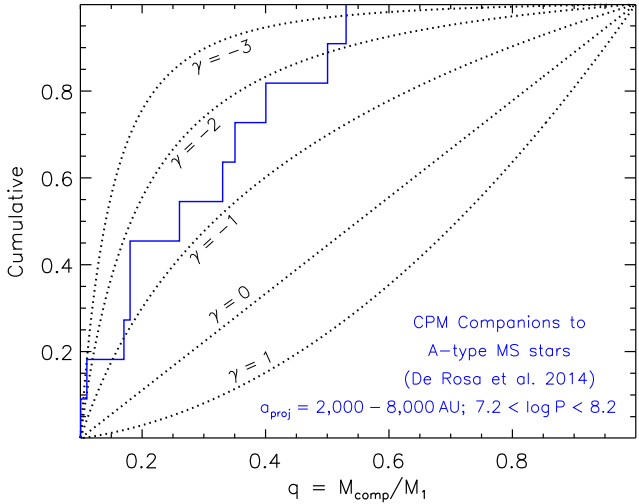}}
\caption{Cumulative distribution of mass ratios $q$ for the 11 VBs in the \citet{DeRosa2014} CPM survey with $q$~$>$~0.1 and projected separations $a_{\rm proj}$~=~2,000\,-\,8,000\,AU (blue). Similar to the VBs found at shorter separations (see Fig.~18), we measure a two-component power-law distribution with slopes {\large $\gamma$}$_{\rm largeq}$~=~$-$2.2\,$\pm$\,0.6 across large mass ratios $q$~$>$~0.3 and {\large $\gamma$}$_{\rm smallq}$~=~$-$1.1\,$\pm$\,0.5 across 0.1~$<$~$q$~$<$~0.3.}
\end{figure}

We display the cumulative distribution of mass ratios for these 11 VBs in Fig.~19.  We measure an excess twin fraction consistent with zero, i.e., ${\cal F}_{\rm twin}$~$<$~0.10 at the 1$\sigma$ confidence level.  We again find that the power-law slope $\gamma$ is steeper across larger mass ratios and then flattens toward $q$ = 0.1.  Statistically, we measure {\large $\gamma$}$_{\rm largeq}$~=~$-$2.2\,$\pm$\,0.6 and {\large $\gamma$}$_{\rm smallq}$~=~$-$1.1\,$\pm$\,0.5 (Table~8).

\subsection{Late-B Primaries}

For a sample of ${\cal N}_{\rm prim}$ = 115 B-type stars in the Sco~OB2 association ($d$ $\approx$ 145 pc), \citet{Shatsky2002} utilized near-infrared adaptive optics to search for visual companions across $\rho$~=~0.3$''$\,-\,6.4$''$, i.e., $a$~=~45\,-\,900~AU.  Their sample of B0-B9 stars is weighted toward lower primary masses according to the IMF, and so the average primary mass is $\langle M_1 \rangle$~=~5\,$\pm$\,2\,\Msun.  \citet{Shatsky2002} measured the mass ratios $q$ from the observed infrared colors and brightness contrasts.  Adaptive optics are sensitive to $q$ = 0.1 companions for angular separations $\rho$~$>$~0.5$''$ while incompleteness and observational biases become important beyond $\rho$~$>$~4$''$ \citep[][see their Fig.~8]{Shatsky2002}.  We therefore select the ${\cal N}_{\rm comp}$~=~18 companions with angular separations $\rho$~=~0.5$''$\,-\,4.0$''$ and measured mass ratios $q$~$>$~0.1.  These 18 companions  represent a relatively complete subsample across separations $a$~=~70\,-\,600~AU, i.e., 4.9~$<$~log\,$P$\,(days)~$<$ 6.3.

We display in Fig.~20 the cumulative distribution of mass ratios for the 18 companions.  The excess twin fraction is consistent with zero, i.e., the 1$\sigma$ upper limit is ${\cal F}_{\rm twin}$ $<$ 0.05 (Table~9).  If we fit the power-law slope of the mass-ratio distribution across the large range 0.1~$<$~$q$~$<$~1.0, we measure  {\large $\gamma$}~=~$-$0.8\,$\pm$\,0.3.  Our measurement is slightly steeper than the slope {\large $\gamma$}~=~$-$0.5 reported by \citet{Shatsky2002}.  This is primarily because \citet{Shatsky2002} fit the mass-ratio distribution across the entire interval 0.0~$<$~$q$~$<$~1.0.   Like the IMF, the mass-ratio probability distribution $p_q$ cannot be described by a single power-law across all mass ratios 0~$<$~$q$~$<$~1. As we have parameterized in the present study (see \S2), the mass-ratio distribution is more accurately described by a broken power-law distribution down to some threshold $q$~$\approx$~0.1, below which the observations are insensitive and/or incomplete.  Allowing for a break in the mass-ratio distribution at $q$ = 0.3, we measure {\large $\gamma$}$_{\rm smallq}$~=~$-$0.7\,$\pm$\,0.5 across $q$~=~0.1\,-\,0.3 and {\large $\gamma$}$_{\rm largeq}$~=~$-$1.0\,$\pm$\,0.5 across $q$~=~0.3\,-\,1.0 (Table~9).

\begin{figure}[t!]
\centerline{
\includegraphics[trim=0.5cm -0.1cm 0.5cm -0.2cm, clip=true, width=3.6in]{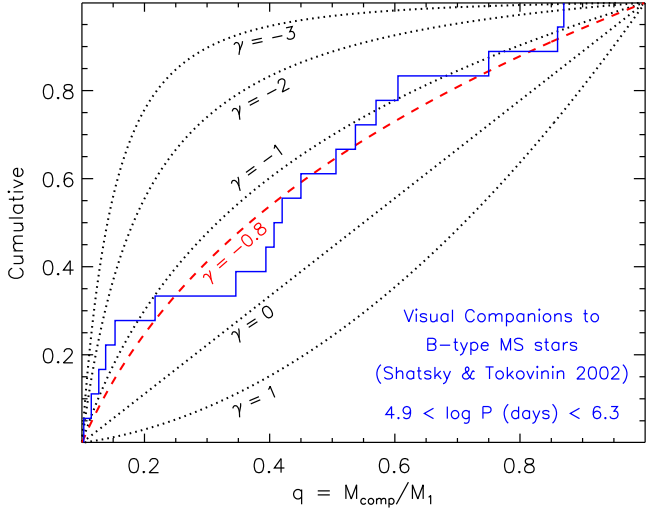}}
\caption{Cumulative distribution of mass ratios $q$ $>$ 0.1 for the 18 visual companions (blue) to 115 B-type MS primaries ($\langle M_1 \rangle$~=~5\,$\pm$\,2\,\Msun) resolved at angular separations 0.5$''$\,-\,4.0$''$  with adaptive optics \citep{Shatsky2002}. For this relatively complete subsample, we measure the power-law component of the mass-ratio probability distribution $p_q$~$\propto$~$q^{\gamma}$ to be {\large $\gamma$}~=~$-$0.8\,$\pm$\,0.3 for 0.1 $<$ $q$ $<$ 1.0 (dashed red).}
\end{figure}

Of the 18 companions, ${\cal N}_{\rm largeq}$ = 12 have large mass ratios $q$ $>$ 0.3.  The subgroups within the Sco-OB2 assocation have a range of ages $\tau$~=~4\,-\,15~Myr \citep{deZeeuw1999,Shatsky2002}, so we account for the correction factor ${\cal C}_{\rm evol}$ = 1.2\,$\pm$\,0.1 due to binary evolution.  The intrinsic frequency of companions with $q$~$>$~0.3 per decade of orbital period is $f_{\rm logP;q>0.3}$ = ${\cal N}_{\rm largeq}{\cal C}_{\rm evol}$/${\cal N}_{\rm prim}$/$\Delta$log$P$ = (12\,$\pm$\,$\sqrt{12}$)(1.2\,$\pm$\,0.1)/115/(6.3\,$-$\,4.9) = 0.09\,$\pm$\,0.03 for late-B MS stars and wide orbital periods (Table~9).

\subsection{Mid-B Primaries}

For a sample of ${\cal N}_{\rm prim}$ = 109 B2-B5 primaries ($\langle M_1 \rangle$~=~7\,$\pm$\,2\,\Msun), \citet{Abt1990} identified 49 VB companions.  These 49 VBs exhibit CPM, have orbital solutions, and/or have sufficiently small angular separations $\rho$~$\lesssim$~5$''$ to ensure the systems are physically associated.  Their sample is relatively complete down to $\Delta$V~$<$~7.0~mag ($q$~$\gtrsim$~0.13) for angular separations $>$0.65$''$ (see their Table 5).  We therefore select the 10 VBs from \citet{Abt1990} with listed brightness constrasts $\Delta$V~$\le$~7.0~mag and angular separations $\rho$~=~0.65$''$\,-\,6.5$''$, i.e., 5.4~$\lesssim$~log\,$P$\,(days)~$\lesssim$~6.9. Beyond $\rho$~$>$~6.5$''$,  the binaries may be spurious associations or dynamicaly unstable, even if they exhibit CPM \citep{Abt2000,Shatsky2002,Duchene2013}. 

To estimate the mass ratios $q$ of these 10 VBs from the observed brightness contrasts $\Delta$V, we utilize the empirical relations between spectral type, bolometric corrections, absolute magnitudes, and masses provided in \citet[][and references therein]{Pecaut2013}.  We fit the following relation:

\begin{align}
  {\rm log}~q~=~&-0.155\Delta{\rm V}~~{\rm for}~0.0\,<\,\Delta{\rm V}\,({\rm mag})\,\le \,4.0, \nonumber \\
          -0.32 &-0.075\Delta{\rm V}~~{\rm for}~4.0\,<\,\Delta{\rm V}\,({\rm mag})\,<\,8.0,
\end{align}

\noindent applicable only for B2-B5 MS primaries with $M_1$~$\approx$~5\,-\,9\,\Msun.  The break in the equation above models how the slope of the mass-luminosity relation for K-A type secondaries with $M_2$ $\approx$ 0.5\,-\,2\,\Msun\ is steeper than the
 slope of the mass-luminosity relation for late/mid B-type secondaries with $M_2$~$\approx$~3\,-\,9\,\Msun\ (see also \S5).  Depending on the precise age and spectral type of the primary, the uncertainty in the mass ratio according to Eqn.~4 is $\delta$q~$\approx$~0.04.  

The 10 VBs we selected from \citet{Abt1990} have brightness contrasts $\Delta$V = 1.3\,-\,7.0 mag, which correspond to mass ratios $q$ = 0.14\,-\,0.63 according to Eqn.~4.  In Fig.~21, we display the cumulative distribution of mass ratios for these 10 VBs.  The excess twin fraction is consistent with zero, i.e., we measure ${\cal F}_{\rm twin}$~$<$~0.12 at the 1$\sigma$ confidence level (Table~9).  Assuming the sample is complete down to $q$~=~0.13, we measure  {\large $\gamma$}~=~$-$2.0\,$\pm$\,0.4 across the entire interval $q$~=~0.13\,-\,1.00.  This indicates wide binaries with mid-B primaries have mass ratios $q$~$\gtrsim$~0.13 that are consistent with random pairings drawn from a Salpeter IMF ({\large $\gamma$} = $-$2.35).   \citet{Abt1990} examined all visual and common proper motion binaries across 5~$\lesssim$~log\,$P$\,(days)~$\lesssim$~9 in their sample, and also concluded that the mass ratios are consistent with random pairings from a Salpeter IMF.  The widest systems, however,  may be contaminated by faint spurious associations.  Nevertheless, we have shown that {\large $\gamma$} $\approx$ $-$2.0\,$\pm$\,0.4 applies for relatively close VBs (5.4~$<$~log\,$P$~$<$~6.9) that are most probably gravitationally bound and dynamically stable.

\begin{figure}[t!]
\centerline{
\includegraphics[trim=0.5cm -0.2cm 0.5cm -0.3cm, clip=true, width=3.6in]{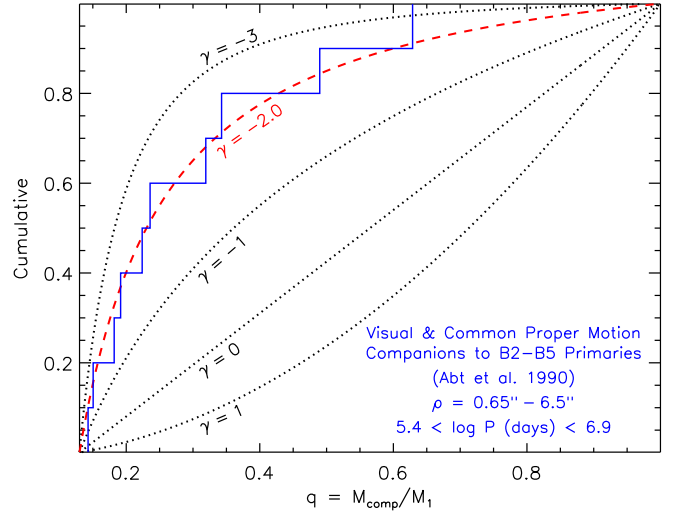}}
\caption{Cumulative distribution of mass ratios $q$ $>$ 0.13 for the 10 visually resolved companions (blue) to 109 B2\,-\,B5 MS primaries ($\langle M_1 \rangle$~=~7\,$\pm$\,2\,\Msun) with angular separations 0.65$''$\,-\,6.5$''$  \citep{Abt1990}. We measure the power-law component of the mass-ratio distribution to be {\large $\gamma$}~=~$-$2.0\,$\pm$\,0.4 (dashed red).  For a broad range of mass ratios 0.1~$\lesssim$~$q$~$<$~1.0, companions to mid-B MS stars with long orbital periods have a mass-ratio distribution consistent with random pairings drawn from a Salpeter IMF ({\large $\gamma$}~=~$-$2.35).}
\end{figure}

Allowing for a break in the mass-ratio distribution at $q$ = 0.3, we measure {\large $\gamma$}$_{\rm smallq}$~$\approx$~$-$1.3 and {\large $\gamma$}$_{\rm largeq}$~$\approx$~$-$2.7.  Given the small sample size, the uncertainties in these two statistics are quite large and significantly correlated.  It is physically unlikely that either of these slopes is steeper than the the value {\large $\gamma$} = $-$2.35 implied by random pairings from a Salpeter IMF.  If we impose the additional constraint that {\large $\gamma$}~$>$~$-$2.5, then we measure {\large $\gamma$}$_{\rm smallq}$~$=$~$-$1.7\,$\pm$\,0.6 and {\large $\gamma$}$_{\rm largeq}$~$=$~$-$2.2\,$\pm$\,0.6 (Table~9).  

Of the 10 VBs, ${\cal N}_{\rm largeq}$~=~4 have mass ratios $q$~$>$~0.3.  As done in \S3, we account for the correction factor ${\cal C}_{\rm evol}$ = 1.2\,$\pm$\,0.1 due to binary evolution within the volume-limited sample of \citet{Abt1990}.  For mid-B primaries and wide separations, the intrinsic frequency of companions with $q$ $>$ 0.3 per decade of orbital period is $f_{\rm logP;q>0.3}$ = ${\cal N}_{\rm largeq} {\cal C}_{\rm evol}$/${\cal N}_{\rm prim}$/$\Delta$log$P$ = (4\,$\pm$\,$\sqrt{4}$)(1.2\,$\pm$\,0.1)/109/(6.9\,$-$\,5.4) = 0.03\,$\pm$\,0.02 (Table~9).  

The binary statistics for wide companions to mid-B primaries ({\large $\gamma$}$_{\rm largeq}$~$=$~$-$2.2\,$\pm$\,0.6, ${\cal F}_{\rm twin}$ $<$ 0.12, $f_{\rm logP;q>0.3}$ = 0.03\,$\pm$\,0.02) nearly matches the statistics we measured in \S6.1 for wide companions to Cepheids ({\large $\gamma$}$_{\rm largeq}$~$=$~$-$2.1\,$\pm$\,0.5, ${\cal F}_{\rm twin}$ $<$ 0.07, $f_{\rm logP;q>0.3}$ = 0.04\,$\pm$\,0.02).  Recall that Cepheids evolve from mid-B primaries with $\langle M_1 \rangle$~=~6\,$\pm$\,2\,\Msun.  The similarity in the statistical parameters  validates our ability to measure the properties of mid-B MS binaries with intermediate and wide separations based on observations of Cepheids.

\renewcommand{\arraystretch}{1.6}
\setlength{\tabcolsep}{10pt}
\begin{figure}[t!]\footnotesize
{\small {\bf Table 9:} Companion statistics based on visually resolved companions to B-type and O-type MS stars.}  \\
\vspace*{-0.45cm}
\begin{center}
\begin{tabular}{|c|c|}
\hline
 Reference, Primary Mass,  &  \\
 \& Period Interval & Statistic \\
\hline
 \citet{Shatsky2002}; & {\large $\gamma$}$_{\rm smallq}$  = $-$0.7\,$\pm$\,0.5 \\
\cline{2-2}
  $\langle M_1 \rangle$ = 5\,$\pm$\,2\,\Msun\,;  & {\large $\gamma$}$_{\rm largeq}$  = $-$1.0\,$\pm$\,0.5  \\
\cline{2-2}
 log\,$P$\,(days) = 5.6\,$\pm$\,0.7   & ${\cal F}_{\rm twin}$ $<$ 0.05 \\
\cline{2-2}
  & $f_{\rm logP;q>0.3}$ = 0.09\,$\pm$\,0.03 \\
\hline
 \citet{Abt1990}; & {\large $\gamma$}$_{\rm smallq}$ = $-$1.7\,$\pm$\,0.6 \\
\cline{2-2}
  $\langle M_1 \rangle$ = 7\,$\pm$\,2\,\Msun\,;  & {\large $\gamma$}$_{\rm largeq}$ = $-$2.2\,$\pm$\,0.6 \\
\cline{2-2}
 log\,$P$\,(days) = 6.15\,$\pm$\,0.75   & ${\cal F}_{\rm twin}$ $<$ 0.12 \\
\cline{2-2}
 & $f_{\rm logP;q>0.3}$ = 0.03\,$\pm$\,0.02 \\
\hline
\citet{Duchene2001} and & {\large $\gamma$}$_{\rm smallq}$ = $-$1.3\,$\pm$\,0.5 \\
\cline{2-2}
  \citet{Peter2012}    & {\large $\gamma$}$_{\rm largeq}$ = $-$1.9\,$\pm$\,0.5 \\
\cline{2-2}
  $\langle M_1 \rangle$ = 12\,$\pm$\,3\,\Msun\,;  & ${\cal F}_{\rm twin}$ $<$ 0.10 \\
\cline{2-2}
 log\,$P$\,(days) = 6.4\,$\pm$\,0.7    & $f_{\rm logP;q>0.3}$ = 0.08\,$\pm$\,0.04 \\
\hline
\citet{Sana2014}; & {\large $\gamma$}$_{\rm smallq}$ = $-$1.5\,$\pm$\,0.5 \\
\cline{2-2}
  $\langle M_1 \rangle$ = 28\,$\pm$\,8\,\Msun\,;  & {\large $\gamma$}$_{\rm largeq}$ = $-$2.1\,$\pm$\,0.5 \\
\cline{2-2}
 log\,$P$\,(days) = 6.6\,$\pm$\,0.4   & ${\cal F}_{\rm twin}$ $<$ 0.10 \\
\cline{2-2}
 & $f_{\rm logP;q>0.3}$ = 0.07\,$\pm$\,0.03 \\
\hline
\citet{Aldoretta2015}; & {\large $\gamma$}$_{\rm largeq}$ = $-$1.8\,$\pm$\,0.4 \\
\cline{2-2}
  $\langle M_1 \rangle$ = 28\,$\pm$\,8\,\Msun\,;  & ${\cal F}_{\rm twin}$ $<$ 0.03 \\
\cline{2-2}
 log\,$P$\,(days) = 5.95\,$\pm$\,0.75   & $f_{\rm logP;q>0.3}$ = 0.10\,$\pm$\,0.03 \\
\hline
\end{tabular}
\end{center}
\end{figure}
\renewcommand{\arraystretch}{1.0}

\subsection{Early-B Primaries}

\citet{Duchene2001} utilized adaptive optics to search for visual companions to massive stars in the young open cluster NGC\,6611 (the Eagle Nebula; $\tau$~$\approx$~2\,-\,6 Myr; $d$~$\approx$~2.1\,kpc).  More recently, \citet{Peter2012} performed lucking imaging of intermediate-mass and massive stars within several subgroups of the Cep~OB2/3 association ($\tau$~$\approx$~3\,-\,10~Myr; $d$ $\approx$ 0.8\,kpc).  A significant fraction of both samples contain B0-B3 primaries.  In addition, both surveys are sensitive to low-mass companions across similar ranges of projected separation, and both studies determined the binary mass ratios from the observed brightness contrasts.  To more accurately measure the binary statistics, we combine the VB data on the B0-B3 primaries from the \citet{Duchene2001} and \citet{Peter2012} surveys.  In the following, we designate the two samples with subscripts A and B, respectively.

\citet{Duchene2001} observed ${\cal N}_{\rm prim, A}$ = 30 B0-B3 stars in NGC\,6611, 10 of which were resolved with adaptive optics to have visually resolved companions (see their Tables~1, 3, and~4).  Three of these VBs have large near-infrared brightness contrasts $\Delta$K~$>$~3.5~mag ($\lambda$~$\approx$~2.2\,$\mu$m), indicating very extreme mass ratios $q$~$<$~0.1.  Meanwhile, two other companions have large dust reddenings compared to their primaries, suggesting they are are background giants.  We select the ${\cal N}_{\rm comp,A}$~=~5 companions with projected separations 0.5$''$\,-\,1.5$''$, B0-B3 primaries, and estimated mass ratios $q$~$>$~0.1 (Walker designations 25, 188, 254, 267, and 311). For Walker 311, we adopt the average mass ratio $q$ = 0.63 of the listed range $q$ = 0.47\,-\,0.78 in Table~3 of \citet{Duchene2001}.  Of the five companions, ${\cal N}_{\rm largeq,A}$~=~2 have large mass ratios $q$~$>$~0.3.  

From the \citet{Peter2012} survey, we select the ${\cal N}_{\rm prim, B}$ = 56 primaries with spectral types B0-B3 and luminosity classes III-V (see their Table A.1).  Of these primaries, ${\cal N}_{\rm comp,B}$~=~8 have projected separations 0.45$''$\,-\,1.5$''$ and brightness contrasts $\Delta$z$'$~$<$~4.3~mag ($\lambda$~$\approx$~900\,nm). We adopt the mass ratios presented in Fig.~10 of \citet{Peter2012} for our nine selected VBs with B0-B2 primaries.  For the two systems with B3 primaries, we utilize the relation between $\Delta$z$'$ and $q$ calculated in \citet{Peter2012}.  Specifically, we estimate $q$ $=$0.10 for BD+61~2218 ($\Delta$z$'$~=~4.3~mag) and $q$ $=$0.12 for HD~239649 ($\Delta$z$'$~=~4.2~mag).  Of the eight companions, ${\cal N}_{\rm largeq,B}$~=~3 have large mass ratios $q$~$>$~0.3.

\begin{figure}[t!]
\centerline{
\includegraphics[trim=0.5cm 0.1cm 0.5cm 0.2cm, clip=true, width=3.55in]{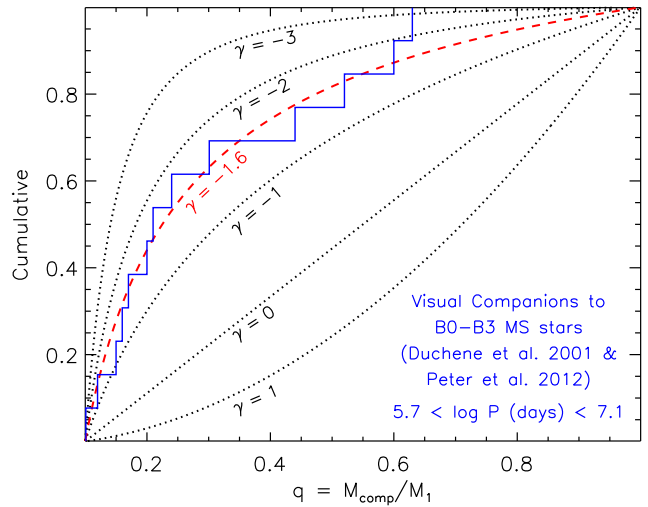}}
\caption{Cumulative distribution of mass ratios $q$ $>$ 0.1 for 13 visual binaries (blue) with B0-B3 primaries ($\langle M_1 \rangle$~=~12\,$\pm$\,3\,\Msun) using a combined sample from  \citet{Duchene2001} and \citet{Peter2012}. The power-law component {\large $\gamma$}~=~$-$1.6\,$\pm$\,0.3 (dashed red) of the mass-ratio distribution is moderately discrepant with random pairings drawn from a Salpeter IMF, but is still weighted toward extreme mass ratios $q$ = 0.1\,-\,0.3.}
\end{figure}

In Fig.~22, we display the cumulative distribution of mass ratios for the ${\cal N}_{\rm comp}$ = ${\cal N}_{\rm comp,A}$\,+\,${\cal N}_{\rm comp,B}$ = 5\,$+$\,8 = 13 binaries we selected from \citet{Duchene2001} and \citet{Peter2012}.  We measure an excess twin fraction ${\cal F}_{\rm twin}$~$<$~0.10 that is consistent with zero (Table~9).  By fitting a single power-law slope across the entire interval 0.1~$<$~$q$~$<$~1.0, we determine {\large $\gamma$}~=~$-$1.6\,$\pm$\,0.4.  This is moderately discrepant with random pairings drawn from a Salpeter IMF ({\large $\gamma$}~=~$-$2.35) at the 1.9$\sigma$ significance level.  We note that \citet{Duchene2001} conclude that their data alone, which is smaller than our combined sample, is consistent with {\large $\gamma$}~=~$-$2.35.  Meanwhile, \citet{Peter2012} report {\large $\gamma$}~$\approx$~$-$1.0 that is inconsistent with random pairings drawn from a Salpeter IMF. However, \citet{Peter2012} fit this exponent down to extreme mass ratios $q$ $\lesssim$ 0.05 where the secondary masses $M_2$~$\lesssim$~0.5\,\Msun\ are no longer expected to follow a Salpeter-like slope. In any case, our value of {\large $\gamma$}~=~$-$1.6\,$\pm$\,0.4 after combining the two samples is between the two independent measurements.  Allowing for a break at $q$~=~0.3, we measure {\large $\gamma$}$_{\rm smallq}$~=~$-$1.3\,$\pm$\,0.5 and {\large $\gamma$}$_{\rm largeq}$~=~$-$1.9\,$\pm$\,0.5 (Table~9).

Given the distance $d$~$\approx$~2.1\,kpc to NGC~6611, the ${\cal N}_{\rm largeq,A}$~=~2 companions with $\rho$~=~0.5$''$\,-\,1.5$''$ and $q$~$>$~0.3 from the \citet{Duchene2001} sample have separations $a$~$\approx$~1,000\,-\,3,000\,AU, i.e., 6.4~$<$~log\,$P$\,(days)~$<$~7.1. Meanwhile, the ${\cal N}_{\rm largeq,B}$~=~3 companions with $\rho$~=~0.45\,-\,1.5$''$ and $q$~$>$~0.3 from the \citet{Peter2012} sample have separations $a$~$\approx$~360\,-\,1,200\,AU, i.e., 5.7~$<$~log\,$P$~$<$~6.5, given the distance $d$~$\approx$~0.8\,kpc to the Cep~OB2/3 association.  The open cluster NGC~6611 is sufficiently young that the correction factor ${\cal C}_{\rm evol,A}$ = 1.0\,$\pm$\,0.1 due to binary evolution is negligible. For the slightly older Cep~OB2/3 association, we adopt ${\cal C}_{\rm evol,B}$~=~1.1\,$\pm$\,0.1.  Combining the statistics, we measure $f_{\rm logP;q>0.3}$ = (${\cal N}_{\rm largeq,A}{\cal C}_{\rm evol,A}$/$\Delta$log$P_{\rm A}$~+~${\cal N}_{\rm largeq,B}{\cal C}_{\rm evol,B}$/$\Delta$log$P_{\rm B}$)\,/ (${\cal N}_{\rm prim,A}$~+~${\cal N}_{\rm prim,B}$)  = 0.08\,$\pm$\,0.04 for early-B primaries and log\,$P$\,(days) = 6.4\,$\pm$\,0.7 (Table~9).

\subsection{O-type Primaries}

In addition to long-baseline interferometry and sparse aperture masking (see \S5.2), \citet{Sana2014} utilized near-infrared adaptive optics to search for visual companions to the O-type stars in their sample.  \citet{Aldoretta2015} recently performed high-contrast imaging in the optical ($\Delta$m $\lesssim$ 5.0 mag; $\lambda$~$\approx$~583\,nm) of O-type stars with the Fine Guidance Sensor aboard the {\it Hubble Space Telescope}. \citet{Mason2009} and \citet{Caballero2014} have discovered visual companions to other O-type stars, but these two surveys are limited to small optical brightness contrasts $\Delta$V~$\lesssim$~2\,-\,3~mag and therefore large mass ratios $q$~$\gtrsim$~0.4\,-\,0.5.  In the following, we analyze the VB statistics of O-type primaries based on the \citet{Sana2014} and \citet{Aldoretta2015} samples.  These two studies cover slightly different angular separations, incorporate different bandpasses, and report only the observed brightness contrasts.  We therefore treat these two surveys independently and model our own values of the mass ratios according to the listed brightness contrasts.  

For the \citet{Sana2014} survey, we follow the same procedure as in \S5.2.  We select the ${\cal N}_{\rm prim}$~=~106 O-type primaries with luminosity classes II.5-V that they observed with near-infrared adaptive optics.  This includes the 76 such systems in the \citet{Sana2014} main target list (see their Fig.~1), and the 30 additional objects in their supplementary target lists.  As shown below and in Fig.~23, VBs with brightness contrasts $\Delta$K$_{\rm s}$~$<$~5.8~mag ($\lambda$ $\approx$~2.2\,$\mu$m) are complete down to $q$~=~0.1.  The \citet{Sana2014} adaptive optics survey is sensitive to $\Delta$K$_{\rm s}$~$\approx$~5.8~mag beyond $\rho$~$>$~0.6$''$ (see their Fig.~7), while confusion with background/foreground stars becomes non-negligible beyond $\rho$~$>$~2.0$''$ (see their Fig.~8).  From our selected sample of 106 O-type MS primaries,   \citet{Sana2014} identified 18 companions with angular separations $\rho$~=~0.6$''$-2.0$''$ and brightness contrasts $\Delta$K$_{\rm s}$~$<$~5.8~mag.  Based on the observed surrounding stellar densities, \citet{Sana2014} estimate extremely small probabilities ${\cal P}_{\rm spur}$ $\le$ 2\% that each of these 18 VBs are spurious associations.  The total probability that any of the 18 VBs are optical doubles is $\sim$5\%.  Our subsample of 18 VBs is therefore relatively unbiased and complete down to $q$ = 0.1.  

Incorporating the same stellar isochrones, relations, and methods adopted in \S5.2, we calculate the brightness contrasts $\Delta$K$_{\rm s}$ as a function of mass ratio $q$. In Fig.~23, we compare $q$ and $\Delta$K$_{\rm s}$ for different combinations of primary spectral type and luminosity class. A sample of binaries with O-type primaries and luminosity classes III-V is complete down to $q$~=~0.1 if the observations are sensitive to  $\Delta$K$_{\rm s}$~$\approx$~5.8~mag.  The $\Delta$K$_{\rm s}$ ($\lambda$ $\approx$~2.2\,$\mu$m) relations shown in Fig.~23 are nearly indistinguishable from the near-infrared $\Delta$H ($\lambda$~$\approx$~1.6\,$\mu$m) curves presented in Fig.~13. We estimate $\Delta$K$_{\rm s}$ = 0.96$\Delta$H for a broad range of binary masses and ages.

\begin{figure}[t!]
\centerline{
\includegraphics[trim=0.6cm 0.0cm 0.5cm 0.1cm, clip=true, width=3.55in]{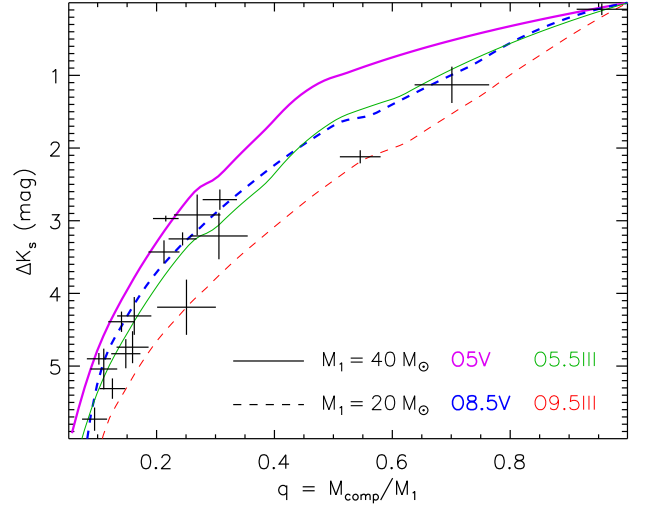}}
\caption{Similar to Fig.~13, but for the 18 VBs in the \citet{Sana2014} adaptive optics survey with O-type MS primaries,  projected separations 0.6$''$\,-\,2.0$''$, and brightness contrasts $\Delta$K$_{\rm s}$ $<$ 5.8 mag.  This VB subsample is complete down to $q$ = 0.1 and does not contain any spurious optical doubles at the 95\% confidence level. }
\end{figure}

 \citet{Sana2014} measured both brightness contrasts $\Delta$K$_{\rm s}$ and $\Delta$H for 14 of our 18 selected VBs.  For these systems, we first use the relation $\Delta$K$_{\rm s}$ = 0.96$\Delta$H to convert the near-infrared brightness contrasts, and then we calculate a weighted average and uncertainty for $\Delta$K$_{\rm s}$.  For the four remaining VBs, we adopt the values of and uncertainties in $\Delta$K$_{\rm s}$ reported by \citet{Sana2014}. We calculate the mass ratios $q$ of the 18 VBs based on the relations presented in Fig.~23.  As done in \S5.2, we propagate the measurement uncertainty in $\Delta$K$_{\rm s}$ as well as uncertainties of 0.5 in both the primary's spectral subtype and luminosity class. 

 In Fig.~23, we compare the estimated mass ratios $q$ to the observed brightness contrasts $\Delta$K$_{\rm s}$ for our 18 VBs. The majority of the VBs have $q$ $<$ 0.35.  One system, HD\,93161, has a mass ratio $q$ $\approx$ 0.96 near unity according to the observed brightness contrast $\Delta$K$_{\rm s}$~=~0.07~mag.  The fainter component HD\,93161\,B is a O6.5IV-V star while the brighter component HD\,93161\,A  has a spectral classification of O7.5-8V \citep{Naze2005,Sana2014}.  The fainter component HD\,93161\,B should therefore be more massive and luminous.  This apparent discrepancy is because component A is itself an SB2 with an orbital period of $P$~$\approx$~8.6~days and estimated masses of $M_{\rm Aa}$~$=$~22.2\,$\pm$\,0.6\,\Msun\ and $M_{\rm Ab}$~$=$~17.0\,$\pm$\,0.4\,\Msun\ \citep{Naze2005}.  This short-period SB2 with $q$ = $M_{\rm Ab}$/$M_{\rm Aa}$ = 0.77 contributes to the statistics of close companions to early-type stars in \S3.  Using the O-type stellar relations provided in \citet{Martins2005}, we estimate HD\,93161\,B to have a mass of $M_B$~$=$~30\,$\pm$\,3\,\Msun\ according to its observed spectral classification of O6.5IV-V.  For the VB HD\,93161, we adopt a mass ratio of $q$ $\equiv$ $M_{\rm comp}$/$M_1$ = $M_{\rm Aa}$/$M_{\rm B}$ = 22.2/30 = 0.74. For the 17 remaining VBs with  $\Delta$K$_{\rm s}$~$<$~1.1~mag, the presence of an additional companion to the O-type star will not significantly affect the mass ratio inferred from the brightness contrast.   We adopt the mass ratios presented in Fig.~23 for the 17 other VBs.

\begin{figure}[t!]
\centerline{
\includegraphics[trim=0.5cm 0.1cm 0.5cm 0.2cm, clip=true, width=3.45in]{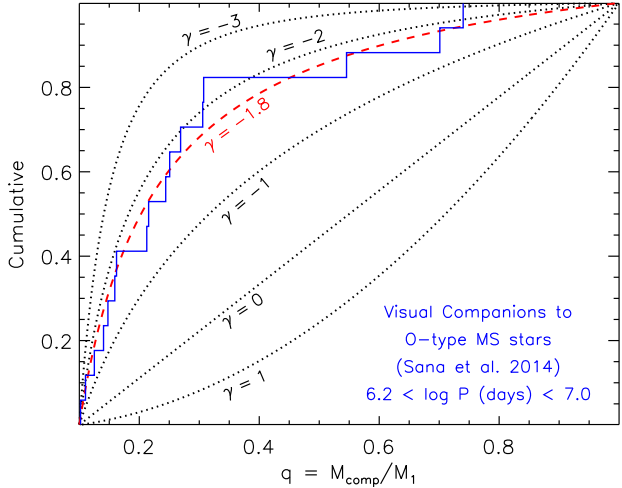}}
\caption{Cumulative distribution of mass ratios for the 17 VBs (blue) with O-type MS primaries, $q$ $>$ 0.1, and projected separations 0.6$''$\,-\,2.0$''$ (6.2~$<$~log\,$P$~$<$~7.0) from \citet{Sana2014}.  For this broad interval, we find the slope {\large $\gamma$}~=~$-$1.8\,$\pm$\,0.3 (dashed red) of the mass-ratio probability distribution is weighted toward small values $q$~=~0.1\,-\,0.3.}
\end{figure}

Of our 18 selected VBs,  ${\cal N}_{\rm smallq}$ = 12 have $q$~=~0.1\,-\,0.3 and ${\cal N}_{\rm largeq}$ = 5 have  $q$~=~0.3\,-\,1.0.  The one remaining VB has a large brightness contrast $\Delta$K$_{\rm s}$~=~5.7~mag near our selection limit and a mass ratio $q$~$\approx$~0.09 just below our statistical cut-off.  In Fig.~24, we display the cumulative distribution of mass ratios for the 17 VBs with $q$ $>$ 0.1.   We measure an excess twin fraction of  ${\cal F}_{\rm twin}$ $<$ 0.10 (Table 9). By fitting a single-power law mass-ratio distribution across $q$~=~0.1\,-\,1.0, we determine  {\large $\gamma$}~=~$-$1.8\,$\pm$\,0.3 (see Fig.~24).  Allowing for a break in the mass-ratio distribution at $q$ = 0.3, the slopes are {\large $\gamma$}$_{\rm smallq}$ = $-$1.5\,$\pm$\,0.5 and {\large $\gamma$}$_{\rm largeq}$ = $-$2.1\,$\pm$\,0.5 (Table~9).  As found for mid-B (\S7.3) and early-B (\S7.4) VB samples, the component {\large $\gamma$}$_{\rm largeq}$ across large mass ratios $q$~=~0.3\,-\,1.0 is consistent with random pairings drawn from the Salpeter IMF  while the slope {\large $\gamma$}$_{\rm smallq}$ across small mass ratios $q$~=~0.1\,-\,0.3 is slightly flatter.

 As in \S5.2, we adopt an average distance $\langle d \rangle$~=~1.5~kpc for the O-type binaries in the \citet{Sana2014} survey.  The angular separations $\rho$~=~0.6$''$-2.0$''$ correspond to separations $a$~=~900\,-\,3,000~AU, i.e., 6.2~$<$~log\,$P$\,(days)~$<$~7.0.  We adopt the same correction factor ${\cal C}_{\rm evol}$~=~1.1\,$\pm$\,0.1  for O-type stars with luminosity classes II.5-V.  The intrinsic companion frequency is $f_{\rm logP;q>0.3}$ = ${\cal N}_{\rm largeq} {\cal C}_{\rm evol}$/${\cal N}_{\rm prim}$/$\Delta$log$P$ = (5\,$\pm$\,$\sqrt{5}$)(1.1\,$\pm$\,0.1)/106/(7.0\,$-$\,6.2) = 0.07\,$\pm$\,0.03 (Table~9).

We next analyze the statistics of VBs in the \citet{Aldoretta2015} catalog.  They utilized the high-resolution of {\it Hubble}'s Fine Guidance Sensor with the F583W filter ($\lambda$~$\approx$~460\,-\,700\,nm) to identify 74 close companions to 224 O and early-B stars. Their sample is complete down to $\Delta$m = 4.0 mag beyond $>$0.1$''$, while incompleteness and confusion with background/foreground objects become important at large separations $>$1.0$''$.  We do not consider their supplementary list of 81 additional targets, as these systems were observed with lower sensitivity $\Delta$m~$<$~3.0~mag \citep{Nelan2004,Caballero2014}.  From the  \citet{Aldoretta2015} main sample, we select the ${\cal N}_{\rm prim}$~=~128 galactic O-type stars with luminosity classes II.5-V.  From these primaries, \citet{Aldoretta2015} identified ${\cal N}_{\rm comp}$ = 21 VB companions with projected separations 0.1$''$\,-\,1.0$''$ and optical brightness contrasts $\Delta$m~$<$~4.0~mag.  

We implement the same technique as above to estimate the mass ratios $q$.  To model the VB brightness contrasts $\Delta$m in the broadband F583W filter ($\lambda$~$\approx$~460\,-\,700\,nm), we average the $\Delta$V and $\Delta$R$_{\rm C}$ relations for $q$.  In Fig.~25, we display $\Delta$m as a function of $q$ for the same primary masses and luminosity classes as in Figs.~13 and 23.  In the optical, the sensitivity limit of $\Delta$m~$<$~4.0~mag is complete down to $q$~$>$~0.3.  For the 21 VBs we selected from \citet{Aldoretta2015}, we measure the mass ratios $q$ from $\Delta$m and the spectral types and luminosity classes of the primaries.  We show our results in Fig.~25, where we find ${\cal N}_{\rm largeq}$ = 18 VBs have mass ratios $q$ $>$ 0.3.  The three remaining VBs have large brightness contrasts $\Delta$m~=~3.6\,-\,4.0~mag near the sensitivity limit and mass ratios $q$ = 0.25\,-\,0.30 just below our statistical cut-off.

\begin{figure}[t!]
\centerline{
\includegraphics[trim=0.6cm 0.1cm 0.5cm 0.2cm, clip=true, width=3.5in]{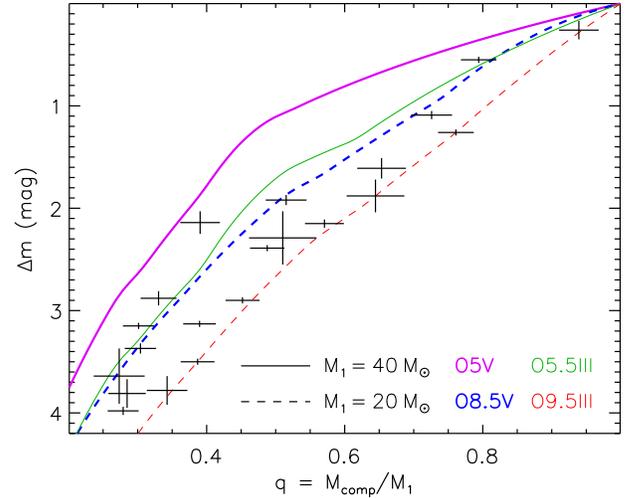}}
\caption{Similar to Figs.~13 and 23, but for the 21 VBs in the \citet{Aldoretta2015} main survey with O-type MS primaries,  projected separations 0.1$''$\,-\,1.0$''$, and optical brightness contrasts $\Delta$m~$<$~4.0~mag ($\lambda$~$\approx$~460\,-\,700\,nm).  This VB subsample is complete down to $q$ = 0.3.}
\end{figure}

Given the average distance $\langle d \rangle$~$\approx$~2.0~kpc to the ${\cal N}_{\rm largeq}$ = 18 VBs we selected from \citet{Aldoretta2015}, the projected angular separations $\rho$~=~0.1$''$\,-\,1.0$''$ correspond to orbital periods 5.2~$<$~log\,$P$\,(days)~$<$~6.7.  The selected VBs have primary luminosity classes II.5-V, and so the correction factor ${\cal C}_{\rm evol}$ = 1.1\,$\pm$\,0.1 due to binary evolution is relatively small.  The intrinsic frequency of companions with $q$ $>$ 0.3 per decade of orbital period is $f_{\rm logP;q>0.3}$ = ${\cal N}_{\rm largeq} {\cal C}_{\rm evol}$/${\cal N}_{\rm prim}$/$\Delta$log$P$ = (18\,$\pm$\,$\sqrt{18}$)(1.1\,$\pm$\,0.1)/128/(6.7\,$-$\,5.2) = 0.10\,$\pm$\,0.03 for O-type stars and wide orbital periods (Table~9). 

In Fig.~26, we display the cumulative mass-ratio distribution for the ${\cal N}_{\rm largeq}$~=~18 VBs from the \citet{Aldoretta2015} main sample with O-type MS primaries, angular separations $\rho$~=~0.1$''$\,-\,1.0$''$, and mass ratios $q$ $>$ 0.3.  We measure a negligible excess twin fraction ${\cal F}_{\rm twin}$~$<$~0.03 and a power-law component {\large $\gamma$}$_{\rm largeq}$~=~$-$1.8\,$\pm$\,0.4 that is weighted toward small mass ratios $q$~=~0.3\,-\,0.5 (Table~9).  Our measurement of {\large $\gamma$}$_{\rm largeq}$~=~$-$1.8\,$\pm$\,0.4 based on the \citet{Aldoretta2015} O-type VB sample is consistent with our measurements for VB companions to Cepheids ({\large $\gamma$}$_{\rm largeq}$~=~$-$2.1\,$\pm$\,0.5), mid-B stars ({\large $\gamma$}$_{\rm largeq}$~=~$-$2.2\,$\pm$\,0.6), early-B stars ({\large $\gamma$}$_{\rm largeq}$~=~$-$1.9\,$\pm$\,0.5), and O-type stars in the \citet{Sana2014} survey ({\large $\gamma$}$_{\rm largeq}$~=~$-$2.1\,$\pm$\,0.5).   The weighted average and uncertainty of these five values is {\large $\gamma$}$_{\rm largeq}$~=~$-$2.0\,$\pm$\,0.2.  For a broad range of primary masses $M_1$~$\approx$~6\,-\,40\,\Msun\ and orbital periods 5~$\lesssim$~log\,$P$\,(days)~$\lesssim$~7, the power-law slope of the binary mass-ratio distribution across $q$~=~0.3\,-\,1.0 nearly matches the prediction from random pairings drawn from a Salpeter IMF ({\large $\gamma$}$_{\rm largeq}$~=~$-$2.35).  

\begin{figure}[t!]
\centerline{
\includegraphics[trim=0.5cm 0.1cm 0.5cm 0.2cm, clip=true, width=3.4in]{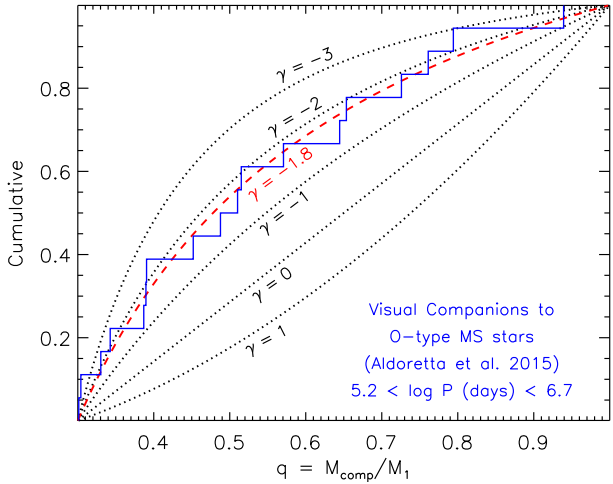}}
\caption{Cumulative distribution of mass ratios for the 18 VBs (blue) with O-type MS primaries, $q$ $>$ 0.3, and projected separations 0.1$''$\,-\,1.0$''$ (5.2~$<$~log\,$P$~$<$~6.7) from \citet{Aldoretta2015}.  The power-law slope {\large $\gamma$}$_{\rm largeq}$~=~$-$1.8\,$\pm$\,0.3 across $q$~=~0.3\,-\,1.0 (dashed red) favors small mass ratios $q$~=~0.3\,-\,0.5. }
\end{figure}

\subsection{Eccentricity Distribution}

As discussed in \S2, catalogs of VBs with reliable orbital solutions are biased against systems with large eccentricities \citep{Harrington1977,Tokovinin2015}.  We can nonetheless use VB orbit catalogs to measure a lower limit to the power-law slope $\eta$ of the eccentricity distribution. We also fit $\eta$ across eccentricities 0.0~$<$~$e$~$<$~0.8 that are not as severely affected by this selection effect.  \citet{Malkov2012} compiled a set of VBs with known distances, measured primary spectral types, and orbits of high quality from both the Catalog of Orbits and Ephemerides of Visual Double Stars \citep{Docobo2001} and the Sixth Catalog of Visual Binary Stars \citep{Hartkopf2001}.  We consider the early-type VBs in their sample with measured orbital periods $P$~=~10\,-\,100~years.  The majority of early-type binaries with shorter orbital periods $P$~$<$~10~years remain unresolved and are unsuitable for orbit determinations.  VBs with longer periods $P$~$>$~100~years generally have incomplete orbits, and so the selection bias against eccentric systems becomes too important.  

In the top panel of Fig.~27, we show the cumulative distribution of eccentricities 0~$<$~$e$~$<$~1 for the early-type VBs with $P$~=~10\,-\,100~years from the \citet{Malkov2012} catalog.  We compare the 18 VBs with primary spectral types O5-B5 (blue) to the 101 VBs with spectral types B6-A5 (red).  The O/early-B subsample is weighted toward larger eccentricities, where we measure $\eta$~=~0.4\,$\pm$\,0.3.  For the late-B/early-A subsample, we find $\eta$~=~0.0\,$\pm$\,0.2.   \citet{Abt2005} also reports a uniform eccentricity distribution ($\eta$ = 0.0) based on a subsample of VBs with orbital solutions, $P$~$>$~1,000~days, and B0-F0 primary spectral types.  In all these cases, however, the estimates for $\eta$ are biased toward smaller values due to the observational selection effects. Our measurements of $\eta$~$>$~0.4 and $\eta$~$>$~0.0 for the O/early-B and late-B/early-A subsamples, respectively, are lower limits.

To account for the selection bias against VBs with $e$~$>$~0.8, we now fit the power-law slope $\eta$ across 0.0~$<$~$e$~$<$~0.8. In the bottom panel of Fig.~27, we display the cumulative distribution of eccentricities 0.0~$<$~$e$~$<$~0.8 for our same two subsamples.  The power-law slope increases to $\eta$~=~0.3\,$\pm$\,0.3 for the late-B/early-A VBs.  For the O/early-B subsample, we measure $\eta$~=~0.8\,$\pm$\,0.3, which is consistent with a thermal eccentricity distribution ($\eta$~=~1). The systematic uncertainties in $\eta$ due to the selection biases are comparable to the measurement uncertainties.  We therefore adopt $\eta$~=~0.8\,$\pm$\,0.4 and $\eta$~=~0.3\,$\pm$\,0.4 for the O/early-B and late-B/early-A VB subsamples, respectively (Table~10).

\renewcommand{\arraystretch}{1.6}
\setlength{\tabcolsep}{10pt}
\begin{figure}[t!]\footnotesize
{\small {\bf Table 10:} Measurements of the eccentricity distribution based on early-type VBs with log\,$P$\,(days) = 3.6\,$\pm$\,0.5 from the \citet{Malkov2012} visual orbit catalog.}  \\
\vspace*{-0.45cm}
\begin{center}
\begin{tabular}{|c|c|}
\hline
 Primary Mass  & $\eta$ \\
\hline
$\langle M_1 \rangle$ =  3\,$\pm$\,1\,\Msun\ & 0.3\,$\pm$\,0.4 \\
\hline
$\langle M_1 \rangle$ = 11\,$\pm$\,4\,\Msun\ & 0.8\,$\pm$\,0.4 \\
\hline
\end{tabular}
\end{center}
\end{figure}
\renewcommand{\arraystretch}{1.0}

\begin{figure}[t!]
\centerline{
\includegraphics[trim=3.9cm 0.1cm 5.0cm 0.1cm, clip=true, width=2.9in]{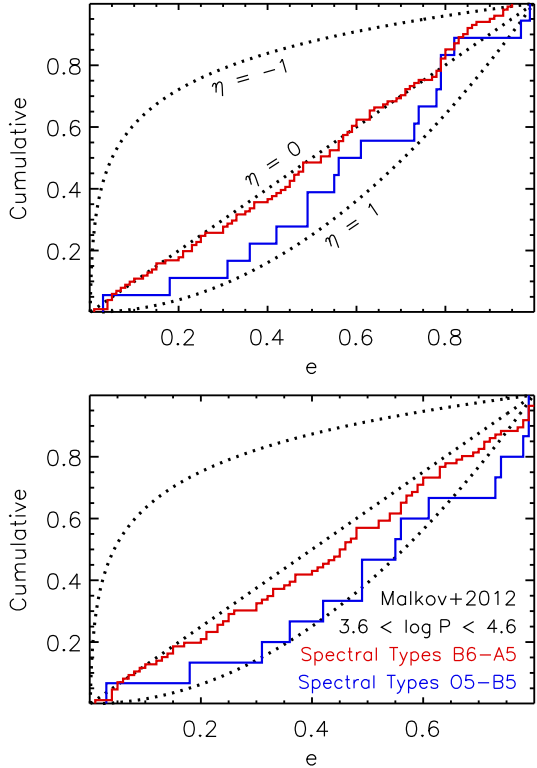}}
\caption{Cumulative distribution of eccentricities across 0.0~$<$~$e$~$<$~1.0 (top) and across 0.0~$<$~$e$~$<$~0.8 (bottom) for early-type VBs with $P$~=~10\,-\,100~years (3.6~$<$~log\,$P$\,(days)~$<$4.6) from the \citet{Malkov2012} visual orbit catalog. We compare the eccentricity distributions of VBs with primary spectral types B6-A5 (red) and O5-B5 (blue).  After accounting for the selection bias against VBs with $e$~$>$~0.8, we find the O/early-B subsample has an eccentricity distribution ($\eta$~=~0.8\,$\pm$\,0.4) consistent with a thermal distribution ($\eta$~=~1).   }
\end{figure}

\section{Solar-type Binaries}

\subsection{Sample Selection}

To extend the baseline toward smaller primary masses, we now investigate the companion properties to solar-type MS primaries ($M_1$~=~1.00\,$\pm$\,0.25\,\Msun).  The most complete solar-type MS binary sample derives from \citet{Raghavan2010}, who updated and extended the sample of \citet{Duquennoy1991} (see \S8.6 for comparison of these two samples of solar-type MS binaries).  \citet{Raghavan2010} combined various observational techniques to search for companions around  454 F6\,-\,K3 type stars located within 25 pc.  The companion properties to solar-type primaries may differ in young star-forming environments \citep{Duchene2007,Connelley2008b,Kraus2011,Tobin2016}, dense open clusters \citep{Patience2002,Kohler2006,Geller2012,King2012}, or at extremely low metallicities \citep{Abt2008,Gao2014,Hettinger2015}.  In \S10, we compare the corrected solar-type binary statistics between the field MS population to younger MS and pre-MS populations in open clusters and stellar associations. For now and for the purposes of binary population synthesis studies, we are mostly interested in the overall companion statistics of typical primaries with average ages.  Most solar-type stars are near solar-metallicity, in the galactic field, and are several Gyr old.   The volume-limited sample of solar-type MS primaries in \citet{Raghavan2010} is therefore most representative of the majority of solar-type MS stars.

We display in Fig.~\ref{solarq.fig} the 168 confirmed companions from \citet{Raghavan2010} with measured orbital periods 0.0 $<$ log\,$P$\,(days) $<$ 8.0 and mass ratios 0.1~$<$~$q$~$<$~1.0.  We utilize the same methods as in \citet{Raghavan2010} to estimate the orbital periods $P$ from projected separations and the stellar masses from spectral types.  Our Fig.~\ref{solarq.fig}  is similar to Figs.~11~\&~17 in \citet{Raghavan2010}.  However, in the cases of triples and higher-order multiples, we always define the period and mass ratio $q$~$\equiv$~$M_{\rm comp}$/$M_1$ with respect to the solar-type primary (see \S2), which differs slightly from the definitions adopted in \citet{Raghavan2010}.

 For example, first consider a triple in an (Aa,\,Ab)\,-\,B hierarchical configuration. A $M_{\rm Aa}$~=~$M_1$~=~1.0\,\Msun\ primary and $M_{\rm Ab}$~=~0.5\,\Msun\ companion are in a short-period orbit of $P$~=~100~days. A long-period tertiary component with $M_{\rm B}$~=~0.4\,\Msun\, orbits the inner binary with a period of $P$~=~10$^5$~days.  This system would contribute two data points in Fig.~\ref{solarq.fig}: one with $q$~=~$M_{\rm Ab}$/$M_{\rm Aa}$~=~0.5 and log\,$P$ = 2.0, and one with $q$ =  $M_{\rm B}$/$M_{\rm Aa}$~=~0.4 and log\,$P$~=~5.0.  We do {\it not} define the mass ratio of the wide system to be $q$~=~$M_{\rm B}$/($M_{\rm Aa}$\,$+$\,$M_{\rm Ab}$) as done in \citet{Raghavan2010}.

 Next, consider a triple in an A\,-\,(Ba,\,Bb) hierarchical configuration.  A solar-type $M_{\rm A}$ =~$M_1$~=~1.0\,\Msun\ primary is in a long-period $P$~=~10$^5$~day orbit around a close, low-mass binary with $M_{\rm Ba}$~=~0.5\,\Msun, $M_{\rm Ba}$~=~0.4\,\Msun, and $P$~=~100~days.  In this situation, only the wide system with $q$~=~$M_{\rm Ba}$/$M_{\rm A}$~=~0.5 and log\,$P$~=~5.0 would contribute to our Fig.~\ref{solarq.fig}.  We do {\it not} consider the low-mass inner binary with log $P$~=~2.0 and $M_{\rm Ba}$/$M_{\rm Bb}$~=~0.8, as done in \citet{Raghavan2010},  because neither component Ba nor Bb is a solar-type star.  Only if component Ba itself has a F6\,-\,K3 spectral type do we include the close (Ba,\,Bb) pair in our sample.  Nearly half of the twins with $q$~$>$~0.95 in Figs.~11~\&~17 of \citet{Raghavan2010} are not solar-type twins.  They instead contain late-K or M-dwarf equal-mass binaries in a long-period orbit with a solar-type primary in a A\,-\,(Ba,\,Bb) hierarchical configuration.  Our Fig.~\ref{solarq.fig} therefore does not contain as many twin components as displayed in Figs.~11~\&~17 of \citet{Raghavan2010}.  

\subsection{Corrections for Incompleteness}

\begin{figure}[t!]
\centerline{
\includegraphics[trim=1.1cm 0.0cm 0.9cm 0.0cm, clip=true, width=3.6in]{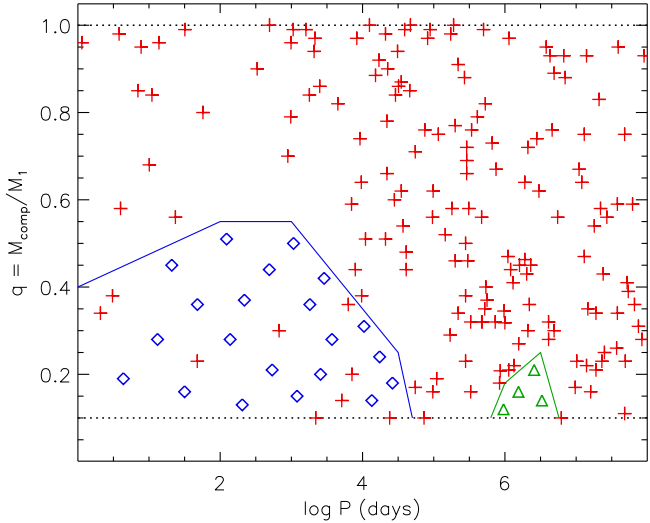}}
\caption{The companions to 454 solar-type stars from the \citet{Raghavan2010} survey as a function of $P$ and $q$~=~$M_{\rm comp}$/$M_1$.  We display the 168 confirmed systems (red pluses) with measured mass ratios 0.1 $<$ $q$ $<$ 1.0 and periods 0 $<$ log\,$P$\,(days) $<$ 8. Two regions (blue and green lines) of this parameter space are incomplete, either because the various observational techniques are insensitive to these systems  and/or the systems in these regions are detectable but have periods and/or mass ratios that cannot be readily measured (e.g., SB1s, radial velocity variables, and companions implied through proper motion acceleration).  We estimate 21 (blue diamonds) and 4 (green triangles) additional stellar MS companions located within these blue and green regions, respectively.}
\label{solarq.fig}
\end{figure}

The \citet{Raghavan2010} sample is relatively complete except for two regions of the parameter space of $P$ versus $q$.  First, the survey is not sensitive to detecting companions with log\,$P$ $\approx$ 5.9\,-\,6.7 and $q$ $\approx$ 0.1\,-\,0.2 (green region in our Fig.~\ref{solarq.fig}).  As shown in Fig.~11 of \citet{Raghavan2010}, companions that occupy this portion in the parameter space are missed by both adaptive optic and common proper motion techniques.  Considering the density of systems in the immediately surrounding regions where the observations are relatively complete, we estimate $\approx$4 additional systems occupy this gap in the parameter space (four green systems in Fig.~\ref{solarq.fig}).  

The second region of incompleteness occurs at $q$ $\lesssim$ 0.5 and log\,$P$ $\lesssim$ 4.5 (blue region in our Fig.~\ref{solarq.fig}).  The optical brightness contrast between binary components is an even steeper function of mass ratio for F\,-\,M type stars.  Solar-type SB2s with sufficiently luminous secondaries are observed only if $q$ $\gtrsim$ 0.40\,-\,0.55 (brightness contrasts $\Delta$V~$<$~5.0~mag, depending on the orbital period).  Spectroscopic binaries with lower-mass companions will generally appear as SB1s and therefore not have mass ratios that can be readily measured.  Of the four spectroscopic binaries with log\,$P$~$<$~3.0 and $q$~$<$~0.4 shown in Fig.~\ref{solarq.fig}, only one is an SB2 with $P$~$\approx$~4~days and a mass ratio $q$ $\approx$ 0.38 close to the detection limit.  The other three systems are SB1s in a hierarchical triple where the tertiary itself has an orbital solution.  The total mass of the inner SB1 is measured dynamically, and so the mass of the companion in the SB1 can be estimated.  The few observed systems with 3.0 $<$ log\,$P$ $<$ 4.5 and $q$ $<$ 0.4 are sufficiently nearby and have favorable orientations for the companions to be resolved with adaptive optics.  In general, companions below the blue line in our Fig.~\ref{solarq.fig} are unresolved, both spatially and spectrally.

We estimate the number of missing systems in the blue region of Fig.~\ref{solarq.fig} as follows.  \citet{Raghavan2010} identified 26 confirmed and candidate binaries that do not have measurable mass ratios, e.g., SB1s, radial velocity variables, and companions implied through proper motion acceleration of the primary.  A few of the seven radial velocity variables may contain substellar companions with $q$ $<$ 0.1.  The SB1s and companions identified through proper motion acceleration, on the other hand, must have $q$ $\gtrsim$ 0.1 to produce the measured signal. There is also a small gap at log\,$P$ = 3.5\,-\,4.5 and $q$ = 0.10\,-\,0.25 where neither spectroscopic radial velocity surveys nor adaptive optic surveys are complete \citep[see Fig.~11 of][]{Raghavan2010}.  Considering the density of low-mass companions at slightly longer orbital periods, we estimate $\approx$3\,-\,5 additional systems that escaped detection in this region.  Finally, only spectroscopic radial velocity surveys are sensitive to closely orbiting low-mass companions, but only $\approx$80\% of the sample of 454 primaries were searched for such radial velocity variations.  We estimate an additional $\approx$20\%, or $\approx$4\,-\,6 SB1s, to be present around primaries that were not surveyed for spectroscopic variability. In total, we estimate $\approx$35 additional unresolved companions with $q$~$>$~0.1 and log\,$P$~$\lesssim$~4.5.  

\subsection{Frequency of White Dwarf Companions}

Like early-type binaries, a certain fraction of unresolved solar-type SB1s likely contain compact remnant companions.  For solar-type binaries, the majority of unresolved compact remnant companions are white dwarfs (WDs) instead of neutron stars or black holes \citep{Hurley2002,Belczynski2008}.  Fortunately, we can estimate the frequency of such close Sirius-like systems using three different methods that we discuss below.

\subsubsection{UV Excess from WD Companions}

  First, we rely on the catalog of \citet{Holberg2013}, who compiled all known Sirius-like binaries with A\,-\,K type primaries and WD companions. Their sample contains ${\cal N}_{\rm hotWD}$~=~7 systems with separations $a$~$<$~25~AU (log\,$P$~$\lesssim$~4.5), distances $d$~$<$~50~pc, and components that were originally identified due to the UV excess from the hot, closely orbiting WD companions.  This subsample is relatively complete as long as the temperatures $T_{\rm WD}$~$>$~15,000~K of the WDs are sufficiently hotter than the temperatures $T_1$~$\approx$~5,000\,-\,10,000~K of the  A\,-\,K type primaries.  According to evolutionary tracks, a WD cools to $T_{\rm WD}$~$\approx$~15,000 K in $t_{\rm cool}$~$\approx$~0.15\,-\,0.60~Gyr, depending on its mass and composition \citep{Fontaine2001}.  There are ${\cal N}_{\rm solar}$~$\approx$~6,000 A\,-\,K type primaries in the {\it Hipparcos} catalog with parallactic distances $d$~$<$~50~pc \citep{Perryman1997}.  The majority of these systems contain G-K type primaries, and so we adopt an average age of $\langle \tau \rangle$~=~5~Gyr. The fraction of solar-type primaries that have WD companions with log\,$P$~$<$~4.5 is ${\cal F}_{\rm solar+WD;logP<4.5}$ = ${\cal N}_{\rm hotWD}\, \langle \tau\rangle$/(${\cal N}_{\rm solar}\,t_{\rm cool}$) = (2.5\,$\pm$\,1.5)\%.  

\subsubsection{Monte Carlo Population Synthesis}
 
 Second, we incorporate the observed MS binary statistics into a Monte Carlo population synthesis technique to estimate the fraction of systems that evolve into solar-type~+~WD binaries.  In Appendix~A.1, we explore the various evolutionary pathways for producing solar-type + WD binaries.  By utilizing the measured distributions of binary properties, we also calculate in Appendix~A.1 the fraction of systems that evolve through these channels as a function of primary mass $M_1$.  In total, we find that (16\,$\pm$\,3)\% of systems with $M_1$~=~1.25\,\Msun\ primaries evolve into solar-type~+~WD binaries.  This fraction increases to (24\,$\pm$\,6)\% for $M_1$~=~5.0\,\Msun.  Similarly, the fraction of systems that evolve into solar-type~+~WD close binaries with log~$P$~$<$~4.5 is (5\,$\pm$\,1)\% for $M_1$~=~1.25\,\Msun\ and (10\,$\pm$\,3)\% for $M_1$~=~5.0\,\Msun.  We interpolate these statistics as continuous functions of log\,$M_1$.

Using a Monte Carlo technique, we now calculate the fraction of solar-type stars that have WD companions as a function of age $\tau$.  We select primary masses across 0.75\,\Msun~$<$~$M_1$~$<$~8.0\,\Msun\ according to a \citet{Kroupa2013} IMF $p_{M_1}$ $\propto$ $M_1^{-\alpha}$ with slope $\alpha$~=~2.3\,$\pm$\,0.3. We choose companion properties (frequency, mass ratios, and orbital periods) based on the statistics above and summarized in Appendix A.1.  Finally, we adopt MS stellar lifetimes according to the solar-metallicity evolutionary tracks of \citet{Bertelli2008,Bertelli2009}.  At each moment in time, we keep track of the number ${\cal N}_{\rm solar}$ of solar-type MS stars ($M$ = 0.75\,-\,1.25\,\Msun) that {\it appear} to be the primaries, i.e., single solar-type stars, solar-type primaries with lower-mass MS companions, and solar-type stars with WD companions.  The number ${\cal N}_{\rm solar}$ does {\it not} include the number of solar-type secondaries with more massive MS primaries that have not yet evolved into WDs.  We also compute the number ${\cal N}_{\rm solar+WD}$ of solar-type~+~WD binaries, including the subset ${\cal N}_{\rm solar+WD;logP<4.5}$ with log~$P$~$<$~4.5.

\begin{figure}[t!]
\centerline{
\includegraphics[trim=0.8cm 0.8cm 0.8cm 0.5cm, clip=true, width=3.4in]{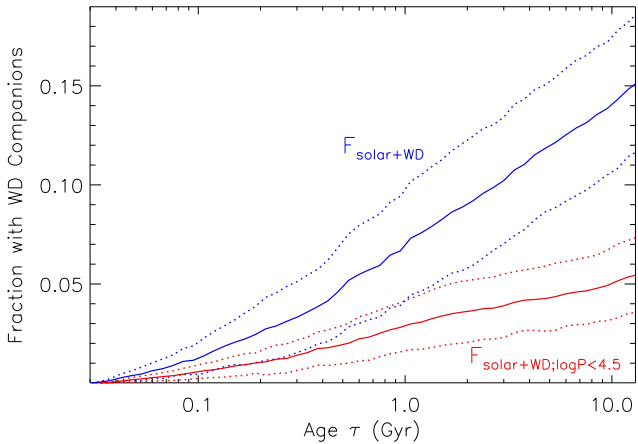}}
\caption{The fraction of solar-type primaries that have WD companions as a function of age $\tau$ (blue), including the subset with orbital periods log\,$P$\,(days) $<$ 4.5 (red).  We indicate the 1$\sigma$ uncertainties with dotted lines.  Approximately ${\cal F}_{\rm solar+WD}$~$=$~14\% of old solar-type primaries with ages $\tau$~$\approx$~10~Gyr are actually the secondaries in binaries in which the initially more massive primaries have already evolved into WDs.  A significant fraction of solar~+~WD binaries have short orbital periods and may appear as SB1s.}
\end{figure}

In Fig.~29 we plot ${\cal F}_{\rm solar+WD}$~=~${\cal N}_{\rm solar+WD}$/${\cal N}_{\rm solar}$ and ${\cal F}_{\rm solar+WD;logP<4.5}$~=~${\cal N}_{\rm solar+WD;logP<4.5}$/${\cal N}_{\rm solar}$ as a function of age $\tau$.  Contamination by WD companions is negligible at young ages $\tau$~$<$~100~Myr because only the most massive primaries $M_1$ $\approx$ 6\,-\,8\,\Msun\ have had enough time to evolve into WDs.  For older populations with ages $\tau$~$\approx$~10~Gyr, ${\cal F}_{\rm solar+WD}$~$\approx$~14\% of solar-type primaries have WD companions, and ${\cal F}_{\rm solar+WD;logP<4.5}$~$\approx$~5\% of solar-type primaries have WD companions with log\,$P$~$<$~4.5.

To assess the degree of WD contamination in the volume-limited \citet{Raghavan2010} sample ($d$~$<$~25~pc), we account for the star formation history in the solar neighborhood.  Observations of both nearby stars and WDs have been utilized to show that the local star formation rate during the past $\approx$\,5 Gyr is approximately twice as high as it was $\approx$\,5\,-\,10 Gyr ago \citep{Bertelli2001,Cignoni2006,Rowell2013}.  We emphasize this is the star formation history within the solar neighborhood $d$~$\lesssim$~50~pc, {\it not} the star formation history integrated through the entire height and/or width of the disk.  For simplicity, we adopt a local star formation history that linearly rises such that the present-day star formation rate is twice the rate it was 10~Gyr ago. Using a Monte Carlo technique and our adopted binary statistics and star formation history, we calculate that ${\cal F}_{\rm solar+WD}$~=~(11\,$\pm$\,4)\% of solar-type primaries in the solar neighborhood have WD companions.  We also find that ${\cal F}_{\rm solar+WD;logP<4.5}$~=~(4.4\,$\pm$\,1.6)\% of solar-type primaries have WD companions with log\,$P$ $<$ 4.5.  

\subsubsection{Barium Stars}

 Finally, we estimate ${\cal F}_{\rm solar+WD,logP<4.5}$ based on the observed frequency of barium stars.  Barium stars are G-K type giants with mild to strong Ba\,\textsc{ii} absorption features \citep{MacConnell1972}.  Only a small fraction ${\cal F}_{\rm Ba}$ = (1.0\,$\pm$\,0.5)\% of G-K type giants are observed to be barium stars \citep{MacConnell1972,Jorissen1998,Karakas2000}. As proof of their binary star origin, more than $\approx$80\% of barium stars are SB1s with companions at intermediate orbital periods $P$~$\approx$~200\,-\,5,000~days, i.e. 2.3 $<$ log $P$ $<$ 3.7 \citep{Boffin1988,Jorissen1998}.  The remaining $\approx$20\% are also expected to be in binaries, but with face-on orientations and/or orbital periods $P$~$>$~5,000~days too long to produce detectable radial velocity variations.  The general consensus is that barium stars were originally solar-type MS stars that accreted s-process rich material from asymptotic giant branch (AGB) donors \citep{Boffin1988,Jorissen1998,Karakas2000}.  The companions to barium stars are therefore the WD remnants of the AGB donors.  Not only did the solar-type MS stars accrete barium, but sufficient mass to become hotter and more massive A-F type MS stars.  Because Ba\,\textsc{ii} absorption features are not easily detected when $T_{\rm eff}$ $\gtrsim$ 6,000~K, the accretors must first evolve into cooler G-K type giants to be readily observed as barium stars. 

  Only a subset of solar-type + WD binaries will appear as barium stars, i.e., those with companions that accreted sufficient material from AGB donors. Binaries with two solar-type MS stars and short orbital periods 0.0~$<$~log\,$P$~$<$~2.3 will fill their Roche lobes when the primary is on the MS, Hertzsprung gap, or red giant branch \citep[][see also Appendix A.1]{Hurley2002,Belczynski2008}.  These systems will leave behind WD remnants with solar-type MS companions that are not chemically enriched with barium.  Similarly, solar-type companions at longer orbital periods 3.7~$<$~log\,$P$~$<$~4.5 are less likely to accrete enough material from the AGB donors to appear as barium stars. If the period distribution of solar-type companions ($M_2$~$\approx$~1\Msun) follows \"{O}pik's law, then the ratio of all close solar-type~+~WD binaries with log\,$P$~$<$~4.5 to those that will appear as barium stars with 2.3~$<$~log~$P$~$<$~3.7 is 
${\cal C}_{\rm Ba}$ = $\Delta$log$P_{\rm total}$/$\Delta$log$P_{\rm Ba}$ = (4.5\,$-$\,0.0)/(3.7\,$-$\,2.3) = 3.2.  In reality, the observed frequency of solar-type companions to solar-type ($M_1$ $\approx$ 1\,\Msun) and  mid-B ($M_1$ $\approx$ 5\,\Msun) primaries increases slightly with increasing logarithmic orbital period (see \S9 and Appendix A.1).  Using the observed period distribution, we calculate the correction factor to be ${\cal C}_{\rm Ba}$~$\approx$~3.1\,$\pm$\,0.8.  Based on the observed  population of barium stars, we estimate that ${\cal F}_{\rm solar+WD,logP<4.5}$ = ${\cal F}_{\rm Ba}\,{\cal C}_{\rm Ba}$ = (1.0\,$\pm$\,0.5)(3.1\,$\pm$\,0.8) = (3.1\,$\pm$\,1.7)\% of solar-type MS stars have WD companions with log~$P$~$<$~4.5.

\subsection{Corrected Population}

The three independent methods described above result in values ${\cal F}_{\rm solar+WD,logP<4.5}$ = (2.5\,$\pm$\,1.5)\%, (4.4\,$\pm$1.6)\%, and (3.1\,$\pm$\,1.7)\% that are consistent with each other.  We adopt a weighted average of ${\cal F}_{\rm solar+WD,logP<4.5}$~=~(3.4\,$\pm$\,1.0)\%.  In the \citet{Raghavan2010} sample of 454 solar-type MS stars, there should be 454\,$\times$\,(0.034\,$\pm$\,0.010) = 15\,$\pm$\,5~WD~companions with log~$P$~$<$~4.5.  Only one of these suspected systems, HD~13445, was barely resolved to have a WD companion with log~$P$~=~4.4.  The remaining 14\,$\pm$\,5 solar-type~+~WD binaries with log $P$~$<$~4.5 remain unresolved, but most likely appear as SB1s and/or systems that exhibit proper motion acceleration.

\renewcommand{\arraystretch}{1.55}
\setlength{\tabcolsep}{10pt}
\begin{figure*}[t!]\footnotesize
{\small {\bf Table 11:} Companion statistics of solar-type primaries ($\langle M_1 \rangle$ = 1.0\,$\pm$\,0.2) based on the \citet{Raghavan2010} survey and corrected sample of ${\cal N}_{\rm prim}$~=~404 primaries and ${\cal N}_{\rm comp}$~=~193 stellar companions with 0.1~$<$~$q$~$<$~1.0 and 0~$<$~log\,$P$\,(days)~$<$~8.}  \\
\vspace*{-0.45cm}
\begin{center}
\begin{tabular}{|c|c|c|c|c|c|c|c|}
\hline
 log $P$ (days) & ${\cal N}_{\rm smallq}$ & ${\cal N}_{\rm largeq}$ & $f_{\rm logP;q>0.3}$ & $\eta$  & {\large $\gamma$}$_{\rm smallq}$ & {\large $\gamma$}$_{\rm largeq}$ & ${\cal F}_{\rm twin}$  \\
\hline
 0.5\,$\pm$\,0.5 &  1 &  7 & 0.017\,$\pm$\,0.007 & $-$0.8\,$\pm$\,0.2 &  
\multirow{2}{*}{0.3\,$\pm$\,0.9} & \multirow{2}{*}{$-$0.6\,$\pm$\,0.7} & \multirow{2}{*}{0.29\,$\pm$\,0.11}  \\
\cline{1-5}
 1.5\,$\pm$\,0.5 &  3 &  8 & 0.020\,$\pm$\,0.007 & $-$0.4\,$\pm$\,0.3 &                     &                    &                    \\
\hline
 2.5\,$\pm$\,0.5 &  3 & 10 & 0.025\,$\pm$\,0.008 &    0.2\,$\pm$\,0.3 &  
\multirow{2}{*}{$-$0.1\,$\pm$\,0.7} & \multirow{2}{*}{$-$0.5\,$\pm$\,0.4} & \multirow{2}{*}{0.20\,$\pm$\,0.07}  \\
\cline{1-5}
 3.5\,$\pm$\,0.5 &  6 & 18 & 0.045\,$\pm$\,0.011 &    0.6\,$\pm$\,0.4 &                     &                    &                    \\
\hline
 4.5\,$\pm$\,0.5 &  7 & 27 & 0.067\,$\pm$\,0.013 &    0.3\,$\pm$\,0.3 &     
\multirow{2}{*}{0.4\,$\pm$\,0.6} & \multirow{2}{*}{$-$0.4\,$\pm$\,0.3} & \multirow{2}{*}{0.10\,$\pm$\,0.04}  \\
\cline{1-5}
 5.5\,$\pm$\,0.5 &  9 & 31 & 0.077\,$\pm$\,0.014 &           -        &                     &                    &                    \\
\hline
 6.5\,$\pm$\,0.5 &  9 & 23 & 0.057\,$\pm$\,0.012 &           -        &    
\multirow{2}{*}{0.5\,$\pm$\,0.5} & \multirow{2}{*}{$-$1.1\,$\pm$\,0.3} &       \multirow{2}{*}{0.02$_{-0.02}^{+0.03}$}     \\
\cline{1-5}
 7.5\,$\pm$\,0.5 & 10 & 21 & 0.052\,$\pm$\,0.011 &           -        &                     &                    &                    \\
\hline
\end{tabular}
\end{center}
\end{figure*}
\renewcommand{\arraystretch}{1.0}

\citet{Raghavan2010} identified 11 companions with measurable mass ratios 0.1 $<$ $q$ $<$ 0.5 and orbital periods 0.0 $<$ log $P$ $<$ 4.5.  In \S8.2, we estimated $\approx$35 additional companions across the same period range based on the \citet{Raghavan2010} detection efficiencies and their statistics of SB1s, systems exhibiting proper motion acceleration, etc. Of the $\approx$35 additional companions, 14\,$\pm$\,5 are most likely WDs.  We conclude that (14\,$\pm$\,5)/(11\,$+$\,35)~=~(30\,$\pm$\,10)\% of  nearby solar-type primaries which appear as SB1s and/or exhibit proper motion acceleration actually contain WD companions.  

The remaining 35\,$-$\,14~=~21 additional binaries contain M-dwarf companions with $q$ $\approx$ 0.1\,-\,0.5 and 0.0 $\lesssim$ log $P$ $\lesssim$ 4.5. The blue region in Fig.~\ref{solarq.fig} is quite large, and so we distribute the estimated 21 additional M-dwarf companions based on the nature of the systems.  The SB1s have known orbital periods, generally 1.0~$<$~log\,$P$\,(days)~$<$~3.5.  There are only two additional SB1s with 0~$<$~log\,$P$~$<$~1, and one or both of these systems may contain post-CE WD companions (see Appendix A.1).  The radial velocity variables and companions implied through proper motion acceleration have systematically longer orbital periods 2.5~$<$~log~$P$~$<$~4.7 \citep{Raghavan2010}.  The $\approx$\,3\,-\,5 systems that escaped detection lie in the interval 3.5~$<$~log~$P$~$<$~4.7 (\S8.2).  We therefore expect one, four, six, six, and four additional stellar companions in the logarithmic period intervals log\,$P$ $=$ 0\,-\,1, 1\,-\,2, 2\,-\,3, 3\,-\,4, and 4.0\,-\,4.7, respectively.

  In terms of mass ratios $q$, we assume the 21 additional systems are evenly distributed between $q$ = 0.1 and the detection limit at $q$ = 0.40\,-\,0.55 indicated by the blue line in Fig.~\ref{solarq.fig}.  Weighting the additional systems toward smaller or larger mass ratios in this interval does not significantly affect our statistical measurements.  We display the 21 additional M-dwarf companions as the blue diamonds in Fig.~\ref{solarq.fig}.

In \S8.3.2, we determined that ${\cal F}_{\rm solar+WD}$~=~(11\,$\pm$\,4)\% of solar-type stars in the local solar-neighborhood have WD companions.  In other words, the correction factor due to binary evolution is ${\cal C}_{\rm evol}$ = 1.11\,$\pm$\,0.04 for solar-type primaries  (see \S2).  Of the 454 solar-type systems in the \citet{Raghavan2010} sample, we estimate that 454\,$\times$\,0.11~$\approx$~50 have WD companions.  Some of the brighter WD companions have been identified \citep{Raghavan2010}, but the majority are probably too faint with binary brightness contrasts $\Delta$V $\gtrsim$~10~mag that are too large to be detected.  The true number of solar-type primaries in the \citet{Raghavan2010} sample, i.e., those that have always been the most massive components of their respective systems, is  ${\cal N}_{\rm prim}$ = 454\,$-$\,50 = 404.  The WD contamination to the total sample ($\approx$11\%) is significantly smaller than the WD contamination to SB1s ($\approx$30\%).  In conclusion, the corrected solar-type sample contains ${\cal N}_{\rm prim}$~=~404 primaries and ${\cal N}_{\rm comp}$~=~193 stellar companions with 0.1~$<$~$q$~$<$~1.0 and 0.0~$<$~log\,$P$\,(days)~$<$~8.0.

\begin{figure}[t!]
\centerline{
\includegraphics[trim=0.8cm 0.2cm 0.3cm 0.1cm, clip=true, width=3.6in]{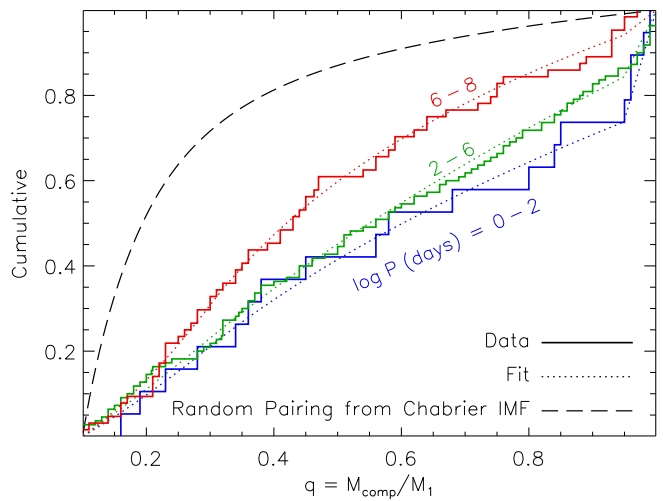}}
\caption{Cumulative distribution of mass ratios $q$ for solar-type binaries divided into three logarithmic period intervals.  We compare the data (solid lines) after correcting for incompleteness to the fits (dotted; parameters presented in Table~11).  The short-period systems (blue) with 0~$<$~log~$P$~(days)~$<$~2 are weighted toward large mass ratios, primarily due to a large excess fraction ${\cal F}_{\rm twin}$ $\approx$ 0.29 of twins with $q$ = 0.95\,-\,1.00. The intermediate-period systems (green) with 2~$<$~log\,$P$~$<$~6 have power-law slopes {\large $\gamma$}$_{\rm smallq}$~$\approx$~{\large $\gamma$}$_{\rm largeq}$~$\approx$~0.0 close to uniform, but with an excess twin fraction ${\cal F}_{\rm twin}$ $\approx$ 0.14 that is half the value of the short-period sample.  Meanwhile, the long-period systems (red) with 6~$<$~log~$P$~$<$~8 have a negligible excess twin fraction ${\cal F}_{\rm twin}$ $\approx$ 0.02 and a power-law component {\large $\gamma$}$_{\rm largeq}$~$\approx$~$-$1.1 that favors small mass ratios $q$ $\approx$ 0.2\,-\,0.5. As found for early-type binaries, solar-type binaries become weighted toward smaller mass ratios with increasing orbital period. Unlike the early-type binaries, however, even the widest solar-type MS binaries are markedly inconsistent with random pairings drawn from a \citet{Chabrier2003} IMF (dashed line). }
\label{solarcumq.fig}
\end{figure}

\subsection{Intrinsic Binary Statistics}

In Table~11, we provide the statistics based on the corrected population of solar-type binaries.  We initially divide the sample into eight intervals of $\Delta$log$P$~=~1 across 0~$<$~log\,$P$\,(days)~$<$~8.  For each logarithmic period interval, we list the number ${\cal N}_{\rm smallq}$ of companions with 0.1~$<$~$q$~$<$~0.3 and the number ${\cal N}_{\rm largeq}$ of companions with 0.3~$<$~$q$~$<$~1.0.  The frequency $f_{\rm logP;q>0.3}$ of companions with $q$ $>$ 0.3 per decade of orbital period simply derives from dividing each value of ${\cal N}_{\rm largeq}$ by ${\cal N}_{\rm prim}$ = 404.  

To analyze the mass-ratio distribution, we divide the sample into four intervals of $\Delta$log$P$~=~2 across 0~$<$~log\,$P$\,(days)~$<$~8.  For each of these four subsamples, we measure {\large $\gamma$}$_{\rm smallq}$, {\large $\gamma$}$_{\rm largeq}$, and ${\cal F}_{\rm twin}$ (Table~11). In Fig.~30, we display the cumulative distribution of mass ratios 0.1~$<$~$q$~$<$~1.0 for three logarithmic period intervals: 0~$<$~log\,$P$~$<$~2, 2~$<$~log\,$P$~$<$~6, and 6~$<$~log\,$P$~$<$~8.  

The most noticeable trend in the mass-ratio distribution of solar-type binaries is that the excess twin fraction ${\cal F}_{\rm twin}$ significantly decreases with orbital period.  At short periods 0~$<$~log\,$P$~$<$~2, we measure a large excess twin fraction ${\cal F}_{\rm twin}$ = 0.29\,$\pm$\,0.11.  This is consistent with the conclusions of \citet{Halbwachs2003}, who also find a substantial excess twin population among solar-type spectroscopic binaries.  As can be seen in Figs.~28 and~30, 5 of the 15 solar-type binaries with 0~$<$~log\,$P$~$<$~2 and $q$~$>$~0.3 have $q$ = 0.95\,-\,1.00, implying ${\cal F}_{\rm twin}$~$\approx$~5/15~$\approx$~0.3.  At intermediate orbital periods 2~$<$~log\,$P$~$<$~6, the excess twin fraction ${\cal F}_{\rm twin}$~=~0.1\,-\,0.2 is smaller but definitively non-zero at a statistically significant level (see Figs.~28 and 30).  Only at the widest orbital separations, i.e. 6~$<$~log\,$P$~$<$~8, is the excess twin fraction ${\cal F}_{\rm twin}$ $<$ 0.05 negligible.  

For solar-type binaries, the power-law components of the mass-ratio distribution exhibit only minor variations with orbital period.  In fact, the slopes {\large $\gamma$}$_{\rm smallq}$ $\approx$ 0.3 and {\large $\gamma$}$_{\rm largeq}$ $\approx$ $-$0.5 are relatively constant and mildly consistent with a uniform distribution {\large $\gamma$}~=~0.0 for 0~$<$~log\,$P$~$<$~6.  Only at the longest orbital periods 6~$<$~log\,$P$~$<$~8 does the power-law slope {\large $\gamma$}$_{\rm largeq}$ $\approx$ $-$1.1 become weighted toward smaller mass ratios.  As can be seen in Fig.~28, there is an enhanced concentration of companions with $q$~=~0.1\,-\,0.5 across log $P$~=~5.5\,-\,8.0. As found for early-type binaries, the excess twin fraction ${\cal F}_{\rm twin}$ and power-law component {\large $\gamma$}$_{\rm largeq}$  both decrease with increasing orbital period.

 Using a Monte Carlo technique, we generate a binary star population based on random pairings drawn from a \citet{Chabrier2003} IMF (Salpeter power-law slope above 1\,\Msun\ and log-normal distribution below 1\,\Msun).  From the simulated population, we select the binaries with solar-type primaries $M_1$ = 0.75\,-\,1.25\,\Msun.  We display the cumulative distribution of mass ratios for these solar-type binaries in Fig.~30.  By selecting a narrow interval of primary masses $M_1$~=~0.75\,-\,1.25\,\Msun, the distribution of mass ratios across $q$~=~0.1\,-\,1.0 is nearly indistinguishable from the distribution of masses $M$~=~0.1\,-\,1.0\,\Msun\ according to the IMF \citep[see also][]{Tout1991,Kouwenhoven2009}.  For all orbital periods 0~$<$~log~$P$~$<$~8, the observed population of solar-type  MS binaries is inconsistent with random pairings drawn from the IMF.   Despite the enhancement of $q$~=~0.1\,-\,0.5 companions across log $P$ = 6.0\,-\,8.0, the widest solar-type MS binaries have a top-heavy mass-ratio distribution relative to the IMF random pairing prediction (Fig.~30). Our results are consistent with \citet{Lepine2007}, who also find that wide solar-type binaries identified via common proper motion do not match random pairings from the field population.

\subsection{Comparison with Duquennoy \& Mayor (1991)}

Knowing their survey was partially incomplete, especially toward smaller mass ratios $q$~$<$ ~0.3, \citet{Duquennoy1991} took great lengths to correct for incompleteness in their sample.  The more recent analysis by \citet{Raghavan2010} nearly tripled the sample of solar-type primaries. \citet{Raghavan2010} also employed new observational techniques, e.g., adaptive optics and long-baseline interferometry,  to discover stellar companions in portions of the $f(P,\,q)$ parameter space that eluded \citet{Duquennoy1991}.  The raw sample of \citet{Raghavan2010} is certainly larger and more complete than the raw sample of \citet{Duquennoy1991}.  However, believing their sample was mostly complete, \citet{Raghavan2010} did not conduct as detailed an incompleteness study of their survey.  Because of this, there has been tension reported in the literature as to whether the corrected sample of \citet{Duquennoy1991} or the raw sample of \citet{Raghavan2010} is a better representation of the true statistics of solar-type binaries \citep{Marks2011,Geller2012}.  Because we conducted our own incompleteness study of \citet{Raghavan2010}, we can finally address this concern.

After correcting for incompleteness down to $q$~$\approx$~0.1,  \citet{Duquennoy1991} estimated there to be 91 companions with 0~$<$~log\,$P$\,(days)~$<$~8 in their sample of 164 solar-type primaries (see their Fig.~7). We show their corrected frequency of companions per decade of orbital period per primary in Fig.~31 (blue histogram).  Based on their corrected population, the multiplicity frequency of solar-type primaries is 0.55\,$\pm$\,0.06. However, \citet{Duquennoy1991} included WD companions in their corrected sample.  In fact, in their effort to account for incompleteness, \citet{Duquennoy1991} artificially added 2 WDs per decade of orbital period across intermediate separations.  We estimate there to be 13 WD companions total in the corrected sample of \citet{Duquennoy1991}.  After removing these 13 systems with WDs, there are 78 companions across periods log\,$P$~=~0\,-\,8 to 151 solar-type primaries.  We show in Fig.~31 the frequency of companions per decade of orbital period after applying our correction of removing the WDs from the \citet{Duquennoy1991} sample (green histogram).  By removing the WDs, the multiplicity frequency decreases slightly to 0.52\,$\pm$\,0.06.

\begin{figure}[t!]
\centerline{
\includegraphics[trim=0.6cm 0.0cm 0.3cm -0.1cm, clip=true, width=3.55in]{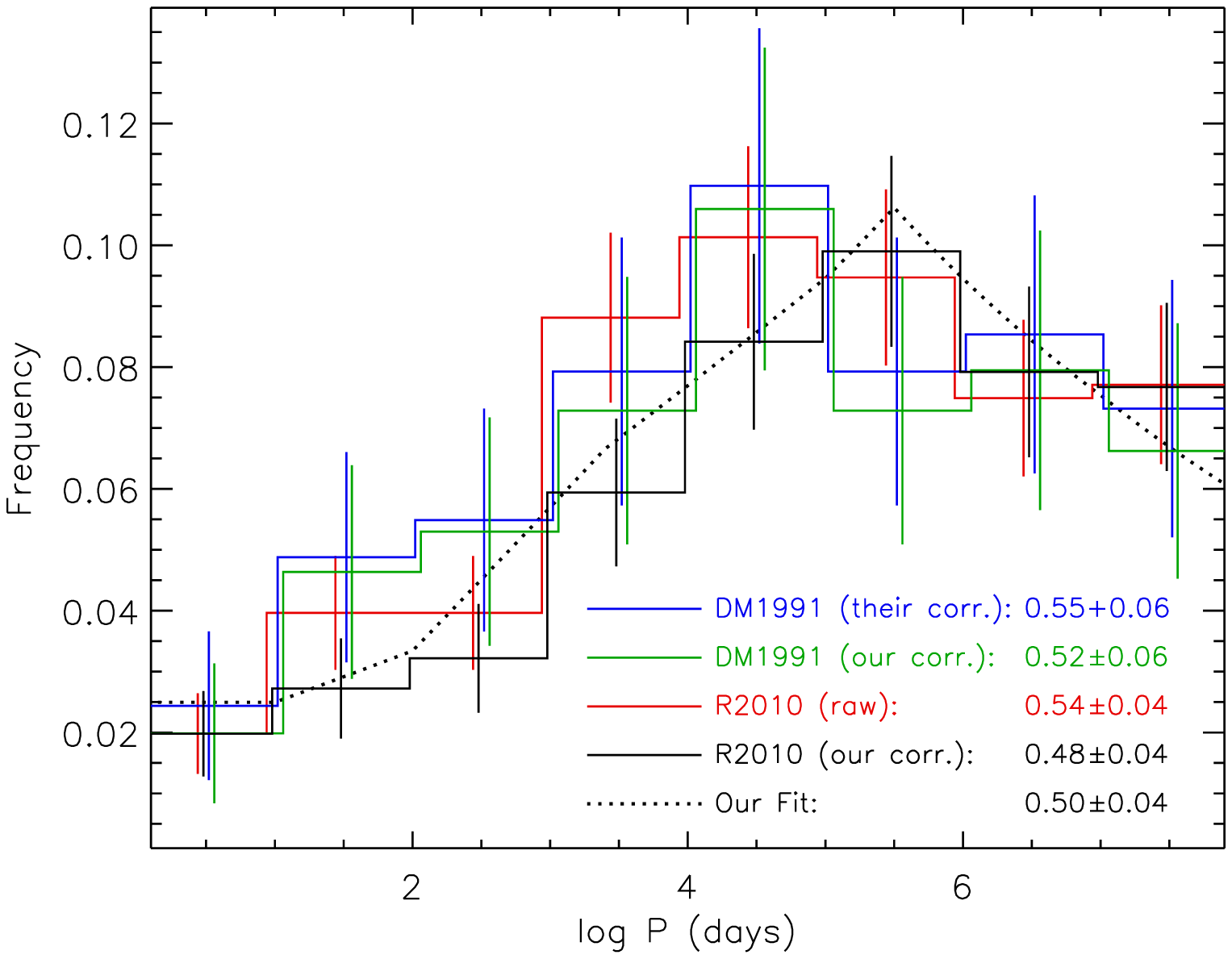}}
\caption{The frequency of companions per decade of orbital period per solar-type MS primary based on different surveys and corrections applied to those surveys.  We compare the \citet{Duquennoy1991} sample based on their corrections for incompleteness (blue), the corrected \citet{Duquennoy1991} sample after removing systems with WD companions (green), the raw \citet{Raghavan2010} sample (red), and the \citet{Raghavan2010} sample after applying our corrections for selection biases (black).  For each sample, we list the multiplicity frequency of companions per primary.  Our analytic fit (dotted) adopted in the present study results in a multiplicity frequency of 0.50\,$\pm$\,0.04 MS companions with $q$ $>$ 0.1 and 0~$<$~log\,$P$\,(days)~$<$~8 per solar-type MS primary.  }
\end{figure}

\citet{Raghavan2010} identified 243 confirmed companions with log\,$P$~=~0\,-\,8 in their sample of 454 solar-type primaries (see their Fig. 13).  This results in a multiplicity frequency of 0.54\,$\pm$\,0.04, and we show the period distribution of the raw \citet{Raghavan2010} sample in Fig.~31 (red histogram).  After accounting for incompleteness and removing extreme mass-ratio companions $q$~$<$~0.1, companions to M-dwarf and late-K secondaries in A\,-\,(Ba,\,Bb) hierarchical triples, and systems with WD companions, we find 193 companions with $q$ $>$ 0.1 and log\,$P$~=~0\,-\,8 to 404 solar-type primaries (Table 11; black histogram in our Fig.~31).  The multiplicity frequency of the \citet{Raghavan2010} sample after accounting for these various selection effects is 0.48\,$\pm$\,0.04. 

The largest difference in the multiplicity frequency stems not from comparing \citet{Duquennoy1991} and \citet{Raghavan2010}, but by comparing the statistics before and after removing WD companions.  Nevertheless, all four different versions of the multiplicity frequency and period distribution displayed in Fig.~31 are consistent with each other within the 2$\sigma$ uncertainty levels. Our analytic fit to the period distribution of companions to solar-type primaries (see \S9.3) is between the \citet{Duquennoy1991} and \citet{Raghavan2010} distributions after applying our corrections and removing WD companions.  This fit results in a multiplicity frequency of 0.50\,$\pm$\,0.04 companions with $q$ $>$ 0.1 and log\,$P$~=~0\,-\,8 per solar-type primary.  We conclude that the differences between the corrected \citet{Duquennoy1991} sample and raw \citet{Raghavan2010} sample are negligible, and, most importantly, smaller than the systematic uncertainties in {\it how} corrections for incompleteness and WD companions are applied.

\subsection{Eccentricity Distribution}

\begin{figure}[t!]
\centerline{
\includegraphics[trim=0.6cm 0.0cm 0.3cm -0.1cm, clip=true, width=3.55in]{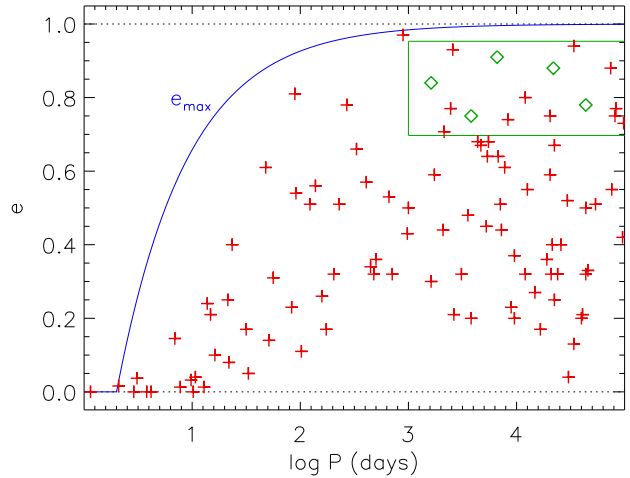}}
\caption{Eccentricities $e$ versus orbital periods $P$ for solar-type binaries from the \citep{Raghavan2010} sample.  We display the 97 binaries (red) with spectroscopic and/or visual orbit solutions.  Five additional detected systems with 3 $<$ log\,$P$\,(days) $<$ 5 do not have visual orbits and most likely have $e$ $\approx$ 0.70\,-\,0.95 (five green systems within green region).  Solar-type binaries with log\,$P$~$<$~1 have been tidally circularized, while longer period systems are weighted toward large eccentricities.}
\label{solare.fig}
\end{figure}

 We next measure the eccentricity probability distribution $p_e$ $\propto$ $e^{\eta}$ of solar-type binaries as a function of orbital period $P$.  In Fig.~\ref{solare.fig}, we plot $e$ versus $P$ for the 97 solar-type binaries in the \citet{Raghavan2010} sample with spectroscopic and/or visual orbit solutions and 0~$<$~log\,$P$\,(days)~$<$~5.  Our Fig.~\ref{solare.fig} is quite similar to Fig.~14 in \citet{Raghavan2010}.  However, we do not include the data points that actually represent the orbits of late-K and M-dwarf binaries.  For example, the two systems near $e$ = 0.12 and log\,$P$ $\approx$~3.9 in Fig.~14 of \citet{Raghavan2010} are the orbits of low-mass binaries with tertiary solar-type primaries in a A\,-\,(Ba,\,Bb) hierarchical configuration (see \S8.1).  We remove these two systems and four additional low-mass binaries  with 4~$<$~log\,$P$\,(days)~$<$~5 in the \citet{Raghavan2010} survey from our sample.

All of the 44 detected companions to solar-type primaries with 0~$<$ log\,$P$\,(days)~$<$~3 have measured eccentricities according to spectroscopic and/or visual orbit solutions.  At 3 $<$ log $P$ $<$ 4 and 4 $<$ log $P$ $<$ 5, however, three and two detected companions, respectively, do not have measured eccentricities.  Visual binaries which do not have reliable orbital solutions generally have large eccentricities $e$ $\gtrsim$ 0.7 \citep{Harrington1977,Tokovinin2015}.   We assume the five visual binaries with intermediate periods 3~$<$~log~$P$~$<$~5 but without orbital solutions are evenly distributed across $e$ $\approx$ 0.70\,-\,0.95 (green systems in Fig.~\ref{solare.fig}). 

We also display in Fig.~\ref{solare.fig} the maximum eccentricity $e_{\rm max}$ as a function of $P$ according to Eqn.~3. As expected, all detected systems have $e$ $<$ $e_{\rm max}$.  In fact, the majority of systems with log $P$ $<$ 1 have been tidally circularized.  In this short-period interval, we measure the power-law component $\eta$ = $-$0.8\,$\pm$\,0.2 of the eccentricity distribution to be weighted toward small values (Table~11).  Solar-type binaries at longer orbital periods log $P$ $>$ 1 not only contain systems with large eccentricities, but exhibit a deficit of binaries with small eccentricities $e$~$\lesssim$~0.15.  

\begin{figure}[t!]
\centerline{
\includegraphics[trim=4.2cm 0.0cm 4.8cm 0.0cm, clip=true, width=3.1in]{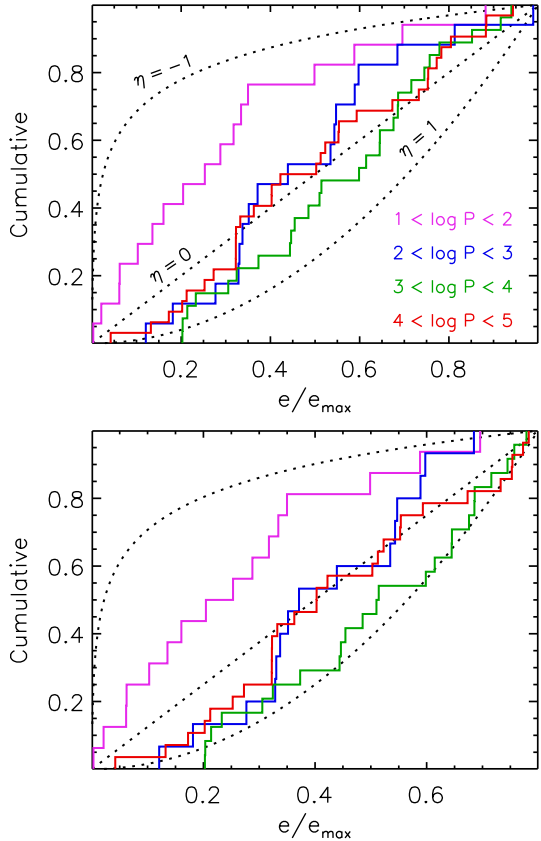}}
\caption{Cumulative distribution of eccentricities $e$/$e_{\rm max}$ for solar-type binaries with log\,$P$\,(days) = 1\,-\,2 (magenta), 2\,-\,3 (blue), 3\,-\,4 (green), and 4\,-\,5 (red).  Top panel: by fitting all solar-type binaries, we measure power-law slopes $\eta$ = $-$0.4, 0.1, 0.4, and 0.2 for log $P$ = 1\,-\,2, 2\,-\,3, 3\,-\,4, and 4\,-\,5, respectively.  Bottom panel: by fitting only those systems with $e$ $<$ 0.8$e_{\rm max}$ that have not been severely affected by tidal evolution, we measure $\eta$ = $-$0.4, 0.3, 0.8, and 0.4 for log $P$ = 1\,-\,2, 2\,-\,3, 3\,-\,4, and 4\,-\,5, respectively.}
\label{solarcume.fig}
\end{figure}

In the top panel of Fig.~\ref{solarcume.fig}, we display the cumulative distributions of $e$/$e_{\rm max}$ for four different logarithmic period intervals.  We measure $\eta$ = $-$0.4, 0.1, 0.4, and 0.2 for log $P$ = 1\,-\,2, 2\,-\,3, 3\,-\,4, and 4\,-\,5, respectively. At these wide separations, the tidal circularization timescales are longer than the ages $\langle \tau \rangle$ $\approx$ 5 Gyr of solar-type binaries \citep{Zahn1977,Hut1981}.  As expected, we do not find any circularized solar-type MS binaries beyond $P$~$>$~20~days \citep[see also][]{Meibom2005}.  However, solar-type binaries initially born with $e$~$>$~0.8 and log\,$P$~$>$~2 tidally evolve toward smaller eccentricities $e$~$<$~0.8 on significantly shorter timescales $\tau_{\rm tide}$~$\propto$~(1$-e^2$)$^{\nicefrac{13}{2}}$ \citep{Zahn1977,Hut1981}.  In the bottom panel of Fig.~\ref{solarcume.fig}, we display the cumulative distribution of eccentricities for only those systems with $e/e_{\rm max}$~$<$~0.8 that are not as severely affected by tidal effects.  By fitting the power-law component $\eta$ across 0~$<$~$e$/$e_{\rm max}$~$<$~0.8, we measure $\eta$ = $-$0.4, 0.3, 0.8, and 0.4 for the intervals log $P$ = 1\,-\,2, 2\,-\,3, 3\,-\,4, and 4\,-\,5, respectively.  We average the two methods for determining $\eta$, and report  $\eta$ = $-$0.4\,$\pm$\,0.3, 0.2\,$\pm$\,0.3, 0.6\,$\pm$\,0.4, and 0.3\,$\pm$\,0.3 for log $P$ = 1\,-\,2, 2\,-\,3, 3\,-\,4, and 4\,-\,5, respectively, in Table 11.  For solar-type binaries, the power-law component $\eta$ of the eccentricity distribution increases with orbital period.  Nonetheless, the measured values of $\eta$~$\approx$~0.2\,-\,0.6 at intermediate periods log~$P$~=~2\,-\,5 are mildly discrepant with a thermal eccentricity distribution ($\eta$ = 1).  

Finally, to investigate the eccentricities $e$ of solar-type binaries as a function of age $\tau$, we examine the data set shown in Fig.~8 of \citet{Meibom2005}.  In their Fig.~8, \citet{Meibom2005} compare $e$ versus $\tau$ for solar-type binaries in eight different environments with various ages.  They find the tidal circularization period increases from $P_{\rm circ}$~$\approx$~7~days for pre-MS binaries up to $P_{\rm circ}$~$\approx$~15 days for the halo population.   In our present study, we are most interested in the eccentricity distributions of binaries with periods log~$P$~$>$~1.2 beyond the circularization period. Observational selection biases become too important in the \citet{Meibom2005} sample toward long orbital periods log~$P$~$>$~2.4 and large eccentricities $e$~$>$~0.6.  We therefore select the 110 solar-type binaries from the \citet{Meibom2005} data set with orbital periods 1.2~$<$~log\,$P$~$<$~2.4 and eccentricities $e$~$<$~0.6.  We divide our sample into two subsets: the 49 systems with ages $\tau$~$<$~700~Myr contained in panels a\,-\,d on the left side of Fig.~8 in \citet{Meibom2005}, and the 61 binaries with ages $\tau$~$\approx$~3\,-\,10~Gyr contained in panels e\,-\,h on the right side of their Fig.~8.

In Fig.~34, we display the cumulative distribution of eccentricities 0.0~$<$~$e$~$<$~0.6 of solar-type binaries with intermediate periods for our young and old populations.  Even with two large subsamples, the eccentricity distributions of young and old solar-type binaries are surprisingly similar.  A K-S test shows that the two populations are consistent with each other at the 92\% confidence level.  This demonstrates that tidal evolution during the MS is inappreciable for solar-type binaries with intermediate periods 1.2~$<$~log\,$P$~$<$~2.4 and modest eccentricities $e$~$<$~0.6.

By fitting our power-law distribution $p_e$~$\propto$~$e^{\eta}$ to the data, we measure $\eta$~=~0.2\,$\pm$\,0.3 and $\eta$~=~0.1\,$\pm$\,0.3 for the young and old populations, respectively (Table~12).  These values are consistent with our measurements of $\eta$~=~-0.4\,$\pm$0.3 at slightly shorter periods log\,$P$~=~1.0\,-\,2.0 and of $\eta$~=~0.2\,$\pm$0.3 at slightly longer periods log\,$P$~=~2.0\,-\,3.0 based on the \citet{Raghavan2010} survey (see Table~11).  Upon visual inspection of Fig.~34, we find it possible that a single-parameter distribution may not necessarily fully describe the data.  For example, both the young and old cumulative eccentricity distributions lie systematically below the $\eta$~=~0 curve across $e$~=~0.10\,-\,0.25, and then they cross above the $\eta$~=~0 distribution near $e$~$\approx$~0.45.  A two-parameter eccentricity distribution, e.g., a Gaussian, may better fit the data.  However, a K-S test reveals that both the old and young eccentricity distributions are consistent with a power-law distribution at the $\approx$\,90\% significance level.  Moreover, selection biases and tidal evolution at large eccentricities $e$~$\gtrsim$~0.8$e_{\rm max}$ are not fully understood, and so the initial zero-age MS eccentricity distribution is not well constrained in this portion of the parameter space.  With the current samples, we do not find sufficient discrepancy between our adopted single-component power-law distribution and the data to warrant a different probability distribution and/or additional parameters.

\renewcommand{\arraystretch}{1.6}
\setlength{\tabcolsep}{10pt}
\begin{figure}[t!]\footnotesize
{\small {\bf Table 12:} Measurements of the eccentricity distribution based on solar-type binaries ($M_1$ = 1.0\,$\pm$\,0.3\,\Msun) with log\,$P$\,(days)~=~1.8\,$\pm$\,0.6 contained within the \citet{Meibom2005} data set.}  \\
\vspace*{-0.45cm}
\begin{center}
\begin{tabular}{|c|c|}
\hline
 Age  & $\eta$ \\
\hline
$\tau$ $<$ 700 Myr & 0.2\,$\pm$\,0.3 \\
\hline
$\tau$ $>$ 3 Gyr & 0.1\,$\pm$\,0.3 \\
\hline
\end{tabular}
\end{center}
\end{figure}
\renewcommand{\arraystretch}{1.0}

\begin{figure}[t!]
\centerline{
\includegraphics[trim=0.6cm 0.2cm 0.1cm 0.3cm, clip=true, width=3.3in]{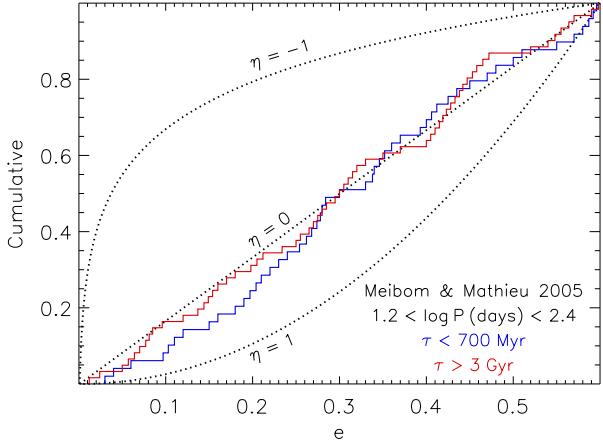}}
\caption{Cumulative distribution of eccentricities 0.0~$<$~$e$~$<$0.6 for solar-type binaries with orbital periods 1.2~$<$~log\,$P$\,(days)~$<$~2.4 contained within the \citet{Meibom2005} data set.  We compare the 49 binaries with ages $\tau$~$<$~700~Myr (blue) to the 61 systems in older environments with $\tau$~$>$~3~Gyr (red).  The two subsamples are both relatively unbiased and surprisingly similar, demonstrating that tidal evolution is negligible for MS solar-type binaries with modest eccentricities $e$~$<$~0.6 and orbital periods log\,$P$~$>$~1.2 slightly beyond the circularization period. We measure the power-law component of the eccentricity distribution to be $\eta$~=~0.2\,$\pm$\,0.3 and $\eta$~=~0.1\,$\pm$\,0.3 for the young and old populations, respectively. }
\end{figure}

\section{Probability Density Functions}

In the following, we compare, fit, and analyze the five statistical parameters $f_{\rm logP;q>0.3}$, {\large$\gamma$}$_{\rm smallq}$, {\large$\gamma$}$_{\rm largeq}$, ${\cal F}_{\rm twin}$, and $\eta$ as a function of primary mass $M_1$ and orbital period $P$ (\S9.1-9.3).  Using our relations for $f_{\rm logP;q>0.3}$ and the parameters {\large$\gamma$}$_{\rm smallq}$, {\large$\gamma$}$_{\rm largeq}$, and ${\cal F}_{\rm twin}$ that describe the mass-ratio distribution down to $q$~=~0.1, we compute the frequency $f_{\rm logP;q>0.1}$ of companions with $q$~$>$~0.1 per decade of orbital period.  We also measure the total MS multiplicity frequencies and multiplicity fractions based on these statistics (\S9.4).

\subsection{Mass-ratio Distributions}

\begin{figure*}[t!]
\centerline{
\includegraphics[trim=3.3cm 0.0cm 3.9cm 0.2cm, clip=true, width=6.9in]{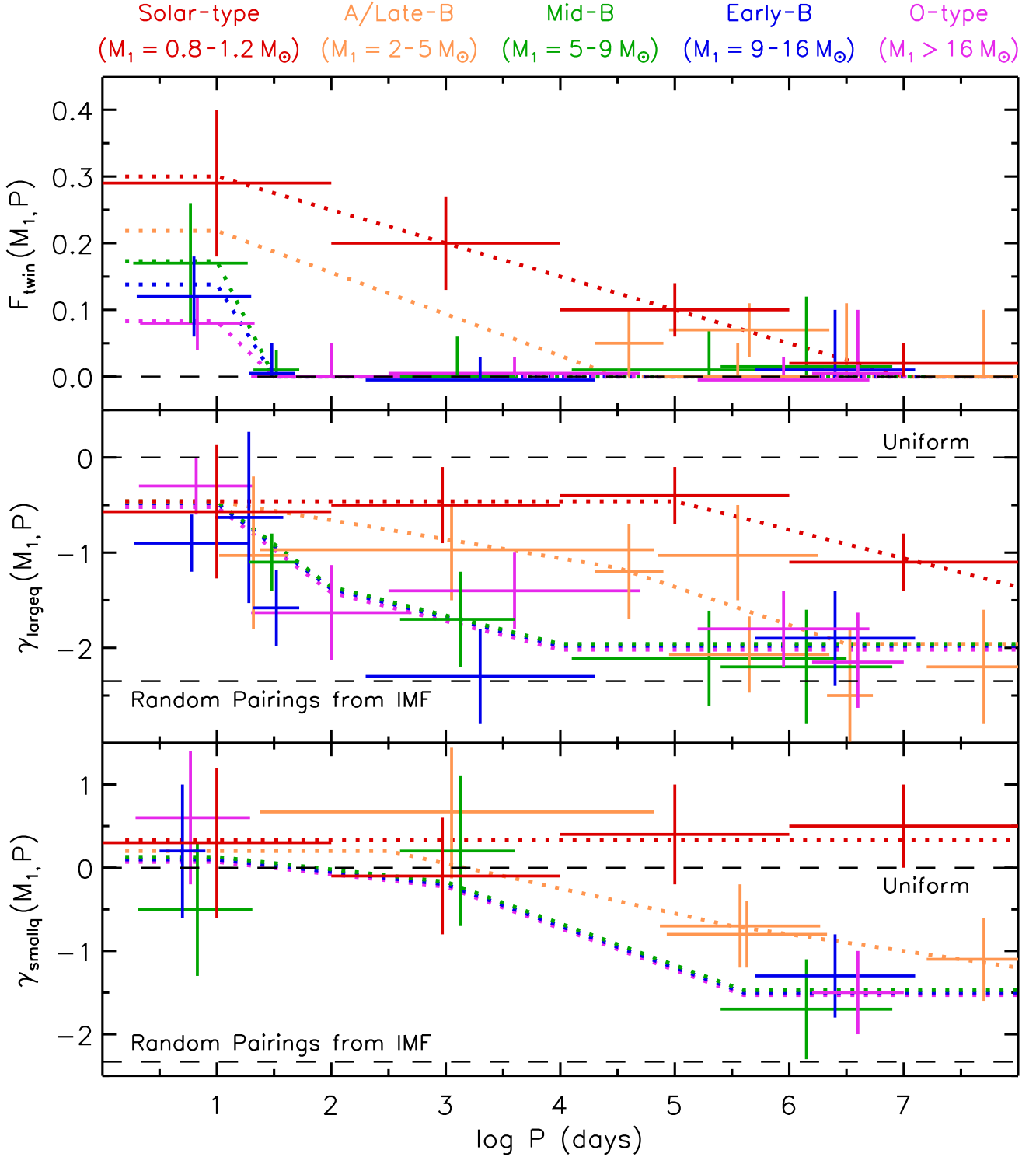}}
\caption{The three statistical parameters that describe the mass-ratio probability distribution $p_q$ (see Fig.~2) as a function of orbital period $P$ and colored according to primary mass $M_1$.  We display the excess fraction ${\cal F}_{\rm twin}$\,($M_1$,\,$P$) of twin components with $q$~$>$~0.95 (top panel), the power-law slope {\large$\gamma$}$_{\rm largeq}$\,($M_1$,\,$P$) of the distribution across large mass ratios $q$~$>$~0.3 (middle panel), and the power-law slope {\large$\gamma$}$_{\rm smallq}$\,($M_1$,\,$P$) of the distribution across small mass ratios $q$~=~0.1\,-\,0.3 (bottom panel).  We display all our measurements after correcting for incompleteness and selection effects.  We group the data into five primary mass / MS spectral type intervals: solar-type ($M_1$~=~0.8\,-\,1.2\,\Msun; red), A/late-B ($M_1$~=~2\,-\,5\,\Msun; orange), mid-B ($M_1$~=~5\,-\,9\,\Msun; green), early-B ($M_1$~=~9\,-\,16\,\Msun; blue), and O-type ($M_1$~$>$~16\,\Msun; magenta). We fit analytic functions (dotted) to the observations.  For comparison, ${\cal F}_{\rm twin}$ = 0.0 and {\large$\gamma$}$_{\rm largeq}$ = {\large$\gamma$}$_{\rm smallq}$ = 0.0 provide a uniform mass-ratio distribution, while ${\cal F}_{\rm twin}$ = 0.0 and {\large$\gamma$}$_{\rm largeq}$ = {\large$\gamma$}$_{\rm smallq}$ = $-$2.35 imply random pairings drawn from a Salpeter IMF (dashed lines).  Qualitatively, for all primary masses $M_1$, binaries become weighted toward smaller mass ratios $q$ with increasing orbital period $P$.  Quantitatively, the variations in ${\cal F}_{\rm twin}$, {\large$\gamma$}$_{\rm largeq}$, and {\large$\gamma$}$_{\rm smallq}$ according to $M_1$ and $P$ are rather complex.  These variations in the mass-ratio distribution provide important clues and diagnostics for binary star formation and evolution (see text for details). }
\end{figure*}

  After combining all our measurements in \S3-8 and Tables 2\,-\,12, we display in Fig.~35 the excess twin fraction ${\cal F}_{\rm twin}$ (top panel) as a function of orbital period $P$ and colored according to primary mass $M_1$.  We group the data points into five spectral subtype intervals: solar-type ($M_1$~=~0.8\,-\,1.2\,\Msun; red), A/late-B ($M_1$~=~2\,-\,5\,\Msun; orange), mid-B ($M_1$~=~5\,-\,9\,\Msun; green), early-B ($M_1$~=~9\,-\,16\,\Msun; blue), and O-type ($M_1$~$>$~16\,\Msun; magenta).  The excess twin fraction ${\cal F}_{\rm twin}$ exhibits three main aspects.  First, for solar-type primaries, the excess twin fraction decreases linearly with respect to log~$P$.  Second, for mid-B, early-B, and O-type primaries, the excess twin fraction is zero for $P$~$\gtrsim$~20~days.  Finally, at shorter periods, the excess twin fraction is anti-correlated with respect to log~$M_1$.  We find the following relation fits these three aspects of the data:

\begin{align}
{\cal F}_{\rm twin}\,(M_1,P)~=&~{\cal F}_{\rm twin;logP<1}~~{\rm for}~{\rm log}\,P\,<\,1, \nonumber \\
 &~{\cal F}_{\rm twin;logP<1}\Big(1 - \frac{{\rm log}\,P - 1}{{\rm log}\,P_{\rm twin}-1}\Big) \nonumber \\
 &~~~~~~~{\rm for}~1\,\le\,{\rm log}\,P\,<\,{\rm log}\,P_{\rm twin},  \nonumber \\
 &~0.0~~{\rm for}~{\rm log}\,P\,\ge\,{\rm log}\,P_{\rm twin},
\end{align}

\noindent where the excess twin fraction at short orbital periods is:

\begin{equation}
 {\cal F}_{\rm twin;logP<1}\,(M_1)~=~0.30 - 0.15{\rm log}\Big(\frac{M_1}{{\rm M}_{\odot}}\Big),
\end{equation}

\noindent and the maximum orbital period of the observed excess twin population is:

\begin{align}
 {\rm log}\,P_{\rm twin}\,(M_1)= &~8.0-\frac{M_1}{{\rm M}_{\odot}}\,&\,{\rm for}~M_1\,\le\,6.5\,{\rm M}_{\odot}, \nonumber \\
                                &~1.5\,&\,{\rm for}~M_1\,>\,6.5\,{\rm M}_{\odot}. 
\end{align}

We compare the analytic fit (Eqn.~5; dotted lines) to the data for ${\cal F}_{\rm twin}$($M_1$,\,$P$) in the top panel of Fig.~35.  We evaluate the fit according to the average primary mass within each spectral subtype interval, i.e.,  $M_1$ = 1, 3.5, 7, 12, and 28\,\Msun\ for solar-type, A/late-B, mid-B, early-B, and O-type primaries, respectively. The fit adequately describes the excess twin fraction as continuous functions of $M_1$ and $P$.  In fact, the fit nearly coincides with the majority of the observed values. The two non-zero ${\cal F}_{\rm twin}$ measurements for A-type stars (orange data points at log\,$P$~$\approx$~4.6 and 5.6 in top panel of Fig.~35) correspond to $\langle M_1 \rangle$~=~2.0\,\Msun, and therefore lie just above the evaluation of the fit at $M_1$~=~3.5\,\Msun\ (orange dotted line).  

There are still uncertainties $\delta {\cal F}_{\rm twin}$ in the excess twin fraction according to the precision of the observational measurements. For all periods and primary masses, we find the 1$\sigma$ uncertainty in the excess twin fraction to be:

\begin{equation}
 \delta {\cal F}_{\rm twin}\,(M_1,\,P) = {\rm max}\{0.03,\,0.3{\cal F}_{\rm twin}\}.
\end{equation}

In the middle panel of Fig.~35, we display the power-law slope {\large$\gamma$}$_{\rm largeq}$ of the mass-ratio distribution across large mass ratios $q$~=~0.3\,-\,1.0.  For all primary masses $M_1$ and for short orbital periods log\,$P$\,(days)~$<$~1.0, the slope is {\large$\gamma$}$_{\rm largeq}$~$\approx$~$-$0.5.  At longer orbital periods, the power-law component {\large$\gamma$}$_{\rm largeq}$ decreases, but the break toward steeper slopes depends significantly on $M_1$. For solar-type primaries, we find:

\begin{align}
 \mbox{{\large $\gamma$}}_{\rm largeq}\,(0.8\,{\rm M}_{\odot}<M_1<1.2&\,{\rm M}_{\odot},\,P)= \nonumber \\
-0.5                         &~{\rm for}~0.2\,\le\,{\rm log}\,P\,<\,5.0, \nonumber \\
-0.5 - 0.3({\rm log}\,P - 5) &~{\rm for}~5.0\,\le\,{\rm log}\,P\,<\,8.0. 
\end{align}

\noindent For the midpoint of A/late-B primaries, we measure:

\begin{align}
 \mbox{{\large $\gamma$}}_{\rm largeq}\,(M_1=3.5\,{\rm M}_{\odot}&,\,P)= \nonumber \\
 -0.5                           &~{\rm for}~0.2\,\le\,{\rm log}\,P\,<\,1.0, \nonumber \\
 -0.5 - 0.2({\rm log}\,P - 1.0) &~{\rm for}~1.0\,\le\,{\rm log}\,P\,<\,4.5, \nonumber \\
 -1.2 - 0.4({\rm log}\,P - 4.5) &~{\rm for}~4.5\,\le\,{\rm log}\,P\,<\,6.5, \nonumber \\
 -2.0                           &~{\rm for}~6.5\,\le\,{\rm log}\,P\,<\,8.0.  
\end{align}

\noindent Finally, for mid-B, early-B, and O-type stars, we fit:

\begin{align}
 \mbox{{\large $\gamma$}}_{\rm largeq}\,(M_1 > 6.0\,{\rm M}_{\odot}&,\,P)= \nonumber \\
 -0.5                          &~{\rm for}~0.0\,\le\,{\rm log}\,P\,<\,1.0, \nonumber \\
 -0.5 - 0.9({\rm log}\,P - 1)  &~{\rm for}~1.0\,\le\,{\rm log}\,P\,<\,2.0, \nonumber \\
 -1.4 - 0.3({\rm log}\,P - 2)  &~{\rm for}~2.0\,\le\,{\rm log}\,P\,<\,4.0, \nonumber \\
 -2.0                          &~{\rm for}~4.0\,\le\,{\rm log}\,P\,<\,8.0.  
\end{align}

\noindent  For primary stars with masses $M_1$~=~1.2\,-\,3.5\,\Msun, we interpolate between Eqn.~9 and Eqn.~10 according to $M_1$.  Similarly, for $M_1$~=~3.5\,-\,6.0\,\Msun, we interpolate between Eqn.~10 and Eqn.~11.  In this manner, Eqns.~9\,-\,11 provide an analytic fit to {\large $\gamma$}$_{\rm largeq}$ as continuous functions of $M_1$ and $P$.  

We compare our fit (dotted lines) to the data for {\large $\gamma$}$_{\rm largeq}$ in the middle panel of Fig.~35.  The analytic functions fit the data reasonably well across all primary masses and orbital periods.  There is scatter, but none of the individual observations deviate more than $\Delta${\large $\gamma$}$_{\rm largeq}$~=~0.7 from the fit.  In fact, the weighted average of observations across a narrow period and primary mass interval match the analytic functions within $\Delta${\large $\gamma$}$_{\rm largeq}$~$\approx$~0.2.  For example, the three measurements of {\large $\gamma$}$_{\rm largeq}$ for early-type binaries at intermediate periods log\,$P$~$\approx$~3.4 are $-$1.7\,$\pm$\,0.5 (mid-B), $-$2.3\,$\pm$\,0.5 (early-B), and $-$1.4\,$\pm$\,0.4 (O-type).  The weighted average and uncertainty is {\large $\gamma$}$_{\rm largeq}$~=~$-$1.7\,$\pm$\,0.3, while evaluation of Eqn.~11 at log\,$P$~$\approx$~3.4 yields {\large $\gamma$}$_{\rm largeq}$~=~$-$1.8. In this parameter space, the analytic function and average of the observations differ by only $\Delta${\large $\gamma$}$_{\rm largeq}$~=~0.1. Similarly, the two A/late-B measurements at log\,$P$~$\approx$~5.6 are {\large $\gamma$}$_{\rm largeq}$~$-$1.0\,$\pm$\,0.5 and $-$2.1\,$\pm$\,0.4.  In this case, the weighted average of {\large $\gamma$}$_{\rm largeq}$~=~$-$1.6\,$\pm$\,0.3 matches the value {\large $\gamma$}$_{\rm largeq}$~=~$-$1.6 of Eqn.~10 evaluated at log\,$P$~=~5.6.  For these two examples, the weighted uncertainties $\delta${\large $\gamma$}$_{\rm largeq}$~$\approx$~0.3 of the observations are larger than the differences $\Delta${\large $\gamma$}$_{\rm largeq}$~$<$~0.2 between the fits and averages of the observations. Based on these examples and further comparisons between the measurements and fits, we adopt the 1$\sigma$ uncertainties between the analytic functions and actual values to be:

\begin{equation}
 \delta\mbox{{\large $\gamma$}}_{\rm largeq}\,(M_1,\,P) = 0.3
\end{equation}

\noindent for all primary masses and orbital periods. 

In the bottom panel of Fig.~35, we display the various measurements of the power-law slope {\large $\gamma$}$_{\rm smallq}$ across small mass ratios $q$~=~0.1\,-\,0.3.  For solar-type primaries, the component {\large $\gamma$}$_{\rm smallq}$~$\approx$~0.3 is nearly constant across all orbital periods.  We adopt:

\begin{align}
 \mbox{{\large $\gamma$}}_{\rm smallq}\,(0.8\,{\rm M}_{\odot}<M_1<1.2&\,{\rm M}_{\odot},\,P)~=~0.3 \nonumber \\
   & {\rm for}~0.2\,<\,{\rm log}\,P\,<\,8.0.
\end{align}

The weighted average of the mid-B, early-B, and O-type measurements of {\large $\gamma$}$_{\rm smallq}$ near log\,$P$ $\approx$ 0.8 is {\large $\gamma$}$_{\rm smallq}$~=~0.1\,$\pm$\,0.5.  At log\,$P$ $\approx$ 6.5, the weighted average of the mid-B, early-B, and O-type observations decreases to {\large $\gamma$}$_{\rm smallq}$~=~$-$1.5\,$\pm$\,0.3.  The three A/late-B measurements across log\,$P$~=~5\,-\,8 span $-$1.1~$<$~{\large $\gamma$}$_{\rm smallq}$~$<$~$-$0.7.  These three values for A/late-B binaries are between the solar-type and early-type VB measurements, indicating a natural progression in {\large $\gamma$}$_{\rm smallq}$ as a function of primary mass.  

Unfortunately, for early-B and O-type primaries, there are no direct measurements of {\large $\gamma$}$_{\rm smallq}$ across 2~$<$~log\,$P$~$<$~4 (see bottom panel in Fig.~35).  As shown in Fig.~1, current observations are insensitive to small mass-ratio binaries at intermediate orbital periods.  A simple linear interpolation between the short-period ({\large $\gamma$}$_{\rm smallq}$~=~0.1\,$\pm$\,0.5) and long-period ({\large $\gamma$}$_{\rm smallq}$~=~$-$1.5\,$\pm$\,0.3) values yields {\large $\gamma$}$_{\rm smallq}$~$=$~$-$0.5\,$\pm$\,0.5 at log\,$P$~$\approx$~3.0. Based on the observed frequency of SB1 companions to Cepheids, we estimate {\large $\gamma$}$_{\rm smallq}$~=~0.2\,$\pm$\,0.9 for mid-B primaries and log\,$P$~=~3.1 (\S6.2).  This measurement is quite uncertain, particularly due to the small sample size, but also because some of the SB1 companions may be compact remnants. For early-type binaries and intermediate orbital periods log\,$P$~$\approx$~3, we adopt the average {\large $\gamma$}$_{\rm smallq}$ = $-$0.2\,$\pm$\,0.6 of the linear interpolation estimate and the Cepheid SB1 estimate.

Based on the above considerations, we determine the following analytic relations.  For the midpoint of A/late-B primaries, we use:

\begin{align}
 \mbox{{\large $\gamma$}}_{\rm smallq}\,(M_1=3.5\,{\rm M}_{\odot}&,\,P)= \nonumber \\
   0.2                             &~{\rm for}~0.2\,\le\,{\rm log}\,P\,<\,2.5, \nonumber \\
   0.2 - 0.3({\rm log}\,P - 2.5)   &~{\rm for}~2.5\,\le\,{\rm log}\,P\,<\,5.5, \nonumber \\
  -0.7 - 0.2({\rm log}\,P - 5.5)   &~{\rm for}~5.5\,\le\,{\rm log}\,P\,<\,8.0.
\end{align}

\noindent For mid-B, early-B, and O-type stars, we fit:

\vspace*{-0.2cm}
\begin{align}
 \mbox{{\large $\gamma$}}_{\rm smallq}\,(M_1 > 6.0\,{\rm M}_{\odot}&,\,P)= \nonumber \\
   0.1                           &~{\rm for}~0.2\,\le\,{\rm log}\,P\,<\,1.0, \nonumber \\
   0.1 - 0.15({\rm log}\,P - 1)  &~{\rm for}~1.0\,\le\,{\rm log}\,P\,<\,3.0, \nonumber \\
  -0.2 - 0.50({\rm log}\,P - 3)  &~{\rm for}~3.0\,\le\,{\rm log}\,P\,<\,5.6, \nonumber \\
  -1.5                           &~{\rm for}~5.6\,\le\,{\rm log}\,P\,<\,8.0.  
\end{align}

\noindent  As done previously, we interpolate between Eqn.~13 and Eqn.~14 for primary stars with masses $M_1$~=~1.2\,-\,3.5\,\Msun.  Similarly, for $M_1$~=~3.5\,-\,6.0\,\Msun, we interpolate between Eqn.~14 and Eqn.~15 with respect to $M_1$.  

We compare our fit to the data for {\large $\gamma$}$_{\rm smallq}$ in the bottom panel of Fig.~35. Our fit passes through all the measurements within their respective 1$\sigma$ uncertainties.  The uncertainty in the fit is therefore dominated by the uncertainty in the observations.  The 1$\sigma$ errors in {\large $\gamma$}$_{\rm smallq}$ depend primarily on $P$. They increase from $\delta${\large $\gamma$}$_{\rm smallq}$~$\approx$~0.4 at log\,$P$~$\approx$~1 to $\delta${\large $\gamma$}$_{\rm smallq}$ $\approx$ 0.6 at log\,$P$~$\approx$~3, and then they decrease to $\delta${\large $\gamma$}$_{\rm smallq}$ $\approx$ 0.3 for log\,$P$~$\gtrsim$~6.  For all primary masses $M_1$, we model the 1$\sigma$ uncertainties as:

\vspace*{-0.2cm}
\begin{align}
 \delta\mbox{{\large $\gamma$}}_{\rm smallq}\,(M_1&,\,P) = \nonumber \\
   0.4                          &~{\rm for}~0.2\,\le\,{\rm log}\,P\,<\,1.0, \nonumber \\
   0.4 + 0.1({\rm log}\,P - 1)  &~{\rm for}~1.0\,\le\,{\rm log}\,P\,<\,3.0, \nonumber \\
   0.6 - 0.1({\rm log}\,P - 3)  &~{\rm for}~3.0\,\le\,{\rm log}\,P\,<\,6.0, \nonumber \\
   0.3                          &~{\rm for}~6.0\,\le\,{\rm log}\,P\,<\,8.0.  
\end{align}

\noindent This relation accounts for the larger uncertainties in {\large $\gamma$}$_{\rm smallq}$ at intermediate orbital periods due to the gaps in the observations.

\subsection{Eccentricity Distributions}

In Fig.~36, we display the power-law slope $\eta$ of the eccentricity distribution as a function of log\,$P$ and colored according to spectral type based on all our measurements in Tables 2\,-\,12.  We divide the data into two spectral type intervals: late-type ($M_1$~=~0.8\,-\,5\,\Msun) and early-type ($M_1$~$>$~5\,\Msun). Both late-type \citep{Meibom2005,Raghavan2010} and early-type \citep{Sana2012} populations have similar zero-age MS circularization periods $P$~$\approx$~2\,-\,6~days (log\,$P$~$\approx$~0.5).  In addition, for both late-type and early-type binaries, the eccentricity distribution becomes weighted toward larger values with increasing orbital period $P$.  However, the power-law slope is $\Delta \eta$~$\approx$~0.5 larger for early-type binaries compared to late-type binaries across all orbital periods.

For close binaries with log\,$P$~$\lesssim$~1.2, the differences in the eccentricity distribution between the two spectral types can be explained by tidal evolution.  Tidal damping via convection is more efficient in cool, late-type stars than is radiative damping in hot, early-type stars \citep{Zahn1975,Zahn1977,Hut1981}.  Moreover, late-type MS binaries live orders-of-magnitude longer and have had more time to tidally evolve toward smaller eccentricities.  

 For binaries with intermediate periods 1.2~$\lesssim$~log\,$P$~$\lesssim$~5, however, the differences cannot be explained solely by tidal effects.  In \S8.7, we showed that tidal evolution is negligible for MS solar-type binaries with periods log\,$P$~=~1.2\,-\,2.4 slightly beyond the tidal circularization period.  For late-type binaries, even those with orbital periods 3~$<$~log\,$P$~$<$~5 where tidal effects are even less significant, the power-law slope $\eta$~$\approx$~0.3\,-\,0.6 is discrepant with a thermal eccentricity distribution (see Fig.~36). At wider separations $\langle a \rangle$~$\approx$~120~AU (log\,$P$~$\approx$~5.6), \citet{Tokovinin2015} demonstrate that solar-type binaries have an intrinsic eccentricity probability distribution described by $p_e$~=~1.2$e$+0.4.  Fitting our power-law model to their data that accounts for selection effects (Fig.~7 in \citealt{Tokovinin2015}), we measure $\eta$~=~0.5\,$\pm$\,0.3. We confirm the conclusion of \citet{Tokovinin2015} that the eccentricity distribution of wide solar-type binaries is flatter than a thermal distribution.  Meanwhile, for early-type binaries, the zero-age MS eccentricity distribution quickly asymptotes toward a thermal distribution ($\eta$ = 1) beyond log\,$P$~$\gtrsim$~1 (Fig.~36). Early-type binaries with intermediate orbital periods are born onto the zero-age MS with systematically larger eccentricities than their solar-type counterparts.  

The differences between the early-type and solar-type intermediate-period eccentricity distributions may stem from the differences in dynamical processing during the earlier pre-MS phase of formation.  As noted in \S1, dynamical interactions
and exchanges tend to drive the eccentricities toward a thermal distribution \citep{Ambartsumian1937,Heggie1975,Pringle1989,Turner1995,Kroupa1995,Kroupa2008}.
Considering massive stars exhibit a larger multiplicity frequency and triple star fraction (see \S9.4), massive systems may more efficiently evolve toward $\eta$~$\approx$~1 at early times $\tau$~$\lesssim$~1~Myr \citep[see also][]{Goodwin2005,Pflamm2006,Oh2015}.  Meanwhile, only $\approx$10\% of solar-type MS primaries are observed to be in triple/quadruple systems (\S9.4).  While the triple star fraction of solar-type stars may be slightly higher during the pre-MS phase, it is still substantially smaller than the triple-star fraction measured for O-type MS stars (see \S10).  The smaller triple-star fraction for solar-type stars may lead to the smaller values of $\eta$~$\approx$~0.5 across intermediate periods.

Another possibility is that both early-type and late-type binaries with intermediate periods are born with $\eta$~$\approx$~1 during the early pre-MS, but then, for solar-type binaries, the eccentricities decreases to $\eta$~$\approx$~0.5 by the zero-age MS.  Solar-type binaries have considerably longer pre-MS contraction timescales, and so tidal interactions of much larger, pre-MS components may cause the binaries to evolve toward smaller eccentricities, even at intermediate separations.  Solar-type pre-MS stars also have longer disk lifetimes, and so binary star interactions with the primordial disks could drive the eccentricities downward.  \citet{Kroupa1995b} provides analytic models of pre-MS binaries embedded in disks that incorporate both tidal circularization and the preferred accretion onto the secondary component.  In addition to orbital circularization, the latter process also drives the binary mass ratio towards unity and may explain the excess fraction of twins observed in solar-type binaries, even those at intermediate separations $a$~$\sim$~0.5\,-\,100~AU.  More detailed comparisons with these models as well as future observations of solar-type pre-MS binaries with intermediate orbital periods should help in determining whether they are born with or evolve toward $\eta$~$\approx$~0.5.

\begin{figure}[t!]
\centerline{
\includegraphics[trim=0.7cm 0.35cm 0.8cm 0.6cm, clip=true, width=3.35in]{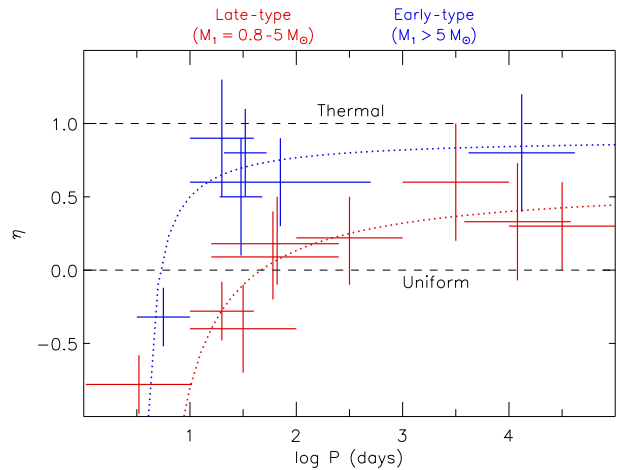}}
\caption{The power-law slope $\eta$ of the eccentricity distribution $p_e$ $\propto$ $e^{\eta}$ across eccentricities 0.0~$<$~$e$~$<$~$e_{\rm max}$ such that the components are not Roche-lobe filling.  We compare our corrected measurements of $\eta$ as a function of orbital period $P$ and colored according to primary spectral type: late ($M_1$~=~0.8\,-\,5\,\Msun; red) and early ($M_1$~$>$~5\,\Msun; blue). We fit the data (dotted) for the late-type and early-type populations.  Although both late-type and early-type MS binaries become weighted toward larger eccentricities with increasing period, early-type binaries quickly asymptote toward a thermal eccentricity distribution ($\eta$~=~1) beyond log\,$P$\,(days)~$\gtrsim$~1.  Meanwhile, solar-type binaries asymptote toward $\eta$~$\approx$~0.5, about halfway between the uniform ($\eta$~=~0) and thermal ($\eta$~=~1) distributions (dashed lines).}
\end{figure}

For zero-age MS binaries, we adopt an analytic function for $\eta$ that approaches circular orbits for short-period binaries and then asymptotes toward a fixed $\eta$ at longer periods.  For late-type MS binaries, we use:

\begin{align}
 \eta\,(0.8\,{\rm M}_{\odot} < M_1 &< 3\,{\rm M}_{\odot},\,0.5 < {\rm log}\,P < 6.0)= \nonumber \\
   & 0.6 - \frac{0.7}{{\rm log}\,P - 0.5}.   
\end{align}

\noindent Meanwhile, for early-type MS binaries, we fit:

\begin{align}
 \eta\,(M_1 &> 7\,{\rm M}_{\odot},\,0.5 < {\rm log}\,P < 5.0)= \nonumber \\
   & 0.9 - \frac{0.2}{{\rm log}\,P - 0.5}.   
\end{align}

\noindent To model a smooth transition between late and early spectral types, we interpolate across $M_1$~=~3\,-\,7\,\Msun\ between Eqns.~17 and 18. We compare our fit to the data in Fig.~36.

 For the shortest orbital periods log\,$P$~$\lesssim$~1, the parameter $\eta$ is not well-defined. This is primarily because the maximum eccentricity $e_{\rm max}$ possible without Roche-lobe filling (Eqn.~3) dramatically decreases toward short periods.  It is difficult to reliably measure the power-law slope $\eta$ of the eccentricity distribution across a narrow interval 0~$<$~$e$~$<$~$e_{\rm max}$.  In any case, the combination of $e_{\rm max}$ provided in Eqn.~3 and our fit to $\eta$ in Eqns.~17 and 18 adequately reproduce the statistics of the observed short-period binary populations.  

For intermediate-period binaries with 1~$\lesssim$~log\,$P$~$\lesssim$~5, our fit to $\eta$ describes the observed variations in the eccentricity distribution.  The fit passes through the measurement uncertainties of all the data points.  For orbital periods $P$ within this interval and for all primary masses $M_1$, the 1$\sigma$ uncertainty between the fit and actual value is:

\begin{equation}
 \delta\eta\,(M_1,\,1\,<\,{\rm log}\,P\,<\,5) = 0.3.
\end{equation}

 For the widest companions with log\,$P$\,(days)~$\approx$~6\,-\,8, we can currently only speculate as to the nature of the eccentricity distribution.  For solar-type binaries, the power-law slope $\eta$~$\approx$~0.5 may continue to the widest separations \citep{Tokovinin2015}.  For early-type primaries, however, wide companions cannot follow a power-law slope $\eta$~$\approx$~0.8 across all eccentricities 0~$<$~$e$~$<$~$e_{\rm max}$~$\approx$~1.  As shown in \S9.4, the majority of wide companions to O-type and B-type stars are tertiary components in hierarchical triples.  Wide tertiary components have been observed to have systematically smaller eccentricities than their wide-binary counterparts \citep{Tokovinin2015}.  The maximum eccentricity $e_{\rm max}$ of tertiary components must not only satisfy Eqn.~3, but also satisfy the dynamical stability criterion (\citealt{Mardling2001}, Eqn.~8 in \citealt{Tokovinin2014}).  For example, if log\,$P_{\rm inner}$~=~3 and log\,$P_{\rm outer}$~=~6, then the eccentricity of the tertiary must be $e_{\rm outer}$ $<$ 0.93 to be dynamically stable. For wide tertiary companions to early-type primaries, the power-law slope $\eta$~$\approx$~0.8 may still continue across 5~$<$~log\,$P_{\rm outer}$~$<$~8, but the domain 0~$<$~$e$~$<$~$e_{\rm max}$ of the distribution must be modified so that $e_{\rm max}$ satisfies the stability criterion.

\subsection{Period Distributions}

We next investigate the period distribution of MS binaries as a function of primary mass $M_1$.  In the top panel of Fig.~37, we display the frequency $f_{\rm logP;q>0.3}$\,($M_1$,\,$P$) of companions with $q$~$>$~0.3 per decade of orbital period based on all our measurements in Tables 2\,-\,12.  We group the data into the same five spectral type intervals as done in \S9.1.  For $M_1$~$\approx$~1\,\Msun, $f_{\rm logP;q>0.3}$ follows a log-normal distribution with a peak of $f_{\rm logP;q>0.3}$~$\approx$~0.08 near log\,$P$~$\approx$~5.  This is consistent with the log-normal period distribution found in previous studies of solar-type MS binaries \citep{Duquennoy1991,Raghavan2010}.  For early-type primaries, the companion frequency $f_{\rm logP;q>0.3}$~$\approx$~0.1\,-\,0.2 is larger at short (log\,$P$~$<$~1.0) and intermediate (1.5~$<$~log\,$P$~$<$~4.0) orbital periods. Meanwhile, at long orbital periods (5~$<$~log\,$P$~$<$~7), our measurements of $f_{\rm logP;q>0.3}$ do not significantly depend on $M_1$.  In the following, we further analyze $f_{\rm logP;q>0.3}$ according to three regimes of orbital period.

First, at short orbital periods log\,$P$~$<$~1, we observe a clear monotonic increase in $f_{\rm logP;q>0.3}$  with respect to $M_1$ (see top panel of Fig.~37).  The close binary frequency increases by more than an order of magnitude from $f_{\rm logP;q>0.3}$~$\approx$~0.02 for solar-type primaries to $f_{\rm logP;q>0.3}$~$\approx$~0.24 for O-type primaries.  By fitting the six data points with log\,$P$~$<$~1.0 in the top panel of Fig.~37, we measure the frequency of short-period binaries with $q$~$>$~0.3 to be:

\begin{align}
f_{\rm logP<1.0;q>0.3}\,(M_1) = & 0.020 + 0.04\,{\rm log}\Big(\frac{M_1}{{\rm M}_{\odot}}\Big) \nonumber \\
                         & +0.07\Big[{\rm log}\Big(\frac{M_1}{{\rm M}_{\odot}}\Big)\Big]^2.
\end{align}

Second, we examine the companion frequency $f_{\rm logP;q>0.3}$ across intermediate periods 1.5~$\lesssim$~log\,$P$~$\lesssim$~4.0.  At these intermediate separations, the observed frequency of companions to mid-B, early-B, and O-type primaries is definitively larger than the solar-type binary frequency (see top panel of Fig.~37).  The weighted average of the two O-type observations in this period interval is $f_{\rm logP;q>0.3}$~=~0.16\,$\pm$\,0.04.  The one early-B measurement is $f_{\rm logP;q>0.3}$~=~0.18\,$\pm$\,0.05, and the weighted average of the three mid-B values is $f_{\rm logP;q>0.3}$~=~0.12\,$\pm$\,0.03. For solar-type binaries, the companion frequency increases from $f_{\rm logP;q>0.3}$~=~0.020\,$\pm$\,0.007 at log\,$P$~=~1.5 to $f_{\rm logP;q>0.3}$~=~0.056\,$\pm$\,0.011 at log\,$P$~=~4.0. At log\,$P$~$\approx$~3, the frequency $f_{\rm logP;q>0.3}$~=~0.06\,$\pm$\,0.02 of companies to A-type / late-B primaries is only slightly larger than the frequency $f_{\rm logP;q>0.3}$~=~0.035\,$\pm$\,0.010 of companions to solar-type primaries.  We fit the weighted averages of the companion frequencies according to primary mass at log\,$P$~$=$~2.7:

\begin{align}
f_{\rm logP=2.7;q>0.3}\,(M_1) = & 0.039 + 0.07\,{\rm log}\Big(\frac{M_1}{{\rm M}_{\odot}}\Big) \nonumber \\
                                & +0.01 \Big[{\rm log}\Big(\frac{M_1}{{\rm M}_{\odot}}\Big)\Big]^2 .
\end{align}

Finally, at wide separations, the binary frequency decreases from $f_{\rm logP;q>0.3}$~$\approx$~0.08 at log\,$P$~=~5 to $f_{\rm logP;q>0.3}$~$\approx$~0.04 at log\,$P$~=~8 (see top panel of Fig.~37). For long orbital periods, there is a slight non-monotonic trend between primary mass $M_1$ and the frequency $f_{\rm logP;q>0.3}$ of companions with $q$~$>$~0.3.  By averaging the observations near log\,$P$~=~6 within each spectral subtype interval, we measure $f_{\rm logP;q>0.3}$~=~0.07\,$\pm$\,0.01, 0.06\,$\pm$\,0.01, 0.05\,$\pm$\,0.01, 0.08\,$\pm$\,0.04, and 0.09\,$\pm$\,0.02 for solar-type, A/late-B, mid-B, early-B, and O-type primaries, respectively.  By fitting these observations and accounting for the period dependence $f_{\rm logP;q>0.3}$~$\propto$~exp[$-$0.3(log\,$P$\,$-$\,5.5)] beyond log\,$P$~$>$~5.5, we find the companion frequency at log\,$P$~=~5.5 to be:

\begin{align}
 f_{\rm logP=5.5;q>0.3}\,(M_1) = & 0.078 - 0.05\,{\rm log}\Big(\frac{M_1}{{\rm M}_{\odot}}\Big) \nonumber \\
                            & +0.04\Big[{\rm log}\Big(\frac{M_1}{{\rm M}_{\odot}}\Big)\Big]^2.
\end{align}

According to the top panel of Fig.~37, the companion frequency $f_{\rm logP;q>0.3}$ for B-type binaries peaks near log\,$P$~$\approx$~3.5 ($a$~$\approx$~10~AU).  This is consistent with \citep{Rizzuto2013}, who find the period distribution of B-type binaries peaks at projected separations log\,$a_{\rm proj}$\,(AU)~$\approx$~0.9.  Interestingly, the magenta data points in the top panel of Fig.~37 suggest the O-type companion frequency is slightly bimodal.  There is a dominant peak at short periods log\,$P$~$<$~1 ($a$~$\lesssim$~0.3~AU). This is consistent with the conclusions of \citet{Sana2012}, who find that spectroscopic binary companions to O-type stars are skewed toward shorter periods.  However, there is also a secondary peak in the period distribution of O-type binaries near log\,$P$~$\approx$~3.5~($a$~$\approx$~10~AU), similar to the peak found in B-type binaries.  The companion frequency $f_{\rm logP;q>0.3}$ definitively increases across log\,$P$~$\approx$~2.0\,-\,3.5 for solar-type and B-type stars, and may also increase for O-type primaries.   We therefore parameterize $f_{\rm logP;q>0.3}$ to increase across these intermediate orbital periods for all spectral types.    We set the slope to be $\alpha$ = $\partial f_{\rm logP;q>0.3}$/$\partial$log$P$ = 0.018 across the interval log\,$P$ = [2.7-$\Delta$log$P$, 2.7+$\Delta$log$P$], where $\Delta$log$P$~=~0.7. As future observations better constrain the functional form of the period distribution across intermediate periods, we can adjust $\alpha$ and $\Delta$log$P$ accordingly.   Based on the above considerations, we can now fit $f_{\rm logP;q>0.3}$ as continuous functions of $M_1$ and $P$:

\onecolumngrid
\begin{align}
 f_{\rm logP;q>0.3}\,&(M_1,\,P) =  \nonumber \\
 & f_{\rm logP<1;q>0.3}~~~~~~~~~~~~~~~~~~~~~~~~~~~~~~~~~~~~~\,\,{\rm for}~0.2\,\le\,{\rm log}\,P\,<\,1.0, \nonumber \\
 & f_{\rm logP<1;q>0.3} + \frac{{\rm log}\,P - 1}{1.7-\Delta{\rm log}P}\times(f_{\rm logP=2.7;q>0.3}-f_{\rm logP<1;q>0.3} - \alpha\,\Delta{\rm log}P) \nonumber \\
 & ~~~~~~~~~~~~~~~~~~~~~~~~~~~~~~~~~~~~~~~~~~~~~~~~~~~~~\,\,\,{\rm for}~1.0\,\le\,{\rm log}\,P\,<\,2.7-\Delta{\rm log}P, \nonumber \\
 & f_{\rm logP=2.7;q>0.3}+\alpha({\rm log}\,P - 2.7) ~~~~~~~~~~~~~~{\rm for}~2.7-\Delta{\rm log}P\,\le\,{\rm log}\,P\,<\,2.7+\Delta{\rm log}P, \nonumber \\
 & f_{\rm logP=2.7;q>0.3} + \alpha\,\Delta{\rm log}P +  \frac{{\rm log}\,P - 2.7 - \Delta{\rm log}P}{2.8-\Delta{\rm log}P}\times(f_{\rm logP=5.5;q>0.3}-f_{\rm logP=2.7;q>0.3} - \alpha\,\Delta{\rm log}P) \nonumber \\
 & ~~~~~~~~~~~~~~~~~~~~~~~~~~~~~~~~~~~~~~~~~~~~~~~~~~~~~\,\,\,{\rm for}~2.7+\Delta{\rm log}P\,\le\,{\rm log}\,P\,<\,5.5, \nonumber \\
 & f_{\rm logP=5.5;q>0.3}\times {\rm exp}\big[-0.3({\rm log}P - 5.5)\big]~~{\rm for}~5.5\,\le\,{\rm log}\,P\,<\,8.0.
\end{align}
\twocolumngrid

\noindent where $\alpha$ = 0.018, $\Delta$log$P$ = 0.7, and the companion frequencies $f_{\rm logP<1;q>0.3}(M_1)$, $f_{\rm logP=2.7;q>0.3}(M_1)$, and $f_{\rm logP=5.5;q>0.3}(M_1)$ are presented in Eqns. 20, 21, and 22, respectively. 

We compare our fit to the data in the top panel of Fig.~37.  The fit passes through the 1$\sigma$ uncertainties of all our measurements.  The relative measurement uncertainties in $f_{\rm logP;q>0.3}$ depend slightly on primary mass and orbital period.  For solar-type primaries, $\delta f_{\rm logP;q>0.3}$/$f_{\rm logP;q>0.3}$ decreases from $\approx$40\% at short periods log\,$P$~$<$~1 to $\approx$20\% beyond log\,$P$~$>$~4. For early-type binaries, $\delta f_{\rm logP;q>0.3}$/$f_{\rm logP;q>0.3}$ initially increases from 25\% at log~$P$~$<$~1 to 35\% at log~$P$~$=$~3, and then decreases to $\approx$30\% beyond log\,$P$~$>$~5.5. We model the 1$\sigma$ uncertainties in $f_{\rm logP;q>0.3}$ for low-mass primaries as:

\begin{align}
 \delta f_{\rm logP;q>0.3}\,(0.8\,{\rm M}_{\odot} < M_1 < 2.0&\,{\rm M}_{\odot},\,P) = \nonumber \\
   (0.40-0.05\,{\rm log}\,P) f_{\rm logP;q>0.3}~~ & {\rm for}~0.2\,\le\,{\rm log}\,P\,<\,4.0, \nonumber \\
                        0.20 f_{\rm logP;q>0.3}~~ & {\rm for}~4.0\,\le\,{\rm log}\,P\,<\,8.0,
\end{align}

\noindent and for early-type primaries as:

\begin{align}
 \delta f_{\rm logP;q>0.3}\,(M_1 > 6.0\,{\rm M}_{\odot},\,P)& = \nonumber \\
                        0.25 f_{\rm logP;q>0.3}~~ &  {\rm for}~0.2\,\le\,{\rm log}\,P\,<\,1.0, \nonumber \\
   (0.20+0.05\,{\rm log}\,P) f_{\rm logP;q>0.3}~~ &  {\rm for}~1.0\,\le\,{\rm log}\,P\,<\,3.0, \nonumber \\
   (0.41-0.02\,{\rm log}\,P) f_{\rm logP;q>0.3}~~ &  {\rm for}~3.0\,\le\,{\rm log}\,P\,<\,5.5, \nonumber \\
                        0.30 f_{\rm logP;q>0.3}~~ &  {\rm for}~5.5\,\le\,{\rm log}\,P\,<\,8.0.
\end{align}

\noindent For primary masses $M_1$~=~2\,-\,6\,\Msun, we interpolate the uncertainties $\delta f_{\rm logP;q>0.3}$ between Eqns.~24 and 25.

\begin{figure*}[t!]
\centerline{
\includegraphics[trim=1.8cm 0.7cm 2.6cm 0.9cm, clip=true, width=6.9in]{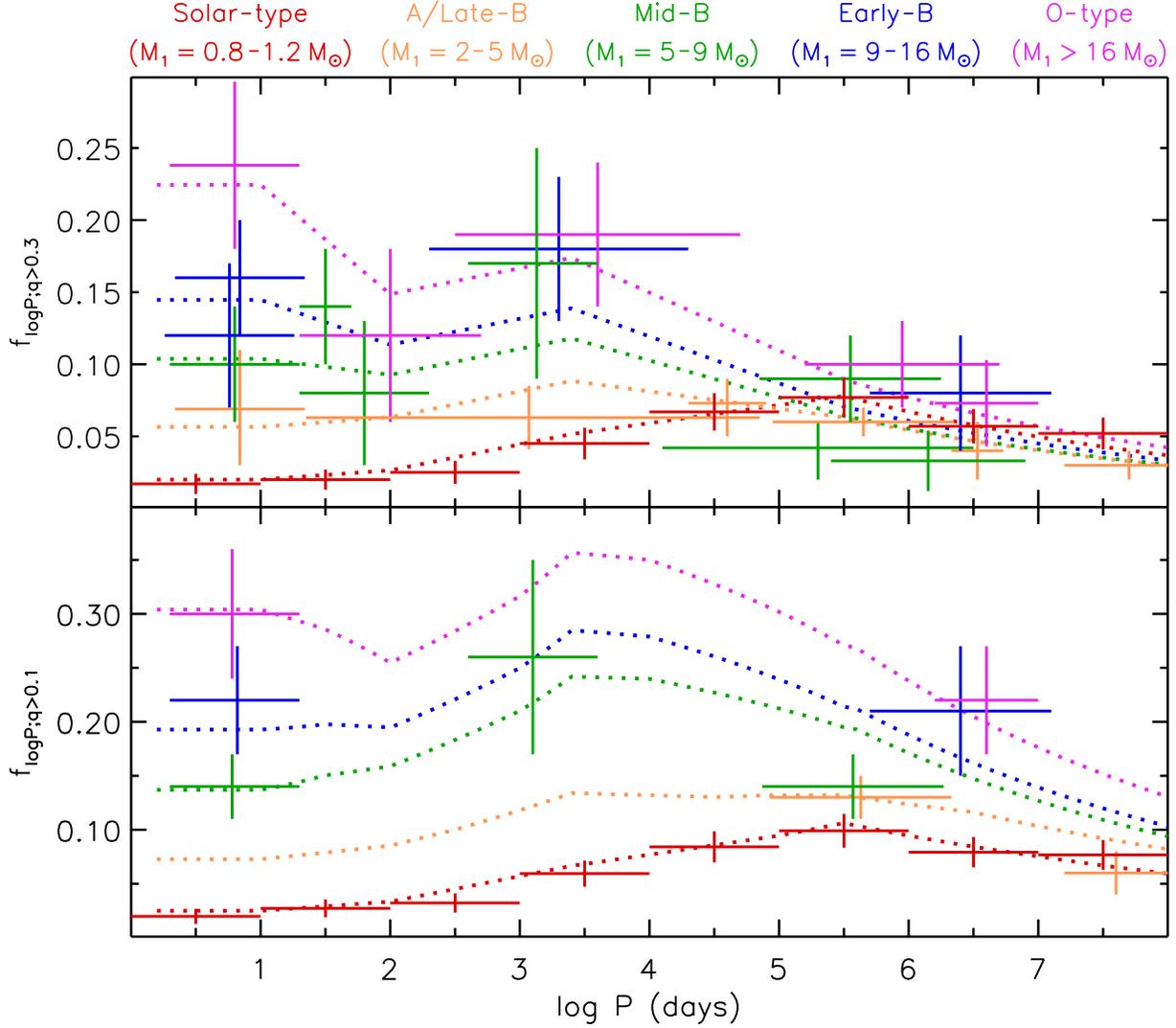}}
\caption{Top panel: Frequency $f_{\rm logP;q>0.3}$\,$(M_1,P)$ of companions with $q$~$>$~0.3 per decade of orbital period.  We display all our measurements after correcting for incompleteness and selection effects, and we group the data into the same five primary mass / MS spectral type intervals as displayed in Fig.~35.  We fit analytic functions (dotted) to the observations. Solar-type MS binaries follow a log-normal period distribution with a peak of $f_{\rm logP;q>0.3}$~$\approx$~0.08 near log\,$P$\,(days)~=~5 ($a$~$\approx$~50\,AU). For early-type primaries, the companion frequencies $f_{\rm logP;q>0.3}$~$\approx$~0.1\,-\,0.2  are substantially larger at short (log\,$P$~$\lesssim$~1) and intermediate (2~$\lesssim$~log\,$P$~$\lesssim$~4) orbital periods.  For mid-B and early-B primaries, the companion frequency peaks at log~$P$~$\approx$~3.5 ($a$~$\approx$~10\,AU).
For O-type MS stars, the orbital period distribution may be slightly bimodal with peaks at short (log\,$P$~$\lesssim$~1) and intermediate (log~$P$~$\approx$~3.5) periods.  Bottom panel: Frequency $f_{\rm logP;q>0.1}$ of companions with $q$~$>$~0.1 per decade of orbital period.  In this case, our model for $f_{\rm logP;q>0.1}$ is completely described by the analytic functions that fit $f_{\rm logP;q>0.3}$ (top panel) and the mass-ratio distribution parameters ${\cal F}_{\rm twin}$, {\large $\gamma$}$_{\rm largeq}$, and {\large $\gamma$}$_{\rm smallq}$ (see Fig.~35). Although we do not directly fit $f_{\rm logP;q>0.1}$, our analytic function matches the data reasonably well.}
\end{figure*}

As shown in the top panel of Fig.~37, our analytic description for $f_{\rm logP;q>0.3}$\,($M_1$,\,$P$) has many breaks and sharp transitions with respect to log\,$P$.   The true period distribution may have smooth transitions better modeled by a log-normal or log-polynomial distribution (but see \citealt{Kobulnicky2014}).  Nevertheless, as future observations become available, we can easily update our coefficients in Eqns.~20\,-\,22 without having to change the functional form of our entire distribution (Eqn.~23).  Most importantly, for all $M_1$ and $P$, our model is consistent with the true values of $f_{\rm logP;q>0.3}$\,($M_1$,\,$P$) within the 1$\sigma$ uncertainties provided in Eqns.~24\,-\,25. 

Given our analytic fits to $f_{\rm logP;q>0.3}$\,($M_1$,\,$P$), {\large $\gamma$}$_{\rm smallq}$\,($M_1$,\,$P$), {\large $\gamma$}$_{\rm largeq}$\,($M_1$,\,$P$), and ${\cal F}_{\rm twin}$\,($M_1$,\,$P$), we calculate the frequency $f_{\rm logP;q>0.1}$\,($M_1$,\,$P$) of companions with $q$~$>$~0.1 per decade of orbital period.  For example,  $f_{\rm logP;q>0.3}$ = 0.14 and a uniform mass-ratio distribution ({\large $\gamma$}$_{\rm smallq}$~=~{\large $\gamma$}$_{\rm largeq}$~${\cal F}_{\rm twin}$~=~0.0) gives $f_{\rm logP;q>0.1}$ = 0.18 (see \S2).  We measure the 1$\sigma$ uncertainties in $f_{\rm logP;q>0.1}$ by propagating in quadrature the 1$\sigma$ uncertainties in the four parameters used to calculate $f_{\rm logP;q>0.1}$.  

In the bottom panel of Fig.~37, we display $f_{\rm logP;q>0.1}$\,($M_1$,\,$P$) as a function of log\,$P$ and colored according to $M_1$.  The frequency $f_{\rm logP;q>0.1}$ of companions with $q$~$>$~0.1 follows a similar functional form as $f_{\rm logP;q>0.3}$, but $f_{\rm logP;q>0.1}$ is larger due to the addition of systems with $q$~=~0.1\,-\,0.3.  For short orbital periods, there is only a slight increase in the companion frequency, i.e., $f_{\rm logP;q>0.1}$~$\approx$~1.3$f_{\rm logP;q>0.3}$.  Meanwhile, for long orbital periods log\,$P$~=~6\,-\,7 and massive primaries $M_1$~$\gtrsim$~6\,\Msun, we measure $f_{\rm logP;q>0.1}$~$\approx$~3.1$f_{\rm logP;q>0.3}$.  This enhanced companion frequency is because the mass-ratio distribution of wider companions to early-type stars is weighted toward smaller mass ratios. A relatively larger fraction of companions with $q$~$>$~0.1 have $q$~=~0.1\,-\,0.3.  This effect is so important at long orbital periods that although $f_{\rm logP;q>0.3}$ may be non-monotonic with respect to $M_1$ (see Eqn.~22 and right side of Fig.~37), $f_{\rm logP;q>0.1}$ is monotonically increasing according to $M_1$ for all orbital periods.  

Not all surveys we have examined in this study are sensitive to  binaries with $q$~$>$~0.1, which is why we have parameterized and measured the frequency $f_{\rm logP;q>0.3}$ of companions with $q$~$>$~0.3 (see Fig.~1 and \S2).  Nonetheless, some samples are complete to $q$~=~0.1 and so we can directly measure the companion frequency $f_{\rm logP;q>0.1}$ down to $q$~=~0.1.  For solar-type primaries, the companion frequency is $f_{\rm logP;q>0.1}$ = (${\cal N}_{\rm largeq}$\,+\,${\cal N}_{\rm smallq}$)/${\cal N}_{\rm prim}$, where ${\cal N}_{\rm prim}$~=~404 and both ${\cal N}_{\rm largeq}$ and ${\cal N}_{\rm smallq}$ are given in Table~11 for each decade of orbital period across 0~$<$~log\,$P$~$<$~8.  Based on the \citet{Sana2012} SB sample containing ${\cal N}_{\rm prim}$ = 71 O-type primaries, we count ${\cal N}_{\rm largeq}$~=~17 companions with $q$~$>$~0.3 and ${\cal N}_{\rm smallq}$~=~4 companions with $q$~=~0.1\,-\,0.3 across $P$~=~2\,-\,20~days (see \S3.5).  This provides $f_{\rm logP;q>0.1}$ = (${\cal N}_{\rm largeq}+{\cal N}_{\rm smallq}$)/${\cal N}_{\rm prim}$ = (17+4)/71 = 0.30\,$\pm$\,0.06.  Similarly, for the combined SB sample of ${\cal N}_{\rm prim}$~=~81\,+\,109\,+\,83~=~273 B-type primaries, we measure $f_{\rm logP;q>0.1}$ = (${\cal N}_{\rm largeq}+{\cal N}_{\rm smallq}$)${\cal C}_{\rm evol}$/${\cal N}_{\rm prim}$ = (22+9)$\times$1.2/273 = 0.14\,$\pm$\,0.03 for log\,$P$~=~0.8\,$\pm$\,0.5.  Based on observations of EBs with early-B primaries (see \S4), we report in \citet{Moe2013} a corrected binary frequency of $f_{\rm logP;q>0.1}$~=~0.22\,$\pm$\,0.05 across $P$~=~2\,-\,20~days.  For the ${\cal N}_{\rm prim}$ = 31 Cepheids brighter than $V$~$<$~8.0~mag that were extensively monitored for radial velocity variations, \citet{Evans2015} find ${\cal N}_{\rm comp}$~=~9 companions with $q$~$>$~0.1 and $P$~=~1\,-\,10~years (see \S6.2).  These statistics provide  $f_{\rm logP;q>0.1}$ = ${\cal N}_{\rm comp} {\cal F}_{\rm Cepheid} {\cal C}_{\rm evol}$/${\cal N}_{\rm prim}$ = 9$\times$0.75$\times$1.2/31 = 0.26\,$\pm$\,0.09.  For a sample of ${\cal N}_{\rm prim,AO}$ = 363 A-type primaries, \citet{DeRosa2014} utilized adaptive optics to detect ${\cal N}_{\rm comp}$ = 54 companions with $q$~$>$~0.1 and projected separations $\rho$~=~0.9$''$\,-\,8.0$''$ (log\,$P$~=~4.9\,-\,6.3; see \S7.1).  The corrected companion frequency is $f_{\rm logP;q>0.1}$ = ${\cal N}_{\rm comp} {\cal C}_{\rm evol}$/${\cal N}_{\rm prim}$/$\Delta$log$P$ = 54$\times$1.2/363/(6.3$-$4.9) = 0.13\,$\pm$\,0.02.  We repeat this exercise for the remaining VB surveys examined in \S7 that are complete to $q$ = 0.1 binaries.  

In the bottom panel of Fig.~37, we present our measured values of $f_{\rm logP;q>0.1}$ based on the samples that are complete down to $q$~=~0.1.  Although we do not fit these data points directly, our analytic function for $f_{\rm logP;q>0.1}$ matches the observed values within their 1$\sigma$ uncertainties.  This demonstrates we can use our fitted functions to $f_{\rm logP;q>0.3}$, {\large $\gamma$}$_{\rm smallq}$, {\large $\gamma$}$_{\rm largeq}$, and ${\cal F}_{\rm twin}$ to reproduce the observed values of $f_{\rm logP;q>0.1}$.  Most importantly, by measuring $f_{\rm logP;q>0.3}$, {\large $\gamma$}$_{\rm largeq}$, and  ${\cal F}_{\rm twin}$ for early-type primaries across all orbital periods and by interpolating {\large $\gamma$}$_{\rm smallq}$  between the short-period (log\,$P$~$<$~1.5) and long-period (log\,$P$~$>$~6) regimes (see \S9.1), we can reliably estimate the total companion frequency $f_{\rm logP;q>0.1}$ across intermediate periods 2~$<$~log\,$P$~$<$~5.  In our analysis, we have always interpolated, never extrapolated, our binary statistics into the regions of the $f$($M_1$,\,$q$,\,$P$) phase space we cannot directly observe.

Based on the observations and our fits to the data points, it is quite evident from Fig.~37 that the period distribution of companions depends critically on the primary mass.  While solar-type MS binaries are weighted toward longer periods log\,$P$\,(days)~$\approx$~5, companions to massive OB stars peak at intermediate periods log\,$P$~$\approx$~3 and are skewed toward even shorter periods log\,$P$~$\lesssim$~1 if we focus only on systems with $q$~$>$~0.3.  As done in \citet{Oh2016}, it is crucial that future population synthesis studies incorporate a primary-mass dependent binary period distribution in order to make meaningful predictions.

\subsection{Multiplicity Frequencies / Fractions}

 We measure the multiplicity frequency $f_{\rm mult;q>0.3}$($M_1$), i.e, the mean number of companions with $q$~$>$~0.3 per primary, by integrating  $f_{\rm logP;q>0.3}$\,($M_1$,\,$P$) across all orbital periods 0.2~$<$~log\,$P$~$<$~8.0 (see Eqn.~1).  For solar-type primaries, for example, we measure $f_{\rm mult;q>0.3}$~=~0.36 by integrating Eqn.~23 with the primary mass set to $M_1$~=~1\,\Msun.  If we were to increase $f_{\rm logP;q>0.3}$ across all orbital periods by the 1$\sigma$ uncertainties $\delta f_{\rm logP;q>0.3}$ provided in Eqn.~24, we would instead measure $f_{\rm mult;q>0.3}$ = 0.44 for solar-type primaries.  This is $\Delta f_{\rm mult;q>0.3}$~=~0.44\,$-$\,0.36~=~0.08 larger than our estimate using the best-fit relation provided in Eqn.~23.  However, the 1$\sigma$ uncertainty $\delta f_{\rm mult;q>0.3}$ in the multiplicity frequency of solar-type stars is {\it not} $\Delta f_{\rm mult;q>0.3}$~=~0.08 because the measurements of $f_{\rm logP;q>0.3}$ at different periods are independent of each other.  In fact, for solar-type binaries, we have ${\cal N}$~=~8 independent measurements of $f_{\rm logP;q>0.3}$ across 0~$\lesssim$~log\,$P$~$\lesssim$~8 (see Table~11 and top panel of Fig.~37). If both $f_{\rm logP;q>0.3}$ and $\delta f_{\rm logP;q>0.3}$ were constant with respect to logarithmic orbital period, then the uncertainty in the multiplicity frequency of solar-type primaries would be $\delta f_{\rm mult;q>0.3}$ = $\Delta f_{\rm mult;q>0.3}/\sqrt{{\cal N}}$ = 0.08/$\sqrt{8}$~=~0.03 given ${\cal N}$~=~8 equally weighted and independent measurements.  In reality, the distributions $f_{\rm logP;q>0.3}$ and $\delta f_{\rm logP;q>0.3}$ are not constant, and so we must weight the measurements accordingly.  

We therefore introduce the coherence length in terms of log\,$P$ and as a function of $M_1$:

\begin{equation}
  l_{\rm logP}\,(M_1) = 1.0 + 0.7\,{\rm log}\Big(\frac{M_1}{{\rm M}_{\odot}}\Big).
\end{equation}

\noindent The coherence length represents the separation with respect to $\delta$log$P$ in which different measurements are used to infer the multiplicity statistics. For example, binaries with log\,$P$\,(days)~=~6 are many coherence lengths away from spectroscopic binaries that probe only short periods log\,$P$~$<$~3, and so we may expect the multiplicity statistics to possibly be different between these two regimes. For solar-type primaries with $M_1$~=~1\,\Msun, the coherence length is $l_{\rm logP}$~=~1.0, and so we have ${\cal N}$~=~$\Delta$log$P$/$l_{\rm logP}$ $\approx$~8.0/1.0~$\approx$~8 independent measurements of $f_{\rm logP;q>0.3}$ across 0~$\lesssim$~log\,$P$~$\lesssim$~8.  Meanwhile, for O-type primaries with $M_1$~$\approx$~28\,\Msun, the coherence length is $l_{\rm logP}$~=~2.0, and so we have only ${\cal N}$~=~4 independent measurements of $f_{\rm logP;q>0.3}$ (see top panel of Fig.~37). To determine the uncertainty in the multiplicity frequency $\delta f_{\rm mult;q>0.3}$, we first divide $\delta f_{\rm logP;q>0.3}$ into ${\cal N}$ equal intervals of length $l_{\rm logP}$.  Within each bin, we integrate $\delta f_{\rm logP;q>0.3}$ to measure the total error in the companion frequency across that interval.  By adding the errors within each interval in quadrature, we calculate $\delta f_{\rm mult;q>0.3}$.  We can write this mathematically as:

\begin{align}
\delta & f_{\rm mult;q>0.3}(M_1) =  \\
 & \Big[ \sum_{i=1}^{\cal N} \Big(\int_{0.2\,+\,(i-1)\,l_{\rm logP}}^{{\rm min}\{8.0,\,0.2\,+\,i\,l_{\rm logP}\}} \delta f_{\rm logP;q>0.3}~d{\rm log}P\Big)^2 \Big]^{\nicefrac{1}{2}} \nonumber
\end{align}

For solar-type primaries with $M_1$~=~1\,\Msun, we measure the value of and uncertainty in the multiplicity frequency to be $f_{\rm mult;q>0.3}$~=~0.36\,$\pm$\,0.03 based on our analytic fits.  This matches the observed value of $f_{\rm mult;q>0.3}$~=$~{\cal N}_{\rm largeq}$/${\cal N}_{\rm prim}$ = (145\,$\pm$\,$\sqrt{145}$)/(404\,$\pm$\,18) = 0.36\,$\pm$\,0.03, where ${\cal N}_{\rm largeq}$~=~145 is the sum of all companions with $q$ $>$ 0.3 presented in Table~11.  The uncertainty $\delta {\cal N}_{\rm prim}$~$\approx$~18 in the number of true primaries derives from the uncertainty in the fraction $\delta {\cal F}_{\rm solar+WD}$~=~4\% of nearby solar-type stars that have WD companions (see \S8.4).

\begin{figure}[t!]
\centerline{
\includegraphics[trim=3.3cm 0.4cm 4.5cm 0.3cm, clip=true, width=3.15in]{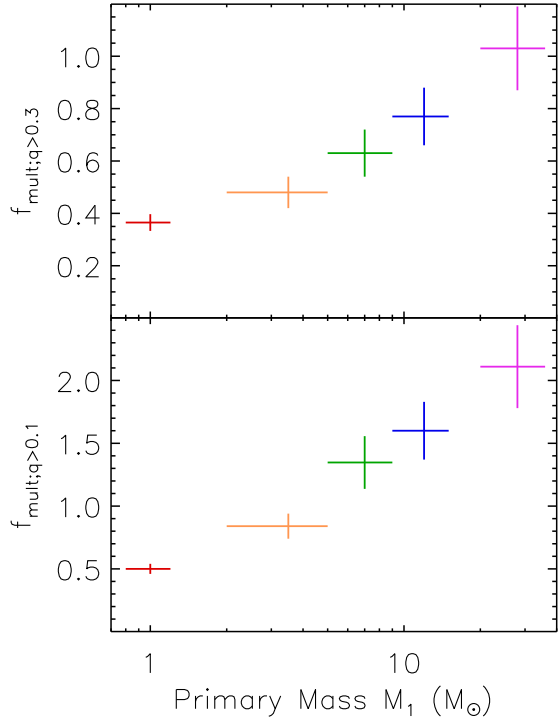}}
\caption{The mean frequency of companions with $q$~$>$~0.3 (top) and $q$~$>$~0.1 (bottom) per primary across orbital periods 0.2~$<$~log\,$P$\,(days)~$<$~8.0 as a function of primary mass $M_1$.  We have colored the data points according to primary spectral type as done in Figs.~34 and 36.  The average solar-type MS primary has $f_{\rm mult;q>0.1}$~=~0.50\,$\pm$\,0.04 companions with $q$~$>$~0.1 (red), while the average O-type MS primary has $f_{\rm mult;q>0.1}$~=~2.1\,$\pm$\,0.3 companions with $q$~$>$~0.1 (magenta).}
\end{figure}

Unlike solar-type binaries, which have continuous measurements of $f_{\rm logP;q>0.3}$ across all orbital periods, there are gaps in our measurements of $f_{\rm logP;q>0.3}$ for early-type binaries (see top panel of Fig.~37).  We cannot simply sum the observed number of companions to early-type stars to calculate $f_{\rm mult;q>0.3}$.  We instead must incorporate our analytic functions to interpolate across the gaps.  For O-type primaries with $M_1$~=~28\,\Msun, for example, we measure $f_{\rm mult;q>0.3}$~=~1.02\,$\pm$\,0.16 according to Eqns.~1, 20\,-\,23, and 25\,-\,27.    Even without considering low-mass companions with $q$~=~0.1\,-\,0.3, an O-type primary already has, on average, one companion with $q$~$>$~0.3.  In the top panel of Fig.~38, we display the multiplicity frequency $f_{\rm mult;q>0.3}$ as a function of primary mass $M_1$. 

Using the same procedure as outlined above, we calculate the total multiplicity frequency $f_{\rm mult;q>0.1}$($M_1$), i.e., the mean number of companions with $q$~$>$~0.1 per primary.  In this case, we integrate $f_{\rm logP;q>0.1}$ across all orbital periods 0.2~$<$~log\,$P$~$<$~8.0 (see \S9.3 and bottom panel of Fig.~37).  We display $f_{\rm mult;q>0.1}$ as function of primary mass $M_1$ in the bottom panel of Fig.~38.  For solar-type primaries with $M_1$~=~1\,\Msun, we measure $f_{\rm mult;q>0.1}$ = 0.50\,$\pm$\,0.04.  This is consistent with the observed value of $f_{\rm mult;q>0.3}$~=(${\cal N}_{\rm largeq}$+${\cal N}_{\rm small}$)/${\cal N}_{\rm prim}$ = (193\,$\pm$\,$\sqrt{193}$)/(404\,$\pm$\,18)~=~0.48\,$\pm$\,0.04, where ${\cal N}_{\rm largeq}$+${\cal N}_{\rm smallq}$~=~193 is the total sum of all companions presented in Table~11.

For early-type primaries with $M_1$~$>$~5\,\Msun, the total multiplicity frequency $f_{\rm mult;q>0.1}$ exceeds unity.  For mid-B primaries with $M_1$~$\approx$~7\,\Msun, we measure $f_{\rm mult;q>0.1}$~=~1.3\,$\pm$\,0.2.  \citet{Kobulnicky2007} and \citet{Kouwenhoven2007} show the total corrected binary fraction approaches 100\% for B-type stars.  These studies modeled all companions as binaries, and so they limited their multiplicity frequency to unity.  In reality, some companions are in triples and/or higher-order multiples, and so the total companion frequency can be $f_{\rm mult;q>0.1}$~$>$~1 as we have measured for massive primaries.  

\citet{Abt1990} report the average mid-B star ($M_1$~$\approx$~8\,\Msun) has 0.8 companions with $M_2$~$>$~2\,\Msun\ ($q$~$\gtrsim$~0.25) and 1.9 companions with $M_2$~$>$~1\,\Msun\ ($q$~$\gtrsim$~0.12).  These statistics translate to $f_{\rm mult;q>0.3}$~$\approx$~0.7 and $f_{\rm mult;q>0.1}$~$\approx$~2.0, respectively.  Although the \citet{Abt1990} estimate of $f_{\rm mult;q>0.3}$~$\approx$~0.7 is consistent with our measurement of $f_{\rm mult;q>0.3}$~$=$~0.63\,$\pm$\,0.09, their total multiplicity frequency of $f_{\rm mult;q>0.1}$~$\approx$~2.0 is discrepant with our estimate of $f_{\rm mult;q>0.1}$ = 1.3\,$\pm$\,0.2 at the 3.2$\sigma$ significance level. For mid-B binaries with intermediate to long orbital periods, \citet{Abt1990} measured the mass-ratio distribution across $q$~=~0.3\,-\,1.0  to be consistent with random pairings drawn from a Salpeter IMF ({\large $\gamma$}$_{\rm largeq}$~=~$-$2.35). Based on their data and more recent observations, we have confirmed this conclusion (see \S9.1).  However, \citet{Abt1990} also assumed this slope could be extrapolated down to $q$~$\approx$~0.1. More recent observations have demonstrated that the power-law component of the mass-ratio distribution flattens toward shallower slopes {\large $\gamma$}$_{\rm smallq}$~$>$~{\large $\gamma$}$_{\rm largeq}$ across smaller mass ratios $q$~=~0.1\,-\,0.3, especially for early-type binaries with intermediate periods (see \S9.1).  For this reason, there are fewer companions with small mass ratios $q$~=~0.1\,-\,0.3 than that predicted by \citet{Abt1990}.

For early-B primaries with $M_1$~$\approx$~12\,\Msun, we measure a slightly larger total companion frequency of $f_{\rm mult;q>0.1}$~$=$~1.6\,$\pm$\,0.2.  \citet{Rizzuto2013} report a corrected multiplicity fraction of $f_{\rm mult}$~$=$~1.35\,$\pm$\,0.25 based on a sample of B-type stars, the majority of which are brighter mid-B and early-B primaries.  This measurement is consistent with and between our mid-B ($f_{\rm mult;q>0.1}$~=~1.3\,$\pm$\,0.2) and early-B ($f_{\rm mult;q>0.1}$~=~1.6\,$\pm$\,0.2) values.

For O-type primaries ($\langle M_1 \rangle$~$\approx$~28\,\Msun), we measure a total multiplicity frequency of $f_{\rm mult;q>0.1}$~$=$~2.1\,$\pm$\,0.3.  This demonstrates that the most massive stars are found almost exclusively in binaries, triples, and quadruples.  Previous studies have also shown the mean multiplicity frequency of O-type stars to be close to two \citep{Preibisch1999,Sana2014}.  We emphasize that in the present study, we have clearly defined the range of binary mass ratios $q$~=~0.1\,-\,1.0 and orbital periods 0.2~$<$~log\,$P$~$<$~8.0 that are incorporated into our measurements of $f_{\rm mult;q>0.1}$.  Hence, the enhanced multiplicity frequency of O-type stars cannot be explained by a larger dynamic range of mass ratios available to more massive stars.  For example, while solar-type primaries with $M_1$~$=$~1.0\,\Msun\ can have stellar-mass companions with $M_2$~$>$~0.08\,\Msun\ only if $q$~$>$~0.08, O-type primaries can have stellar-mass companions down to $q$~$\approx$~0.003. By restricting our analysis to companions with $q$~$>$~0.1 for all spectral types, we can make a more meaningful comparison.  As shown in the bottom panel of Fig.~38, the mean number of companions with $q$~$>$~0.1 per primary increases by a factor of four from $f_{\rm mult;q>0.1}$~=~0.50\,$\pm$\,0.04 for solar-type primaries to $f_{\rm mult;q>0.1}$~=~2.1\,$\pm$\,0.3 for O-type primaries.

Although we cannot fully differentiate between companions in binaries versus those in triples and higher-order multiples (see \S2), we can still use $f_{\rm mult;q>0.1}$($M_1$) to estimate the single ${\cal F}_{\rm n=0;q>0.1}$($M_1$), binary ${\cal F}_{\rm n=1;q>0.1}$($M_1$), triple ${\cal F}_{\rm n=2;q>0.1}$($M_1$), and quadruple ${\cal F}_{\rm n=3;q>0.1}$($M_1$) star fractions as a function of primary mass.  As defined in \S2 and discussed in \S8.1, $f_{\rm mult;q>0.1}$($M_1$) includes only the companions with $q$~$=$~$M_{\rm comp}$/$M_1$~$>$~0.1 that directly orbit the primary of mass $M_1$.  In a (Aa,\,Ab)\,-\,B hierarchical triple configuration, both companions Ab and B would contribute to $f_{\rm mult;q>0.1}$.  Meanwhile, in a A\,-\,(Ba,\,Bb) configuration, only the component Ba is included in $f_{\rm mult;q>0.1}$ unless the secondary itself is comparable in mass to the primary (see \S8.1).  

Most importantly, there is a large phase space of A\,-\,(Ba,\,Bb) triple configurations that completely elude detections, even for nearby solar-type primaries.  For example, suppose adaptive optics and/or long baseline interferometry was utilized to detect an M-dwarf companion (component B) at a separation of $\rho$~=~0.1$''$ ($a$~$\approx$~2\,AU; log\,$P$~$\approx$~3) from a solar-type primary (component A) that is $d$~$\approx$~20~pc away.  Spectroscopic radial velocity observations may reveal that the primary has an additional closer companion (component Ab) in a (Aa,\,Ab)\,-\,B triple configuration.  However, we cannot yet obtain spectroscopic radial velocities of the M-dwarf companion that is only  $\rho$~=~0.1$''$ away from a solar-type primary. If the M-dwarf itself has a close, spectroscopic companion (component Bb) in a  A\,-\,(Ba,\,Bb) configuration, we cannot detect it.  

 Based on our definitions (see also \S2 and Table~1), we can relate the multiplicity frequency and multiplicity fractions:

\begin{equation}
 f_{\rm mult;q>0.1} = {\cal F}_{\rm n=1;q>0.1} + 2 {\cal F}_{\rm n=2;q>0.1} + 3 {\cal F}_{\rm n=3;q>0.1}.
\end{equation}

\noindent The single star fraction is:

\begin{equation}
 {\cal F}_{\rm n=0;q>0.1} = 1 - {\cal F}_{\rm n=1;q>0.1} - {\cal F}_{\rm n=2;q>0.1} - {\cal F}_{\rm n=3;q>0.1}.
\end{equation}

For solar-type primaries, the \citet{Raghavan2010} survey is relatively complete toward binaries, triples, and quadruples as we have defined them.  Although we added 25 companions to account for selection effects (see Fig.~28), the majority of these companions have already been detected but simply have mass ratios and/or orbital periods that cannot be readily measured (see \S8.2-8.4).  \citet{Raghavan2010} present mobile diagrams of triple stars in their Figs.~20\,-\,22, which show the masses and orbital periods / separations of the individual components in the hierarchical triples.  Based on these triple-star mobile diagrams, we count ${\cal N}_{\rm 3A}$~=~24 confirmed and suspected triples in (Aa,\,Ab)\,-\,B configurations that satisfy our selection criteria.  We include an additional ${\cal N}_{\rm 3B}$ = 4 triples in a A\,-\,(Ba,\,Bb) configuration in which the component Ba is itself a solar-type F6\,-\,K3 star.  The remaining 14 triples presented in Figs.~20\,-\,22 of \citet{Raghavan2010} have brown dwarf companions with $q$~$<$~0.1, WD companions, tertiary companions with orbital periods log\,$P$~$>$~8 beyond what we have investigated in this study, and/or A\,-\,(Ba,\,Bb) configurations in which the Ba component is a late-K or M-dwarf star.  

\citet{Raghavan2010} also present mobile diagrams of 14 quadruples and higher-order multiples in their Figs.~23\,-\,24.  Of these systems, we count ${\cal N}_{\rm 3C}$~=~7 as triple systems where the additional fourth component either orbits a late-K/M-dwarf companion in a double-double (Aa,\,Ab)\,-\,(Ba,\,Bb) configuration or has an orbital period log\,$P$~$>$~8.0 too long to be included in our statistical sample.  We find ${\cal N}_{\rm 4A}$~=~3 quadruples in a double-double (Aa,\,Ab)\,-\,(Ba,\,Bb) configuration in which component Ba is itself a solar-type F6\,-\,K3 star.  There is ${\cal N}_{\rm 4B}$~=~1 quadruple in a [(Aa,\,Ab)\,-\,B]\,-\,C configuration.  One of the three remaining quadruples contain a pair of brown dwarfs.  The final two quadruples are in double-double configurations in which the two pairs have extremely wide separations $\rho$~$>$~700$''$ (log\,$P$~$\gtrsim$~8.5).

\begin{figure}[t!]
\centerline{
\includegraphics[trim=0.4cm 0.4cm 0.2cm 0.2cm, clip=true, width=3.45in]{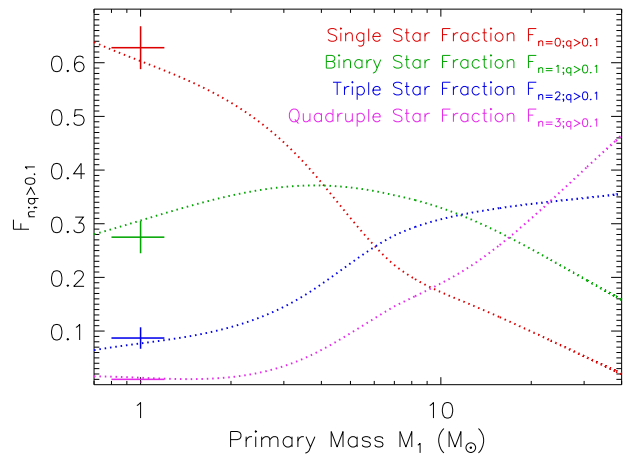}}
\caption{Multiplicity fractions as a function of primary mass (dotted lines), including the single star fraction ${\cal F}_{n=0;q>0.1}$ (red), binary star fraction ${\cal F}_{n=1;q>0.1}$ (green), triple star fraction ${\cal F}_{n=2;q>0.1}$ (blue), and quadruple star fraction ${\cal F}_{n=3;q>0.1}$ (magenta). Given a primary mass $M_1$, our model {\it assumes} the multiplicity fractions follow a Poisson distribution across the interval $n$ = [0,\,3] in a manner that reproduces the measured multiplicity frequency $f_{\rm mult;q>0.1}$~=~$\sum_{n=1}^3$\,$n$\,${\cal F}_{n;q>0.1}$. For solar-type stars, this model matches the measured values (solid) within their uncertainties.  Regardless of the uncertainties in the multiplicity fractions, $\lesssim$10\% of O-type stars are single while $\gtrsim$55\% are born in triples and/or quadruples. }
\end{figure}

Based on the statistics above, we find ${\cal N}_{\rm trip}$ = ${\cal N}_{\rm 3A}$~+~${\cal N}_{\rm 3B}$~+~${\cal N}_{\rm 3C}$ = 35 triples and ${\cal N}_{\rm quad}$~= ${\cal N}_{\rm 4A}$~+~${\cal N}_{\rm 4B}$ =~4 quadruples in which all the companions have mass ratios $q$~$>$~0.1 and directly orbit solar-type stars with periods log\,$P$~$<$~8.  Given the ${\cal N}_{\rm comp}$~=~193 total companions in Table~11, there are ${\cal N}_{\rm bin}$ = ${\cal N}_{\rm comp}$\,$-$\,2${\cal N}_{\rm trip}$\,$-$\,3${\cal N}_{\rm quad}$ = 111 binaries in the corrected \citet{Raghavan2010} sample (Eqn.~28).  Of the ${\cal N}_{\rm prim}$~=~404 solar-type primaries, ${\cal N}_{\rm single}$ = ${\cal N}_{\rm prim}$\,$-$\,${\cal N}_{\rm bin}$\,$-$\,${\cal N}_{\rm trip}$\,$-$\,${\cal N}_{\rm quad}$ = 254 are single stars that do not have any companions with $q$~$>$~0.1 and log\,$P$~$<$~8.0 (Eqn.~29).  We measure the multiplicity fractions of solar-type primaries to be ${\cal F}_{\rm n=0;q>0.1}$ = ${\cal N}_{\rm single}$/${\cal N}_{\rm prim}$ = 254/404 = 0.63\,$\pm$\,0.04 single, ${\cal F}_{\rm n=1;q>0.1}$~=~0.27\,$\pm$\,0.03 binary,  ${\cal F}_{\rm n=2;q>0.1}$~=~0.09\,$\pm$\,0.02 triple, and ${\cal F}_{\rm n=3;q>0.1}$~=~0.010\,$\pm$\,0.005 quadruple. We present the measured multiplicity fractions of solar-type primaries in Fig.~39. These statistics reproduce the observed solar-type multiplicity frequency $f_{\rm mult;q>0.1}$ = 0.27~+~2\,$\times$\,0.09~+~3\,$\times$\,0.01 = 0.48 according to Eqn.~28.  

For early-type systems, we cannot reliably estimate multiplicity fractions without selection biases.  In the following, we {\it assume} the multiplicity fractions ${\cal F}_{\rm n;q>0.1}$ follow a Poisson distribution truncated to the interval $n$~=~[0,\,3] and with a mean that reproduces the measured multiplicity frequency $f_{\rm mult;q>0.1}$ according to Eqn.~28.   We display in Fig.~39 the multiplicity fractions as a function of primary mass based on this model.  For solar-type stars, our Poisson model matches the observed values within their 1$\sigma$ uncertainties. \citet{Kraus2011} and \citet{Duchene2013} also note that the multiplicity fractions of solar-type systems follow a Poisson distribution, concluding that the addition of multiple companions resembles a stochastic process.  The O-type multiplicity fractions are 6$_{-3}^{+6}$\% single, (21\,$\pm$\,7)\% binary, (35\,$\pm$\,3)\% triple, and (38\,$\pm$\,11)\% quadruple, where we have propagated the 1$\sigma$ measurement uncertainties in the multiplicity frequency $f_{\rm mult;q>0.1}$~=~2.1\,$\pm$\,0.3. Systematic errors in our Poisson model likely contribute an additional $\approx$5\% uncertainty in the multiplicity fractions. Given the observational and systematic uncertainties, the single star fraction of O-type stars is consistent with zero. 

No matter how we distribute the multiplicity fractions of O-type stars, the sum of the triple and quadruple star fraction must be ${\cal F}_{\rm n\ge2;q>0.1}$ $>$ 55\% given $f_{\rm mult;q>0.1}$~=~2.1.  For example, by setting the single star and triple star fractions to zero, then the binary star fraction is ${\cal F}_{\rm n=1;q>0.1}$ = 45\% and the quadruple star fraction is ${\cal F}_{\rm n=3;q>0.1}$ = 55\%.  All other distributions of the multiplicity fractions lead to larger values of ${\cal F}_{\rm n\ge2;q>0.1}$ = ${\cal F}_{\rm n=2;q>0.1}$ + ${\cal F}_{\rm n=3;q>0.1}$ $>$ 55\%.   The majority of O-type MS primaries are therefore found in triples and quadruples.  

\section{Binary Star Formation}

Despite their ubiquity, a close stellar companion with $a$~$\lesssim$~1~AU cannot easily form in situ \citep{Tohline2002}.  Instead, the companion most likely forms via fragmentation on large, core scales of $\sim$1,000 AU or within the circumstellar disk at separations $\sim$10\,-\,100~AU \citep{Kratter2006}.  Some mechanism for orbital evolution is required to bring the binary to shorter periods. While the dominant migration mechanism remains unknown, likely candidates include migration through a circumbinary disk due to hydrodynamical forces, dynamical interactions in an initially unstable hierarchical multiple system, or secular evolution in triple stars, such as Kozai cycles, coupled with tidal interactions \citep{Bate1995,Kiseleva1998,Kratter2011}.  

As mentioned in \S1, the measured mass-ratio distribution of binaries offers insight into their formation processes.  For example, if the companion migrates inward through the primordial disk, it most likely accretes additional mass.  Competitive accretion in the circumbinary disk tends to drive the binary mass ratio toward unity  \citep{Kroupa1995,Bate1997, Bate2000, Tokovinin2000, White2001, Marks2011}. While Roche-lobe overflow during the early, fully convective, pre-MS phase may cause the binary component masses to diverge \citep{Tokovinin2000}, mass transfer during the late pre-MS phase may instead increase the mass ratio, possibly contributing to an excess twin population at very short orbital periods $P$~$\lesssim$~10~days.  In both the accretion and mass transfer scenarios, the binary components coevolve during the pre-MS phase, which most likely lead to correlated component masses.  

Early-type binaries with $P$~$<$~20~days exhibit a small but statistically significant excess twin fraction ${\cal F}_{\rm twin}$~=~0.1 (see Fig.~35).  While excess twins are absent among early-type binaries with $P$~$>$~20~days, their mass ratio distribution is measurably discrepant with random pairings of the IMF out to log\,$P$\,(days)~$\approx$~5.5 ($a$~$\approx$~200~AU).  For solar-type binaries, the excess twin fraction ${\cal F}_{\rm twin}$ = 0.3 is measurably larger at short periods $P$~$<$~100~days.  Moreover, the excess twin population of solar-type binaries extends to significantly wider separations $a$~$\approx$~200~AU (log\,$P$~$\approx$~6; Fig.~35). This separation of $a$~$\approx$~200~AU is comparable to the radii of primordial disks observed around young, accreting pre-MS solar-type systems \citep{Andrews2009}.  \citet{White2001} find that the presence of circumprimary and circumsecondary disks are significantly correlated only if the binary separations are $a$~$<$~200~AU.  The lack of disk correlation for wider binaries with $a$~$>$~200~AU indicates the components separately accrete from their own gas reservoirs.  

Based on these various lines of observational evidence, we surmise that wide components with separations $a$~$\gtrsim$~200~AU initially fragmented from molecular cores / filaments and have since evolved relatively independently.  For both solar-type and early-type systems, the mass-ratio distribution of wide companions are weighted toward more extreme mass ratios compared to their counterparts with smaller separations.  For wide early-type systems, we measure {\large $\gamma$}$_{\rm smallq}$~=~$-$1.5\,$\pm$\,0.4 and {\large $\gamma$}$_{\rm largeq}$~=~$-$2.0\,$\pm$\,0.3, which is close to but slightly flatter than that expected from random pairings from a Salpeter IMF ({\large $\gamma$}$_{\rm smallq}$~=~{\large $\gamma$}$_{\rm largeq}$~=~$-$2.35).  Similarly, the widest solar-type binaries investigated in this study are still measurably discrepant from random pairings from the IMF (Fig.~30).  This demonstrates that wide binaries are not perfectly randomly paired based solely on the IMF, possibly suggesting that fragmentation of molecular cores / filaments leads to slightly correlated component masses.  As another possibility, wide companions may be dynamically disrupted and/or captured \citep{Heggie1975}.  Wide systems may therefore still be randomly paired, but where the pairings are modified to include the effects of dynamical processing \citep{Kouwenhoven2010,Marks2011,Perets2012,Thies2015}.  For instance, wide binaries may initially form with mass ratios consistent with random pairings drawn from the IMF, but subsequent dynamical interactions preferentially eject the lower-mass companions with smaller binding energies (see more below). 

Meanwhile, we conclude that closer binaries with $a$~$\lesssim$~200~AU initially fragmented from the disk and subsequently coevolved via accretion.  Utilizing analytic models, \citet{Kratter2006} predict that primordial disks around more massive stars are more prone to fragmentation.  This may explain why the observed frequency $f_{\rm logP;q>0.1}$~$\approx$~0.3 of companions to early-type stars at intermediate separations $a$~$\approx$~20~AU (log\,$P$~$\approx$~4.0) is $\approx$3\,-\,4 times larger than the companion frequency $f_{\rm logP;q>0.1}$ $\approx$ 0.08 to solar-type stars (Fig.~37).  

In addition, \citet{Kratter2006} find that, although the typical fragment mass $M_{\rm frag}$ increases with final primary mass $M_1$, the relation between the two is flatter than linear (see their Fig.~6).  For example, they estimate $q_{\rm frag}$~=~$M_{\rm frag}$/$M_1$~$\approx$~0.08 for $M_1$~=~3\,\Msun\ and $q_{\rm frag}$~=~0.02 for $M_1$~=~40\,\Msun.  The initial fragment can still accrete additional mass from the disk.  Nevertheless, if the companion does not significantly migrate and instead opens a gap in the disk at intermediate separations, the growth of the companion is limited by the amount of mass in the disk within its Hill radius \citep{Goodman2004,Kratter2006}.  \citet{Kratter2006} calculate this so-called isolation mass $M_{\rm iso}$, and also find it to be flatter than linear with respect to $M_1$.  For example, they predict $q_{\rm iso}$~=~$M_{\rm iso}$/$M_1$~$\approx$~0.2 for $M_1$ = 3\,\Msun\ and $q_{\rm iso}$  = 0.1 for $M_1$ = 40\,\Msun.  At intermediate periods log\,$P$~$\approx$~3.5, we observe that early-type binaries are weighted significantly toward smaller mass ratios ({\large $\gamma$}$_{\rm smallq}$~$\approx$~$-$0.5 and {\large $\gamma$}$_{\rm largeq}$~$\approx$~$-$1.8) compared to solar-type binaries ({\large $\gamma$}$_{\rm smallq}$~$\approx$~0.5 and {\large $\gamma$}$_{\rm largeq}$~$\approx$~$-$0.5; Fig.~35).  This trend may be the result of disk fragmentation and subsequent accretion whereby more massive systems produce systematically smaller fragment mass ratios $q_{\rm frag}$, smaller isolation mass ratios $q_{\rm iso}$, and therefore ultimately smaller binary mass ratios $q$.

With decreasing binary separation, the component masses become more highly correlated.  For both early-type and solar-type binaries with $P$~$<$~10~days, we measure {\large $\gamma$}$_{\rm smallq}$~$\approx$~0.4 and {\large $\gamma$}$_{\rm largeq}$~$\approx$~$-$0.5.    This suggests that companions that migrate further inward through the disk accrete substantially more mass.  It is also possible that closer binaries initially fragment on smaller separations scales and with systematically larger fragment mass ratios $q_{\rm frag}$.   For very close binaries with periods $P$~$<$~10~days,  Roche-lobe overflow during the pre-MS phase is a third option to explain the correlated component masses and larger excess twin fraction.  

In any case, we emphasize the transition in the mass-ratio distribution between short ($a$~$\lesssim$~0.3 AU) and intermediate ($a$~$\approx$~50~AU) separations is more pronounced for more massive systems (Fig.~35).  For early-type binaries, the excess twin fraction vanishes beyond $P$~$>$~20~days and the power-law component dramatically decreases from {\large $\gamma$}$_{\rm largeq}$~$\approx$~$-$0.5 to {\large $\gamma$}$_{\rm largeq}$~$\approx$~$-$2.0.  Meanwhile, for solar-type binaries, the excess twin fraction decreases slightly from ${\cal F}_{\rm twin}$~$\approx$~0.3 to ${\cal F}_{\rm twin}$~=~0.1 and the power-law component {\large $\gamma$}$_{\rm largeq}$~$\approx$~$-$0.5 remains relatively constant. This intrinsic variation with respect to primary mass may be due to the scaling of the fragment mass ratio $q_{\rm frag}$ discussed above, but may also be due to the longer formation timescales associated with lower mass primaries.  For example, the average primordial disk lifetimes $\tau_{\rm disk}$~=~3~Myr of solar-type primaries \citep{Mamajek2009} is an order of magnitude longer than the disk photoevaporation timescales $\tau_{\rm disk}$~$\lesssim$~0.3~Myr measured in more massive Herbig Be stars \citep{AlonsoAlbi2009}.  The longer disk lifetimes of solar-type systems may allow a larger fraction of companions to accrete relatively more mass from the disk, possibly toward $q$~$\approx$~1.  Similarly, the pre-MS contraction timescales is significantly longer for solar-type binaries, and so short-period solar-type binaries are more likely to exchange material through Roche-lobe overflow while on the pre-MS.  This may explain the correlation between the excess twin fraction and primary mass at short orbital periods $P$~$<$~10~days.  In summary,  the processes of disk fragmentation, accretion in the disk, and pre-MS mass transfer may all contribute to the larger excess twin fraction and higher degree of correlation between component masses observed in solar-type binaries compared to early-type binaries.

Although the correlation between binary component masses demonstrate they coevolved as they migrated toward shorter separations, they do not reveal precisely {\it how} the companions migrated.  It is possible that companions naturally undergo orbital decay toward smaller separations due to hydrodynamical forces in the disk.  It is also possible that the inner binary requires an outer tertiary to evolve toward shorter periods. After considering selection effects and accounting for incompleteness, \citet{Tokovinin2006} show that $\approx$(70\,-\,90)\% of solar-type binaries with periods $P$~$\approx$~2\,-\,6~days have outer tertiaries with $q$~=~$M_3$/$M_1$~$\gtrsim$~0.2.  Meanwhile, only $\approx$30\% of binaries with $P$~$\approx$~10\,-\,30~days have such tertiary components.  \citet{Tokovinin2006} argue that close binaries form predominantly through Kozai cycles in triples in which the outer tertiary pumps the eccentricity of the inner binary to large values.  The inner binary is subsequently tidally dissipated into a shorter orbit \citep{Kiseleva1998}.  This scenario may also explain the origin of the large eccentricities observed in young early-type close binaries (\S9.2).  

We further investigate the correlation between triples and close binaries as a function of primary spectral type.  Unfortunately, we cannot repeat the \citet{Tokovinin2006} measurement for early-type systems due to the observational selection effects and incompleteness.  We instead compare in Fig.~40 the fraction ${\cal F}_{\rm 0.3<logP<0.8;q>0.1}$ of primaries that have close companions with $P$ = 2\,-\,6 days and $q$~$>$~0.1 to the overall triple/quadruple star fraction ${\cal F}_{n\ge2;q>0.1}$ = ${\cal F}_{n=2;q>0.1}$\,+\,${\cal F}_{n=3;q>0.1}$. Although we cannot directly associate early-type close binaries with tertiary companions on a system by system basis, the correlation between ${\cal F}_{\rm 0.3<logP<0.8;q>0.1}$ and ${\cal F}_{n\ge2;q>0.1}$ is intriguing for three reasons.  

First,  ${\cal F}_{\rm 0.3<logP<0.8;q>0.1}$ is nearly directly proportional to ${\cal F}_{n\ge2;q>0.1}$.  By fitting a linear relation to the four data points in Fig.~40, we measure the y-intercept to be ${\cal F}_{\rm 0.3<logP<0.8;q>0.1}$ = $-$0.005\,$\pm$\,0.007, which is slightly smaller than but consistent with zero.  If a process other than dynamical evolution in triples was the dominant formation mechanism for producing close binaries, then we would expect the y-intercept to be measurably greater than zero, i.e., there would be close binaries even if there were no triple/quadruple systems.

\begin{figure}[t!]
\centerline{
\includegraphics[trim=0.4cm 0.3cm 0.4cm 0.1cm, clip=true, width=3.55in]{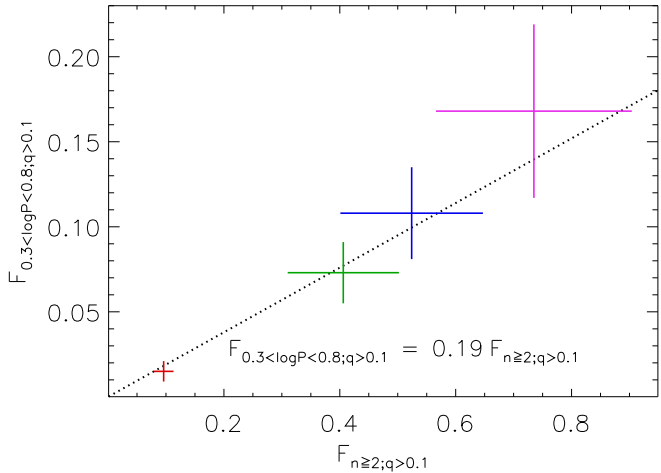}}
\caption{The close companion fraction ${\cal F}_{\rm 0.3<logP<0.8;q>0.1}$ as a function of overall triple plus quadruple star fraction ${\cal F}_{n\ge2;q>0.1}$ colored according to primary spectral type. For solar-type primaries (red), ${\cal F}_{\rm 0.3<logP<0.8;q>0.1}$ = (1.5\,$\pm$\,0.6)\% have close companions with $q$ $>$ 0.1 and $P$~$\approx$~2\,-\,6~days, and ${\cal F}_{n\ge2;q>0.1}$ = (10\,$\pm$\,2)\% are in triple or quadruple systems.  Meanwhile, ${\cal F}_{\rm 0.3<logP<0.8;q>0.1}$ = (17\,$\pm$5)\% of O-type stars have very short-period companions and ${\cal F}_{n\ge2;q>0.1}$ = (73\,$\pm$\,16)\% of O-type stars have $n$~$\ge$~2 companions with $q$~$>$~0.1 (magenta).  If dynamical evolution in triple/quadruple systems is the dominant formation mechanism of close binaries, then it must be a relatively efficient process.  For every triple/quadruple system, (16\,-\,22)\% have inner binaries with short periods $P$~$=$~2\,-\,6~days, irrespective of primary mass (dotted line). }
\end{figure}

Second, the slope of the relation $\epsilon$ = ${\cal F}_{\rm 0.3<logP<0.8;q>0.1}$/${\cal F}_{n\ge2;q>0.1}$ = (19\,$\pm$\,3)\% provides a direct constraint for the efficiency of close binary formation via triple-star dynamical evolution.  If not all close binaries have outer tertiaries, then the efficiency $\epsilon$ would be correspondingly smaller.  For solar-type systems, where we know $\approx$70\,-\,90\% of close binaries with $P$~=~2\,-\,6 days have outer tertiaries, then the efficiency must be $\epsilon$~$\approx$~15\%.  This is substantially larger than that expected if the orbital periods of the inner and outer companions in triples were uncorrelated.  For example, if we randomly select $P_{\rm inner}$ and $P_{\rm outer}$ from the underlying period distribution $f_{\rm logP;q>0.1}$ with the added constraint that $P_{\rm outer}$ $\gtrsim$ 10$P_{\rm inner}$ for long-term dynamical stability, then only 6\% of solar-type triples would have inner binaries with $P_{\rm inner}$ = 2\,-\,6 days.  Given ${\cal F}_{n\ge2;q>0.1}$ = 0.10 for solar-type systems, then the predicted efficiency in this random pairing scenario would be  $\epsilon$ = 0.06\,$\times$\,0.1~=~0.6\%.  This is a factor of $\approx$25 smaller than the measured value.  The hierarchical distributions of triples/quadruples are clearly not uncorrelated, suggesting that dynamical evolution in triples/quadruples is at least partially responsible for the formation of close binaries.

Finally, the correlation between the close binary fraction ${\cal F}_{\rm 0.3<logP<0.8;q>0.1}$ and overall triple/quadruple fraction ${\cal F}_{n\ge2;q>0.1}$ appears to be independent of primary mass $M_1$.  If there is indeed a causal relationship between close binaries and outer tertiaries, then the larger multiplicity frequency $f_{\rm mult;q>0.1}$~=~2.1\,$\pm$\,0.3 and triple/quadruple star fraction ${\cal F}_{n\ge2;q>0.1}$~=~(73\,$\pm$\,16)\% among O-type systems leads to their substantially larger close binary fraction.  The period distribution $f_{\rm logP;q>0.1}$ of companions to O-type stars is slightly bimodal (see Fig.~37), further suggesting that close binaries require a different formation mechanism compared to binaries with intermediate separations.  More precisely, we conclude that both close binaries and binaries with intermediate orbital periods originally fragmented in the disk at intermediate spatial scales.  Only a subset of these primordial binaries, namely those with outer tertiaries in certain orbital configurations, dynamically evolved toward short periods $P$~$\lesssim$~10~days to produce the observed close binary population.

 Dynamical evolution in triples alone cannot explain the observed properties of close binaries.  As discussed above, close binaries exhibit an excess twin fraction ${\cal F}_{\rm twin}$ = 0.1\,-\,0.3 and a correlated mass-ratio distribution  {\large $\gamma$}$_{\rm smallq}$~$\approx$~0.3 and {\large $\gamma$}$_{\rm largeq}$ $\approx$ $-$0.5.  These properties demonstrate that components in close binaries coevolved during their pre-MS formation.  While Kozai cycles in triples may preferentially lead to the formation of close binaries with moderate mass ratios $q$~$\gtrsim$~0.5 \citep{Kiseleva1998,Naoz2014}, it cannot alone reproduce the observed sharp, nearly discontinuous excess fraction of twins with $q$~=~0.95\,-\,1.00. Orbital migration toward shorter periods must generally occur on rapid timescales $\tau$~$\lesssim$~3\,($M_1$/\Msun)$^{-1}$\,Myr during the early pre-MS phase.  Only during these early epochs can the mass ratios of close binaries coevolve toward unity, either through pre-MS Roche lobe overflow and/or shared accretion in the primordial disk.  At later times, the binary components are too small to exchange material and the disk masses have been sufficiently reduced so that coevolution toward $q$~$\approx$~1 is unlikely. As discussed above, we conclude the longer pre-MS lifetimes of solar-type primaries may lead to the larger excess twin fraction at short and intermediate periods compared to their early-type counterparts.  If Kozai cycles in triples produce close binaries, it must operate concurrently with the early pre-MS phase at least ${\cal F}_{\rm twin}$~$\approx$~(10\,-\,30)\% of the time. The processes of disk fragmentation, pre-MS Roche-lobe overflow, accretion and migration in the circumbinary disk, and dynamical perturbations due to outer tertiary companions most likely all contribute to the formation of close binaries.

There is an increasingly popular theory that both solar-type and massive stars may be born with similar binary star fractions and multiplicity statistics during the early pre-MS phase of formation, but then subsequent dynamical processing and ejections dramatically reduce the binary fraction of the longer lived solar-type MS stars \citep{Marks2011,Thies2015}.  In this paradigm, the differences in the binary star properties between O-type MS stars and solar-type MS stars stem primarily from age, not primary mass.  To investigate this hypothesis, we analyze the statistics of younger MS and pre-MS solar-type binaries in open clusters and associations.  In this manner, we account for the compounding age variable so that we can more reliably compare the multiplicity statistics of solar-type and massive stars.

For reference, we display in Fig.~41 the frequency $f_{\rm logP;q>0.1}$ of companions with $q$~$>$~0.1 per decade of orbital period based on our corrections to the \citet{Raghavan2010} survey of solar-type MS stars in the field (Table~11, red data points in bottom panel of Fig.~37). We also show our analytic fits to $f_{\rm logP;q>0.1}$ for solar-type MS (dotted) and O-type MS (dashed) primaries. While the frequency of wide companions to O-type MS stars is twice that measured for solar-type MS stars, the close companion frequency to massive MS stars is an order of magnitude larger than the close companion frequency to solar-type MS stars in the field.

We start by compiling observations of solar-type MS binaries in young open clusters with ages $\tau$~$<$~150~Myr.  At such young ages, we expect only $\lesssim$2\% of solar-type MS stars to be the secondaries in binaries in which the more massive primaries have already evolved into compact remnants (\S8.3; blue curve in Fig.~29).  Similarly, the frequency of closely orbiting solar-type MS + WD binaries in young open clusters with $\tau$~$\approx$~150~Myr is $\approx$4 times smaller than that calculated for the field (red line in Fig.~29).  For the following populations under consideration, the level of contamination by WD companions is smaller than the measurement uncertainties, and so we assume that all observed solar-type primaries are the true primaries of their respective systems.

\citet{Leiner2015} obtained multi-epoch high-resolution spectra of 418 confirmed and likely solar-type MS members of the M35 open cluster ($\tau$~$\approx$~150~Myr).  Within this sample, they identified 9 SB2s with $P$~$<$~100~days and $q$~$>$~0.6, and 35 SB1s with $P$~$<$~1000~days.  The \citet{Leiner2015} binary survey is sensitive to velocity semi-amplitudes $K_1$~$\approx$~3\,km\,s$^{-1}$, and so binaries with $q$~$>$~0.1 and $P$~$<$~200~days are $\gtrsim$95\% complete in their sample assuming random orientations.  Based on the 4 SB2s and 10 SB1s with $P$~=~2\,-\,20~days ($\Delta$\,log\,$P$ = 1.0), we calculate a companion frequency of $f_{\rm logP;q>0.1}$ = (4\,+\,10)/418/1.0 = 0.033\,$\pm$\,0.009 at very short periods.  Similarly, using the 5 SB2s and 11 SB1s with $P$~=~20\,-\,200~days ($\Delta$\,log\,$P$ = 1.0), we measure $f_{\rm logP;q>0.1}$ = (5\,+\,11)/418/1.0 = 0.038\,$\pm$\,0.010 companions with $q$~$>$~0.1 per decade of orbital period.  The remaining 14 of our selected SB1s reside across $P$~=~200\,-\,1,000~days ($\Delta$\,log\,$P$ = 0.7) where the observations are slightly incomplete toward systems with small mass ratios $q$~=~0.1\,-\,0.4 systems.  We estimate a small overall correction factor $\approx$1.2 to account for the incompleteness at these intermediate separations, and so the companion frequency is $f_{\rm logP;q>0.1}$ = 14$\times$1.2/418/0.7 = 0.057\,$\pm$\,0.015.  We display these three data points based on SB companions to solar-type MS stars in M35 as red data points in Fig.~41.

 \begin{figure}[b!]
\centerline{
\includegraphics[trim=0.6cm 0.3cm 0.8cm 0.1cm, clip=true, width=3.65in]{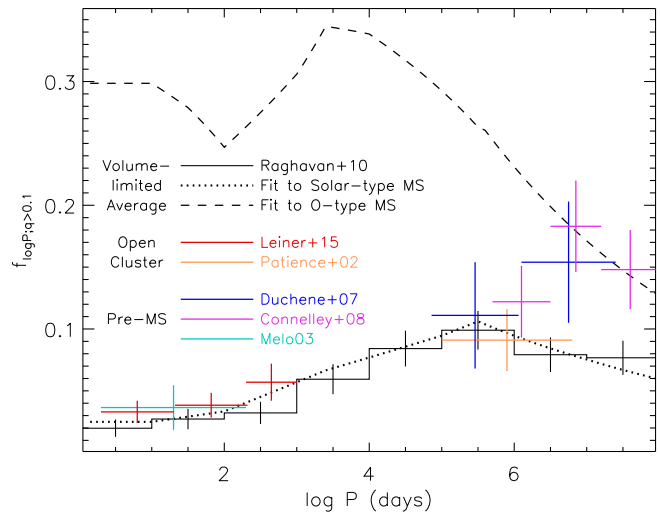}}
\caption{The frequency $f_{\rm logP;q>0.1}$ of companions with $q$~$>$~0.1 per decade of orbital period for various populations of primary stars after correcting for selection effects and incompleteness.  In black, we compare the distribution based on the volume-limited sample of solar-type MS stars in the field (solid) to our analytic fits for the solar-type MS (dotted) and O-type MS (dashed) populations. We also display the binary properties of solar-type stars in young open clusters (red and orange) and of pre-MS solar-type stars (blue, magenta, and cyan).  For wide binaries (log\,$P$~=~6\,-\,8; $a$~=~200\,-\,5,000~AU) that form via fragmentation of cores / filaments, the multiplicity statistics of solar-type pre-MS stars is consistent with that measured for massive MS stars.  Subsequent dynamical processing that preferentially ejects $q$~=~0.1\,-\,0.3 companions reduces the overall companion frequency of solar-type binaries at these separations. For closer binaries (log\,$P$~$<$~6; $a$~$<$~200\,AU) that form via disk fragmentation, the solar-type MS and pre-MS companions frequencies are consistent with each other, both of which are significantly smaller than the measured O-type MS companions frequencies.  A primary-mass dependent process, e.g., more massive protostars are more prone to disk fragmentation, is required to explain why massive stars have larger companion frequencies at short and intermediate separations. }
\end{figure}

\citet{Patience2002} utilized ground-based speckle imaging and direct imaging with {\it HST} to resolve companions to B\,-\,K MS stars in both the $\alpha$~Persei (90~Myr) and Praesepe (660~Myr) open clusters.  We select their 79 systems with K~=~8.5\,-\,10.5\,mag (F5\,-\,K5 primary spectral types) in the younger $\alpha$~Per cluster. For these primaries, \citet{Patience2002} identified  12 companions with $q$~$>$~0.1 ($\Delta$K~$\lesssim$~5.5\,mag) across projected separations $\rho$~=~0.3$''$\,-\,5$''$, i.e., $a$~$\approx$~50\,-\,800\,AU (log\,$P$~=~5.0\,-\,6.8) given the distance $d$~$\approx$~180\,pc to $\alpha$~Per.  Their survey was relatively complete down to $q$~$=$~0.1 ($\Delta$K~$\approx$~5.5\,mag) across this separation range, and so we measure $f_{\rm logP;q>0.1}$ = 13/79/(6.8\,$-$\,5.0) = 0.091\,$\pm$\,0.025 companions with $q$~$>$~0.1 per decade of orbital period across log\,$P$\,(days)~=~5.0\,-\,6.8 (orange data point in Fig.~41).

As can be seen in Fig.~41, the frequency of companions to solar-type MS stars in young open clusters is consistent with that measured in the field population across both short (log~$P$~$<$~3) and long (log~$P$~=~5\,-\,7) orbital periods.  Other surveys have also concluded that the statistics of solar-type binaries in open clusters are indistinguishable from those in the field for a broad range of cluster densities and ages $\tau$~$\approx$~3\,Myr~-~7\,Gyr \citep{Bouvier1997,Kohler2006,Kraus2011,Geller2012,King2012}.  This demonstrates that the formation of solar-type binaries is relatively universal and that there is negligible evolution of the solar-type MS binary statistics for ages $\tau$~$\gtrsim$~3~Myr.

We next turn our attention to wide companions to pre-MS protostars with ages $\tau$~$\lesssim$~3~Myr.  Unlike MS binaries, where the measured brightness contrasts map robustly to mass ratios, accretion luminosity in pre-MS stars can dominate over the photospheric flux.  Accounting for this effect, \citet{Connelley2008a} estimate a near-infrared brightness contrast $\Delta$L~=~4~mag roughly corresponds to $q$~$\approx$~0.1 for coeval pre-MS binaries on the Hayashi track.

\citet{Duchene2007} employed near-IR adaptive optics to search for wide companions to 45 Class~I and flat spectrum protostars embedded in four different molecular clouds.  They identified 15 physically associated companions with separations $\rho$~=~0.2$''$\,-\,10.0$''$ and brightness contrasts $\Delta$L~$<$~4~mag ($q$~$\gtrsim$~0.1).  \citet{Duchene2007} is complete to $\Delta$L~$<$~4.0~mag companions across this separation range, which corresponds to $a$~$\approx$~40\,-\,2,000\,AU (log\,$P$~=~4.9\,-\,7.4) given the average distance $d$~$\approx$~200\,pc to the four molecular clouds.  Of our 15 selected binaries, 6 have $\rho$~=~0.2$''$\,-\,1.2$''$ (log\,$P$ = 4.9\,-\,6.1) and the remaining 9 have 
$\rho$~=~1.2$''$\,-\,10$''$ (log\,$P$~=~6.1\,-\,7.4).  We measure $f_{\rm logP;q>0.1}$ = 6/45/(6.1\,$-$\,4.9) = 0.11\,$\pm$\,0.04 across log\,$P$\,(days)~=~4.9\,-\,6.1 and $f_{\rm logP;q>0.1}$ = 9/45/(7.4\,$-$\,6.1) = 0.15\,$\pm$\,0.05 across log\,$P$\,(days)~=~6.1\,-\,7.4 (blue data points in Fig.~41).

\citet{Connelley2008a} observed a much larger sample of 189 Class I young stellar objects (YSOs) in the near-infrared.  They identified a total of 65 companions with separations $\rho$~=~0.3$''$\,-\,10$''$ and brightness contrasts $\Delta$L~$<$~4~mag ($q$~$\gtrsim$~0.1).  We note that 13 of their YSOs have two or even three resolved companions that reside in the narrow interval $\rho$~=~0.3$''$\,-\,10$''$, contributing 31 of the 65 total companions in our statistic.  A significant fraction of these triples and quadruples in which the companions all have similar separations are most likely gravitationally unstable in their current configurations.  If so, either they will dynamically evolve into stable hierarchical configurations or one of the components will get ejected (see more below).  The \citet{Connelley2008a} survey is complete to $\Delta$L~$=$~4~mag ($q$~$\approx$~0.1) across our selected interval $\rho$~=~0.3$''$\,-\,10$''$, which corresponds to $a$~=~150\,-\,5,000\,AU (log\,$P$ = 5.7\,-\,8.0) given the average distance $d$~$\approx$~500~pc to the YSOs.  We divide our 65 companions across three intervals: 18 with $\rho$~=~0.3$''$\,-\,1.0$''$ ($f_{\rm logP;q>0.1}$ = 0.12\,$\pm$\,0.03 across log\,$P$~=~5.7\,-\,6.5), 25 companions with $\rho$~=~1.0$''$\,-\,3.0$''$ ($f_{\rm logP;q>0.1}$ = 0.18\,$\pm$\,0.04 across log\,$P$~=~6.5\,-\,7.2), and 22 companions with $\rho$~=~3.0$''$\,-\,10.0$''$ ($f_{\rm logP;q>0.1}$ = 0.15\,$\pm$\,0.03 across log\,$P$~=~7.2\,-\,8.0).  We display these three data points in magenta in Fig.~41.

 Finally, we analyze the spectroscopic binary survey of pre-MS T Tauri stars conducted by \citet{Melo2003}, who updated and extended the sample of \citet{Mathieu1992,Mathieu1994}.  \citet{Melo2003} identified 4 SBs with $P$~=~2\,-\,200 days ($\Delta$\,log\,$P$~=~2.0) within their sample of 65 T~Tauri stars.  According to their Fig.~2, the \citet{Melo2003} survey is relatively complete toward binaries with $q$~$>$~0.3 and $P$~$<$~200~days, but rather insensitive to companions with $q$~=~0.1\,-\,0.3.  We adopt a correction factor of 1.2 to account for incompleteness toward small mass ratios.  We find $f_{\rm logP;q>0.1}$ = 4$\times$1.2/65/2.0 = 0.037\,$\pm$0.018 companions with $q$~$>$~0.1 per decade of orbital period across $P$~=~2\,-\,200 days (cyan data point in Fig.~41).

 As shown in Fig.~41, the frequency of wide companions (log\,$P$~=~6\,-\,8; $a$~=~200\,-\,5,000~AU) to solar-type pre-MS stars is larger than that measured for solar-type MS stars in the field.  Indeed, \citet{Ghez1993}, \citet{Duchene2007}, \citet{Connelley2008b}, and \citet{Tobin2016} all measure the frequency of wide companions to solar-type pre-MS stars to be $\approx$\,2\,-\,3 times higher than that measured for the field population. Moreover, the frequency of wide companions to solar-type pre-MS stars is consistent with the frequency of wide companions to O-type MS stars.

These observations of wide companions are largely in agreement with theoretical models of multiple star formation via fragmentation of molecular cores / filaments on large spatial scales \citep{Goodwin2005,Marks2011,Offner2012,Thies2015}.  Namely, the observations are consistent with the notion that wide binaries are born during the early pre-MS phase with (1) the same companion frequency independent of primary mass, and (2) mass ratios consistent with random pairings drawn from the IMF. In fact, if we focus solely on wide binaries with modest mass ratios $q$ $>$ 0.3, the frequency of such wide companions $q$~$>$~0.3 to O-type MS stars already matches the frequency of wide companions $q$~$>$~0.3 to solar-type MS stars in the field (see top panel of Fig.~37).  As mentioned above, a significant fraction of the wide companions to pre-MS solar-type stars are dynamically unstable.  Dynamical processing tends to eject the lower mass companions with $q$~$\approx$~0.1\,-\,0.3 due to their lower binding energies \citep{Marks2011,Kouwenhoven2010,Perets2012,Thies2015}.  This explains why O-type MS stars, which are only a few Myr old, exhibit a larger frequency of wide companions with mass ratios that are nearly consistent with random pairings drawn from the IMF ({\large $\gamma$}$_{\rm largeq}$~=~$-$2.0 and {\large $\gamma$}$_{\rm smallq}$~=~$-$1.5).  Meanwhile, solar-type MS stars in the field, which are several Gyr old, have a smaller frequency of wide companions and show a large deficit of $q$~=~0.1\,-\,0.3 companions relative to random pairings from the IMF (see Fig.~30).

However, the theory that the multiplicity statistics at the time of pre-MS formation are independent of primary mass holds true {\it only} for wide binaries ($a$~$>$~200~AU) that derive from fragmentation of cores / filaments.  At smaller separations $a$~$\lesssim$~200~AU, where we believe binaries formed predominantly via disk fragmentation, a different picture emerges.  As shown in Fig.~41, the frequency of companions to pre-MS solar-type stars across short (log\,$P$~$<$~2.3; $a$~$<$~0.7\,AU) and intermediate (log\,$P$~=~5\,-\,6; $a$ ~=~50\,-\,200~AU) separations is consistent with that measured for solar-type MS stars in the field.  This demonstrates the properties of solar-type binaries with $a$~$<$~200~AU are set very early $\tau$~$\lesssim$~2~Myr in the formation process.  By integrating $f_{\rm logP;q>0.1}$, we measure a multiplicity frequency of $f_{\rm mult;q>0.1}$~$\approx$~0.7\,-\,0.8 companions per primary for solar-type pre-MS systems.  This is only slightly larger than the multiplicity frequency $f_{\rm mult;q>0.1}$~$=$~0.5 of solar-type MS stars in the field and open clusters.  Many of the observed wide companions to solar-type pre-MS primaries are actually tertiaries (see discussion of the \citealt{Connelley2008a} survey above), and so we estimate that $\approx$(30\,-\,40)\% of pre-MS solar-type stars are single.  These apparently single pre-MS solar-type stars may still have companions with $q$~$<$~0.1 and/or log\,$P$\,(days)~$\approx$~8\,-\,10 outside the parameter space investigated in this study.  If these supposed companions are included, then the single star fraction of solar-type pre-MS stars may be closer to zero.  As motivated in \citet{Kroupa1995}, the angular momentum barrier inhibits the formation of single stars, and so the fraction of pre-MS stars with stellar and/or substellar companions should be large.  In any case, even after removing the compounding age variable, the frequency of close and intermediate-period companions with $q$~$>$~0.1 to O-type MS stars is significantly larger than that measured for solar-type pre-MS stars. The dominant reason why the overall multiplicity frequency of O-type MS stars ($f_{\rm mult;q>0.1}$~=~2.1\,$\pm$\,0.3) is larger is due to the substantially larger companion frequency at short and intermediate separations $a$~$<$ 200~AU (see Fig.~41).  This reaffirms our conclusion that a primary-mass dependent process, e.g., the disks around more massive protostars are more prone to fragmentation \citep{Kratter2006}, is the main reason why the binary fraction and multiplicity frequency of massive stars are larger.

\section{Binary Star Evolution}

In the following, we estimate the fraction of systems that will substantially interact with a binary companion $q$ $>$ 0.1 via Roche-lobe overflow (RLOF) as a function of primary mass.  Solar-type primaries with $M_1$~$\approx$~1.0\,\Msun\ expand to $R_1$ $\approx$ 250\,\Rsun\ at the tip of the AGB \citep{Bertelli2008}.  Assuming a uniform eccentricity distribution $\eta$~=~0.0, then the average eccentricity is $\langle e \rangle$~=~0.5.  Meanwhile, a thermal eccentricity distribution provides $\langle e \rangle$~=~0.66.   Both early-type and solar-type binaries with intermediate orbital periods log\,$P$~=~3\,-\,4 have initial eccentricity distributions between these two limits (\S9.2; Fig.~36).  Because the AGB phase is extremely short $\tau_{\rm AGB}$~$\approx$~1~Myr \citep{Bertelli2008}, binaries with intermediate orbital periods log\,$P$~$\approx$~3.5 tidally evolve to only marginally smaller eccentricities prior to Roche-lobe filling. We adopt $\langle e \rangle$ = 0.5, and so solar-type binaries with $a$ $\lesssim$ 2.2$R_1$/(1\,$-$\,$\langle e \rangle$) $\approx$ 1,100\,\Rsun\ $\approx$ 5\,AU will fill their Roche-lobes at periastron \citep{Eggleton1983}.  According to Kepler's laws, solar-type binaries with $P$~$\lesssim$~10~years, i.e., log\,$P$\,(days)~$<$~3.6, undergo RLOF.

Meanwhile, massive stars with $M_1$ $>$ 15\Msun\ expand to much larger radii $R_1$ = 700\,-\,1,200\,\Rsun\ during their red supergiant phase \citep{Bertelli2009}.   Early-type binaries can fill their Roche-lobes across even wider separations $a$ $\lesssim$ 2.2$R_1$/(1\,$-$\,$\langle e \rangle$) $\approx$ 4,000\,\Rsun\ $\approx$ 19\,AU. Assuming $M_1$ = 20\,\Msun\ and the average mass ratio $\langle q \rangle$~=~0.3 of massive binaries with intermediate periods (\S9.1; Fig.~35), then early-type binaries with $P$~$\lesssim$~16~years, i.e., log\,$P$\,(days) $\lesssim$ 3.8, undergo RLOF.  Although more massive stars expand to larger radii, the threshold orbital period log\,$P$\,(days) $\lesssim$ 3.7 below which binaries fill their Roche-lobes is relatively independent of $M_1$ due to Kepler's laws (see Fig.~1).

Slightly wider companions with log\,$P$~$\approx$~3.7\,-\,4.5 may encounter  enhanced wind accretion, especially if the geometry and evolution of the primary's mass loss is affected by the companion \citep{Mohamed2007,Moe2012}.  We nonetheless consider wind-accreting systems with 
log\,$P$~$\gtrsim$~3.7 as weak binary interactions compared to closer binaries with log\,$P$~$\lesssim$~3.7 which experience full RLOF.  In the short-period regime, we have measured the statistics of companions down to log\,$P$ = 0.2. Despite the uncertainty in the functional form of the period distribution below log\,$P$~$<$~0.2, the addition of companions with log\,$P$~$<$~0.2 is negligible compared to the overall companion frequency at longer periods.  The frequency $f_{\rm 0.2<logP<3.7;q>0.1}$ of companions with mass ratios $q$~$>$~0.1 across orbital periods 0.2~$<$~log\,$P$~$<$~3.7 is therefore a reliable indicator of the fraction of primaries that experience significant binary evolution via RLOF.

As a function of primary mass $M_1$, we calculate $f_{\rm 0.2<logP<3.7;q>0.1}$ by integrating $f_{\rm logP;q>0.1}$ across 0.2~$<$~log\,$P$~$<$~3.7.  We measure the uncertainties as done in \S9.4, and we display our results in Fig.~42.  Only (15\,$\pm$\,3)\% of solar-type primaries experience RLOF with companions $q$ $>$ 0.1.  Meanwhile, the frequency $f_{\rm 0.2<logP<3.7;q>0.1}$ = 1.0\,$\pm$\,0.2 of companions with $q$ $>$ 0.1 and 0.2~$<$~log\,$P$~$<$~3.7 to O-type primaries is nearly an order of magnitude larger.  
Essentially all O-type primaries undergo RLOF with companions $q$~$>$~0.1. In fact, the measured frequency $f_{\rm 0.2<logP<3.7;q>0.1}$~=~1.0\,$\pm$\,0.2 is quite close to and may exceed unity.  This suggests that $\approx$(10\,-\,20)\% of O-type primaries are in compact triple configurations in which the outer tertiary has $q$~$>$~0.1 and log\,$P_{\rm outer}$~$<$~3.7 ($a_{\rm outer}$~$\lesssim$~10\,AU).  Close tertiaries can induce Kozai oscillations and may cause the inner binary to merge while still on the MS, thereby producing a blue straggler \citep{Perets2009}. If instead the inner binary first evolves into a pair of compact remnants, for example, the tertiary may accelerate the merger of the two compact objects and lead to the formation of a Type Ia supernova or short gamma-ray burst \citep{Thompson2011}.  The evolution of compact triples should be studied in more detail, especially if they are relatively more common among massive stars.

\begin{figure}[t!]
\centerline{
\includegraphics[trim=0.4cm 0.4cm 0.3cm 0.0cm, clip=true, width=3.6in]{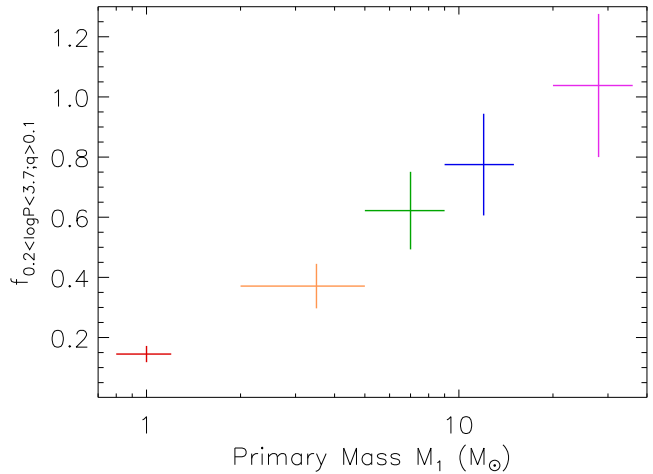}}
\caption{The frequency of companions with $q$~$>$~0.1 and 0.2~$<$~log\,$P$\,(days)~$<$~3.7 per primary as a function of primary mass $M_1$.  Only (15\,$\pm$\,3)\% of solar-type primaries (red) will experience significant binary evolution via Roche-lobe overflow (RLOF).  Meanwhile, essentially all O-type primaries (magenta) will undergo RLOF with companions $q$~$>$~0.1. About (10\,-\,20)\% of O-type primaries are in compact triples in which the outer tertiary has log\,$P$~$<$~3.7 and may therefore significantly affect the evolution of the inner binary. }
\end{figure}

\citet{Sana2012} report that (71\,$\pm$\,8)\% of O-type stars will interact  with companions $q$~$>$~0.1 via RLOF.  Our estimate of $f_{\rm 0.2<logP<3.7;q>0.1}$~=~1.0\,$\pm$\,0.2 is consistent with this estimate but slightly larger for two reasons.  First, \citet{Sana2012} consider only binaries with $P$ $<$ 1,500 days, i.e., log\,$P$ $<$ 3.2, to experience significant binary evolution.  This is primarily because they measure the power-law slope $\eta$ = $-$0.4\,$\pm$\,0.2 of the eccentricity distribution to be weighted toward small values.  Although $\eta$ = $-$0.4 describes the eccentricity distribution of short-period binaries with $P$~$<$~20~days, we find that massive binaries with intermediate periods 2~$<$~log\,$P$~$<$~4 are weighted toward larger eccentricities ($\eta$ $\approx$ 0.8; Fig.~36). Early-type binaries with slightly longer orbital periods log\,$P$ $\approx$ 3.2\,-\,3.7 undergo RLOF at periastron given $\langle e \rangle$ $\approx$ 0.5.  This effect increases the fraction of O-type stars that will interact with a companion by $f_{\rm logP;q>0.1} \Delta$log$P$ $\approx$ 0.3\,$\times$\,0.5 = 0.15 (see bottom panel of Fig.~37).  

Second,  while \citet{Sana2012} assume the distributions of mass ratios and orbital periods are independent, we find that early-type binaries with intermediate periods are weighted toward smaller mass ratios.  There are more companions with $q$~$\approx$~0.1\,-\,0.4 and log\,$P$~$\approx$~2\,-\,3 to O-type stars than that predicted by \citet{Sana2012}. More recent observations with long-baseline interferometry confirm an enhanced companion frequency at intermediate periods log\,$P$~=~3.5 \citep[][see Fig.~37]{Rizzuto2013,Sana2014}.  This second effect increases the fraction of O-type primaries that will interact with a binary companion by an additional $\approx$15\%.  

Because we find early-type binaries with intermediate orbital periods are weighted toward larger eccentricities and smaller mass ratios, the frequency of companions that will interact with a massive primary increases by $\approx$30\%.  We still reaffirm the overall conclusion of \citet{Sana2012} that massive stars are dominated by interactions with binary companions.  We simply find that the fraction is even larger if we account for the variations between $P$, $q$, and $e$.  Moreover, the \citet{Sana2012} spectroscopic binary sample contains only companions that are members of the inner binary.  Meanwhile, long-baseline interferometry is sensitive to all companions $q$~$>$~0.3 across intermediate orbital periods, regardless if the companions are outer tertiaries or members of the inner binaries.  In fact, LBI surveys have detected outer tertiaries at log\,$P_{\rm outer}$\,(days)~$\approx$~3\,-\,4 to massive stars in compact triple configurations \citep{Rizzuto2013,Sana2014}.  This is why we estimate that $\approx$(80\,-\,90)\% of massive stars will interact with a companion, and $\approx$(10\,-\,20)\% of massive primaries are in compact triple configurations with log\,$P_{\rm outer}$\,(days)~$<$~3.7 ($a_{\rm outer}$~$\lesssim$~10~AU).  Combining these two statistics brings the total close companion frequency to our measured value $f_{\rm 0.2<logP<3.7;q>0.1}$~=~1.0\,$\pm$\,0.2 for massive O-type MS primaries.

We next utilize the measured multiplicity statistics to estimate the fraction ${\cal F}_{\rm evol}$ of early-type primaries that are actually the products of binary evolution.  The fraction ${\cal F}_{\rm evol}$ not only includes close binaries that merge or experience stable mass transfer,  but also wide companions in binaries in which the true primaries have already evolved into compact remnants.  Using a Monte Carlo technique, we simulate a large population of single and binary early-type stars (similar to our methods in \S8.3.2 for solar-type systems).  We first select primaries across 4\,\Msun~$<$~$M_1$~$<$~40\,\Msun\ from a Salpeter IMF. Given $M_1$, we then determine the properties of the companions, i.e., intrinsic frequency, period, and mass ratio, based on our probability distributions $f(P,\,q\,|\,M_1)$ measured in \S9.

Once we generate our initial population, we evolve each binary according to the stellar evolutionary tracks of \citet{Bertelli2008,Bertelli2009} and the following assumptions regarding binary interactions.  We assume wide companions with log\,$P$~$>$~3.7 experience negligible mass accretion ($\Delta M_2$~$\approx$~0).  The predicted evolutionary pathways of closer binaries with log\,$P$~$<$~3.7 depend not only on their physical properties $M_1$, $M_2$, and $P$, but also on the still uncertain prescriptions for the physics that describes the interaction \citep{Hurley2002,Belczynski2008}.  For close binaries with log\,$P$~$<$~3.7 which undergo RLOF, we assume for simplicity that all systems either (1) survive common envelope (CE) evolution with negligible mass transfer ($\Delta M_2$~$\approx$~0), (2) undergo stable mass transfer (MT) with $\approx$40\% efficiency ($\Delta M_2$~$\approx$~0.4$M_1$), or (3) merge ($\Delta M_2$~$\approx$~$M_1$). These three scenarios encompass the full range of binary mass transfer efficiency 0~$<$~$\Delta M_2$~$<$~$M_1$ (see below).  For the stable MT systems and mergers, we estimate visibility times on the MS based on the rejuvenated properties of the mass gainers.  For the  post-CE systems and wide binaries, the secondaries are unaffected and evolve according to their birth MS masses.

As a function of age $\tau$, we count the number ${\cal N}_{\rm prim}$ of true primaries with $M_1$ = 8\,-\,12\,\Msun\ that are still on the MS.  We also keep track of the number ${\cal N}_{\rm evol}$ of evolved systems with MS secondaries $M_2$ = 8\,-\,12\,\Msun.  This number includes both MS merger products and MS secondaries in which the primaries have already evolved into compact remnants.  The fraction of apparent primaries that are actually products of binary evolution is  ${\cal F}_{\rm evol}$ = ${\cal N}_{\rm evol}$/(${\cal N}_{\rm evol}$\,+\,${\cal N}_{\rm prim}$).

In Fig.~43, we display ${\cal F}_{\rm evol}$($\tau$) as a function of age for our three scenarios of close binary evolution.  Systems as young as $\tau$~=~3~Myr are already contaminated by products of binary evolution.  This is because massive primaries $M_1$~$\approx$~30\,-\,40\,\Msun\ with short-period companions $P$~$\lesssim$~5~days fill their Roche lobes within $\tau$~$\lesssim$~3~Myr.  Some high-mass X-ray binaries, such as Cyg~X-1 for example, are expected to be only $\tau$~$\approx$~4\,-\,7~Myr old \citep{Mirabel2003}.

At ages $\tau$ $\approx$ 15\,-\,18 Myr, none of the true primaries in our mass interval $M_1$ = 8\,-\,12\,\Msun\ of interest have yet evolved off the MS.  Nevertheless, ${\cal F}_{\rm evol}$ = (12\,-\,28)\% of the systems are actually products of binary evolution that derived from initially more massive primaries (see Fig.~43).  The range in ${\cal F}_{\rm evol}$ is dominated by the uncertainty in the mass transfer efficiencies of close binaries.  For example, mergers contribute the least to ${\cal F}_{\rm evol}$ at these ages. In fact, the majority of systems represented by the green line in Fig.~43 are wide binaries that avoid RLOF, {\it not} mergers.   
Alternatively, binaries that evolve via stable MT produce the largest fraction ${\cal F}_{\rm evol}$ at $\tau$~=~15\,-\,18~Myr (Fig.~43).  We vary the mass transfer efficiency parameter in our population synthesis models, and find that $\Delta M_2$~$\approx$~0.4$M_1$ maximizes ${\cal F}_{\rm evol}$ at these ages. The mass gainers in these stable MT systems are the massive counterparts to blue stragglers \citep{Schneider2014}.  

\begin{figure}[t!]
\centerline{
\includegraphics[trim=0.4cm 0.0cm 0.3cm 0.1cm, clip=true, width=3.5in]{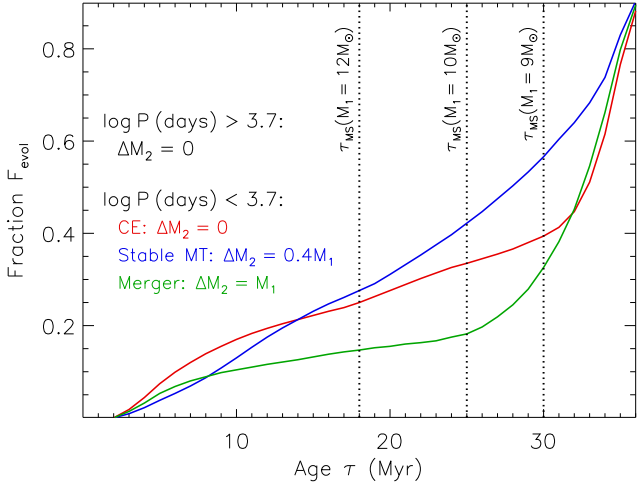}}
\caption{The fraction ${\cal F}_{\rm evol}$ of primaries with $M_1$~=~8\,-\,12\,\Msun\ that are actually the secondaries in binaries in which the true primaries have already evolved into neutron stars or black holes via core-collapse supernovae.  We assume wide companions with log\,$P$~$>$~3.7 do not interact, while all close binaries with log\,$P$~$<$~3.7 experience Roche-lobe overflow and either survive common envelope evolution (red), undergo stable mass transfer (blue), or merge (green). A population of early-type primaries is already contaminated by products of binary evolution by $\tau$~$\gtrsim$~3\,Myr.  All MS stars with masses $M$~=~8\,-\,12\,\Msun\ and ages $\tau$~$>$~$\tau_{\rm MS}$(8\,\Msun)~$=$~37~Myr must be rejuvenated secondaries that gained mass from the primaries, i.e., the massive counterparts to blue stragglers.  Assuming a constant star formation rate, we measure an average fraction $\langle {\cal F}_{\rm evol}\rangle$ = 12\,-\,25\%.  The ratio of true primaries to observed primaries in a volume-limited sample of early-type systems is the correction factor ${\cal C}_{\rm evol}$~=~1.2\,$\pm$\,0.1.}
\end{figure}

With increasing age, a larger fraction of apparent primaries are actually the original secondaries in which the true primaries have evolved into compact remnants.  At $\tau$~=~25~Myr, for example, ${\cal F}_{\rm evol}$~$\approx$~(20\,-\,40)\% of systems are the products of binary evolution while the remaining (60\,-\,80)\% are true primaries with $M_1$~=~8\,-\,10\,\Msun\ (see Fig.~43).  All the true primaries with $M_1$~=~10\,-\,12\,\Msun\ have already evolved off the MS by $\tau$~=~25~Myr.  Similarly, at $\tau$~=~30~Myr, ${\cal F}_{\rm evol}$~$\approx$~(35\,-\,55)\% of systems are secondaries while the remaining (45\,-\,65)\% are true primaries with $M_1$~=~8\,-\,9\,\Msun.   By construction, all systems with ages $\tau$~$>$~37~Myr beyond the MS lifetimes of $M_1$~$>$~8\,\Msun\ primaries must be the original secondaries. In all three scenarios of close binary evolution, the fraction ${\cal F}_{\rm evol}$~=~1 must approach unity by $\tau$~=~37~Myr. 

Assuming a constant star formation rate, we calculate the average fraction $\langle{\cal F}_{\rm evol} \rangle$ of primaries that are actually the products of binary evolution.  We measure $\langle{\cal F}_{\rm evol} \rangle$ = 0.21, 0.25, and 0.12 for the CE, stable MT, and merger scenarios, respectively.  In a volume-limited sample of early-type stars, $\langle{\cal F}_{\rm evol} \rangle$ = (12\,-\,25)\% of the apparent primaries are actually the secondaries in evolved binaries. The correction factor due this effect of binary evolution is ${\cal C}_{\rm evol}$ = 1/(1-$\langle{\cal F}_{\rm evol} \rangle$) = 1.2\,$\pm$\,0.1.  If there is no prior information on the ages of the systems, we have incorporated this correction factor of ${\cal C}_{\rm evol}$ = 1.2\,$\pm$\,0.1 into our statistical analysis.  If instead the observational sample contains only systems in young open clusters $\tau$~$\lesssim$~5~Myr \citep[e.g.,][]{Sana2012} or if we limit our sample to systematically younger primaries with luminosity classes III-V (see \S5 and \S6), then we have adopted a slightly smaller correction factor ${\cal C}_{\rm evol}$ $\approx$ 1.0\,-\,1.1.

\citet{deMink2014} report that $\langle{\cal F}_{\rm evol} \rangle$ = 30$_{-15}^{+10}$\% of early-type systems are the products of binary interactions.   This is slightly larger than but consistent with our estimate of $\langle{\cal F}_{\rm evol} \rangle$ = (12\,-\,25)\%.  The main reason for the difference is because \citet{deMink2014} utilized the measured binary properties of O-type primaries with $\langle M_1 \rangle$~$\approx$~28\,\Msun\ to estimate $\langle{\cal F}_{\rm evol} \rangle$.  We find the multiplicity frequencies of early/mid B-type stars to be 60\,-\,80\% the O-type values (\S9.4 and Fig.~38), and so $\langle{\cal F}_{\rm evol} \rangle$ for $M_1$ = 8\,-\,12\,\Msun\ primaries is correspondingly smaller.  For a volume-limited sample of solar-type primaries $M_1$ $\approx$ 1\Msun, we measure an even smaller fraction $\langle{\cal F}_{\rm evol} \rangle$ = (11\,$\pm$\,4)\% (\S8.4-8.5).   In a volume-limited sample, more massive primaries are more likely to be the products of binary evolution due to their intrinsically larger multiplicity frequencies.  

About (10\,-\,30)\% of O/early-B stars are ``runaway stars" that have large peculiar velocities $v$~$\approx$~50\,-\,200~km~s$^{-1}$ and are found in isolation removed from young clusters/associations \citep{Gies1987,Stone1991,Hoogerwerf2001}.  \citet{Hoogerwerf2001} discuss the two preferred formation channels of runaway OB stars.  First, gravitational interactions between massive stars in dense open clusters can lead to dynamical ejections of a small fraction of OB stars \citep{Poveda1967,Oh2015}.  Second, OB companions in massive binaries receive large recoil kicks when the primaries explode as supernovae \citep{Blaauw1961}.  We measure that (12\,-\,25)\% of early-type stars in a volume-limited sample are the secondaries in binaries in which the original primaries have already collapsed into neutron stars or black holes via core-collapse supernovae.  This happens to match the observed fraction of OB stars that are runaway. However, we do not know precisely what fraction of OB companions to massive stars will receive large kicks and become runaway stars after the primaries explode. If this fraction is large, then in principle, there are sufficient numbers of OB companions to massive stars to explain runaway stars via the binary star supernova-kick scenario.  Detailed population synthesis models utilizing our updated multiplicity statistics for massive stars are needed to confirm this conclusion.

Given the uncertainties in CE evolution, stable mass transfer, and supernova kick velocities, we cannot robustly predict via population synthesis models the fraction of early-type stars with closely orbiting compact remnant companions.  Nonetheless, the observed ratio of early-type SB1s to SB2s increases dramatically with age, indicating that 30$_{-15}^{+10}$\% of early-type SB1s in a volume-limited sample actually contain compact remnant companions (see \S3.5).  Only a certain fraction of compact remnant companions will exhibit signs of accretion, e.g., X-ray emission, UV excess, emission lines, etc.  For example, Cyg X-1 is a short-period $P$~=~5.6~day binary with a relatively large $R_{\rm donor}$~$\approx$~20\,\Rsun, evolved,  blue supergiant donor of type O9.7Iab \citep{Mirabel2003,Ziolkowski2005}. If instead the orbital period $P$~$\approx$~15~days was slightly longer and if the early-type donor was a less evolved, non-rotating B2V dwarf star with $R_{\rm donor}$~$\approx$~6\Rsun, then the mass accretion rate onto the compact object would be $\approx$\,3\,-\,4 orders of magnitude smaller according to binary evolution models \citep{Hurley2002,Belczynski2008}.  A non-negligible fraction of early-type SB1s likely contain neutron star and black hole companions in such orbital configurations and phases of evolution that do not produce detectable emission lines or X-rays.

\section{Summary}

{\it Mind your Ps and Qs:} The distributions of primary masses~$M_1$, mass ratios~$q$, orbital periods~$P$, eccentricities~$e$, and multiplicity fractions are not independent.  We have compiled $\approx$30 separate surveys and subsamples of binaries that have been identified through spectroscopy, eclipses, long-baseline interferometry, adaptive optics, common-proper motion, etc. (Fig.~1).  By correcting for their respective selection effects (\S3-8), we have modeled the intrinsic joint probability density function $f$($M_1$,\,$q$,\,$P$,\,$e$) $\ne$ $f$($M_1$)$f$($q$)$f$($P$)$f$($e$) and its uncertainty (\S9).  We summarize our main results and statistics in Table~13 according to the five primary mass / MS spectral type intervals investigated in \S9.  

{\it Multiplicity Frequencies and Fractions:}  Counting only stellar companions with $q$~=~$M_{\rm comp}$/$M_1$~$>$~0.1 that directly orbit the primary with orbital periods 0.2~$<$~log\,$P$\,(days)~$<$~8.0 (see \S2 and \S8.1), a solar-type MS primary with $M_1$ = 1\,\Msun\ has, on average, $f_{\rm mult;q>0.1}$ = 0.50\,$\pm$\,0.04 companions.  Meanwhile, the average O-type MS primary has $f_{\rm mult;q>0.1}$ = 2.1\,$\pm$\,0.3 companions with $q$~$>$~0.1 and 0.2~$<$~log\,$P$~$<$~8.0 (\S9.4 and Fig.~38).  We measure the multiplicity fractions of solar-type MS primaries to be ${\cal F}_{n=0;q>0.1}$ = 0.60\,$\pm$\,0.04 single, ${\cal F}_{\rm n=1;q>0.1}$~=~0.30\,$\pm$\,0.04 binary,  ${\cal F}_{\rm n=2;q>0.1}$~=~0.09\,$\pm$\,0.02 triple, and ${\cal F}_{\rm n=3;q>0.1}$~=~0.010\,$\pm$\,0.005 quadruple.  We note the single star fraction for solar-type MS primaries will be lower, probably close to half, if extreme mass-ratio companions $q$~$<$~0.1 and/or extremely wide companions with log\,$P$\,(days)~$>$~8 are considered. For O-type MS primaries, we estimate that only ${\cal F}_{n=0;q>0.1}$~$\lesssim$~10\% are single stars while ${\cal F}_{n\ge2;q>0.1}$~$=$~(73\,$\pm$\,16)\% are in triple or quadruple systems.  The multiplicity fractions ${\cal F}_{n;q>0.1}$ are consistent with a Poisson distribution across $n$ = [0,\,3], suggesting the addition of multiple companions broadly resembles a stochastic process (Fig.~39).  The close binary fraction ${\cal F}_{\rm 0.3<logP<0.8;q>0.1}$ is nearly proportional to the overall triple plus quadruple star fractions ${\cal F}_{n\ge2;q>0.1}$ across all primary spectral types (\S10, Fig.~40).  If dynamical evolution in triples and quadruples is the dominant formation mechanism of close binaries, then the process must have a relatively high efficiency $\epsilon$~=~${\cal F}_{\rm 0.3<logP<0.8;q>0.1}$/${\cal F}_{n\ge2;q>0.1}$~$\approx$~15\%.  In other words, 15\% of triples, regardless of primary mass, have $P_{\rm inner}$~$\lesssim$~6~days.

\renewcommand{\arraystretch}{1.6}
\setlength{\tabcolsep}{2.5pt}
\begin{figure*}[t!]\footnotesize
{\small {\bf Table 13:} Multiplicity statistics as a function of primary mass / spectral type after correcting for observational selection effects.} \\
\vspace*{-0.3cm}
\begin{center}
\begin{tabular}{|c|c|c|c|c|c|c|}
\hline
                          &             &      Solar-type               &  A / late-B                &   mid-B                   &      early-B               &     O-type   \\
Statistic                 & Explanation & $M_1$\,=\,0.8\,-\,1.2\,\Msun\ &  $M_1$\,=\,2\,-\,5\,\Msun\ & $M_1$\,=\,5\,-\,9\,\Msun\ & $M_1$\,=\,9\,-\,16\,\Msun\ & $M_1$\,$>$\,16\,\Msun\ \\
\hline
\hline
$f_{\rm mult;q>0.1}$        &  Total multiplicity frequency       & 0.50\,$\pm$\,0.04      & 0.84\,$\pm$\,0.11    & 1.3\,$\pm$\,0.2     & 1.6\,$\pm$\,0.2      & 2.1\,$\pm$\,0.3    \\
\hline
$f_{\rm logP<3.7;q>0.1}$    &  Close binary frequency             & 0.15\,$\pm$\,0.03      & 0.37\,$\pm$\,0.08    & 0.63\,$\pm$\,0.13   & 0.8\,$\pm$\,0.2      & 1.0\,$\pm$\,0.2    \\ 
\hline
\hline
${\cal F}_{n=0;q>0.1}$      &  Single star fraction               & 0.60\,$\pm$\,0.04      & 0.41\,$\pm$\,0.08    & 0.24\,$\pm$\,0.08   & 0.16\,$\pm$\,0.09    & 0.06\,$\pm$\,0.06  \\
\hline
${\cal F}_{n=1;q>0.1}$      &  Binary star fraction               & 0.30\,$\pm$\,0.04      & 0.37\,$\pm$\,0.06    & 0.36\,$\pm$\,0.08   & 0.32\,$\pm$\,0.10    & 0.21\,$\pm$\,0.11 \\
\hline
${\cal F}_{n\ge2;q>0.1}$    &  Triple + quadruple star fraction   & 0.10\,$\pm$\,0.02      & 0.22\,$\pm$\,0.07    & 0.40\,$\pm$\,0.10   & 0.52\,$\pm$\,0.13    & 0.73\,$\pm$\,0.16 \\
\hline
\hline
                            & Companion frequency across:         &                        &                      &                     &                      &   \\
\hline
$f_{\rm logP=1;q>0.1}$      & log\,$P$\,=\,0.5\,-\,1.5       &  0.027\,$\pm$\,0.009   & 0.07\,$\pm$\,0.02    & 0.14\,$\pm$\,0.04   & 0.19\,$\pm$\,0.06    & 0.29\,$\pm$\,0.08 \\
\hline
$f_{\rm logP=3;q>0.1}$      & log\,$P$\,=\,2.5\,-\,3.5       &  0.057\,$\pm$\,0.016   & 0.12\,$\pm$\,0.04    & 0.22\,$\pm$\,0.07   & 0.26\,$\pm$\,0.09    & 0.32\,$\pm$\,0.11 \\ 
\hline
$f_{\rm logP=5;q>0.1}$      & log\,$P$\,=\,4.5\,-\,5.5       &  0.095\,$\pm$\,0.018   & 0.13\,$\pm$\,0.03    & 0.20\,$\pm$\,0.06   & 0.23\,$\pm$\,0.07    & 0.30\,$\pm$\,0.09 \\
\hline
$f_{\rm logP=7;q>0.1}$      & log\,$P$\,=\,6.5\,-\,7.5       &  0.075\,$\pm$\,0.015   & 0.09\,$\pm$\,0.02    & 0.11\,$\pm$\,0.03   & 0.13\,$\pm$\,0.04    & 0.18\,$\pm$\,0.05 \\
\hline
\hline
                            & Excess twin fraction at:         &                        &                      &                     &                      &   \\
\hline
${\cal F}_{\rm twin}$       & log\,$P$\,=1                     &  0.30\,$\pm$\,0.09     & 0.22\,$\pm$\,0.07    & 0.17\,$\pm$\,0.05   & 0.14\,$\pm$\,0.04    & 0.08\,$\pm$\,0.03 \\
\hline
${\cal F}_{\rm twin}$       & log\,$P$\,=3                     &  0.20\,$\pm$\,0.06     & 0.10\,$\pm$\,0.04    &  $<$\,0.03          & $<$\,0.03            & $<$\,0.03         \\
\hline
${\cal F}_{\rm twin}$       & log\,$P$\,=5                     &  0.10\,$\pm$\,0.03     & $<$\,0.03            &  $<$\,0.03          & $<$\,0.03            & $<$\,0.03         \\ 
\hline
${\cal F}_{\rm twin}$       & log\,$P$\,=7                     &  $<$\,0.03             & $<$\,0.03            &  $<$\,0.03          & $<$\,0.03            & $<$\,0.03         \\ 
\hline
           & Power-law slope of $p_q$\,$\propto$\,$q^{\gamma}$ &                        &                      &                     &                      &   \\
                            & across $q$~=~0.3\,-\,1.0 at:     &                        &                      &                     &                      &   \\
\hline
{\large $\gamma$}$_{\rm largeq}$    & log\,$P$\,=1             &  $-$0.5\,$\pm$\,0.3    & $-$0.5\,$\pm$\,0.3   & $-$0.5\,$\pm$\,0.3  & $-$0.5\,$\pm$\,0.3   & $-$0.5\,$\pm$\,0.3 \\
\hline
{\large $\gamma$}$_{\rm largeq}$    & log\,$P$\,=3             &  $-$0.5\,$\pm$\,0.3    & $-$0.9\,$\pm$\,0.3   & $-$1.7\,$\pm$\,0.3  & $-$1.7\,$\pm$\,0.3   & $-$1.7\,$\pm$\,0.3 \\
\hline
{\large $\gamma$}$_{\rm largeq}$    & log\,$P$\,=5             &  $-$0.5\,$\pm$\,0.3    & $-$1.4\,$\pm$\,0.3   & $-$2.0\,$\pm$\,0.3  & $-$2.0\,$\pm$\,0.3   & $-$2.0\,$\pm$\,0.3 \\
\hline
{\large $\gamma$}$_{\rm largeq}$    & log\,$P$\,=7             &  $-$1.1\,$\pm$\,0.3    & $-$2.0\,$\pm$\,0.3   & $-$2.0\,$\pm$\,0.3  & $-$2.0\,$\pm$\,0.3   & $-$2.0\,$\pm$\,0.3  \\
\hline
           & Power-law slope of $p_q$\,$\propto$\,$q^{\gamma}$ &                        &                      &                     &                      &   \\
                            & across $q$~=~0.1\,-\,0.3 at:     &                        &                      &                     &                      &   \\
\hline
{\large $\gamma$}$_{\rm smallq}$    & log\,$P$\,=1             &  0.3\,$\pm$\,0.4       & 0.2\,$\pm$\,0.4      &  0.1\,$\pm$\,0.4    &  0.1\,$\pm$\,0.4     &  0.1\,$\pm$\,0.4   \\
\hline
{\large $\gamma$}$_{\rm smallq}$    & log\,$P$\,=3             &  0.3\,$\pm$\,0.6       & 0.1\,$\pm$\,0.6      & $-$0.2\,$\pm$\,0.6  & $-$0.2\,$\pm$\,0.6   & $-$0.2\,$\pm$\,0.6 \\
\hline
{\large $\gamma$}$_{\rm smallq}$    & log\,$P$\,=5             &  0.3\,$\pm$\,0.4       & $-$0.5\,$\pm$\,0.4   & $-$1.2\,$\pm$\,0.4  & $-$1.2\,$\pm$\,0.4   & $-$1.2\,$\pm$\,0.4 \\
\hline
{\large $\gamma$}$_{\rm smallq}$    & log\,$P$\,=7             &  0.3\,$\pm$\,0.3       & $-$1.0\,$\pm$\,0.3   & $-$1.5\,$\pm$\,0.3  & $-$1.5\,$\pm$\,0.3   & $-$1.5\,$\pm$\,0.3 \\
\hline
\hline
            & Power-law slope of $p_e$\,$\propto$\,$e^{\eta}$  &                        &                      &                     &                      &   \\
                   & across $e$~=~0.0\,-\,0.8$e_{\rm max}$ at:    &                        &                      &                     &                      &   \\
\hline
$\eta$                                   & log\,$P$\,=2        &  0.1\,$\pm$\,0.3      & 0.3\,$\pm$\,0.3      &  0.6\,$\pm$\,0.3    &  0.7\,$\pm$\,0.3     &  0.7\,$\pm$\,0.3   \\
\hline
$\eta$                                   & log\,$P$\,=4        &  0.4\,$\pm$\,0.3      & 0.5\,$\pm$\,0.3      &  0.7\,$\pm$\,0.3    &  0.8\,$\pm$\,0.3     &  0.8\,$\pm$\,0.3 \\
\hline
\end{tabular}
\end{center}
\end{figure*}

{\it Period Distributions:} The period distribution of solar-type MS binaries peaks at log\,$P$\,(days)~$\approx$~5.0 ($a$~$\approx$~50~AU).  The peak in the O-type and B-type MS companion distribution occurs at slightly smaller scales log\,$P$~$\approx$~3.5 ($a$~$\approx$~10 AU) and is $\approx$3\,-\,4 times larger (\S9.3 and Fig.~37).  This indicates that disk fragmentation during the binary formation process is significantly more efficient for more massive systems (\S10).  The O-type companion period distribution is slightly bimodal with a secondary peak at $P$~$\lesssim$~20~days ($a$~$<$~0.5~AU).  This suggests intermediate-period and close binaries require two different formation mechanisms, e.g., disk fragmentation and dynamical triple-star evolution coupled with tidal interactions, respectively (\S10).  

{\it Mass-ratio Distributions:} To accurately model the data, we require a three-parameter mass-ratio distribution: a power-law slope {\large $\gamma$}$_{\rm smallq}$ across small mass ratios $q$~=~0.1\,-\,0.3, a power-law slope {\large $\gamma$}$_{\rm largeq}$ across large mass ratios $q$~=~0.3\,-\,1.0, and an excess fraction ${\cal F}_{\rm twin}$ of twins with $q$ = 0.95\,-\,1.00 (\S2; Fig.~2).  For early-type binaries, we measure a small non-zero excess twin fraction ${\cal F}_{\rm twin}$~$\approx$~0.1 only at short orbital periods $P$~$<$~20~days ($a$~$\lesssim$~0.4\,AU; Fig.~35), indicating the twin components coevolved via Roche-lobe overflow and/or shared accretion in the circumbinary disk during their pre-MS formation (\S10).  The excess twin fraction of solar-type binaries is substantially larger ${\cal F}_{\rm twin}$~$\approx$~0.3 at short periods and is measurably non-zero up to log\,$P$~$\approx$~6.0 ($a$~$\approx$~200\,AU).  This is possibly due to the longer primordial disk lifetimes associated with solar-type primaries (\S10).  

The power-law components {\large $\gamma$}$_{\rm smallq}$~$\approx$~0.3 and {\large $\gamma$}$_{\rm largeq}$~$\approx$~$-$0.5 of solar-type binaries with short and intermediate orbital periods log\,$P$~$\lesssim$~6 are consistent with a uniform mass-ratio distribution {\large $\gamma$}~=~0 (Fig.~35).  While short-period early-type binaries with $P$~$<$~20~days ($a$~$\lesssim$~0.4\,AU) are also consistent with a uniform mass-ratio distribution, the companions to massive primaries quickly become weighted toward smaller mass ratios $q$ = 0.2\,-\,0.3 with increasing period, e.g., {\large $\gamma$}$_{\rm smallq}$~$\approx$~$-$0.5 and {\large $\gamma$}$_{\rm largeq}$ $\approx$ $-$1.8 at log\,$P$\,(days)~$\approx$~3.5 ($a$~$\approx$~10\,AU).   We explain the differences between solar-type and early-type binaries with intermediate separations in the context of disk fragmentation, whereby the ratio $q_{\rm frag}$ = $M_{\rm frag}$/$M_1$ of the fragment mass to the final primary mass is expected to decrease with increasing $M_1$ (\S10).  

Finally, at long orbital periods log\,$P$\,(days)~$\approx$~6\,-\,8 ($a$~$\approx$~500\,-\,10$^4$~AU), the power-law components {\large $\gamma$}$_{\rm smallq}$~$\approx$~$-$1.5 and {\large $\gamma$}$_{\rm largeq}$ $\approx$ $-$2.0 of early-type systems are nearly consistent with random pairings drawn from the Salpeter IMF ({\large $\gamma$} = $-$2.35).  This demonstrates that wide companions to massive stars form relatively independently, most likely through fragmentation of molecular cores / filaments.  Even at these wide separations, the power-law slopes {\large $\gamma$}$_{\rm smallq}$~$\approx$~0.3 and {\large $\gamma$}$_{\rm largeq}$ $\approx$ $-$1.1 for solar-type MS binaries are still measurably discrepant with random pairings from the IMF, indicating some degree of correlation between the component masses (\S8.5 and Fig.~30).  

However, the frequency of wide companions to solar-type {\it pre-MS} stars is twice that measured for solar-type MS stars and actually consistent with the frequency of wide companions to O-type MS stars (\S10 and Fig.~41). Wide binaries are likely born during the early pre-MS phase with the same companion frequency independent of primary mass and with a mass-ratio distribution consistent with random pairings drawn from the IMF.  Subsequent dynamical processing that preferentially ejects $q$~=~0.1\,-\,0.3 companions leads to the smaller wide companion frequency observed in the longer lived solar-type MS population (\S10).  We emphasize this statement applies only to wide binaries that likely form via fragmentation of cores/filaments on large spatial scales.  At short and intermediate separations, the frequency of companions to solar-type pre-MS stars matches the solar-type MS value, both of which are substantially smaller than that measured for O-type MS stars (\S10 and Fig.~41).  A primary-mass dependent physical process, e.g., more massive protostars are more prone to disk fragmentation, is required to explain the significantly larger frequency of close and intermediate-period companions to massive stars.

{\it Eccentricity Distributions:} Samples of spectroscopic binaries, eclipsing binaries, and visual binaries (with orbital solutions) can be substantially biased against systems with large eccentricities $e$~$\gtrsim$~0.8. Moreover, highly eccentric binaries tidally evolve toward smaller eccentricities on more rapid timescales $\tau_{\rm tide}$~$\propto$~(1$-e^2$)$^{\nicefrac{13}{2}}$. We can reliably model the power-law slope $\eta$ of the initial eccentricity distribution $p_e$~$\propto$~$e^{\eta}$ only up to $e$~$\approx$~0.8$e_{\rm max}$, where $e_{\rm max}$ is the maximum eccentricity possible without substantially filling the primary's Roche-lobe at periastron (Eqn.~3).  While short-period binaries have tidally circularized, we measure $\eta$~$\approx$~0.4 for solar-type binaries with intermediate periods log\,$P$\,(days) = 1.5\,-\,5 (\S9.2 and Fig.~36).  Alternatively, we find a larger $\eta$~$\approx$~0.8 for early-type binaries across log\,$P$\,(days) = 1.5\,-\,4.5, which is consistent with a Maxwellian "thermal" eccentricity distribution ($\eta$ = 1).  This suggests that dynamical interactions, possibly with an outer tertiary, may play a larger role in the formation of early-type binaries with short and intermediate orbital periods.  It is also possible that solar-type binaries are born with initially larger eccentricities, but then experience more efficient tidal circularization and/or dampening in the disk during their longer lived pre-MS phase.  In any case, for a zero-age MS population, early-type binaries with intermediate periods log\,$P$~$\approx$~1.5\,-\,5.0 are weighted toward larger eccentricities compared to their solar-type counterparts.  

{\it Binary Evolution:}  Only $f_{\rm 0.2<logP<3.7;q>0.1}$ = (15\,$\pm$\,3)\% of solar-type primaries will undergo RLOF with stellar companions $q$ $>$ 0.1 and log\,$P$\,(days) $<$ 3.7, while essentially all O-type primaries will experience significant binary evolution (\S11 and Fig.~42).  In fact, we estimate that (10\,-\,20)\% of O-type primaries are in compact triple configurations in which the outer tertiary $q$ = $M_3$/$M_1$ $>$ 0.1 has log\,$P_{\rm outer}$~$<$~3.7 and may significantly affect the evolution of the inner binary.  Utilizing the measured multiplicity statistics and a binary population synthesis technique, we calculate the fraction of primaries within a volume-limited sample that are actually the secondaries in which the true primaries have already evolved into compact remnants.  We measure that (11\,$\pm$\,4)\% of solar-type primaries in the solar neighborhood contain WD companions (\S8.3.2 and Fig.~29).  Similarly, we calculate that $\approx$(10\,-\,30)\% of early-type primaries are actually the secondaries in which the true primaries have evolved into neutron stars or black holes via core-collapse supernovae (\S11 and Fig.~43).  This measurement matches the fraction of OB stars that are runaway, indicating that supernova kicks in massive binaries may significantly contribute to the formation of runaway OB stars.  It is imperative that future population synthesis studies of binary evolution incorporate the corrected joint probability distribution function $f$($M_1$,\,$q$,\,$P$,\,$e$) $\ne$ $f$($M_1$)$f$($q$)$f$($P$)$f$($e$).  Independently adjusting the individual distribution functions to the extremes will still not encompass the true nature of the population because $M_1$, $q$, $P$, and $e$ are all highly correlated with each other. 

{\it Nature of SB1s:} We estimate the fraction ${\cal F}_{\rm solar+WD;logP<4.5}$ = (3.4\,$\pm$\,\,1.0)\% of solar-type stars in the solar-neighborhood that contain close WD companions with log\,$P$\,(days)~$<$~4.5 ($a$~$\lesssim$~30~AU) using three independent methods (\S8.3). Because $\approx$10\% of solar-type stars contain close companions with log\,$P$\,(days)~$<$~4.5 in which the nature of the secondary cannot be determined, we find that (30\,$\pm$\,10)\% of solar-type SB1s actually contain WD companions. The remaining (70\,$\pm$\,10)\% of solar-type SB1s have M-dwarf secondaries with mass ratios $q$~$\approx$~0.1\,-\,0.5.  For early-type binaries, the observed ratio of SB1s to SB2s increases by an order of magnitude between young and old populations  (\S3.5).  Selection effects alone cannot explain this discrepancy, and so we conclude that 30$_{-15}^{+10}$\% of early-type SB1s in a volume-limited sample contain neutron star and/or black hole companions.  Only a fraction of SB1s with compact remnant companions will exhibit UV excess, X-rays, or emission lines (\S11), so it is important to never assume that the companions in SB1s are stellar in nature.

  This research was funded by the National Science Foundation under grant AST-1211843. M.M. acknowledges additional financial support from NASA's Einstein Fellowship program PF5-160139. 
We thank N.~Evans, H.~Sana, S.~de Mink, and I.~Czekala for useful discussions regarding statistics and observational selection biases.  We thank the anonymous referee, whose comments helped to improve the quality of the manuscript.  We also thank A.~Pr{\v s}a, K.~Kratter, H.~Perets, J.~Grindlay, R.~Narayan, and R.~Kirshner for their helpful recommendations and suggestions.

\section*{Appendix A.1}

 In this section, we examine the evolutionary pathways for producing solar-type~+~WD binaries.  The progenitors of these systems can be mid-B stars with solar-type companions (Channel 1), solar-type stars with near-equal mass companions (Channel 2), and/or primary masses and mass ratios that are between these two limits.  For each of the two specified channels, we calculate the occurrence rates based on the observed MS multiplicity statistics and distributions.  We implement these rates in \S8.3.2 to calculate the fraction of solar-type stars that have WD companions.  

{\it Channel 1: B-type + solar-type.} A $M_1$~$\approx$~5\,\Msun\ B-type MS star must initially have a $M_2$~$<$~1.25\,\Msun\ ($q$~$<$~0.25) companion to be capable of evolving into a binary with a solar-type star ($M$ = 0.75\,-\,1.25\Msun) and WD companion.  Mass transfer via Roche-lobe overflow (RLOF) is unstable if $q$~$\lesssim$~0.3, and so extreme mass-ratio binaries with short periods log\,$P_{\rm initial}$\,(days)~$<$~1.5 are likely to merge \citep{Hurley2002,Belczynski2008}.  In order to emerge as a solar-type~+~WD binary, the system must either survive common envelope (CE) evolution (1.5~$<$~log\,$P_{\rm initial}$~$\lesssim$~3.7) or avoid RLOF altogether (log\,$P_{\rm initial}$~$\gtrsim$~3.7; see \S11).  Binaries with 4.0~$<$~log~$P_{\rm initial}$~$<$~4.5 expand beyond log~$P_{\rm final}$~$>$~4.5 as the $M_1$~$\approx$~5\,\Msun\ B-type MS primary loses most of its mass via winds. Mass transfer through CE evolution and wind accretion is negligible for extreme mass-ratio binaries \citep{Hurley2002,Belczynski2008}.  The original secondary must therefore be a solar-type star with 0.75~$\lesssim$~$M_2$~$\lesssim$~1.25\,\Msun\ (0.15~$<$~$q$~$<$~0.25). 

In summary, systems with $M_1$~$\approx$~5\,\Msun\ MS primaries, $q$ = 0.15\,-\,0.25 companions, and log~$P_{\rm initial}$~$>$~1.5 produce solar-type~+~WD binaries.  The subset with 1.5~$<$~log~$P_{\rm initial}$~$<$~4.0 emerge as closely orbiting solar-type~+~WD binaries with log~$P_{\rm final}$~$<$~4.5. Based on the measured multiplicity statistics $f$($M_1$,\,$q$,\,$P$) in \S9.1-9.4, we find that (24\,$\pm$\,6)\% of mid-B $M_1$~=~5\,\Msun\ MS primaries have solar-type $q$~=~0.15\,-\,0.25 companions with 1.5~$<$~log\,$P_{\rm initial}$~$<$ 7.5. About half of these binaries, i.e., (10\,$\pm$\,3)\% of systems with mid-B MS primaries, have 1.5~$<$~log\,$P_{\rm initial}$~$<$ 4.0 and are expected to become closely orbiting solar-type~+~WD binaries with log~$P_{\rm final}$~$<$~4.5.

{\it Channel 2: solar-type + solar-type}. Solar-type binaries with log $P_{\rm initial}$ $>$ 3.7 are too wide to experience significant mass transfer (see \S11).  In addition, wide solar-type + solar-type binaries undergo only minor orbital widening as the system loses $\approx$25\% of its total mass. A system with a $M_1$~=~1.25\,\Msun\ mid-F primary, a $M_2$~=~0.75\,-\,1.25\,\Msun\ ($q$ = 0.6\,-\,1.0) companion, and a long orbital period log~$P_{\rm initial}$~$\gtrsim$~3.7 evolves into a solar-type~+~WD binary with log~$P_{\rm final}$~$\gtrsim$~3.7. For solar-type binaries with shorter orbital periods 0.0~$\lesssim$~log~$P_{\rm init}$~$\lesssim$~3.7 and moderate mass ratios $q$ $\gtrsim$ 0.5, mass transfer via RLOF and/or wind accretion can be significant.  For example, a $M_1$~=~1.25\,\Msun\ mid-F primary with a $M_2$~=~1.1\,\Msun\ companion ($q$~$\approx$~0.9) in a short to intermediate orbital period log~$P_{\rm init}$~$\lesssim$~3.7 transfers at least $\Delta$M~$>$~0.15\,\Msun\ of its mass to the secondary \citep{Hurley2002,Belczynski2008}.  This system therefore evolves into a  WD~+~late-A/early-F close binary.  In this case, the mass-gainer may appear as a blue straggler \citep{Geller2011}.  To emerge as a WD~+~solar-type close binary, the $M_1$~=~1.25\,\Msun\ mid-F progenitor must initially have a lower-mass $q$~$\approx$~0.5\,-\,0.9 companion with 0.0~$<$~log~$P_{\rm initial}$~$<$~3.7.  

 The \citet{Raghavan2010} sample of F6\,-\,K3 primaries is sufficiently representative of the binary statistics of $M_1$~=~1.25\,\Msun\ primaries. This sample is also relatively complete for solar-type binaries with $q$ $>$ 0.5.  We count 12 companions with $q$ $=$ 0.5\,-\,0.9 and 0.0~$<$~log~$P$~$<$~3.7, and 54 systems with $q$ $=$ 0.6\,-\,1.0 and 3.7~$<$~log~$P$~$<$~8.0.  Approximately one-quarter of the wide companions with $q$ $=$ 0.6\,-\,1.0 and 3.7~$<$~log~$P$~$<$~8.0 are twins with $q$ $>$ 0.95. In total, we expect (12+54)/404 $\approx$ (16\,$\pm$\,3)\% of systems with $M_1$ = 1.25 primaries to evolve into solar-type~+~WD binaries. About one-third of these binaries, i.e., (5\,$\pm$\,1)\% of systems with $M_1$~=~1.25 primaries, emerge as close solar-type~+~WD binaries with log~$P_{\rm final}$~$<$~4.5.

\bibliographystyle{apj}                       
\bibliography{bibliography}

\end{document}